\documentclass[aps,prb,twocolumn,floatfix]{revtex4-1}
\usepackage{amsfonts,bm}
\usepackage{amsmath}
\usepackage{amssymb}
\usepackage{amsthm}
\usepackage{graphicx}
\usepackage{graphics}
\usepackage{subfigure}
\usepackage{color}
\usepackage{bm}
\usepackage{hyperref}
\usepackage{rotating}
\usepackage{comment}

\newcommand{\<}{\langle}
\renewcommand{\>}{\rangle}
\newcommand{\be}{\begin{equation} }
\newcommand{\ee}{\end{equation} }
\newcommand{\ba}{\begin{eqnarray} }
\newcommand{\ea}{\end{eqnarray} }

\newcommand{\bpm}{\begin{pmatrix}}
\newcommand{\epm}{\end{pmatrix}}
\newcommand{\bmm}{\begin{matrix}}
\newcommand{\emm}{\end{matrix}}

\newtheorem{conjecture}{Conjecture}

\begin{document}
\title{Generalizations and limitations of string-net models}
\author{Chien-Hung Lin and Michael Levin}
\affiliation{James Franck Institute and Department of Physics, University of Chicago, Chicago, Illinois 60637, USA}
\begin{abstract}
	We ask which topological phases can and cannot be realized by exactly soluble string-net models. 
We answer this question for the simplest class of topological phases, namely those with abelian braiding statistics. 
Specifically, we find that an abelian topological phase can be realized by a string-net model if and only if (i) it 
has a vanishing thermal Hall conductance and (ii) it has at least one Lagrangian subgroup --- a subset of quasiparticles
with particular topological properties. Equivalently, we find that an abelian topological phase is realizable if 
and only if it supports a gapped edge. We conjecture that the latter criterion generalizes to the non-abelian case. 
We establish these results by systematically constructing all possible abelian string-net models and analyzing the quasiparticle braiding statistics in each model. We show that the low energy effective field theories for these models are multicomponent $U(1)$ Chern-Simons theories, and we derive the $K$-matrix description of each model. An additional feature of this work is that the models we construct are more general than the original string-net models, due to several new ingredients. First, we introduce two new objects $\gamma, \alpha$ into the construction which are related to $\mathbb{Z}_2$ and $\mathbb{Z}_3$ Frobenius-Schur indicators. Second, we do not assume parity invariance. As a result, we can realize topological phases that were not accessible to the original construction, including phases that break time-reversal and parity symmetry.
\end{abstract}

\maketitle

\section{Introduction}
In recent years, it has become clear that the physics of gapped quantum phases of matter is much richer than was previously thought.
One example of this richness is the large class of two dimensional quantum many body systems that support quasiparticle excitations with
fractional statistics. These systems are known as ``topological phases'' of matter.\cite{WenBook} 

Topological phases pose a basic challenge because their properties cannot be understood in terms of symmetry breaking or order parameters.
Therefore, studying them requires new tools and approaches. One approach that proven to be useful is the construction
of exactly soluble lattice models that realize topological phases. One of the simplest examples is the toric code
model of Ref. [\onlinecite{KitaevToric}]. This model is a spin-$1/2$ system where the spins live on the links of the square lattice.
The reason that the toric code is exactly soluble is that the Hamiltonian 
is a sum of commuting projectors: $H = \sum_i P_i$ where $[P_i, P_j] = 0$. 

An interesting aspect of the toric code model is that it can be mapped onto a model of closed loops or strings.
Based on this observation, Ref. [\onlinecite{LevinWenStrnet}]
generalized the toric code to a large class of exactly soluble ``string-net'' models. Like the toric code, string-net models are lattice spin models
whose low energy physics is governed by effective extended objects.

String-net models can realize a large class of topological phases. For example, these models can realize 
all phases whose low energy effective theory is either (a) a gauge theory with finite gauge group or 
(b) a sum of two decoupled Chern-Simons theories with opposite 
chiralities.\cite{LevinWenStrnet} At the same time, string-net models cannot realize \emph{all} topological phases. In particular, they cannot
realize any phase with a nonzero thermal Hall conductance\cite{KaneFisherThermal} (or equivalently, nonzero chiral central charge\cite{KitaevHoneycomb}). This restriction
follows from the fact that any Hamiltonian that is a sum of commuting projectors has a vanishing thermal Hall conductance.
\footnote{
	The fact that the thermal Hall conductance/chiral central charge vanishes for a commuting projector
Hamiltonian follows from the analysis of the chiral central charge given in appendix D.1 in Ref. [\onlinecite{KitaevHoneycomb}].
Specifically, one can see that $f$ in Eq. (159) vanishes for commuting projectors and therefore we can choose $h=0$. It then follows that $c_{-}=0$ in Eq. (160).}

Given these facts, an important question is to determine which topological phases can and cannot be realized by string-net models.
On a mathematical level, the answer to this question is at least partially understood: it has been argued that string-net models, when suitably
generalized from the original construction of Ref. [\onlinecite{LevinWenStrnet}], realize all ``doubled'' phases --- where ``double'' refers to a generalization of 
Drinfeld's quantum double construction.\cite{KitaevKong} However, the physical interpretation of this result is not clear. In other words, what physical property distinguishes 
the phases that can and cannot be realized?

In this paper, we answer this question for a simple case, namely the case of abelian topological phases. Our analysis is based on an explicit
construction: we systematically construct all string-net models that realize abelian topological phases. For each model, 
we compute the quasiparticle braiding statistics and ground state degeneracy, and we derive a low energy effective field theory
that captures these properties. These effective theories are multicomponent $U(1)$ Chern-Simons theories. 

From this analysis, we find necessary and sufficient conditions for 
when an abelian topological phase can be realized by a string-net model: we find that an abelian phase is realizable 
if and only if (i) it has a vanishing thermal Hall conductance and (ii) it has at least one Lagrangian subgroup. 
Here, a ``Lagrangian subgroup''\cite{KapustinTopbc} $\mathcal{M}$ is a subset of quasiparticles with two properties. First,
all the quasiparticles in $\mathcal{M}$ are bosons and have trivial mutual statistics with one another. Second, any quasiparticle 
that is not in $\mathcal{M}$ has nontrivial mutual statistics with at least one particle in $\mathcal{M}$.

Interestingly, the above conditions 
are identical to the conditions for an abelian topological phase to support a \emph{gapped edge}.\cite{LevinProtedge,KapustinTopbc} Thus, an alternative formulation of
the criterion is that an abelian topological phase can be realized by a string-net model if and only if its boundary with the vacuum 
can be gapped by suitable local interactions. 
We conjecture that this criterion generalizes to the non-abelian case. (see section \ref{conclusion})

As we are interested in investigating the scope of string-net models, it is important that we use the most general possible
definition of these models. This issue is particularly relevant since several recent works\cite{KitaevKong,Kong12,LanWen13} have described a modified 
formulation of string-net models which is more general than the original setup of Ref. [\onlinecite{LevinWenStrnet}]. 
Here we use another formulation of these models, which we believe is equally general to the one described in Refs. [\onlinecite{KitaevKong},\onlinecite{Kong12},\onlinecite{LanWen13}], at least for the abelian case we consider here. The main difference
between our construction of string-net models and the original construction of Ref. [\onlinecite{LevinWenStrnet}] is that we introduce two new
ingredients, $\gamma, \alpha$, into the definition of these models. These new objects are related to $\mathbb{Z}_2$ and $\mathbb{Z}_3$ Frobenius-Schur indicators\cite{KitaevHoneycomb,BondersonThesis} respectively,
and they allow us to realize more general topological phases than Ref. [\onlinecite{LevinWenStrnet}]. We note that it is also possible
to define general string-net models without introducing $\gamma, \alpha$, as in Ref. [\onlinecite{Kong12}, \onlinecite{LanWen13}]. The trade-off is that the approaches of 
Ref. [\onlinecite{Kong12} \onlinecite{LanWen13}] explicitly break the rotational symmetry of the lattice since they assume that all links are oriented along a preferred direction. 

The topological phases that we construct are equivalent to the topological gauge theories of Dijkgraaf and Witten\cite{DijkgraafWitten} with finite abelian gauge group $G$.
The braiding statistics and other topological properties of these phases were analyzed previously by Propitius\cite{PropitiusThesis} using the quantum double
construction. Our results for the braiding statistics agree with those of Propitius, but we obtain them using a more concrete approach in which we directly analyze
braiding in our microscopic lattice models. This braiding analysis is similar to that of Mesaros and Ran\cite{MesarosRan13} who derived braiding statistics 
from a lattice Dijkgraaf-Witten model using a ribbon algebra.

Explicit lattice models for Dijkgraaf-Witten gauge theories with general finite gauge group $G$ were constructed in Refs. [\onlinecite{HuWanWu12}, \onlinecite{MesarosRan13}]. We believe 
that the models we discuss here are closely related to the models of Refs. [\onlinecite{HuWanWu12}, \onlinecite{MesarosRan13}]. However, since 
we work in the string-net formalism, our models can be generalized beyond Dijkgraaf-Witten gauge theories.\cite{LevinWenStrnet}

The paper is organized as follows.
In Sec. \ref{summsec}, we outline our analysis and summarize our results.
In Sec. \ref{fixedwf}, we review some basics of string-net models and define ``abelian string-net'' models.
In Secs. \ref{strnetwf},\ref{fixedh}, we construct ground state wave functions and lattice Hamiltonians
for the abelian string-net models. 
We analyze the low energy quasiparticle excitations of these models in Sec. \ref{qpsection}.
In Sec. \ref{qpsection2}, we explicitly compute the quasiparticle braiding statistics for general abelian 
string-net models, and in Sec. \ref{cssec} we derive multicomponent $U(1)$ Chern-Simons theories that
capture these statistics. Finally, we characterize the phases that are realizable by abelian string-net models in
Sec. \ref{phases}. We illustrate our construction with concrete examples in Sec.
\ref{examples}. The mathematical details can be found in the appendices.

\section{Summary of results} \label{summsec}

\subsection{Construction of lattice models}
The first step in our analysis is to systematically construct a large class of exactly soluble lattice models. The models we construct are a subset of string-net models called ``abelian string-net'' models. In these models, the string types are labeled by elements $a,b,c,...$ of a finite abelian group $G$. The allowed branchings are triplets $(a,b,c)$ such that $a+b+c = 0$. We focus on this subset of models because these are the most general string-net models with abelian quasiparticle statistics.

Each abelian string-net model is specified by two pieces of data: (1) a finite abelian group $G$, and (2)
a collection of four complex-valued functions $(F(a,b,c),d_a, \alpha(a,b), \gamma_a)$ defined on $G$,
obeying certain algebraic equations (\ref{selfconseq}). The corresponding Hamiltonian (\ref{h}) is a spin model where the spins live on the links of 
the honeycomb lattice and where each spin can be in $|G|$ states parameterized by elements of the group: $\{|a\>: a \in G\}$. 
Like the toric code,\cite{KitaevToric} the Hamiltonian is exactly soluble because it can be written as a sum of commuting projectors.

\subsection{Relationship with other string-net constructions}
Our construction is more general than the original formalism of Ref. [\onlinecite{LevinWenStrnet}] in two ways. First, we include two 
new objects $\gamma,\alpha$ (related to $\mathbb{Z}_2$ and $\mathbb{Z}_3$ Frobenius-Schur indicators
\cite{KitaevHoneycomb, BondersonThesis}) in the construction of our models. These objects are related to two new structures: a ``dot''
at every vertex with three incoming or three outgoing strings, and a ``null string'' at every vertex with two incoming or two
outgoing strings. 
Ref. [\onlinecite{LevinWenStrnet}] did not include these structures and therefore effectively assumed $\gamma = \alpha = 1$. Here, because we allow for $\gamma, \alpha \neq 1$, we can realize phases that are not accessible to Ref. [\onlinecite{LevinWenStrnet}]. (See section \ref{examples}).

In addition, we do not impose additional symmetry requirements as in Ref. [\onlinecite{LevinWenStrnet}]. In that work, it was assumed
that the ground state and Hamiltonian were parity invariant, and consequently it was assumed that $F$ obeyed reflection symmetry.
Here we do not make any of these assumptions. As a result, our models can realize topological phases that break parity and time reversal symmetry (see section \ref{examples}).

Refs. [\onlinecite{Kong12},\onlinecite{LanWen13}] described another generalization of Ref. [\onlinecite{LevinWenStrnet}] that
does not involve $\gamma$ or $\alpha$, but breaks rotational symmetry by requiring that all strings are oriented along a preferred direction. 
We believe that our construction realizes the same phases as Refs. [\onlinecite{KitaevKong},\onlinecite{Kong12},\onlinecite{LanWen13}], at least
in the abelian case. The main difference is that our formalism does not explicitly break rotational symmetry.

\subsection{Braiding statistics and Chern-Simons description}
The second step in our analysis is to construct the quasiparticle excitations in each of the abelian string-net models.
We find that these models have $|G|^2$ topologically distinct quasiparticle excitations. The excitations can
be labeled by ordered pairs $(s,m)$ where $s \in G$, and $m$ is a $1D$ representation of $G$. We think of the excitations of the form $(0,m)$ as ``pure charges'' 
and the excitations of the form $(s,0)$ as ``pure fluxes.'' General excitations can be thought of as flux/charge composites.

The most important property of the quasiparticle excitations are their braiding statistics. 
We find that the charges braid trivially with one another but have nontrivial mutual statistics with respect to the 
fluxes. Specifically,
the phase associated with braiding a charge $(0,m)$ around a flux $(s,0)$ is $\rho_m(s)$ where $\rho_m$ denotes the $1D$ representation corresponding
to $m$. In addition, we find that the flux excitations have nontrivial statistics with one another (\ref{exzk},\ref{mutzk}). (In fact, in some of the abelian 
string-net models, the fluxes have \emph{non-abelian} statistics\cite{PropitiusThesis}, though we restrict our attention to the subset of models that have only abelian 
quasiparticles). 

We find that the quasiparticle braiding statistics can be described by a multicomponent $U(1)$ Chern-Simons theory of the form
\begin{displaymath}
L=\frac{K_{IJ} }{4\pi }\varepsilon^{\mu \nu \lambda }a_{I\mu }\partial _{\nu }a_{J\lambda }
\end{displaymath}
with a different ``$K$-matrix'' (\ref{kmat2}) for each model. These Chern-Simons theories can be thought of as low energy effective theories for the
abelian string-net models.

\subsection{Characterizing the realizable phases}
The abelian string-net models can realize many $U(1)$ Chern-Simons theories.
For example, our models can realize time-reversal symmetric phases such as
\[K=\begin{pmatrix}
		2 & 0 \\
		0 & -2
	\end{pmatrix}.
\]
In addition, we can also realize some phases that break time-reversal symmetry such as:
\[K=\begin{pmatrix}
		2 & 0 \\
		0 & -8
	\end{pmatrix}.
\]
On the other hand, we find that we cannot realize other time-reversal breaking phases such as: 
\[K=\begin{pmatrix}
		2 & 0 \\
		0 & -4
	\end{pmatrix}.
\] 
Given these examples, it is natural to wonder: what is the physical distinction between the phases that can and cannot be realized by
string-net models? To answer this question, we derive three equivalent criteria for determining whether an abelian topological phase
is realizable: 
\begin{enumerate}
\item{Braiding statistics criterion: an abelian topological phase is realizable if and only if it has a vanishing thermal Hall conductance 
and contains at least one Lagrangian subgroup (see introduction).}
\item{$K$-matrix criterion: an abelian topological phase is realizable if and only if its $K$-matrix has even dimension $2k 
\times 2k$, and there exist $k$ integer vectors $\Lambda_1,...,\Lambda_k$ satisfying $\Lambda_i K \Lambda_j = 0$.}
\item{Edge state criterion: an abelian topological phase is realizable if and only if its boundary with the vacuum can be gapped 
by suitable interactions.}
\end{enumerate}

\begin{figure}[tb]
\begin{center}
\includegraphics[height=1.4in,width=3in]{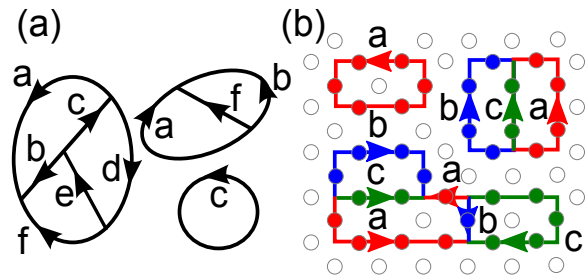}
\end{center}
\caption{(a) String-nets in the continuum. Strings come in different types and carry orientation.
        They can branch and form string-nets.
        (b) String-nets on the lattice. The circles denote spins sitting on the links of the lattice.
        Spins can be in different states indicated by different colors.
They organize to form different types of strings and branch as string-nets.}
        \label{stringnet}
\end{figure}

\section{String-net models}\label{fixedwf}
In this section, we will define string-nets and string-net models and explain their basic structure. This material is mostly a review of Ref. [\onlinecite{LevinWenStrnet}].
We also define ``abelian string-net'' models -- a special class of string-net models which are the main focus of this paper.

\subsection{General string-net models} \label{genstrnetsec}
A \emph{string-net} is a network of strings. The strings that form the edges of the network can come in different ``types'', and carry orientations.
In this paper, we will focus on trivalent networks -- that is, each branch point or node in the network is connected to exactly $3$ strings. Also,
we will assume that the string-nets live in a two-dimensional space. Thus, for the purposes of this paper, string-nets can be thought of
as trivalent graphs with labeled and oriented edges, which live in the plane (see Fig. \ref{stringnet}(a)). These trivalent graphs can live in the continuum,
or (when we want a well-defined quantum theory) on a lattice.  

A \emph{string-net model} is a quantum mechanical model whose basic degrees of freedom are fluctuating string-nets.
To specify a string-net model, one has to provide several pieces of data. First, one needs a finite set of
string types $\{a,b,c \dots \}$. Second, one needs to specify a ``dual'' string type $a^*$ for each string type $a$. The meaning
of the dual string type is related to the string orientations: a string $a$ with a given orientation corresponds to
the same physical state as a string $a^*$ with the opposite orientation -- up to a phase factor which we will specify below.
The final and most important piece of data are the ``branching rules.'' The branching rules are the set of all triplets of
string types $\{(a,b,c) \cdots\}$ which are allowed to meet at a point; these branching rules are specified with the convention that 
the $3$ string types are all oriented away from the point where they meet. 

\begin{figure}[tb]
\begin{center}
\includegraphics[height=0.8in,width=1.6in]{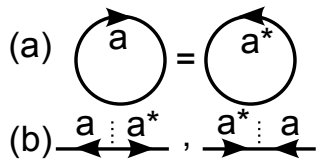}
\end{center}
\caption{(a) String-$a$ and dual string-$a^{\ast}$ have opposite orientations.
        (b) The branching rules associated with null strings are defined as $(a,b,0)$ is allowed iff $b=a^{*}$.}
\label{string0}
\end{figure}

The above data specify the string-net Hilbert space: an orthonormal basis for the string-net Hilbert space 
is given by the set of all string-net configurations which satisfy the above branching rules.
Note that in this Hilbert space, the spatial positioning of the string-net is important: two string-net configurations that are 
\emph{geometrically} distinct correspond to orthogonal states, whether or not the configurations are \emph{topologically} equivalent. 
On the other hand, two string-net configurations that are positioned identically in space, and differ only by reversing string 
orientations and replacing $a \rightarrow a^*$, are regarded as the same physical state up to a phase factor. These phases 
will be defined below. 

As we will see, string-nets and string-net models can be realized in lattice spin-systems (see Fig. \ref{stringnet}(b)). Usually for a weakly interacting spin system, 
the underlying spins can fluctuate independently and the physics is characterized by individual spins. 
However in some spin models, energetic constraints can force the local spin degrees of freedom to organize into effective extended objects.
In this case, the low energy physics of the spin system may be described by a string-net model, where the string-nets live on a lattice.

In order to discuss string-nets on a lattice, and also to simplify some of the mathematics below, it is convenient to include the
``null'' string type into the formalism. The null string type, denoted by $0$, is equivalent to no string at all. This string type is self-dual: $0^{*} = 0$.
The associated branching rule is that $(0,a,b)$ is allowed if $a=b^{\ast}$ (see Fig. \ref{string0}). Unlike the other strings, the orientation of the
null string can be reversed without generating any phase factor. Therefore, we will often neglect the orientation of the null string and draw it as 
an unoriented dotted string.

\begin{figure}[tb]
\begin{center}
\includegraphics[height=0.9in,width=1.4in]{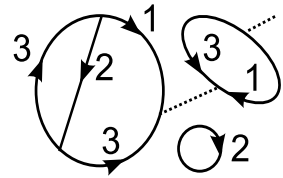}
\end{center}
\caption{A typical string-net configuration for $\mathbb{Z}_{4}$ string-net model. Each string is labeled by the
string types $\{0,1,2,3\}$ and carries an orientation. The dotted line denotes the null string $0$. At each vertex,
strings can branch according to the $\mathbb{Z}_{4}$ branching rules.}
\label{Z4}
\end{figure}

\subsection{Abelian string-net models \label{abstrnet}}
In this paper we focus on a special class of string-net models associated with abelian groups. We call these models
``abelian string-net models.''  To construct an abelian string-net model, one starts with a finite abelian group $G$ 
and then follows a simple recipe. First, one labels the string types by the elements of the group ${a \in G}$, with 
the null string corresponding to the identity element $0$. Second, one defines the dual string $a^*$ using the group 
inverse: $a^{\ast}=-a$. Finally, one defines branching rules by:
 \begin{equation}
	(a,b,c) \text{ is allowed if }a+b+c=0.
	\label{branch}
\end{equation}
(Here we use additive notation for the group operation.) 

We focus on this subset of string-net models because we believe that these are the most general models with abelian quasiparticle statistics --- i.e. other branching rules always give models with at least one non-abelian quasiparticle. Although we do not have a proof of this conjecture, in section \ref{edgechar} we show that even if other branching rules could give abelian topological phases, they could not give any phases beyond those that we realize here. This result justifies our focus on models with the above structure (\ref{branch}).

To see an example of this construction, consider the group $G = \mathbb{Z}_4$. In this case, the corresponding abelian string-net model has four string types, including the null
string: $\{0,1,2,3\}$. The dual string types are $0^* = 0$, $1^* = 3$, $2^* = 2$, and $3^* = 1$.
The branching rules are $\{(0,0,0),(0,1,3),(0,2,2),(1,1,2),(3,3,2)\}$. A typical string-net configuration for this model is shown in Fig. \ref{Z4}.

\subsection{String-net condensation}
To define a string-net model, one needs to specify both the Hilbert space and the Hamiltonian; so far we have focused entirely on the Hilbert space.
Now let us imagine writing down a string-net Hamiltonian. A typical string-net Hamiltonian is a sum of a kinetic energy term and a string tension term.
The kinetic energy term is off-diagonal in the string-net basis. This term gives an amplitude for the string-net states to move. On the other hand, the 
string tension term is diagonal in the string-net basis. This term gives an energy cost to large string-nets.

It is natural to expect that such a Hamiltonian can be in two phases depending on the relative size of the kinetic energy and string tension terms. One phase occurs
when the string tension term dominates over the kinetic energy term. In that case, we expect that the
ground state will contain only a few small strings. The other phase occurs when the kinetic energy term dominates over the string tension term.
In that case, we expect that the ground state will be a superposition of many large string-net configurations. We call the former phase a ``small string'' phase
and the latter phase a ``string-net condensed phase.''

Following the physical picture of Ref. [\onlinecite{LevinWenStrnet}], we expect that string-net condensed phases are topologically ordered -- that is, they support excitations with fractional
statistics -- while the small-string phases do not contain topological order. Therefore, our strategy for constructing topological phases will be to 
construct wave functions for string-net condensed phases. We will then construct exactly soluble Hamiltonians whose ground states are described by these 
wave functions, and we will verify that these exactly soluble models support excitations with fractional statistics.
 
\section{String-net wave functions} \label{strnetwf}
In this section we construct wave functions for abelian string-net condensed phases. As in Ref. [\onlinecite{LevinWenStrnet}], the wave functions that we construct are special: they
describe ``perfect'' string-net condensates with vanishing correlation length. Intuitively, these states can be thought of as fixed points 
under an RG flow. These states capture the universal long distance features of the corresponding phases without any of the complexities of the
short distance physics. In section \ref{fixedh}, we will show that these wave functions are ground states of exactly soluble string-net Hamiltonians, defined on a lattice.

\subsection{Local rules ansatz}
As in Ref. [\onlinecite{LevinWenStrnet}], we will not attempt to construct explicit ground state wave functions for string-net condensed phases. 
Instead, we will define the wave functions \emph{implicitly} using local constraint equations. This approach has the advantage of allowing us to
construct complicated wave functions that would be difficult to write down explicitly. In addition, this approach ensures that 
the wave functions we construct can be realized as ground states of \emph{local} Hamiltonians.

More specifically, we use the following ansatz for constructing abelian string-net wave functions $\Phi$. We assume that $\Phi$ obeys 
local constraint equations that take the following graphical form:
\begin{eqnarray} 
\Phi \left( \raisebox{-0.16in}{\includegraphics[height=0.4in]{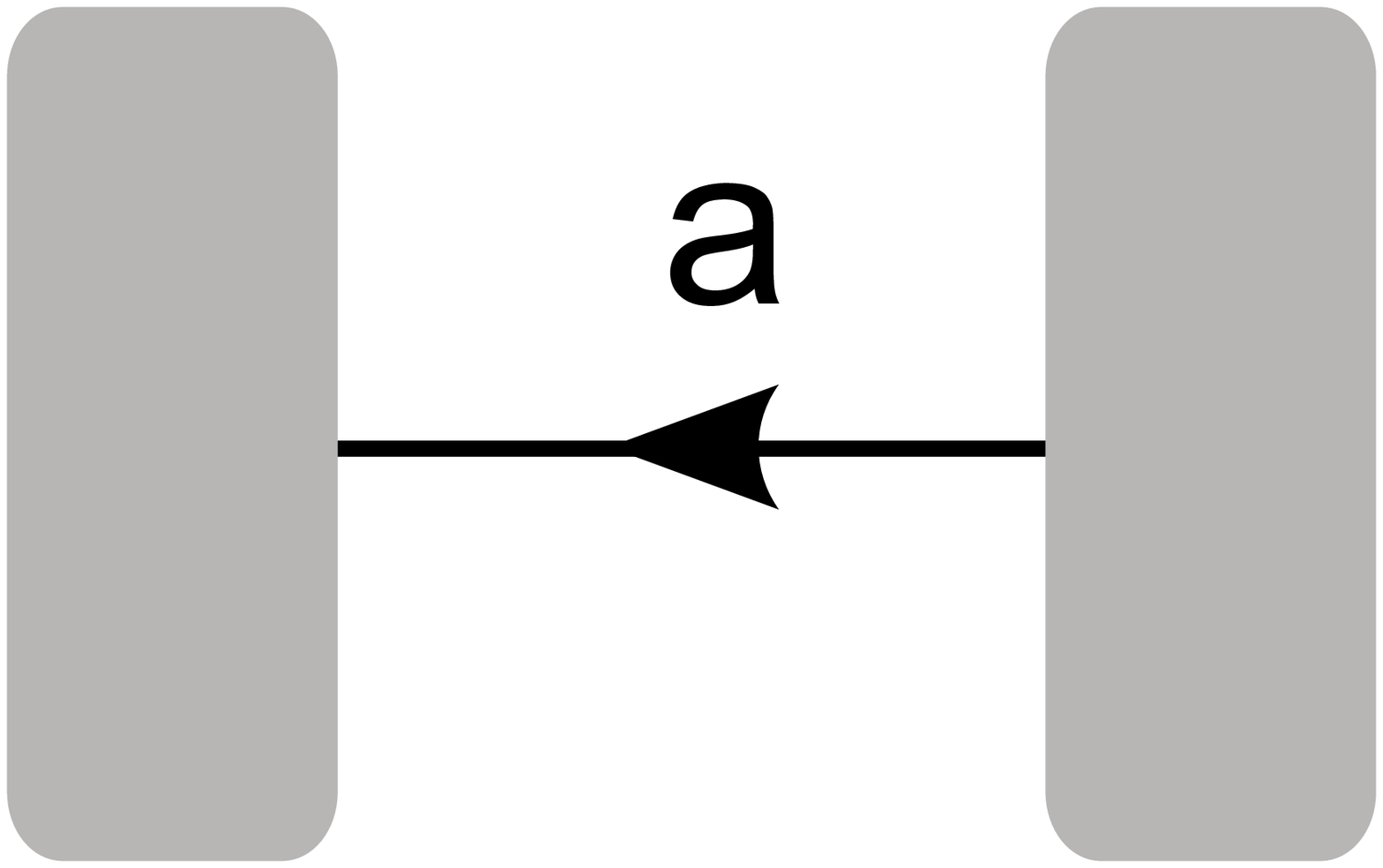}}\right)
&=&
\Phi \left( \raisebox{-0.16in}{\includegraphics[height=0.4in]{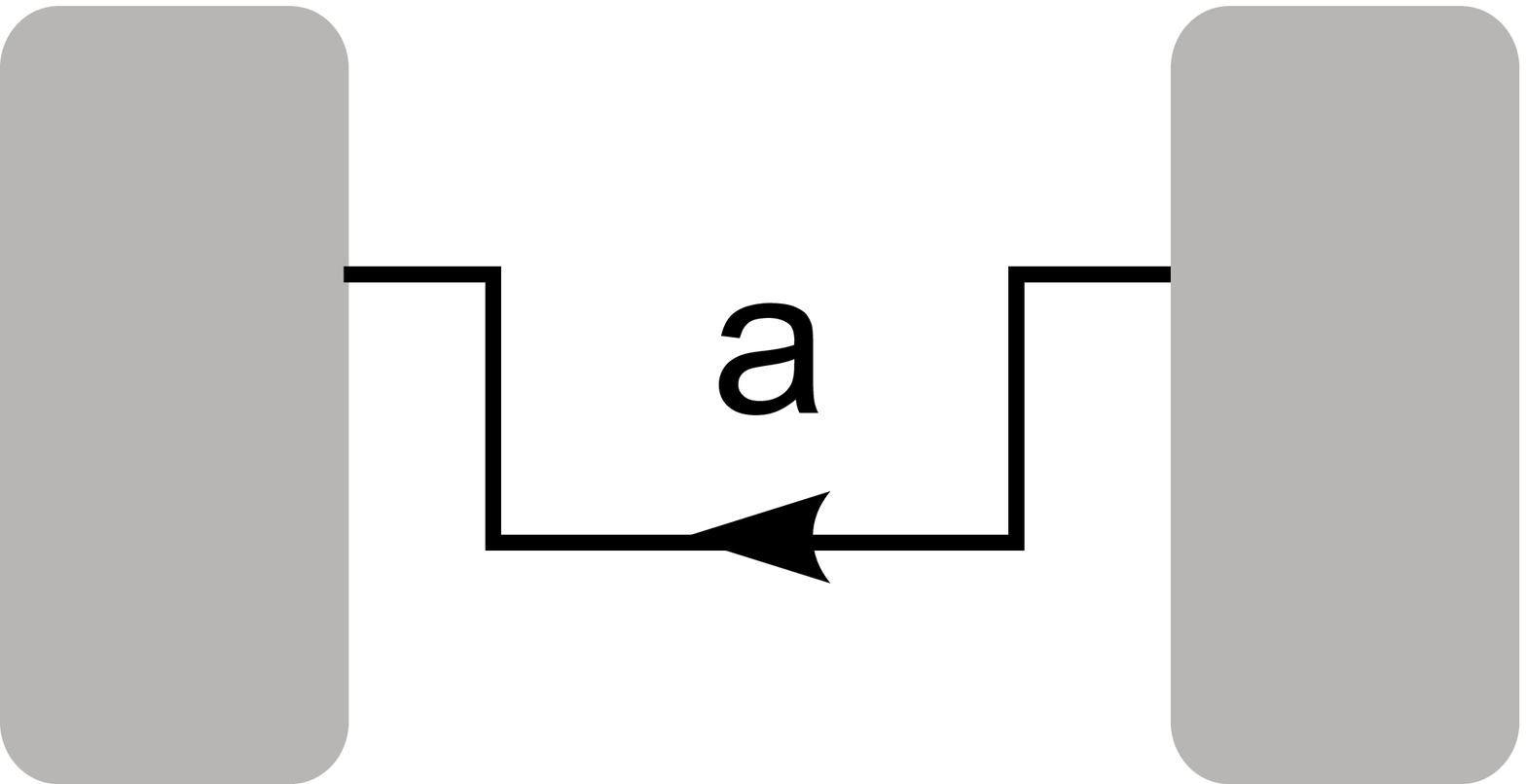}}\right),   \label{rule1} \\
\Phi \left( \raisebox{-0.16in}{\includegraphics[height=0.4in]{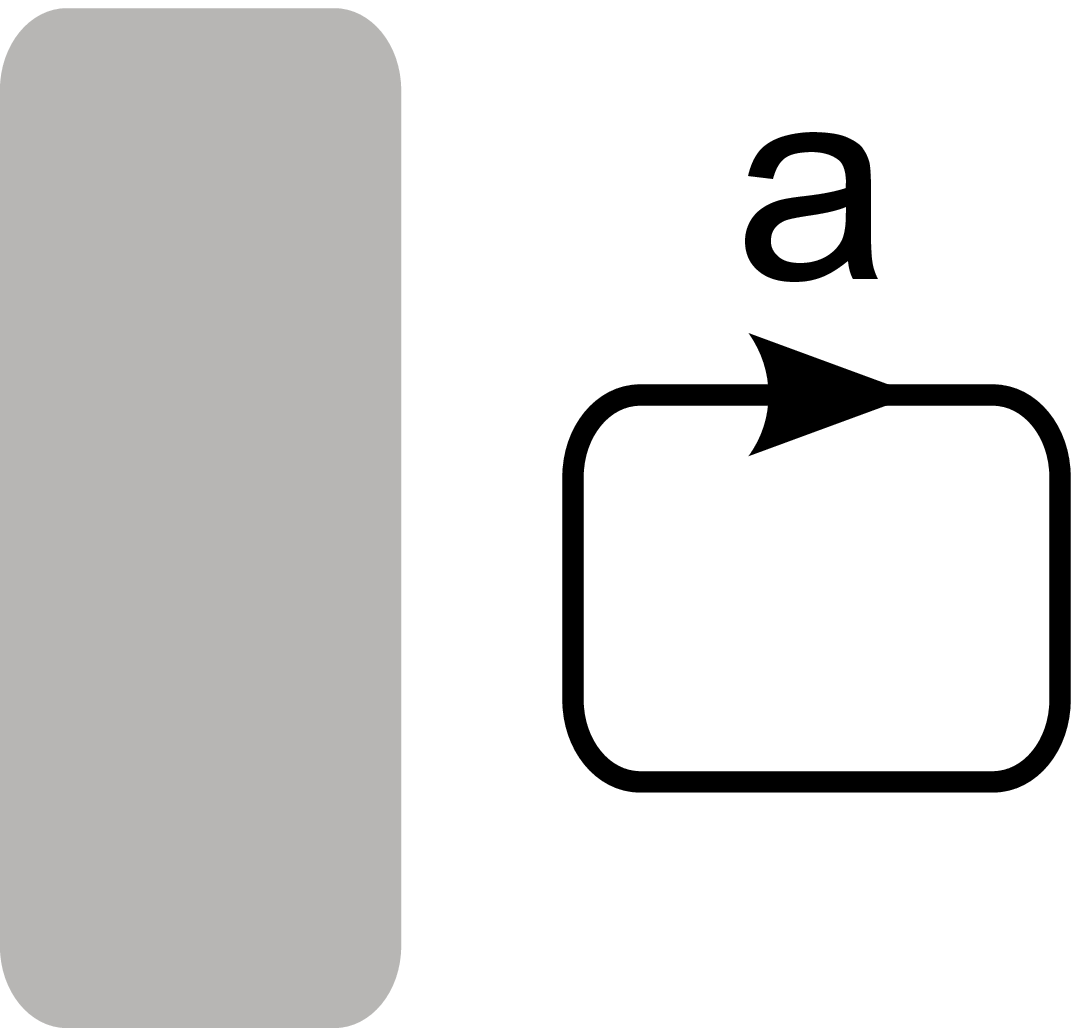}}\right)  
&=&
d_{a}\Phi \left( \raisebox{-0.16in}{\includegraphics[height=0.4in]{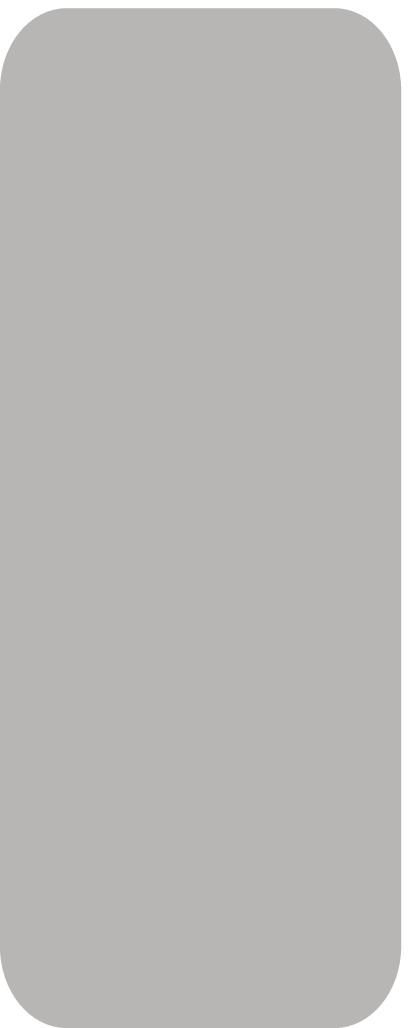}}\right),   \label{rule2} \\
\Phi \left( \raisebox{-0.16in}{\includegraphics[height=0.4in]{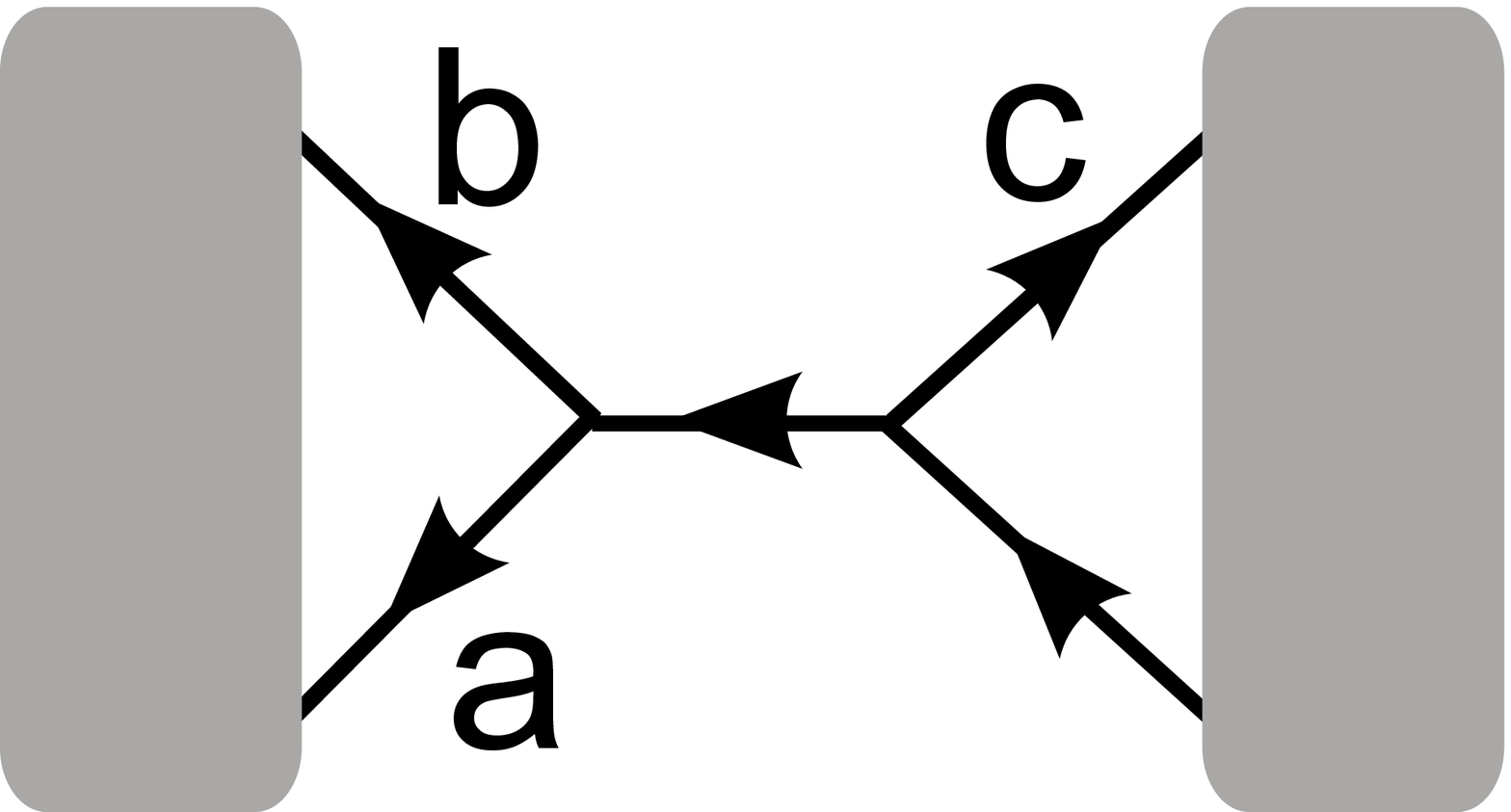}}\right)  
&=&
F(a,b,c)\Phi \left( \raisebox{-0.16in}{\includegraphics[height=0.4in]{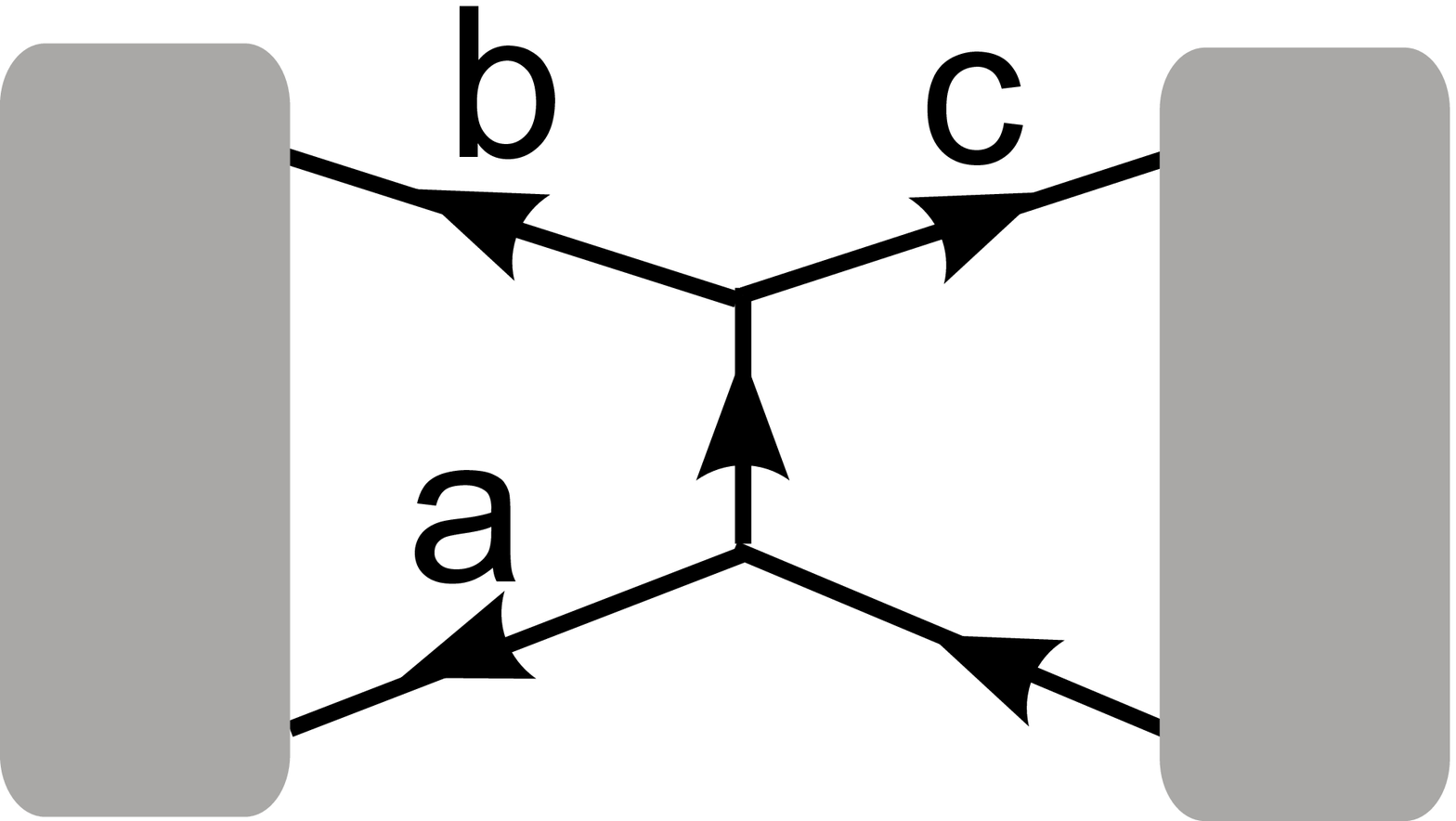}}\right).   \label{rule3} 
\end{eqnarray}
Here $a,b,c$ are arbitrary string types (including the null string type) 
and the shaded regions represent arbitrary string-net configurations which are not changed.
The $d_a$ are complex numbers that depend on the string type $a$, while $F(a,b,c)$ is a complex number that 
depends on $3$ string types $a,b,c$. For the moment, $d_a$ and $F(a,b,c)$ can be arbitrary, but we will soon
see that $d_a, F(a,b,c)$ have to satisfy certain algebraic equations (\ref{selfconseq}-\ref{unit}) in order for
our construction to work.

We now discuss the meaning of these local constraints or local rules. The first rule (\ref{rule1}) has been
drawn schematically. This rule says that two string-net configurations that can 
be continuously deformed into one another must have the same amplitude. Namely, the amplitude of a 
string-net configuration only depends on the topology of the configuration. The second rule (\ref{rule2}) 
says that the amplitude of a string-net configuration containing a closed loop of string type $a$ is equal to
the amplitude of the same configuration \emph{without} the closed loop, multiplied by a factor of $d_a$.

The third rule (\ref{rule3}) is the most important one. This rule relates the amplitude of one string-net
configuration to the amplitude of another configuration that differs from it by recoupling the strings joined at two adjacent vertices.
The reader may notice that two strings have been left unlabeled on both sides of this equation. These labels
are completely determined by the (abelian) branching rules and have been left out due to space constraints. Specifically,
the label on the bottom right hand corner is $a+b+c$, while the middle labels are $a+b$ on the left hand side
and $b+c$ on the right hand side.

The basic idea of equations (\ref{rule1} - \ref{rule3}), is that by applying these local rules multiple times, one
can relate the amplitude of any string-net configuration to the amplitude of the vacuum or ``no-string'' configuration.
Then, using the convention that  
\begin{equation}
\Phi(\text{vacuum}) = 1, 
\label{vacuum}
\end{equation}
the amplitude of every configuration is fully determined. Thus, the rules determine the wave function completely once 
the parameters $d_a, F(a,b,c)$, etc. are given. We will give an example of such a computation below. However, before 
presenting this example, we need to explain our conventions for how to apply these rules, and some additional
structure associated with these conventions.

\subsection{String-net conventions and \texorpdfstring{$\gamma,\alpha$}{gamma,alpha} factors}

First, we discuss our conventions regarding the ``null'' string. In applying the above rules, 
one often encounters string-net configurations containing a null string with label $a=0$. For example, when $b = c^*$, equation (\ref{rule3}) gives:
\begin{equation*}
	\Phi \left(\raisebox{-0.16in}{\includegraphics[height=0.4in]{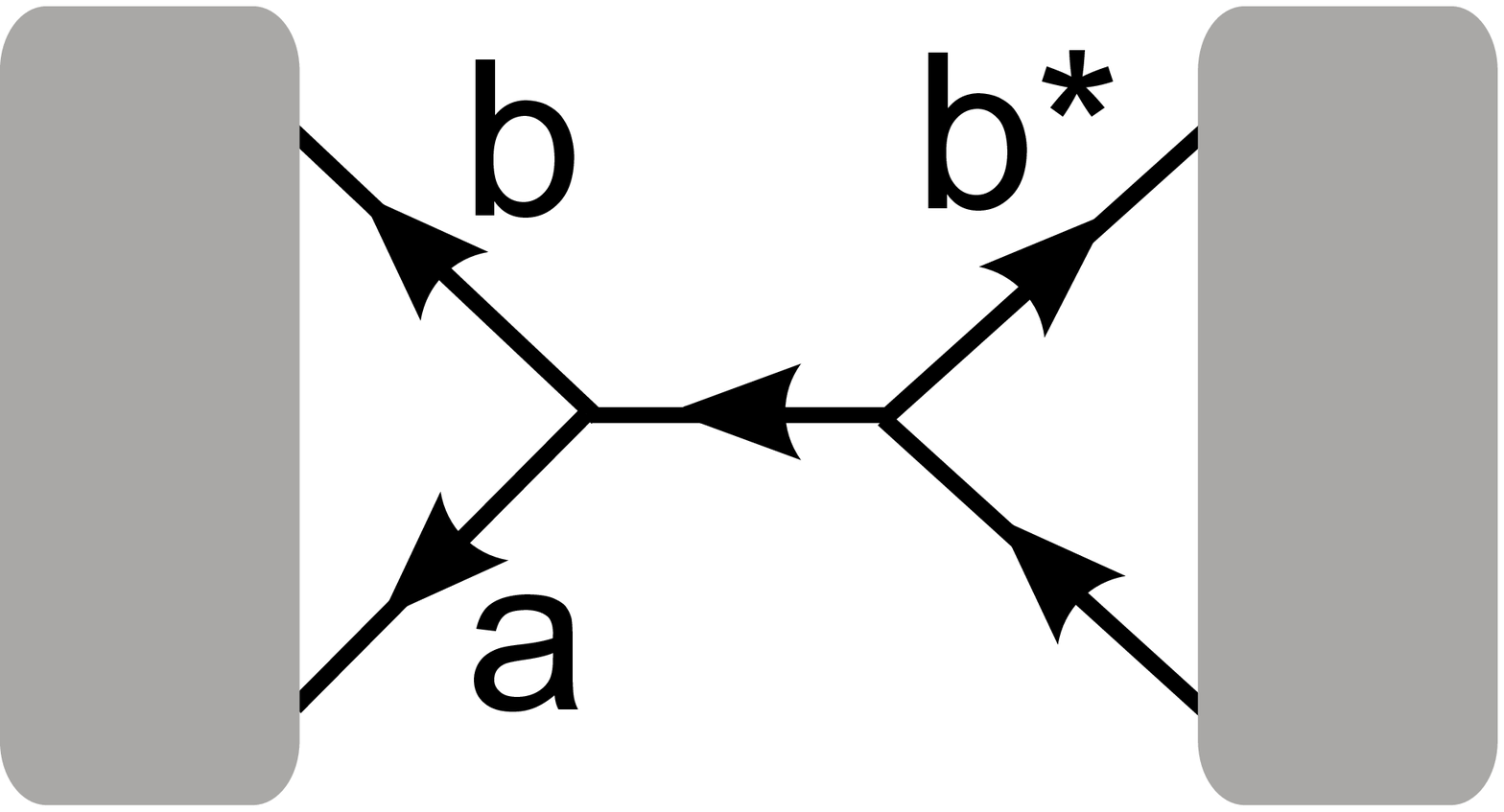}} \right) = F(a,b,b^{*})
\Phi \left(\raisebox{-0.16in}{\includegraphics[height=0.4in]{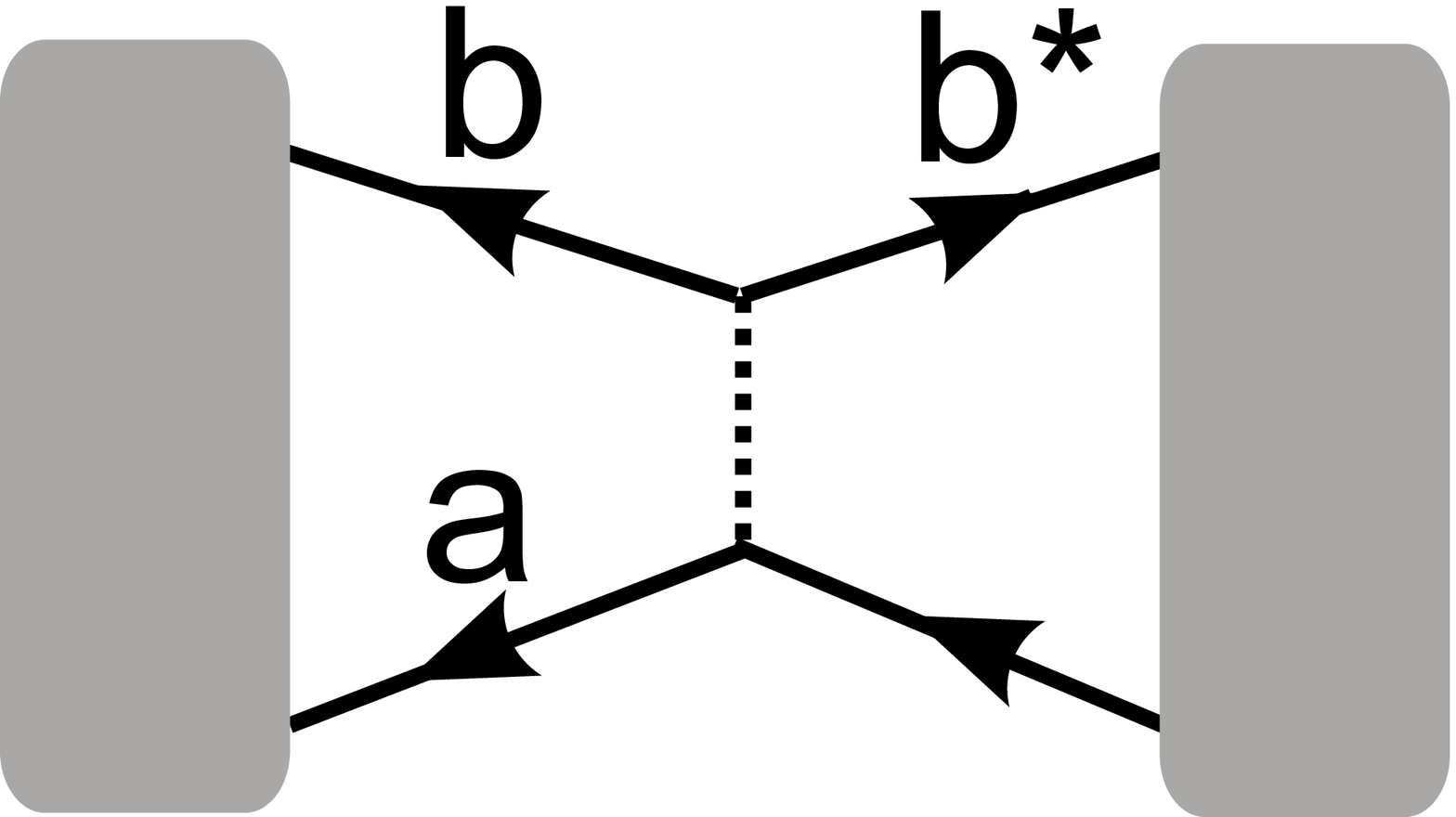}} \right)
\end{equation*}
or in bra-ket notation,
\begin{equation*}
	\left\< 
	\raisebox{-0.16in}{\includegraphics[height=0.4in]{config1a.eps}} 
	\right| \Phi \bigg \rangle = F(a,b,b^{*})
	\left\< \raisebox{-0.16in}{\includegraphics[height=0.4in]{config1b.eps}} 
	\right| \Phi \bigg \rangle
\end{equation*}
where the configuration on the right hand side contains a null string.
In Ref. [\onlinecite{LevinWenStrnet}], it was assumed that these null strings could be freely erased, since the null string
corresponds to the vacuum. This erasing of null strings was a key part of the local rule formalism, since it was what 
allowed us to reduce string-net configurations to the vacuum configuration, and thereby compute their amplitude.
Here, we will also assume that null strings can be erased, but under more restricted circumstances. 

Our rules for dealing with the null string are as follows. We will describe these rules using \emph{bras} $\<X|$ rather than 
\emph{kets} $|X\>$ because it simplifies some of the notation below. First, null strings can be freely erased everywhere except 
near vertices with non-null strings. For example:
\begin{equation}
\left\< \raisebox{-0.16in}{\includegraphics[height=0.4in]{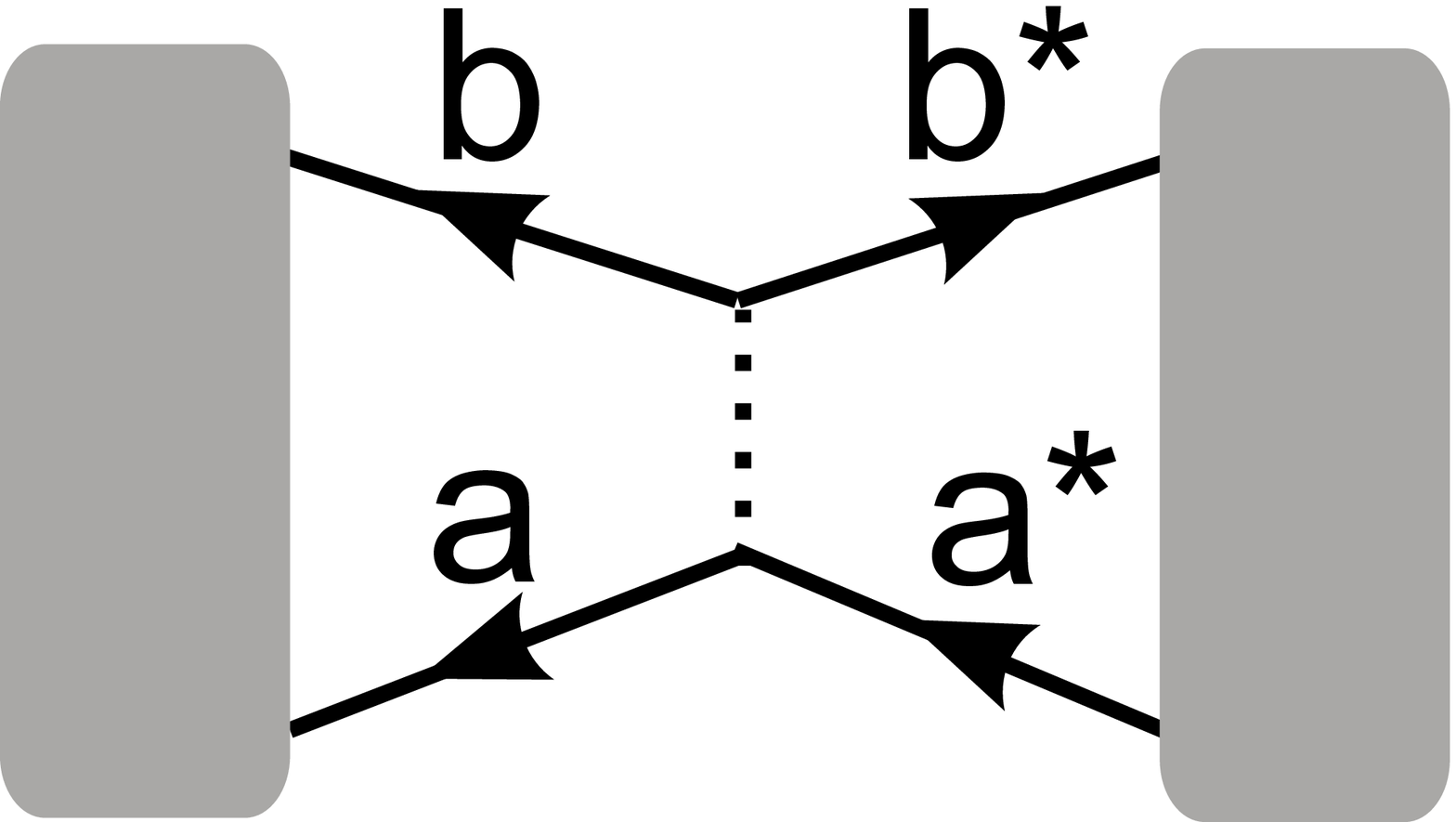}} \right| = 
\left\< \raisebox{-0.16in}{\includegraphics[height=0.4in]{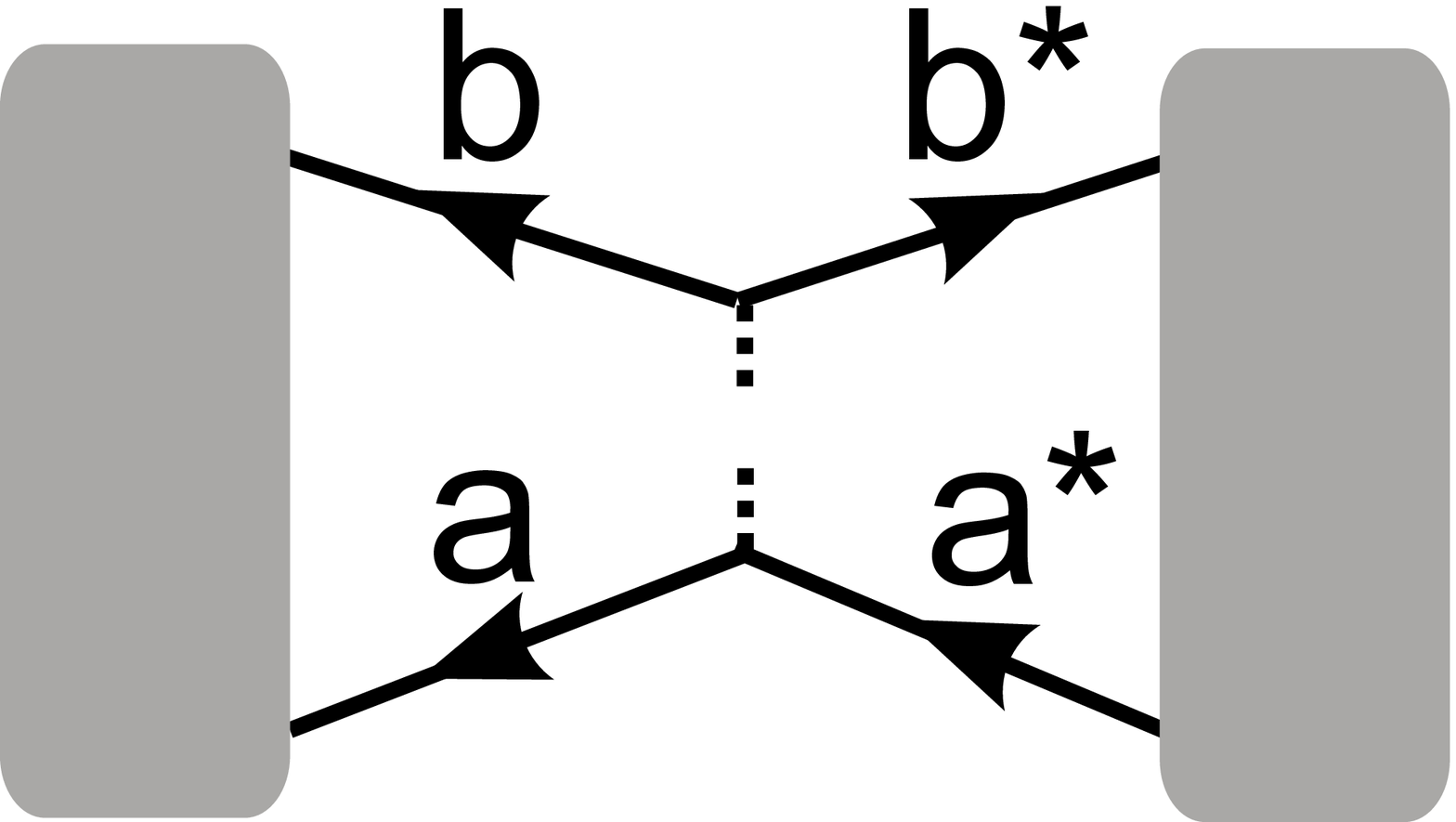}} \right|
\label{nullerase}
\end{equation}
Second, the ``end'' of a null string can be erased at any vertex where the two other strings at the vertex are oriented in the same 
direction:
\begin{equation} 
\left\< \raisebox{-0.16in}{\includegraphics[height=0.4in]{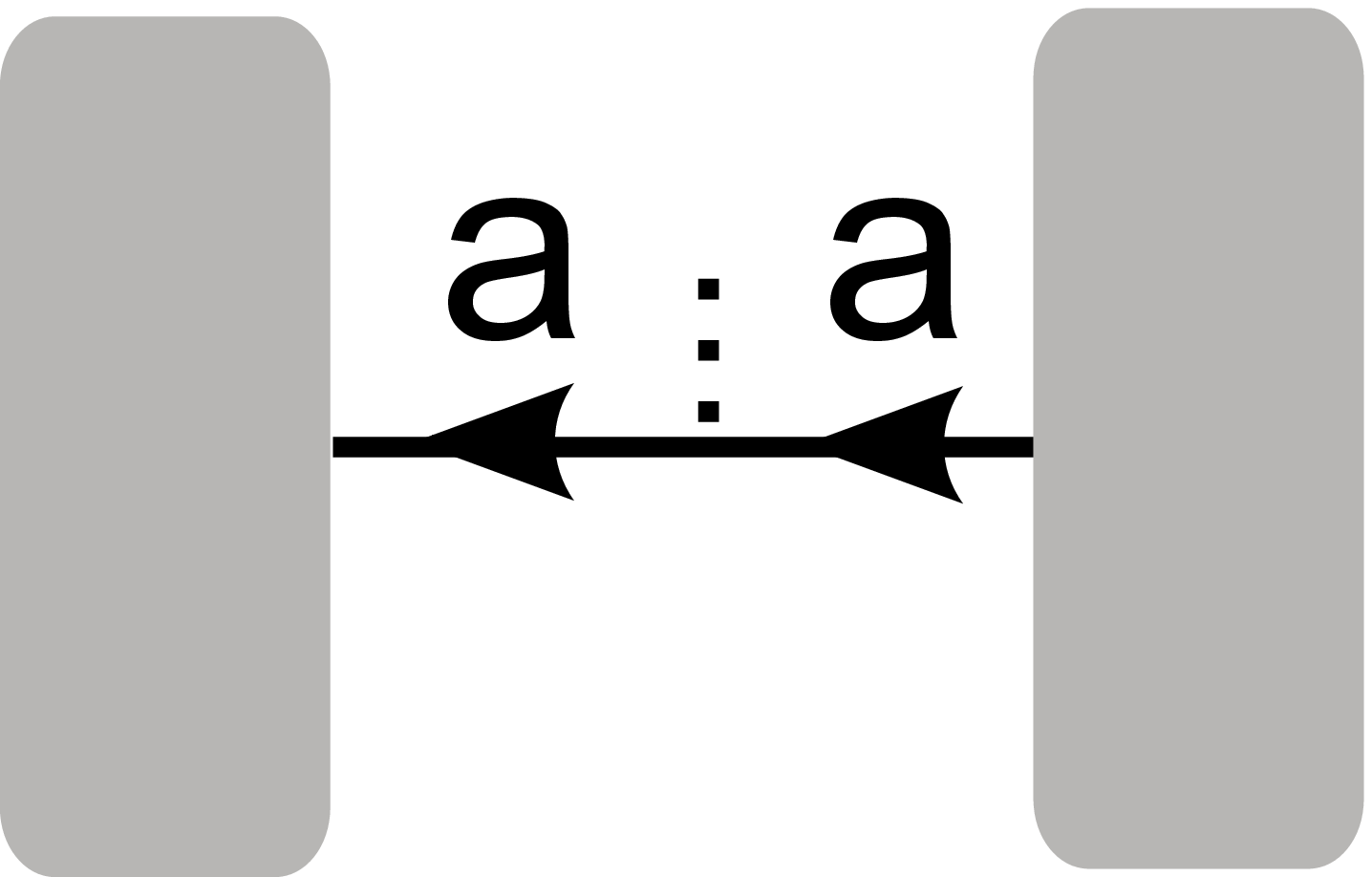}} \right|
= 
\left\< \raisebox{-0.16in}{\includegraphics[height=0.4in]{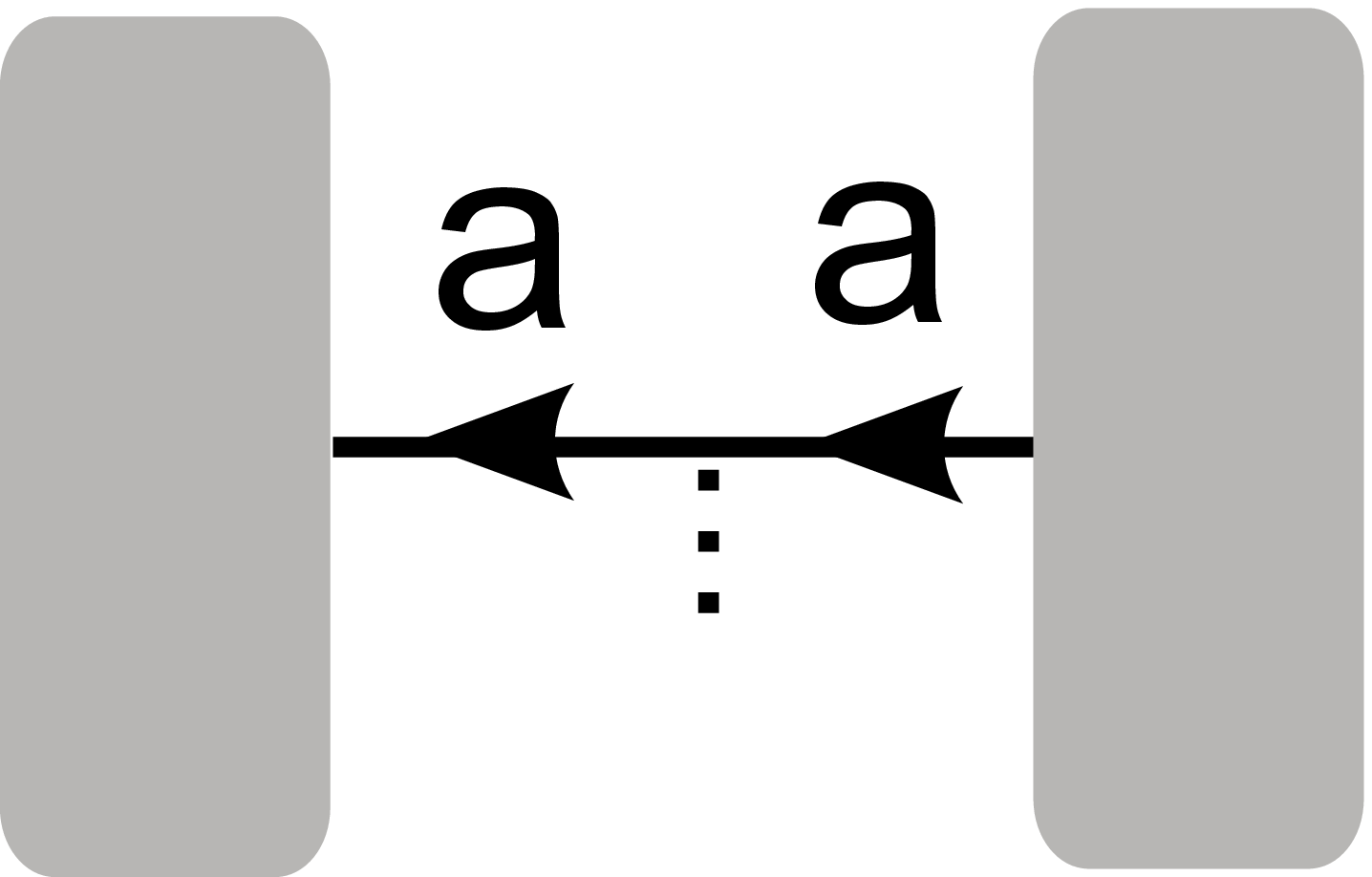}} \right| 
=
\left\< \raisebox{-0.16in}{\includegraphics[height=0.4in]{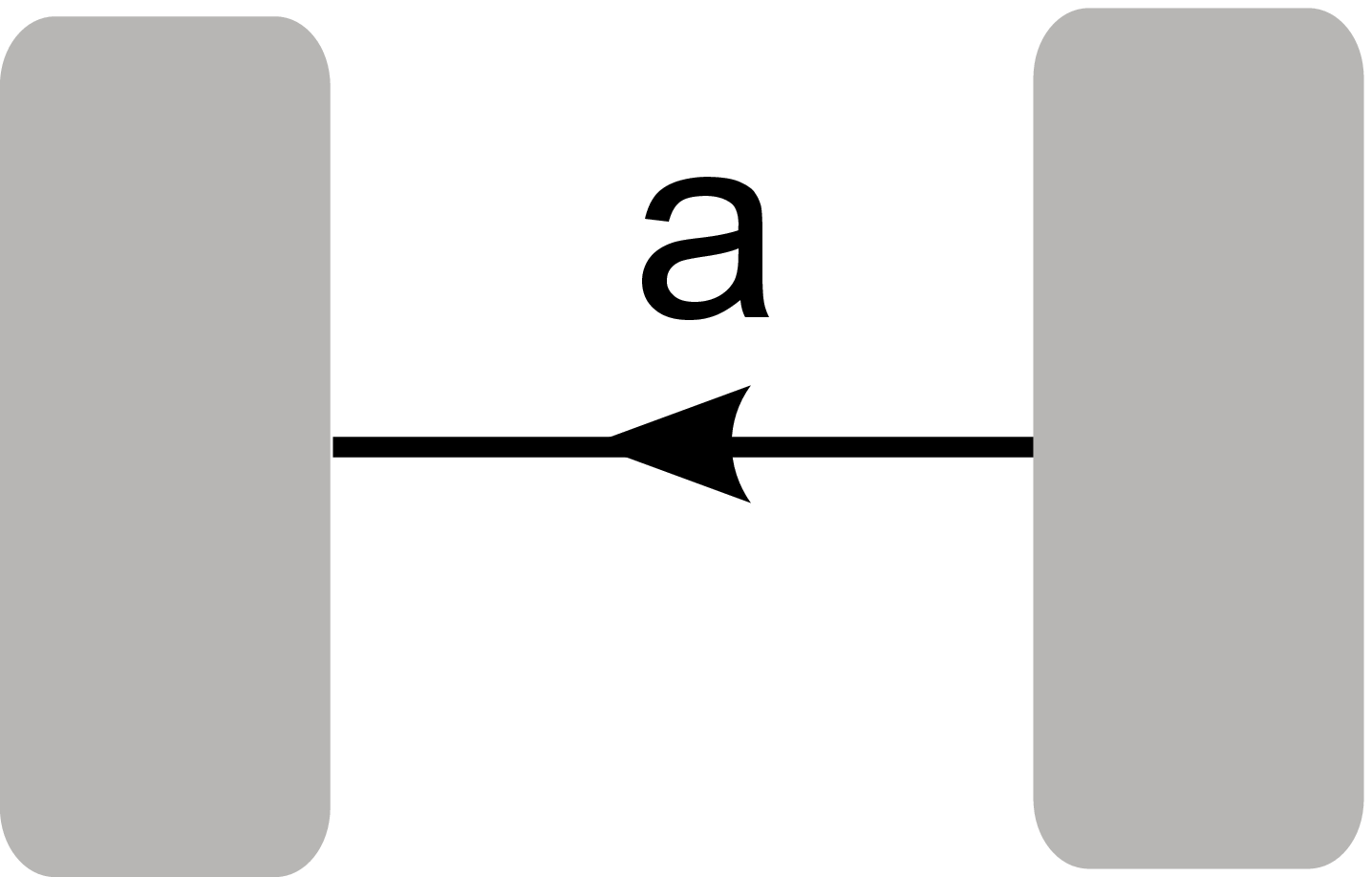}} \right|.
 \label{nullsame}
\end{equation}
On the other hand, the end of the null string \emph{cannot} be erased at vertices where the two other strings are oriented in opposite
directions. Indeed, in this case, we need to keep careful track of the end of the null string, since ``flipping'' the null sting from one side of the vertex to the other introduces a phase factor:
\begin{eqnarray}
\left\< \raisebox{-0.16in}{\includegraphics[height=0.4in]{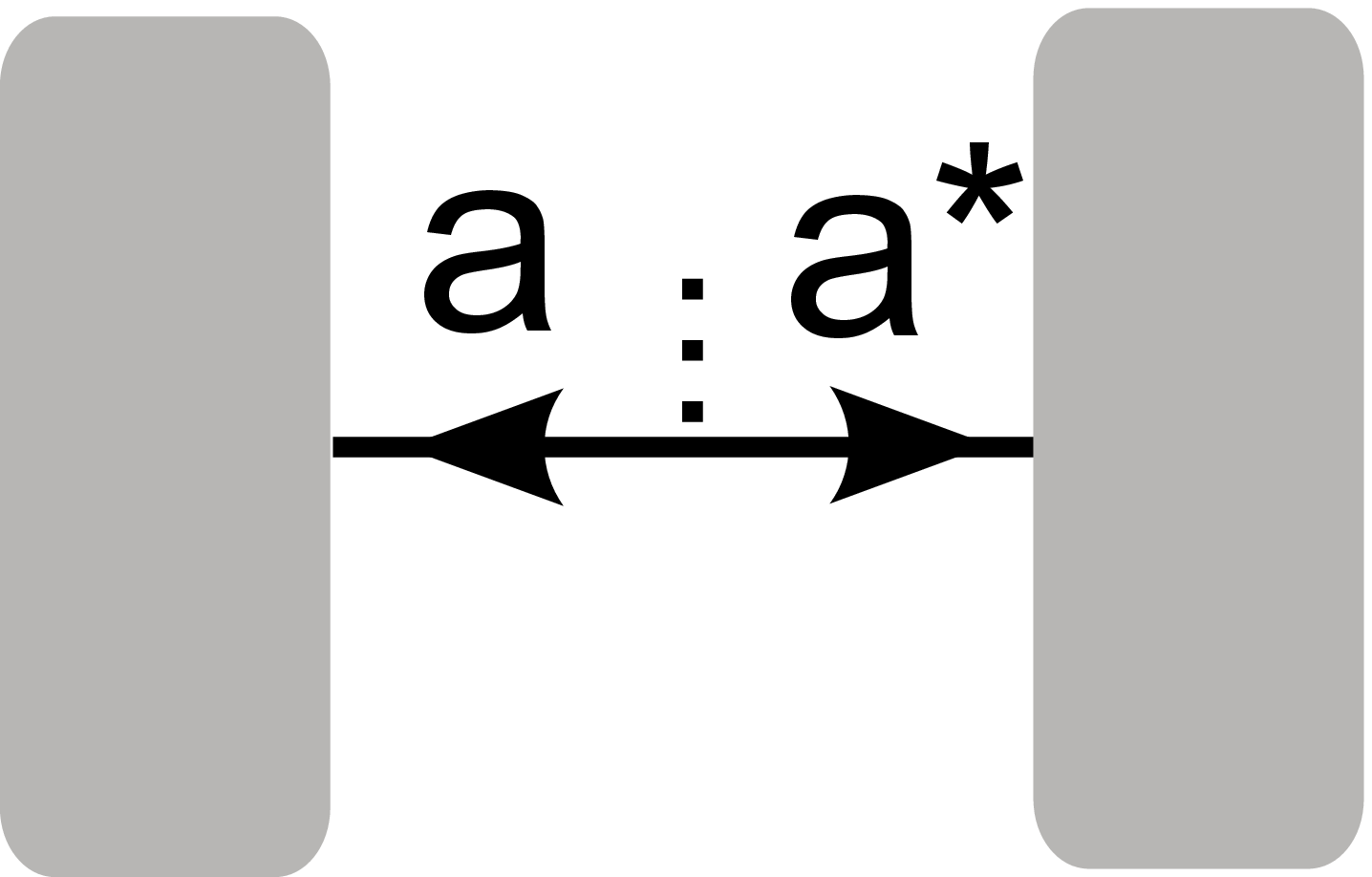}} \right|  
&=&
\gamma_{a} \left\< \raisebox{-0.16in}{\includegraphics[height=0.4in]{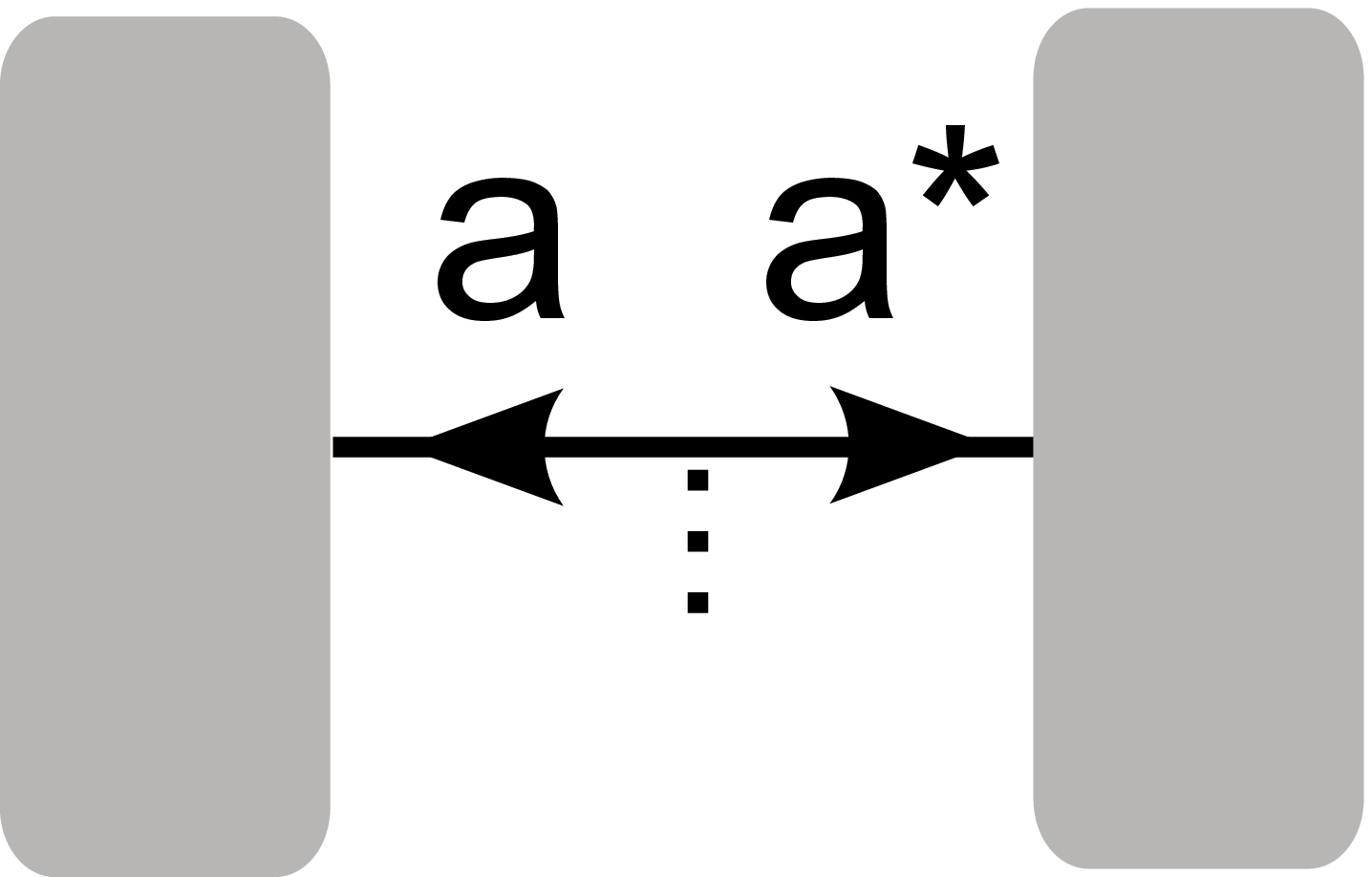}} \right|,   \label{rule5} \\
\left\< \raisebox{-0.16in}{\includegraphics[height=0.4in]{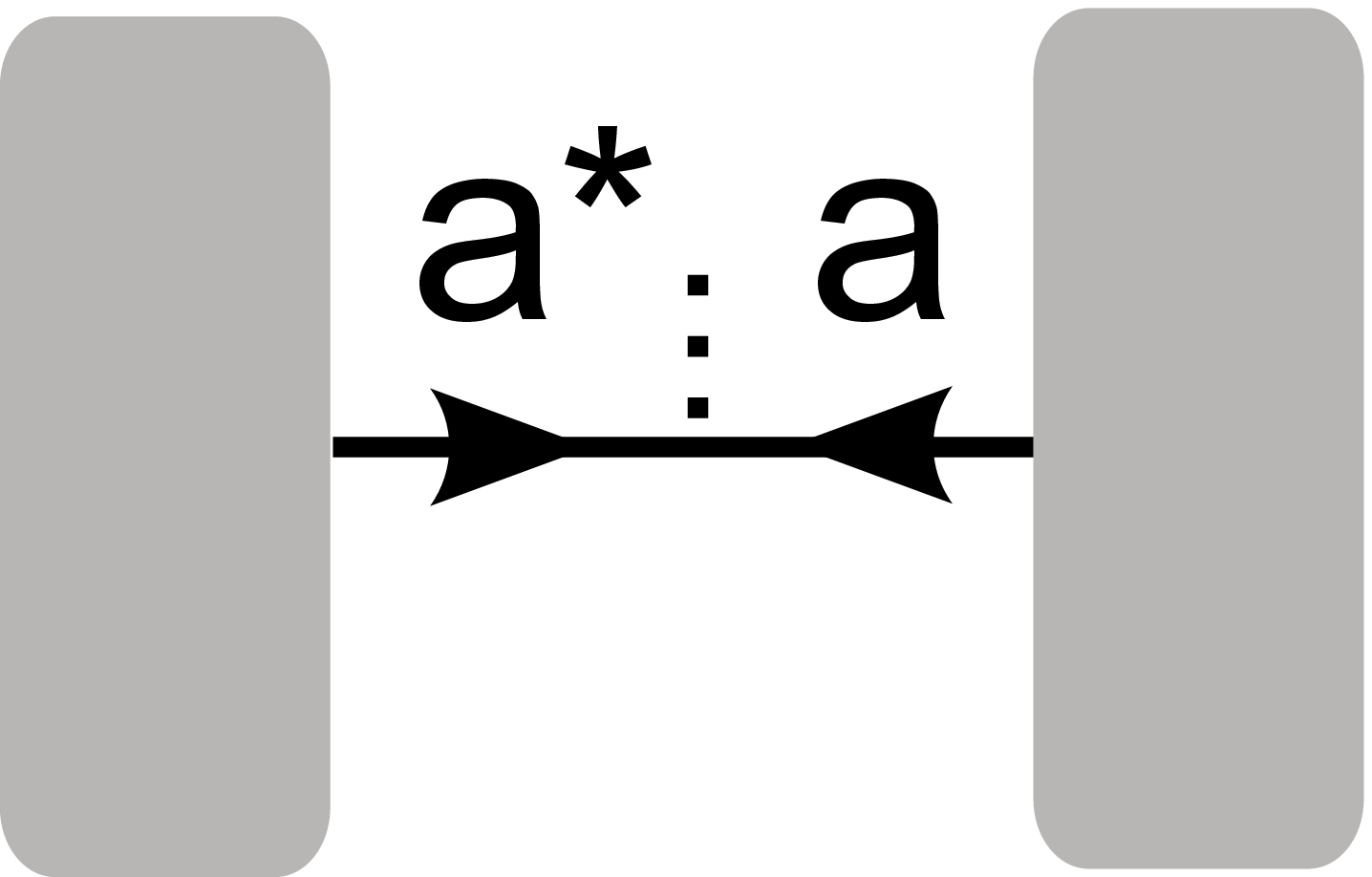}} \right|  
&=&
\gamma_{a} \left\< \raisebox{-0.16in}{\includegraphics[height=0.4in]{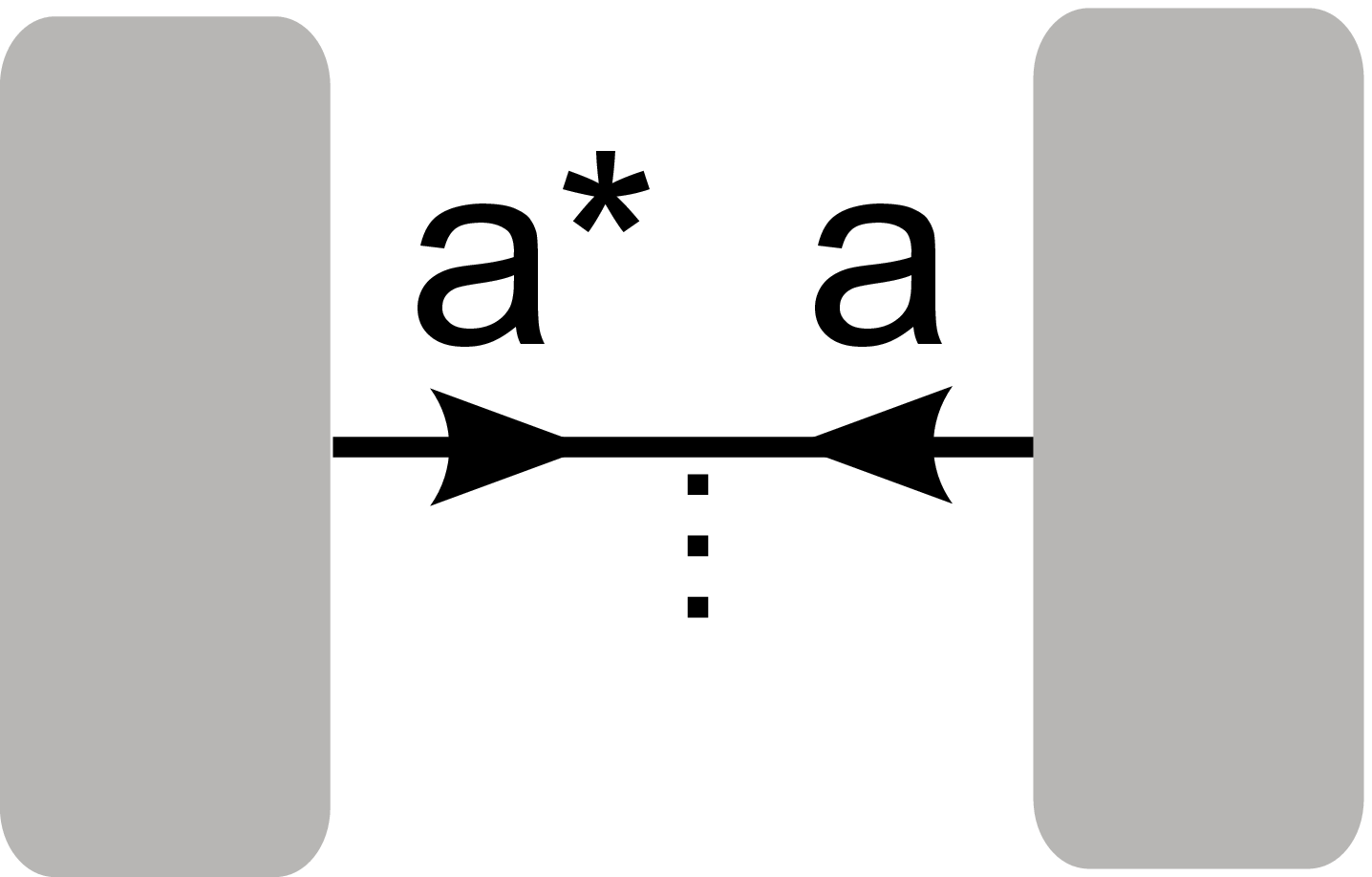}} \right|  \label{rule6}
\end{eqnarray}
where $\gamma_a$ is a complex number with modulus $1$: $|\gamma_a| = 1$. Later we will see that $\gamma_a$ can be
chosen to be $\pm 1$ without loss of generality. (We explain the motivation behind $\gamma$ in appendix \ref{dot}.)

Our fourth rule is that the ends of the null strings can be erased in pairs according to:
\begin{equation}
\left\< \raisebox{-0.16in}{\includegraphics[height=0.4in]{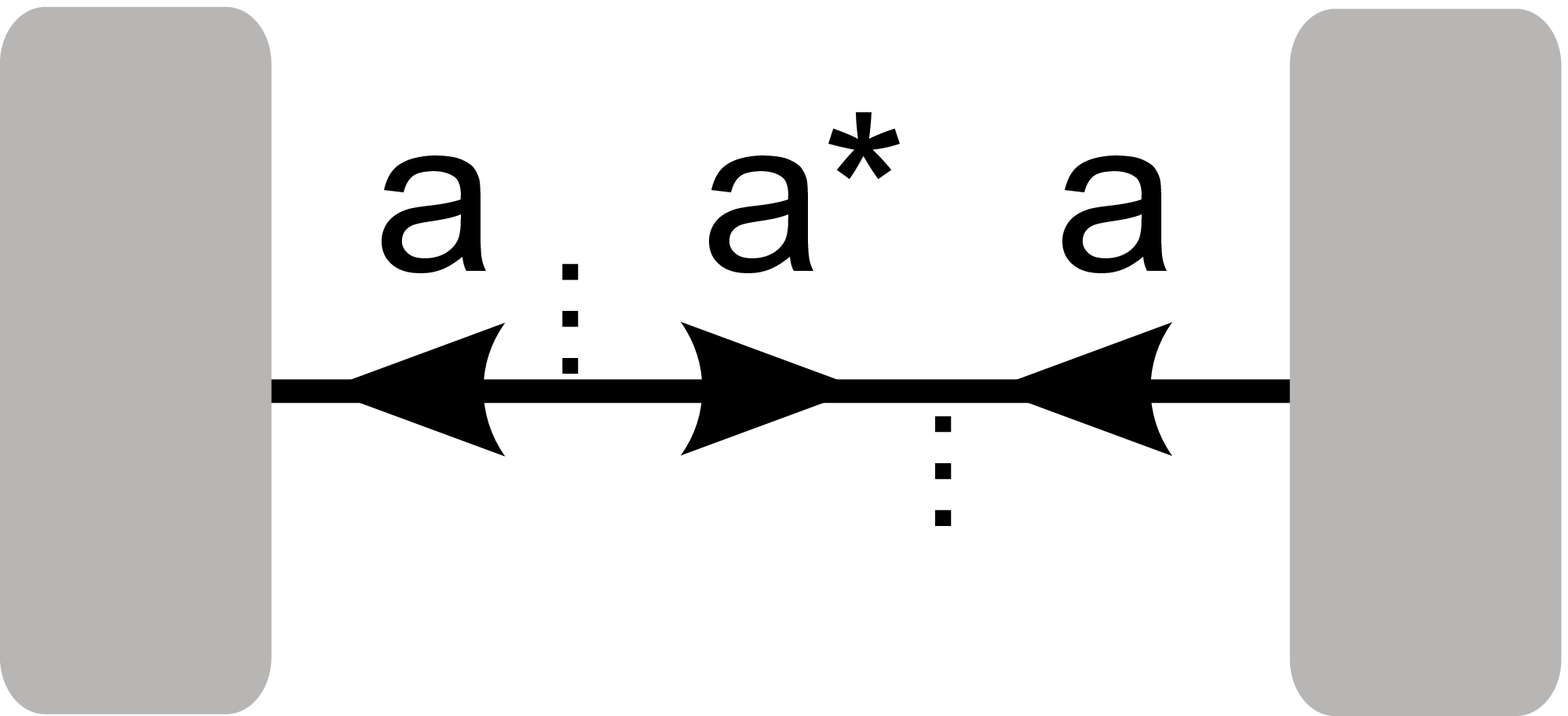}} \right| 
=
\left\< \raisebox{-0.16in}{\includegraphics[height=0.4in]{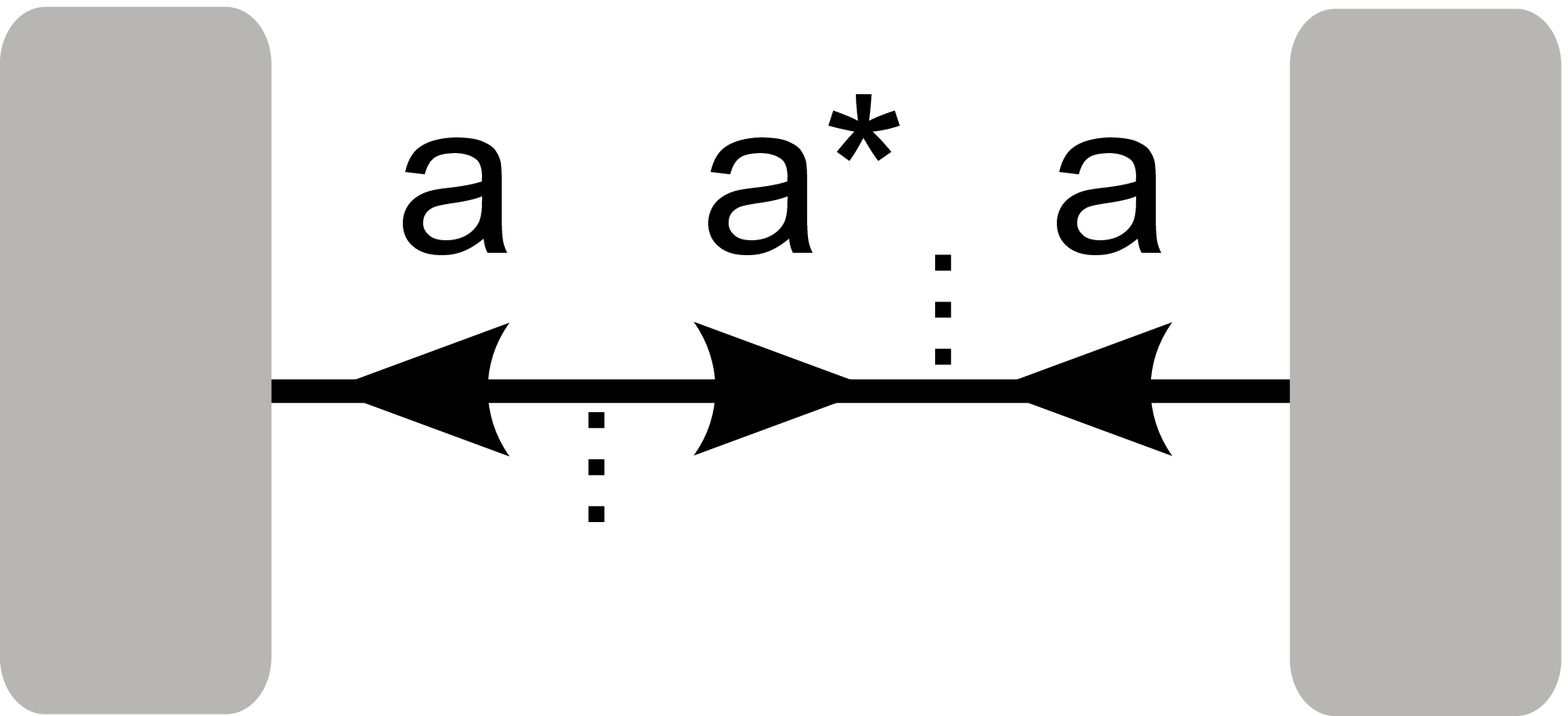}} \right| 
=
\left\< \raisebox{-0.16in}{\includegraphics[height=0.4in]{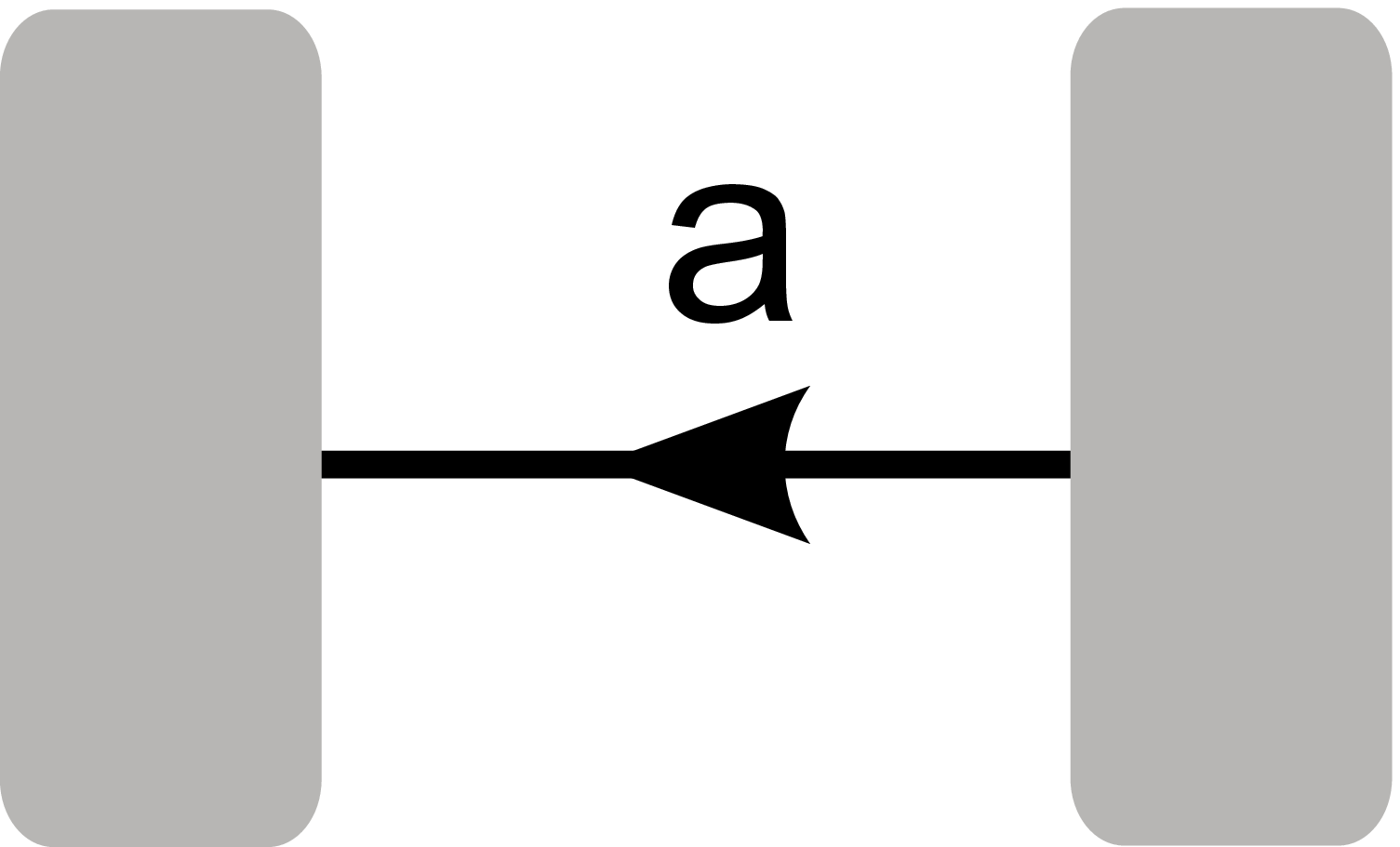}} \right|. 
\label{normal}
\end{equation} 

Finally, the ends of the null strings can be absorbed into vertices as follows:
\begin{eqnarray}
\left\< \raisebox{-0.18in}{\includegraphics[height=0.45in]{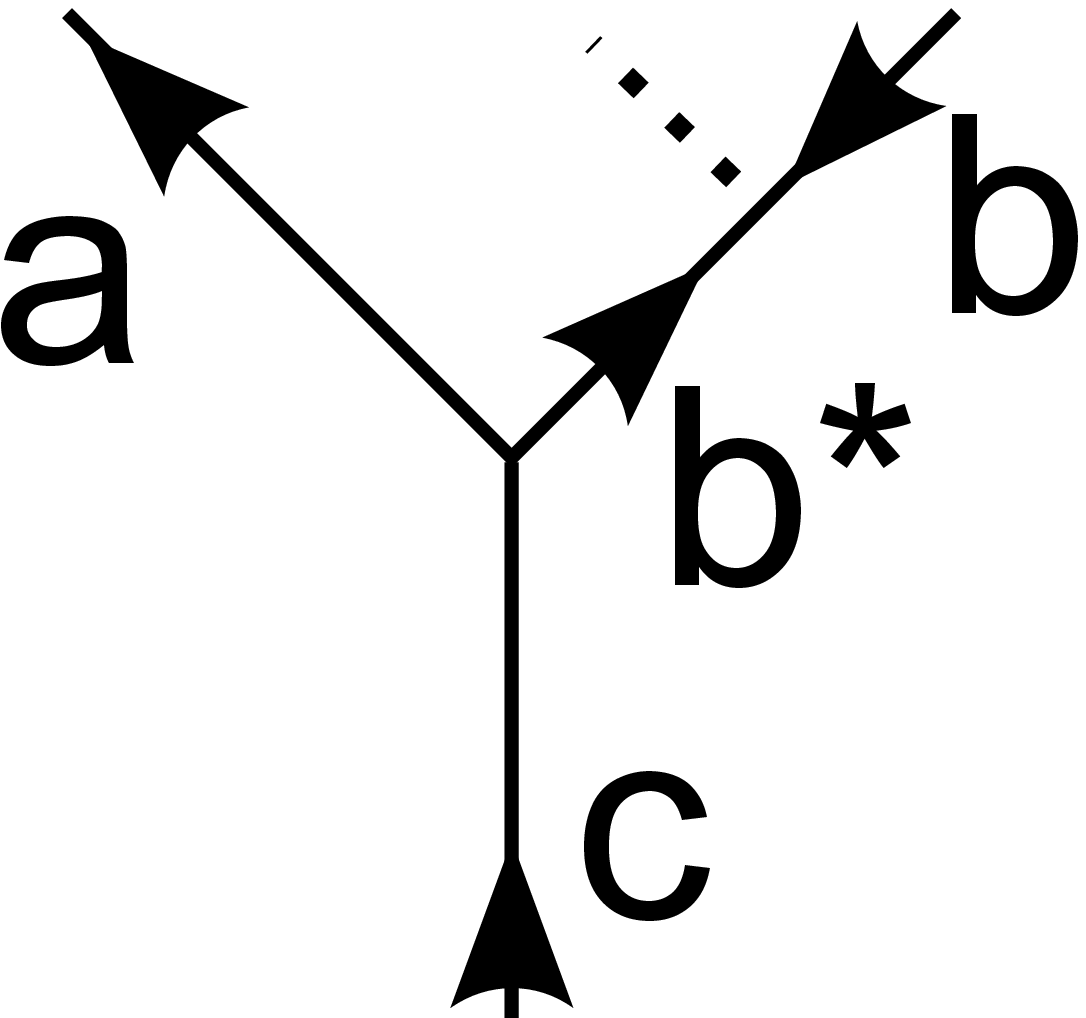}} \right|  
&=&
\left\< \raisebox{-0.18in}{\includegraphics[height=0.45in]{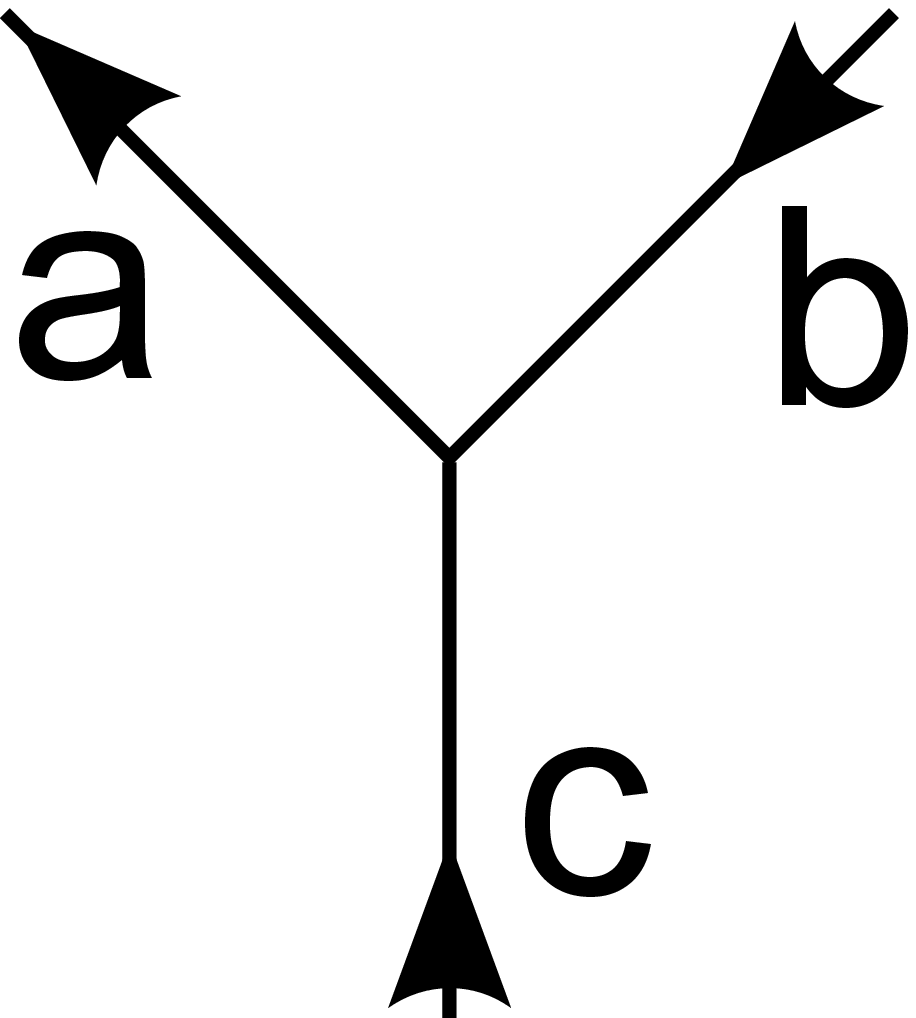}} \right|,   \label{nullabs}\\
\left\< \raisebox{-0.18in}{\includegraphics[height=0.45in]{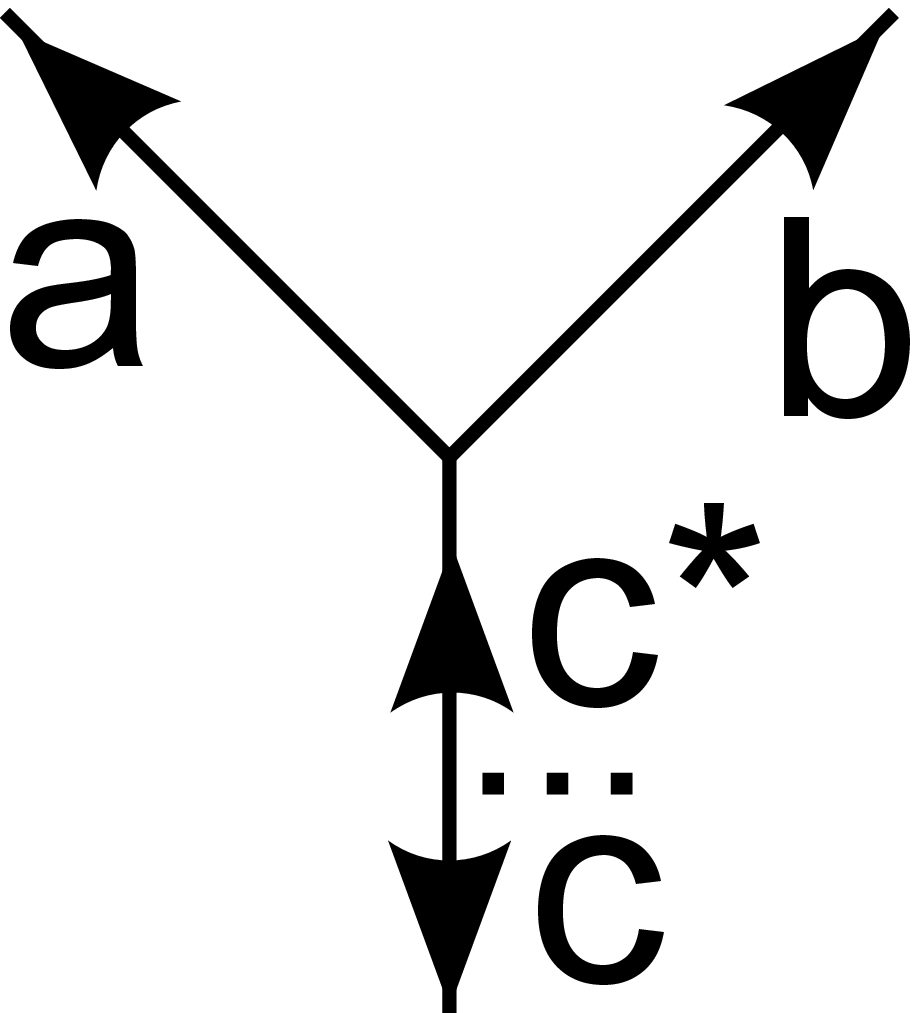}} \right|  
&=&
\left\< \raisebox{-0.18in}{\includegraphics[height=0.45in]{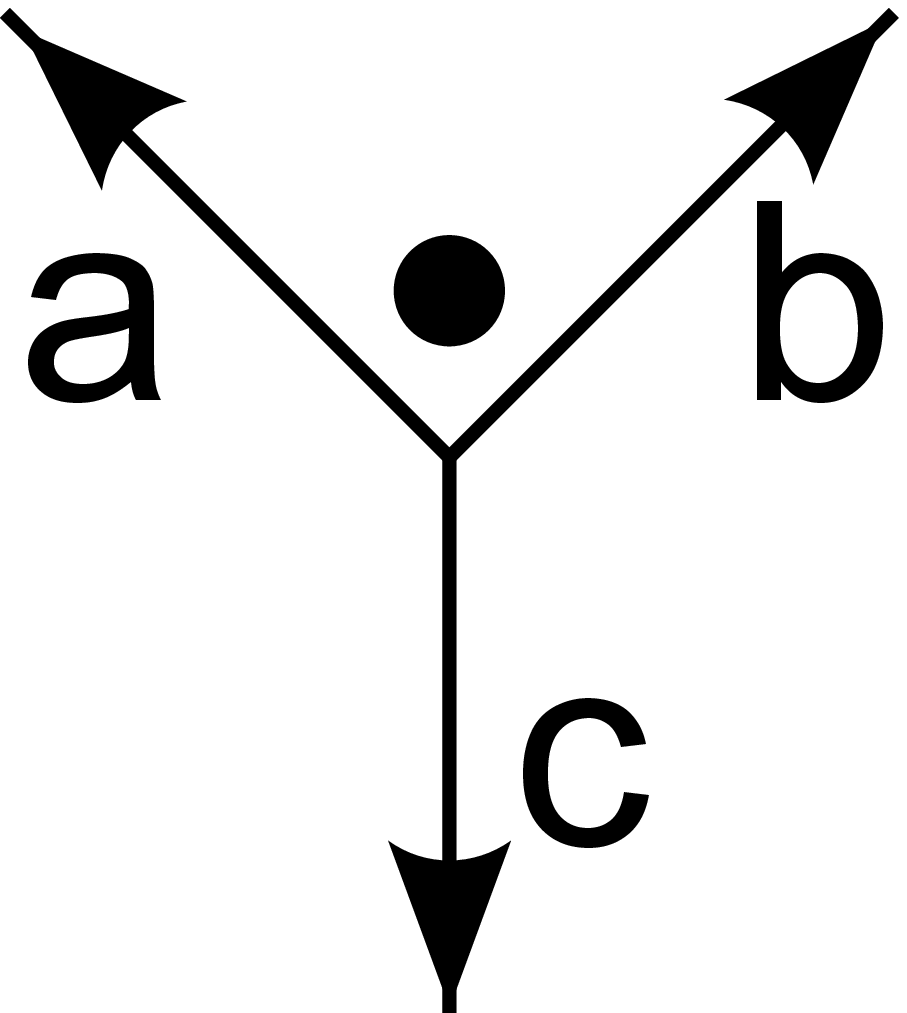}} \right|,   \label{dot1} \\
\left\< \raisebox{-0.18in}{\includegraphics[height=0.45in]{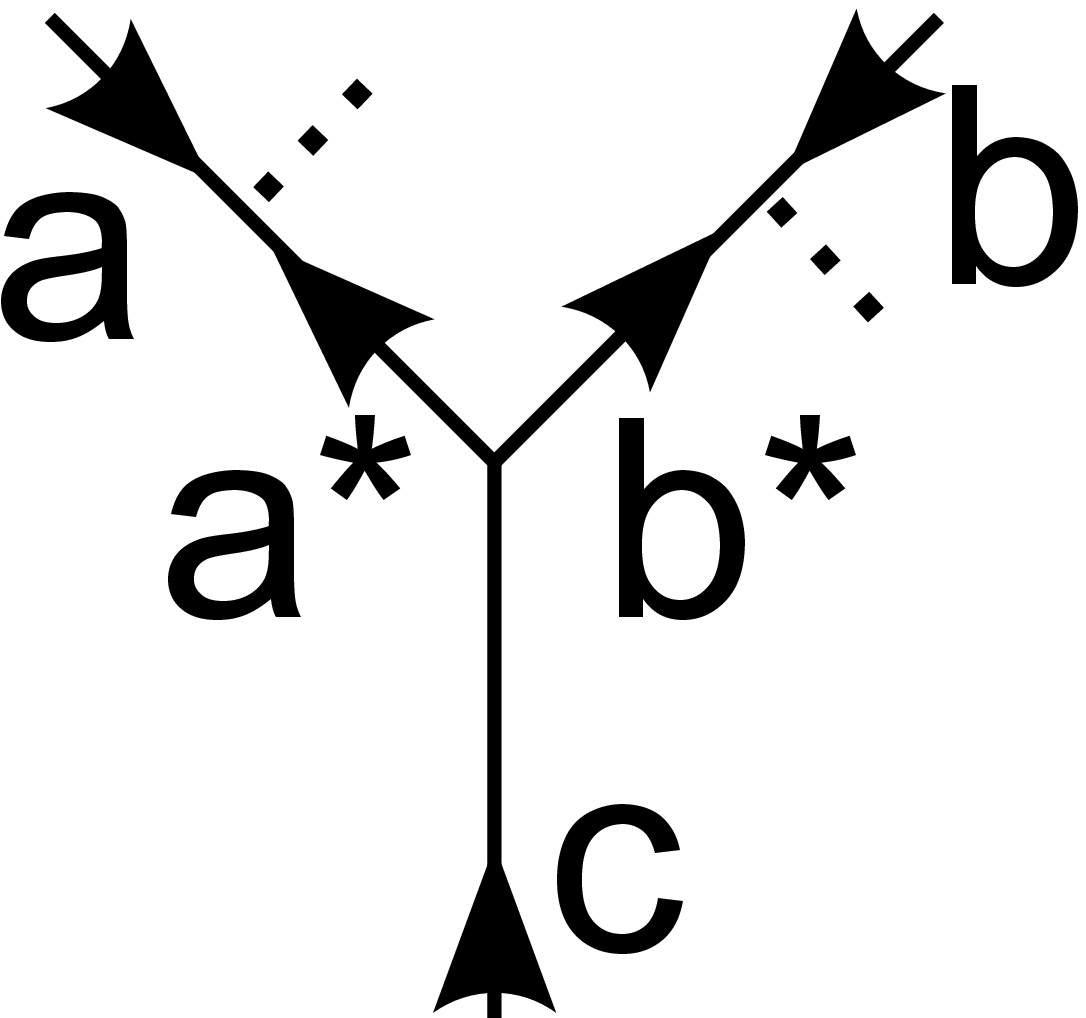}} \right|  
&=&
\left\< \raisebox{-0.18in}{\includegraphics[height=0.45in]{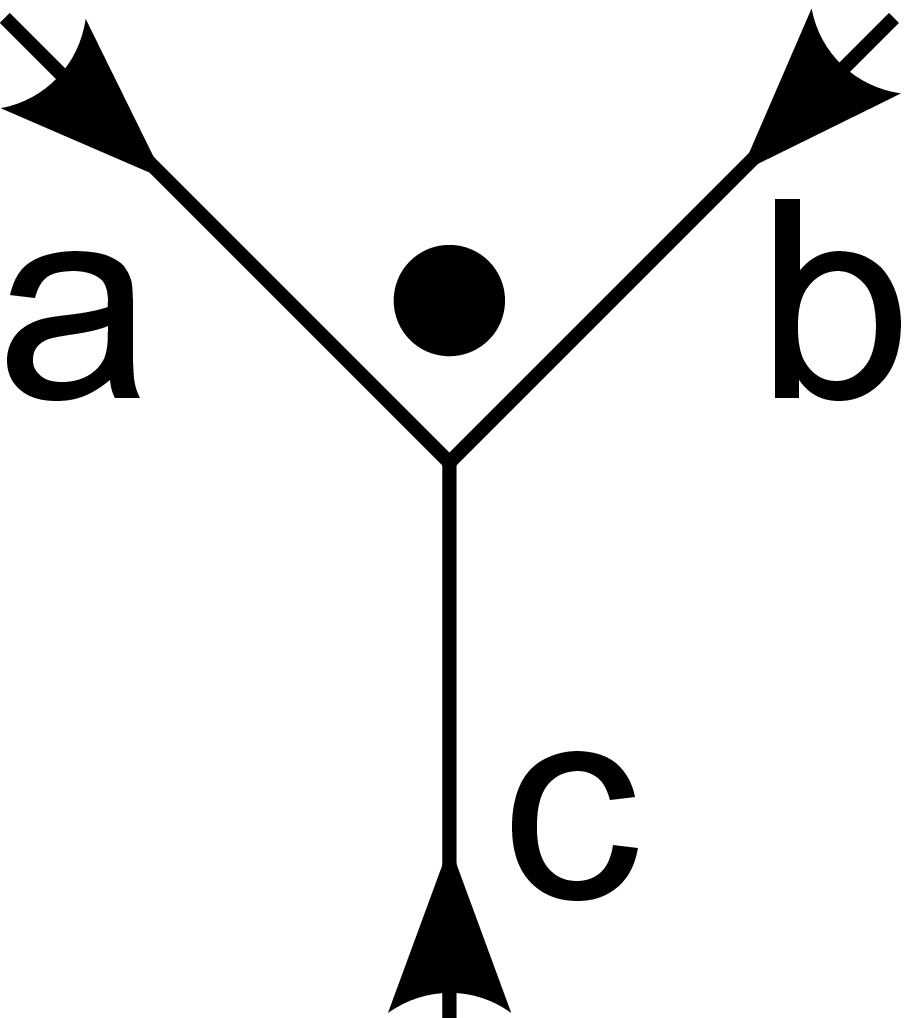}} \right|.	\label{dot2}
\end{eqnarray}

The last two rules (\ref{dot1} - \ref{dot2}) introduce another ingredient into our diagrammatical calculus: we can see that the vertices on
the right hand side of Eq. (\ref{dot1}) and Eq. (\ref{dot2}) are decorated with dots. In general, we decorate all vertices that have three
incoming or three outgoing legs with dots. The dots can be placed in any of the three positions near the vertex. Like the string orientations
or the ends of the null strings, moving the position of the dot does not change the physical state, but it can 
introduce a phase factor (similar to $\gamma_a$). These phase factors are defined by 
\begin{eqnarray}
\left\< \raisebox{-0.16in}{\includegraphics[height=0.4in]{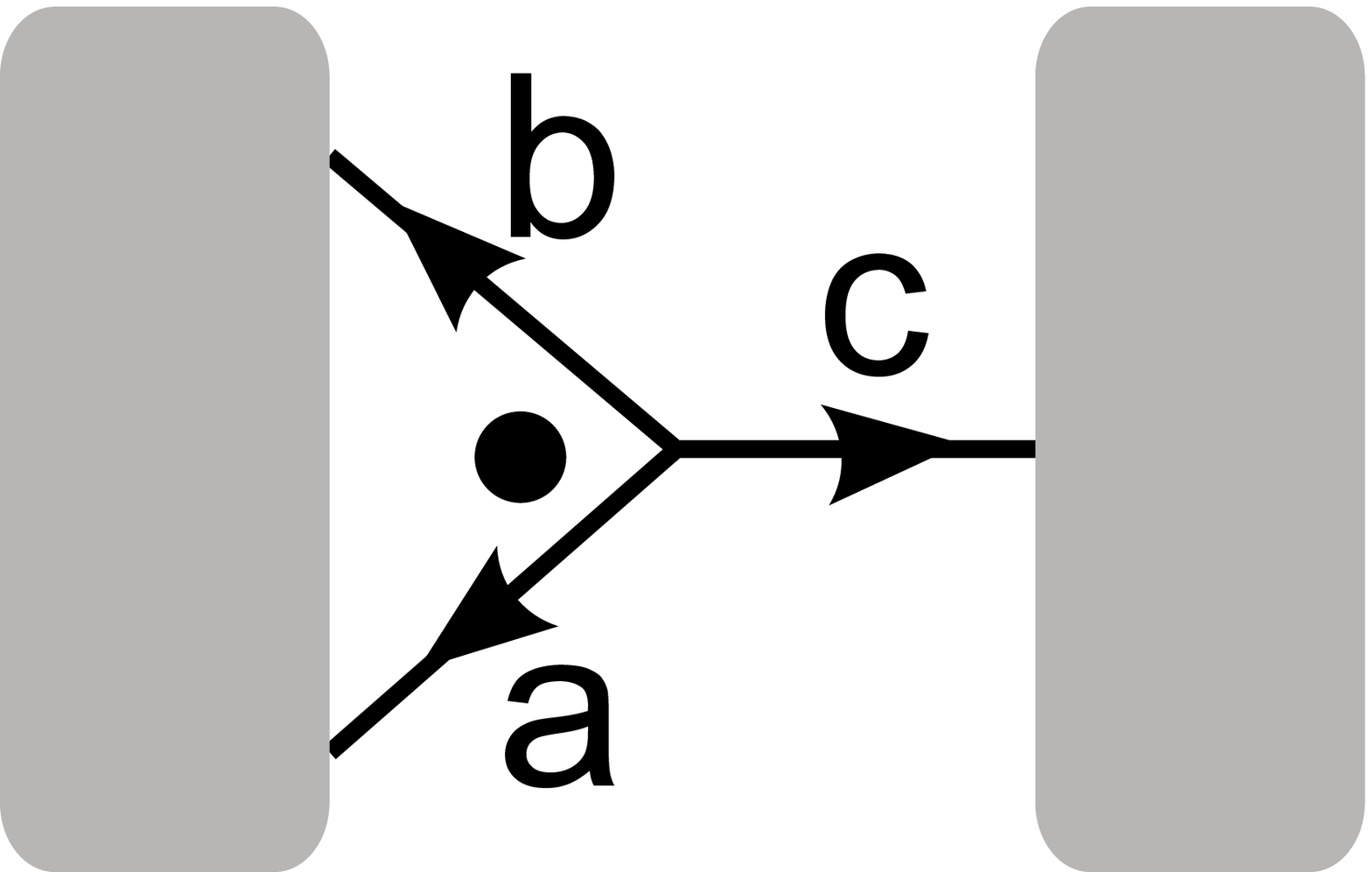}} \right|
&=&
\alpha (a,b) \cdot \left\< \raisebox{-0.16in}{\includegraphics[height=0.4in]{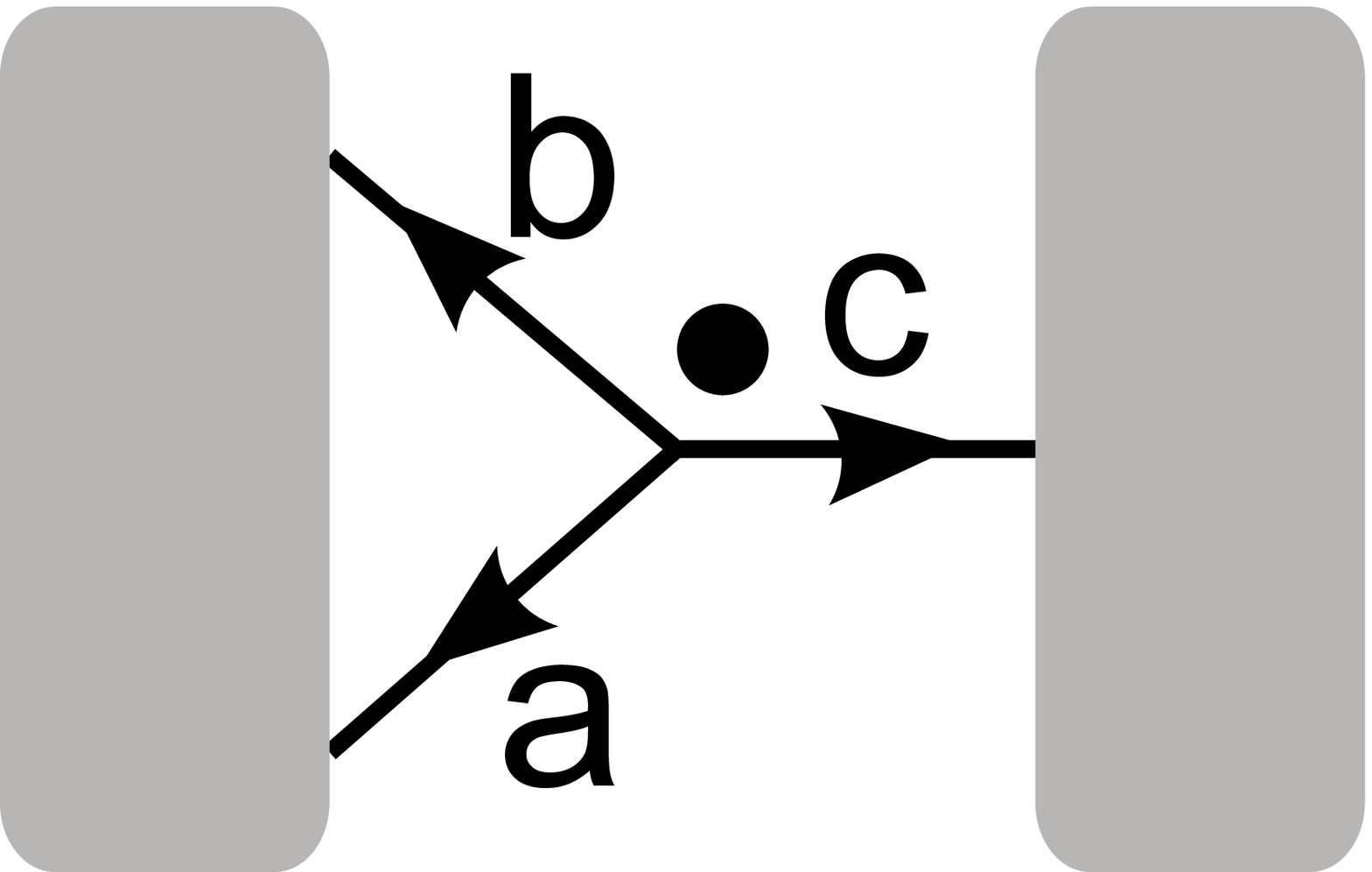}} \right|,   \label{rule4} \\
\left\< \raisebox{-0.16in}{\includegraphics[height=0.4in]{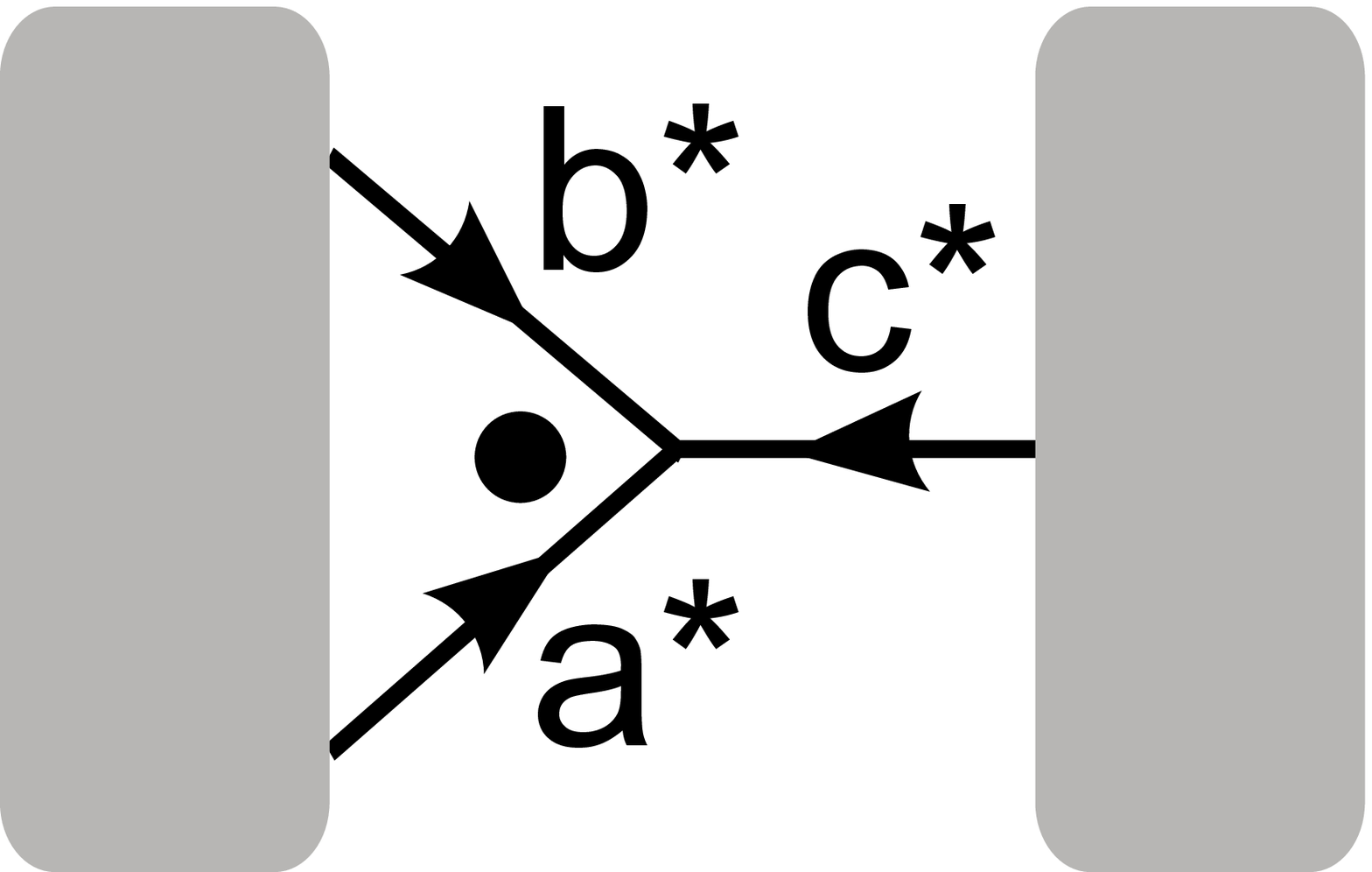}} \right|
&=&
\alpha (a,b) \cdot \left\< \raisebox{-0.16in}{\includegraphics[height=0.4in]{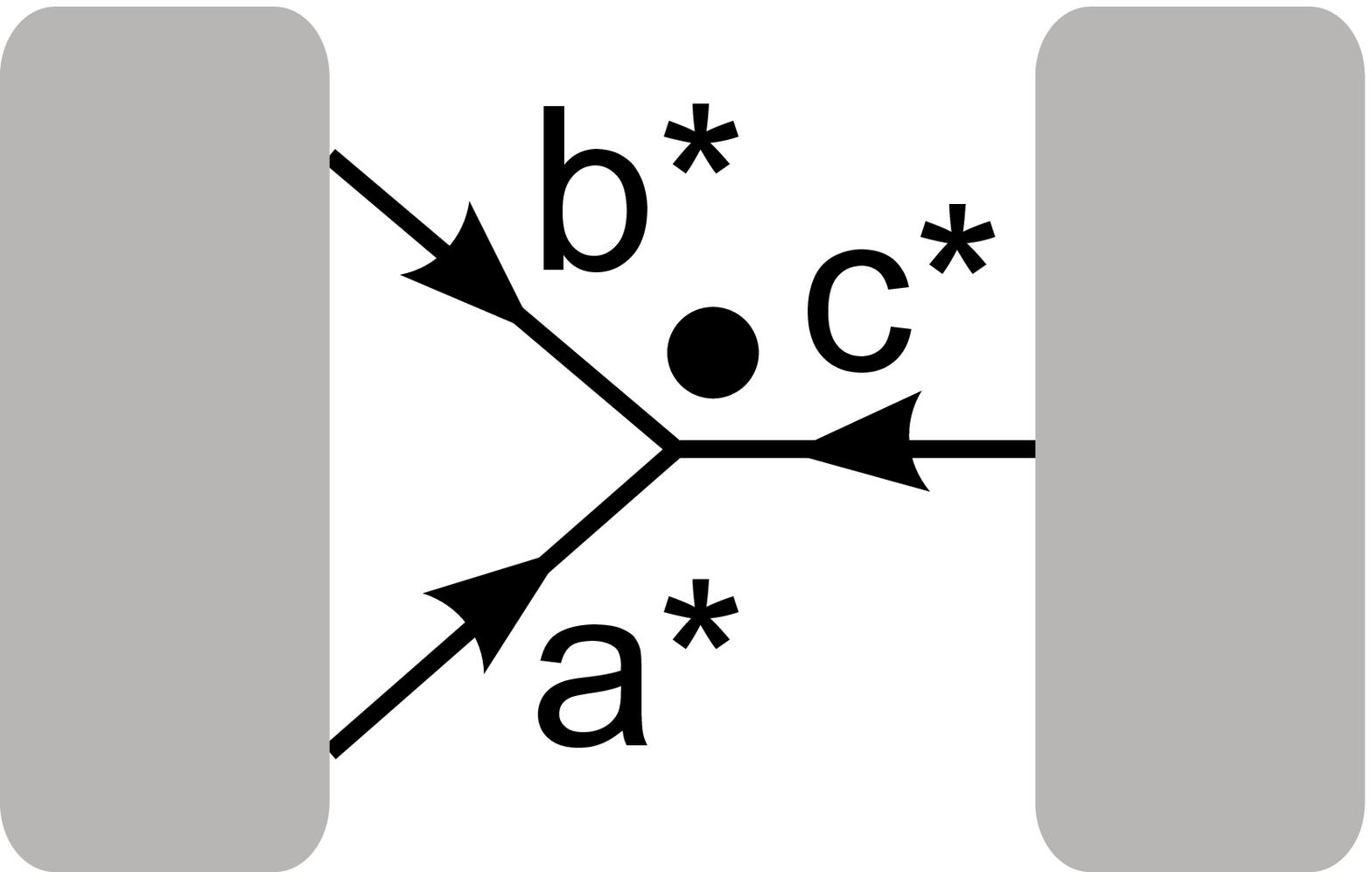}} \right|   \label{rule4'}
\end{eqnarray}
where $\alpha(a,b)$ is a complex number with unit modulus $|\alpha(a,b)| = 1$. Later we will see that $\alpha(a,b)$ can be chosen
to be a third root of unity without loss of generality. (We explain the motivation behind $\alpha$ in appendix \ref{dot}.)

A few comments are in order here. First, we would like to mention that the phases $\gamma_a$ and $\alpha(a,b)$ have an important mathematical meaning 
and are closely related to so-called $\mathbb{Z}_2$ and $\mathbb{Z}_3$ ``Frobenius-Schur indicators'' in tensor category theory\cite{KitaevHoneycomb,BondersonThesis}. One of the main differences
between the formalism in this paper and that of Ref. [\onlinecite{LevinWenStrnet}], is that here we include the phase factors $\gamma_a$, and $\alpha(a,b)$, while 
the construction in Ref. [\onlinecite{LevinWenStrnet}] effectively assumed that $\gamma_a = \alpha(a,b) = 1$. Indeed, Ref. [\onlinecite{LevinWenStrnet}] did not keep track of dots or ends of
null strings at all. Here, by allowing for more general $\gamma_a$ and $\alpha(a,b)$, we are able to construct string-net models and topological phases 
that were inaccessible to Ref. [\onlinecite{LevinWenStrnet}].

Second, we would like to mention that equations (\ref{normal} - \ref{dot2}) are not particularly fundamental and merely represent a 
particular choice of conventions for how to relate different vertices to one another. There are other equally good conventions where these 
rules would include additional phase factors.

Another important point has to do with string orientations. As we mentioned in section \ref{genstrnetsec}, if two string-net configurations differ only by reversing string orientations and replacing labels by $a \rightarrow a^*$, then those two string-net configurations correspond to the same physical state, \emph{up to a phase factor}. 
In Ref. [\onlinecite{LevinWenStrnet}], these phase factors were assumed to vanish. That is, in that work, it was assumed that the string orientations could be changed without introducing any
phases. Here, we allow for nontrivial phase factors, as we find that they are important in realizing more general topological phases. 
In our formalism, the phase factors associated with reversing string orientations are completely determined by the parameters $\gamma_a$ and $\alpha(a,b)$. For example,
we have:
\begin{align*}
\left\< \raisebox{-0.16in}{\includegraphics[height=0.4in]{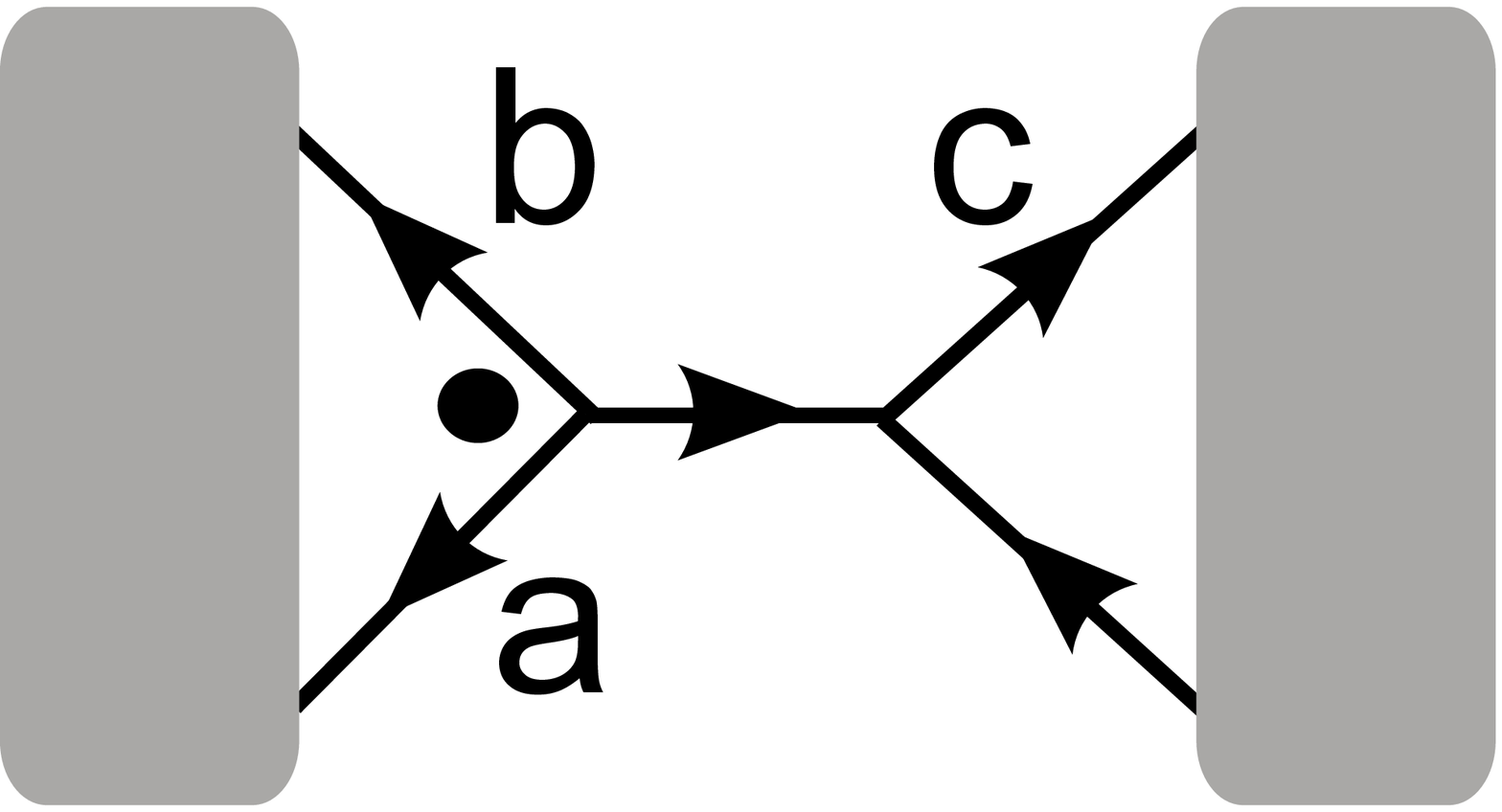}} \right|
&=\left\< \raisebox{-0.16in}{\includegraphics[height=0.4in]{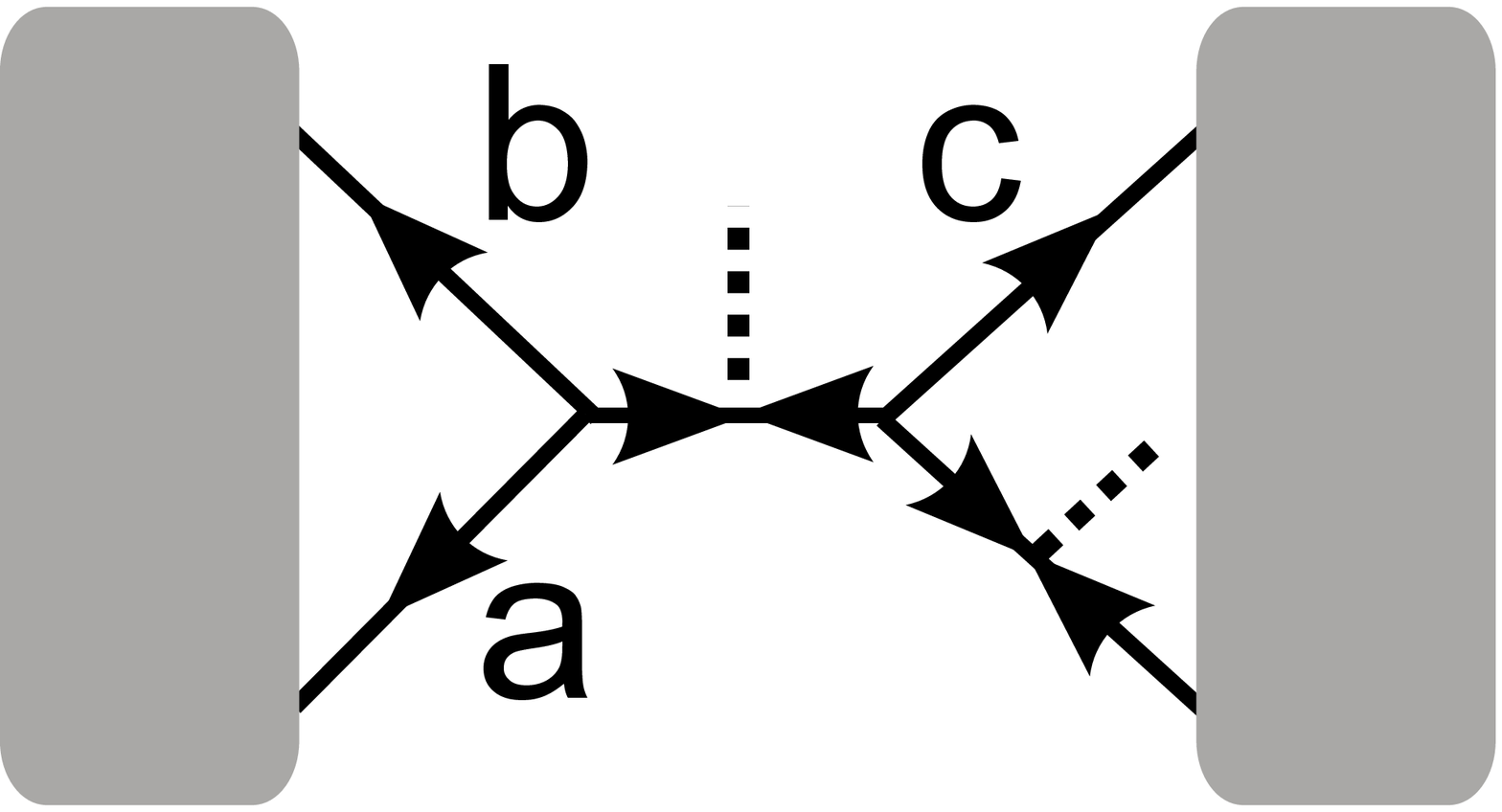}} \right|
=\gamma_{a+b}\left\< \raisebox{-0.16in}{\includegraphics[height=0.4in]{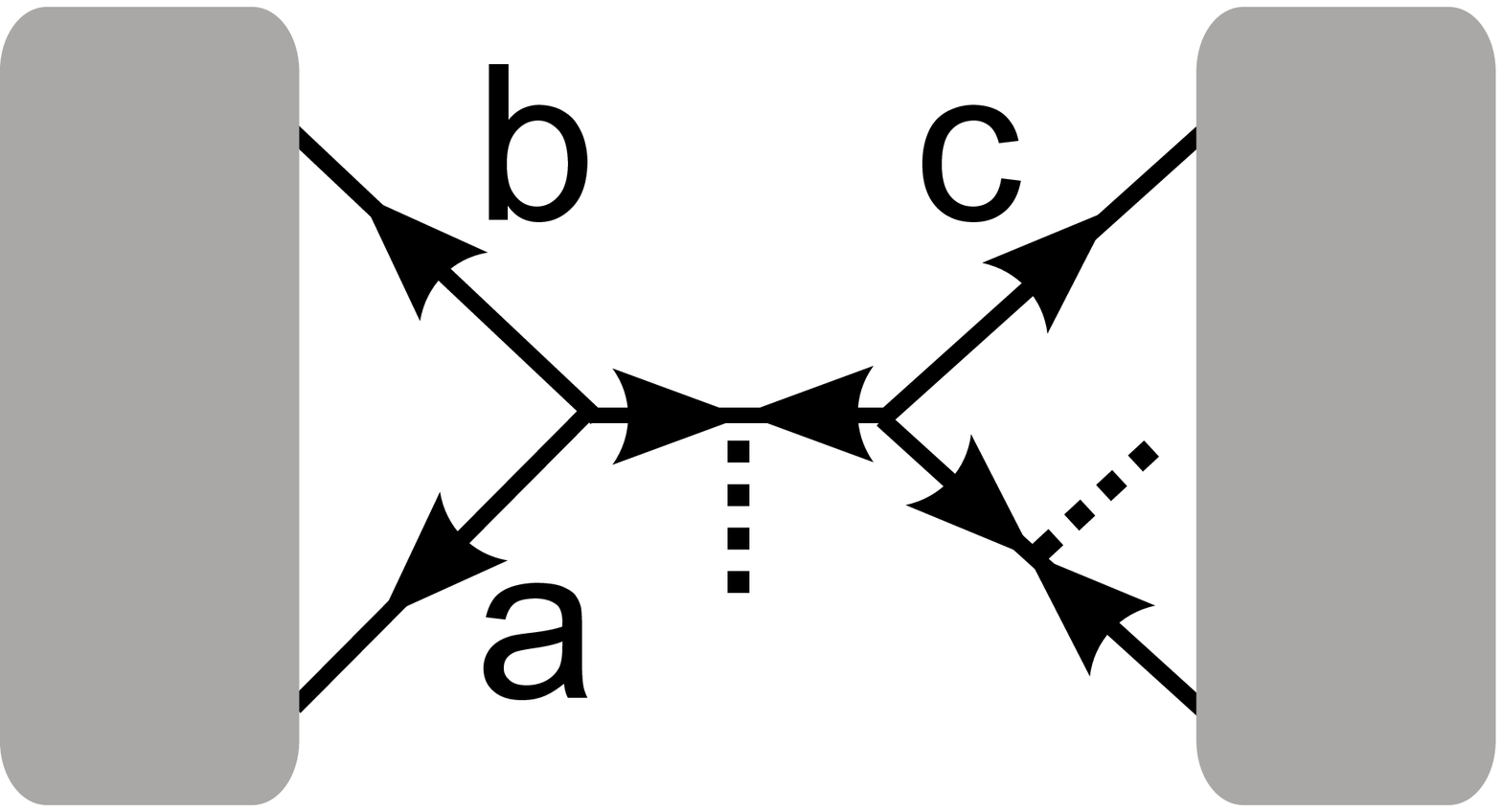}} \right| \nonumber \\
=&\gamma_{a+b}\left\< \raisebox{-0.16in}{\includegraphics[height=0.4in]{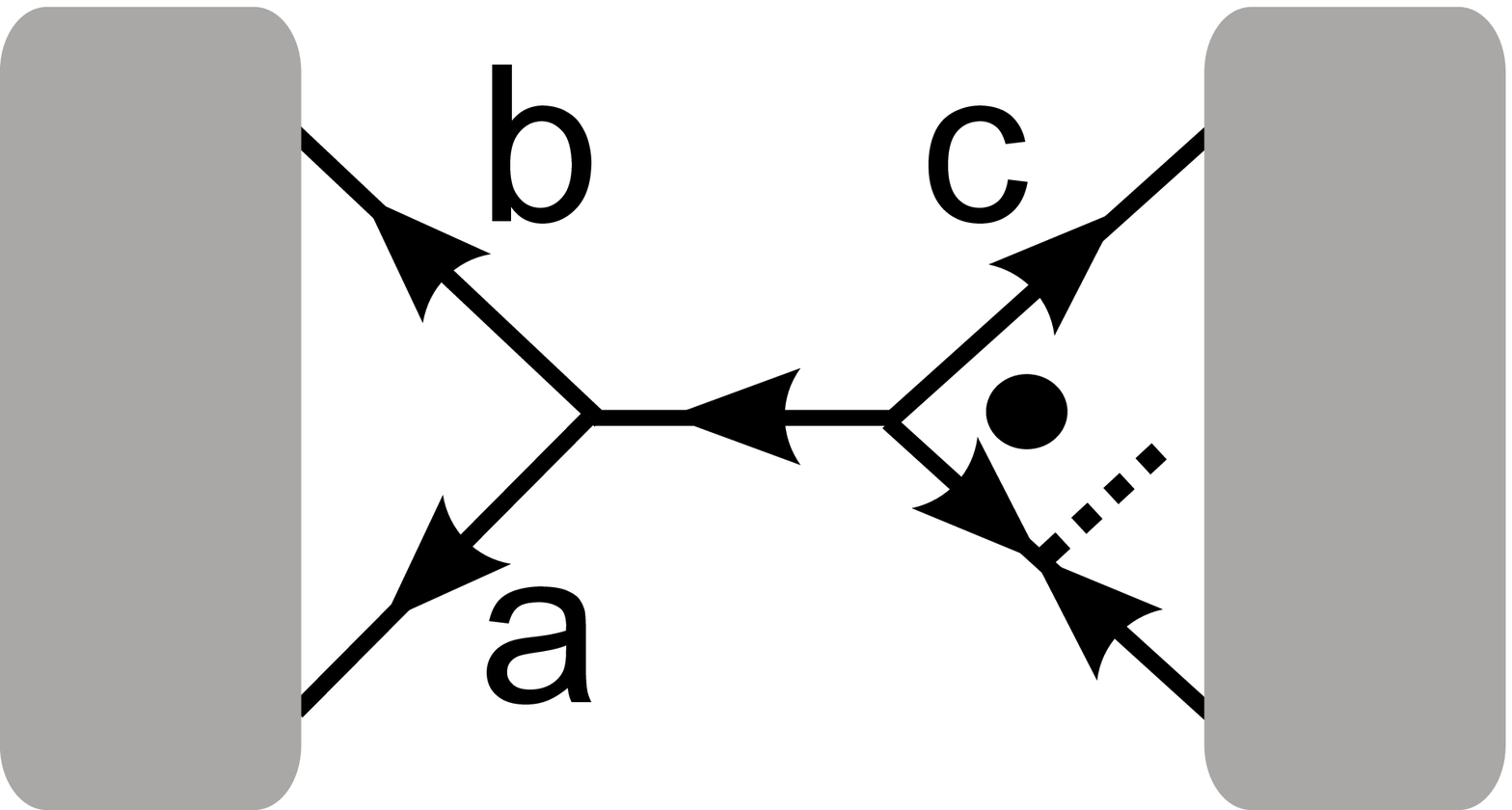}} \right| \nonumber \\
=&\gamma_{a+b}\alpha^{-1}(a+b,c)\left\< \raisebox{-0.16in}{\includegraphics[height=0.4in]{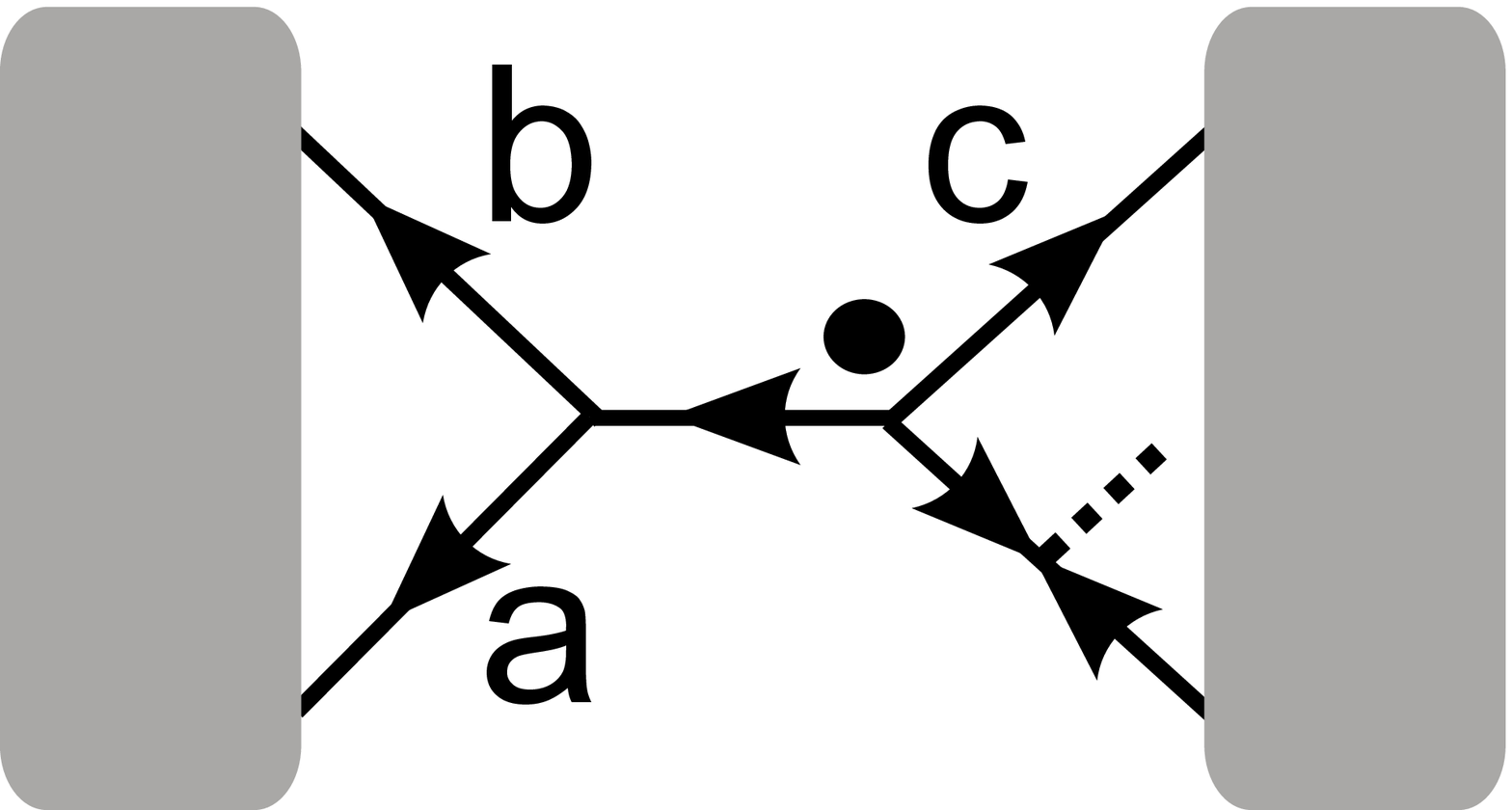}} \right| \nonumber\\
=&\gamma_{a+b}\alpha^{-1}(a+b,c)\gamma_{a+b+c}\left\< \raisebox{-0.16in}{\includegraphics[height=0.4in]{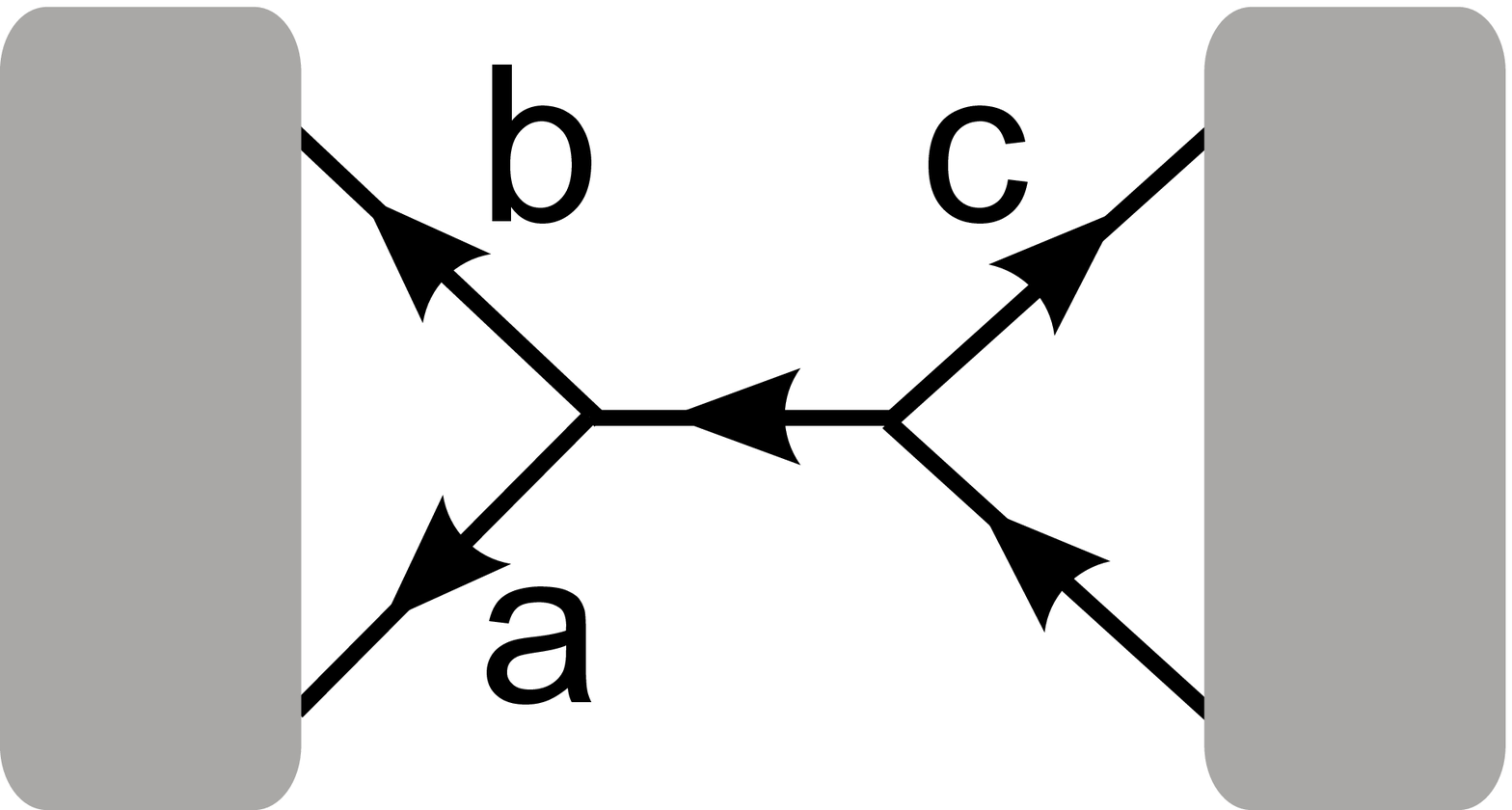}} \right|.
\end{align*}
These phase factors play an important role in our diagrammatical calculus, especially when using (\ref{rule3}). Indeed, in order to implement this rule, the 
string orientations have to match the orientations shown in (\ref{rule3}), and it is often necessary to reverse the orientations of certain strings to achieve 
this matching. In general, this orientation reversal can be accomplished using manipulations similar to those shown above.

\subsection{Example of computing a string-net amplitude}
We now present an example of how the local rules (\ref{rule1} - \ref{rule3}) and the conventions (\ref{nullerase} - \ref{rule4'})
determine the amplitude of general string-net configurations. 
Before discussing the example, we first point out two useful relations:
\begin{align}
	\left\<
	\raisebox{-0.16in}{\includegraphics[height=0.4in]{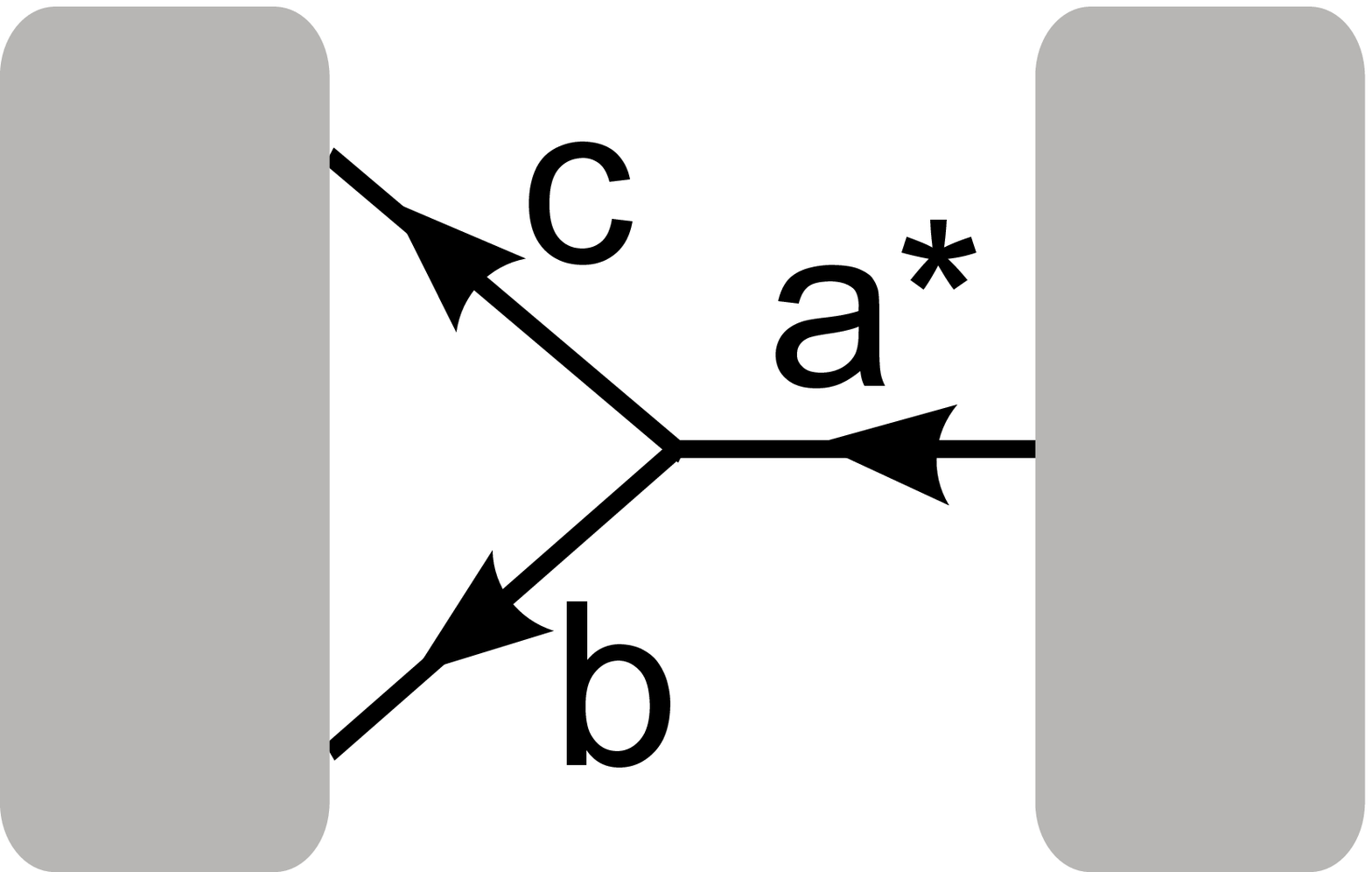}}
	\right|&=\alpha(b,c)
	\left\< 
	\raisebox{-0.16in}{\includegraphics[height=0.4in]{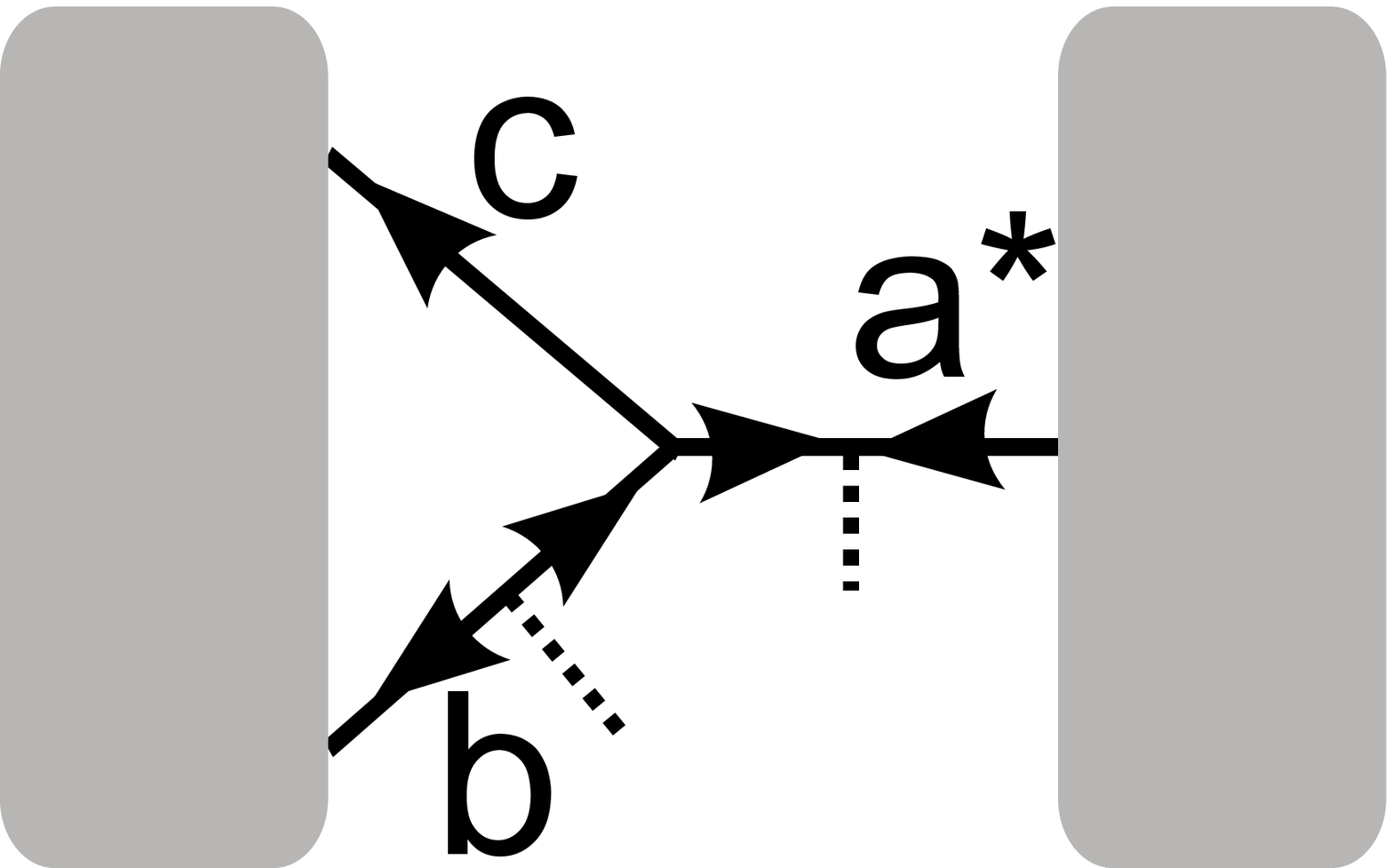}}
	\right| \label{alpha1},\\
	\left\<
	\raisebox{-0.16in}{\includegraphics[height=0.4in]{ex2a.eps}}
	\right|&=\alpha(a,b)^{-1}
	\left\<
	\raisebox{-0.16in}{\includegraphics[height=0.4in]{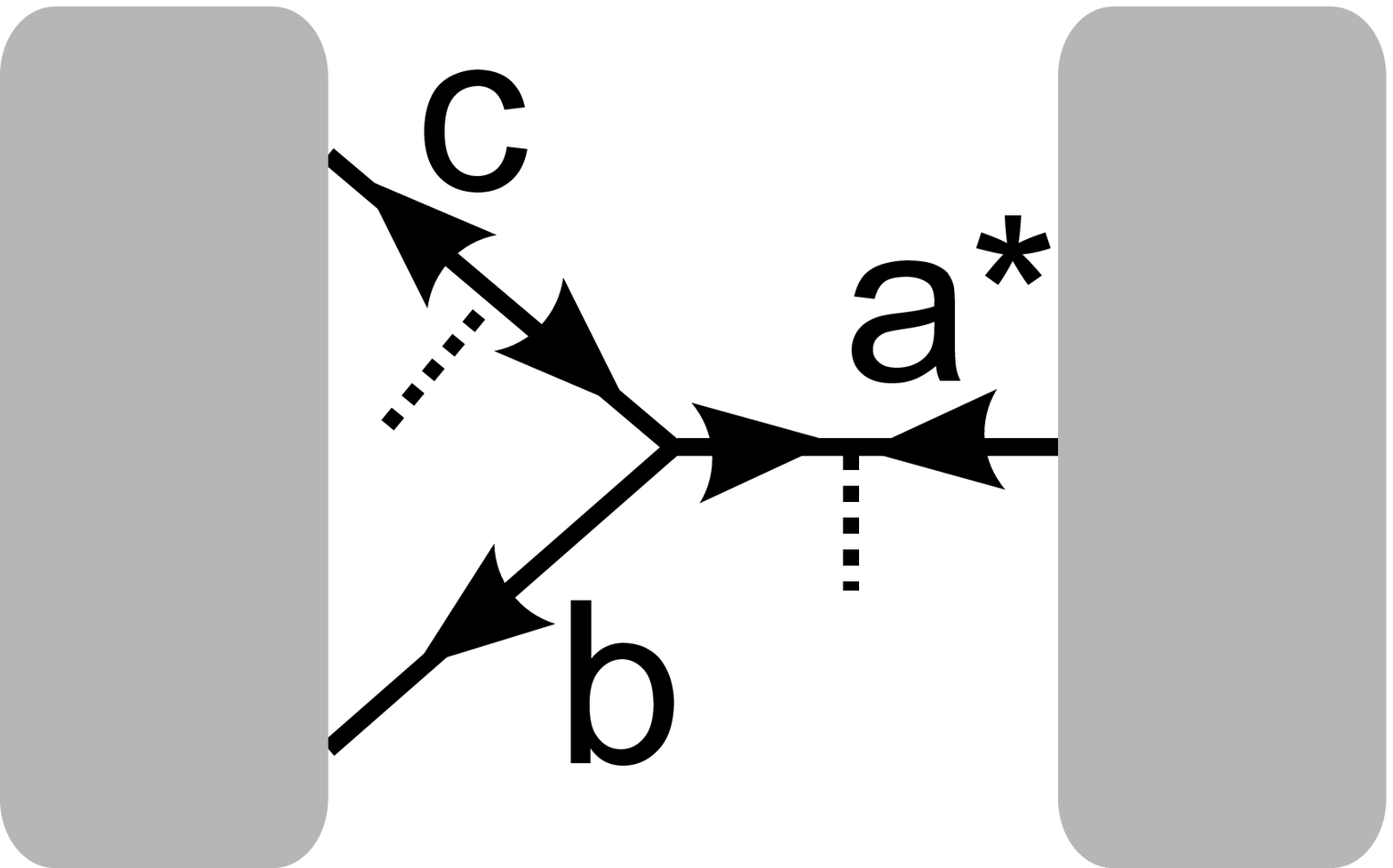}}
	\right|. \label{alpha2}
\end{align}
These relations allow us to change the orientations of vertices with one incoming string and two outgoing strings. Eqs. (\ref{alpha1} - \ref{alpha2}) are often useful when we need to reverse string orientations so that we can apply the local rule (\ref{rule3}). They can be shown by considering
\begin{align*}
	\left\<
	\raisebox{-0.16in}{\includegraphics[height=0.4in]{ex2a.eps}}
	\right|
	&=\left\<
	\raisebox{-0.16in}{\includegraphics[height=0.4in]{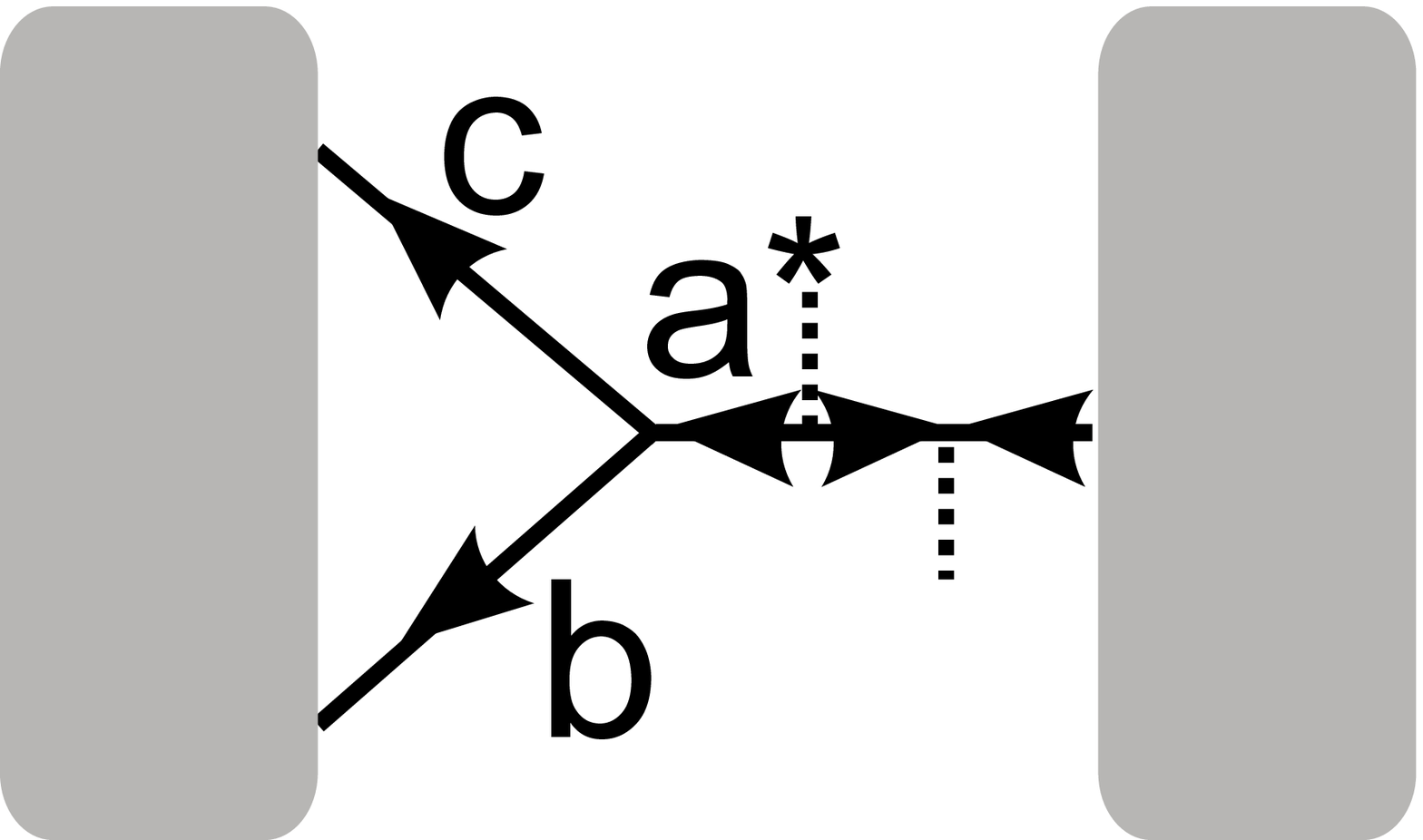}}
	\right|
	=\left\<
	\raisebox{-0.16in}{\includegraphics[height=0.4in]{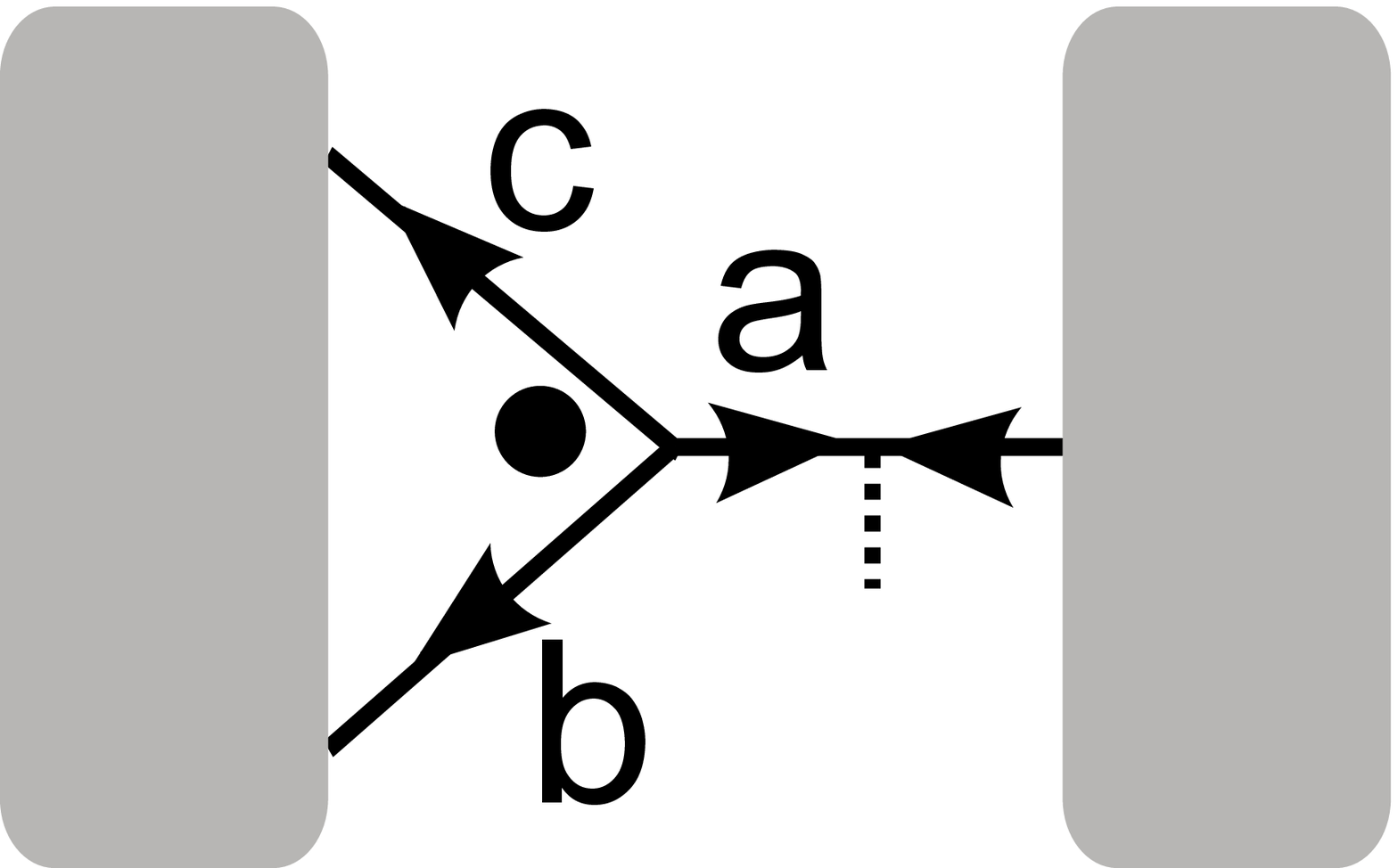}} 
	\right| \\
	&=\alpha(b,c)
	\left\<
	\raisebox{-0.16in}{\includegraphics[height=0.4in]{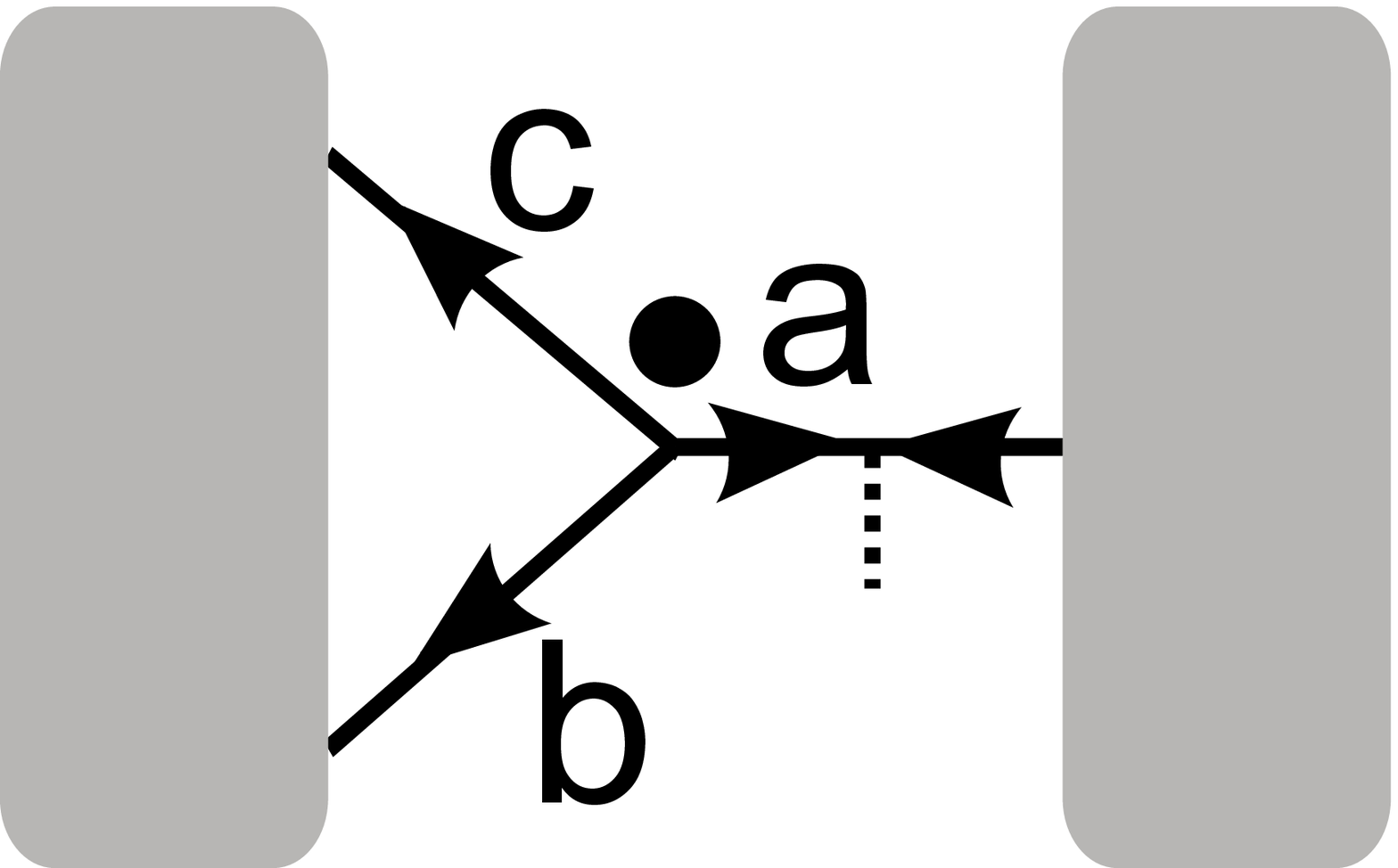}}
	\right| \\ 
	&=\alpha(b,c)
	\left\<
	\raisebox{-0.16in}{\includegraphics[height=0.4in]{ex2b.eps}}.
	\right|
\end{align*}
This shows (\ref{alpha1}). Similarly, by rotating the dot counterclockwise in the second step above, we can show (\ref{alpha2}).

Now let us consider the example:
\begin{align*}
\Phi\left( \raisebox{-0.2in}{\includegraphics[height=0.5in]{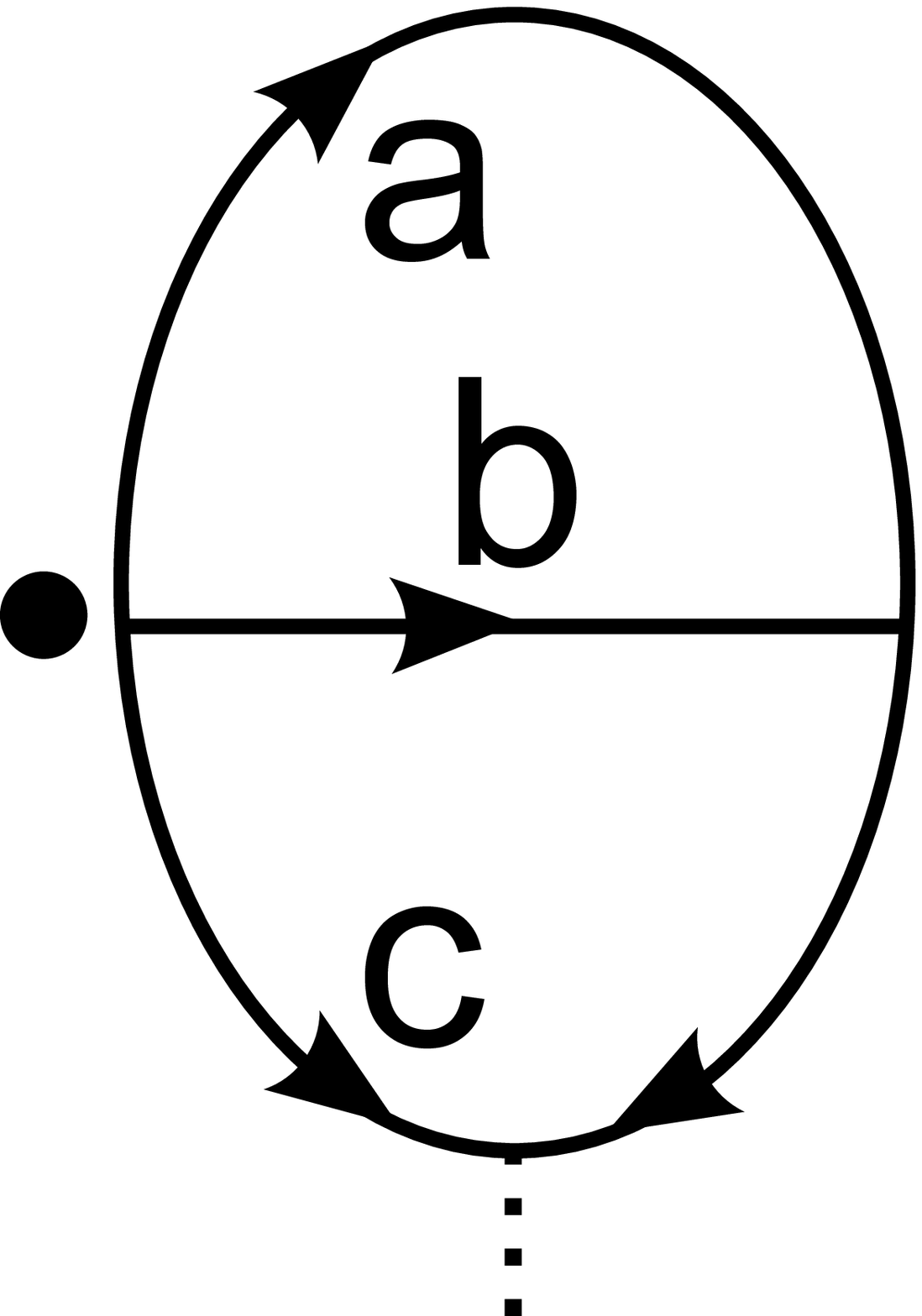}} \right)
&=\Phi\left( \raisebox{-0.2in}{\includegraphics[height=0.5in]{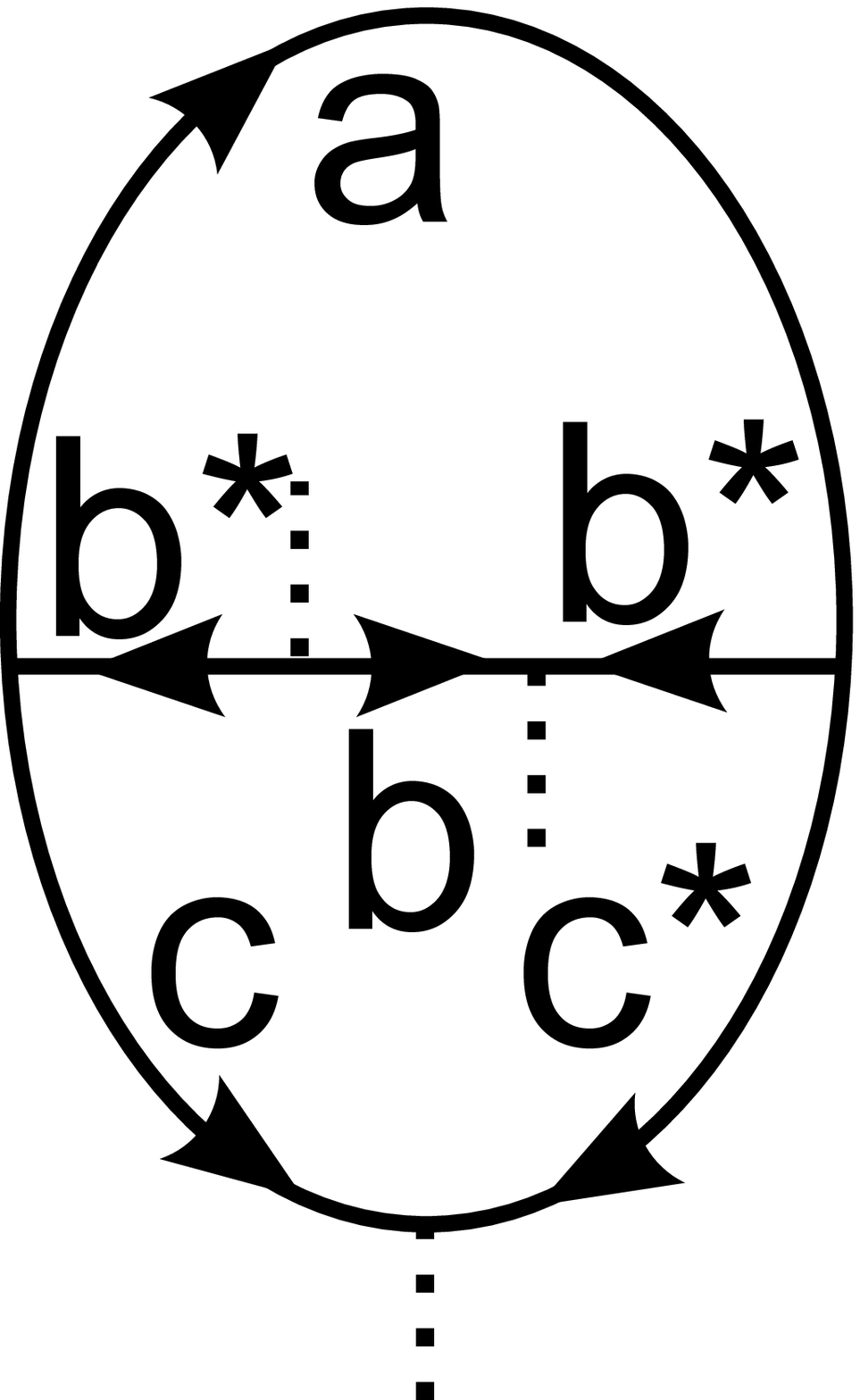}} \right)
=\alpha(c^{*},b^{*})\Phi\left( \raisebox{-0.2in}{\includegraphics[height=0.5in]{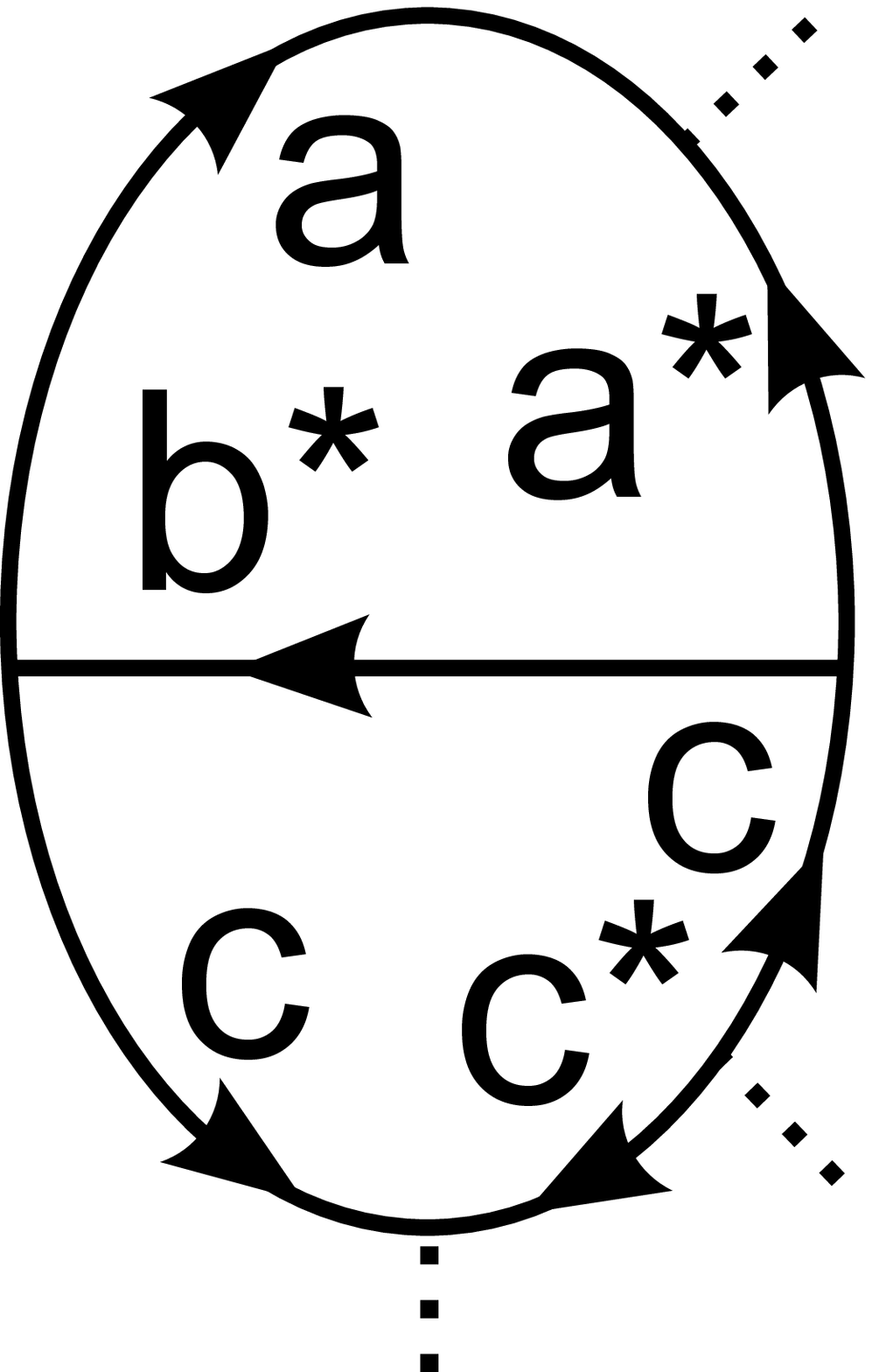}}\right) \\
&=\alpha(c^{*},b^{*})\gamma_{c}\Phi\left( \raisebox{-0.2in}{\includegraphics[height=0.5in]{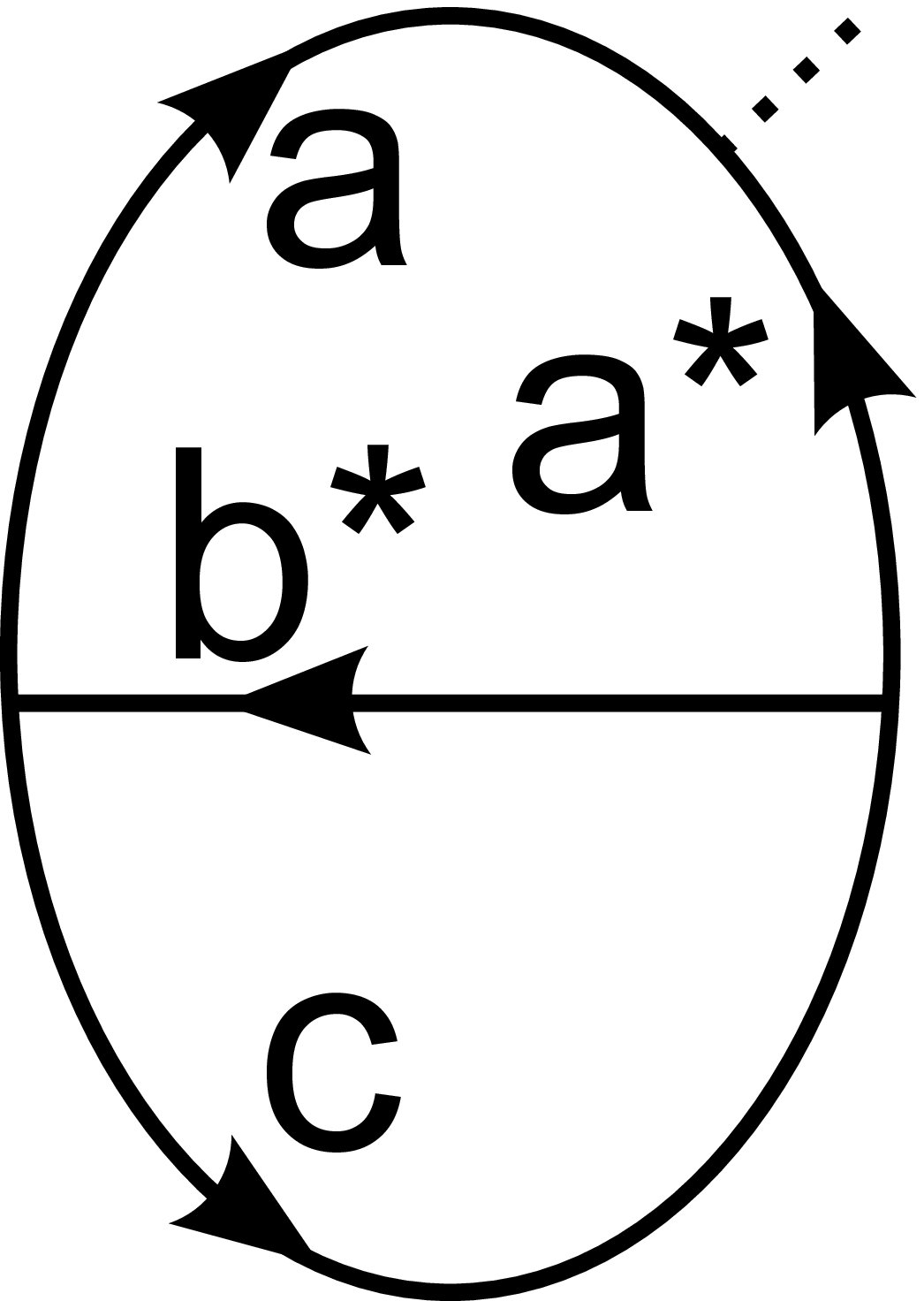}} \right) \\
&=\alpha(c^{*},b^{*})\gamma_{c}F(c,a,a^{*})\Phi\left( \raisebox{-0.2in}{\includegraphics[height=0.5in]{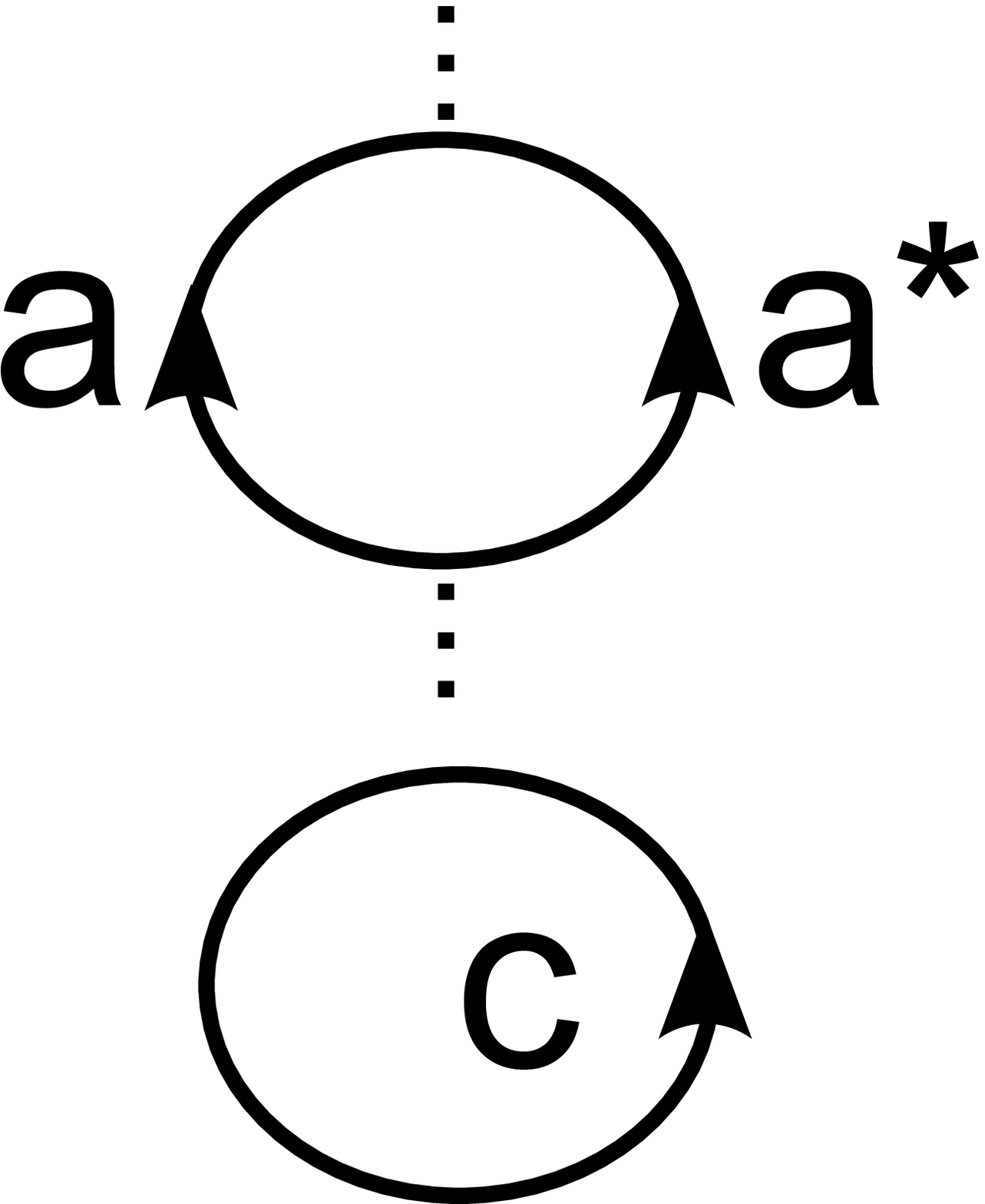}} \right) \\
&=\alpha(c^{*},b^{*})\gamma_{c}F(c,a,a^{*})\gamma_{a^{*}}d_{a}d_{c^{*}} \Phi(\text{vacuum}) \\
&= \alpha(c^{*},b^{*})\gamma_{c}F(c,a,a^{*})\gamma_{a^{*}}d_{a}d_{c^{*}}.
\end{align*}
In the first step we convert both of the vertices to ``basic'' vertices which have one incoming and two outgoing strings. We then use Eq. (\ref{alpha1}) in the second step.
Rules (\ref{rule5}), (\ref{rule3}) and (\ref{rule2}) are then applied in sequence. Finally we use the normalization convention (\ref{vacuum}). 

The above example is typical: in general, any string-net configuration can be reduced to the vacuum configuration by applying the above rules 
and conventions multiple times. In this way, these rules completely determine the wave function $\Phi$.

\subsection{Self-consistency conditions \label{scc}}
\begin{figure}[tb]
\begin{center}
\includegraphics[height=1.5in,width=3in]{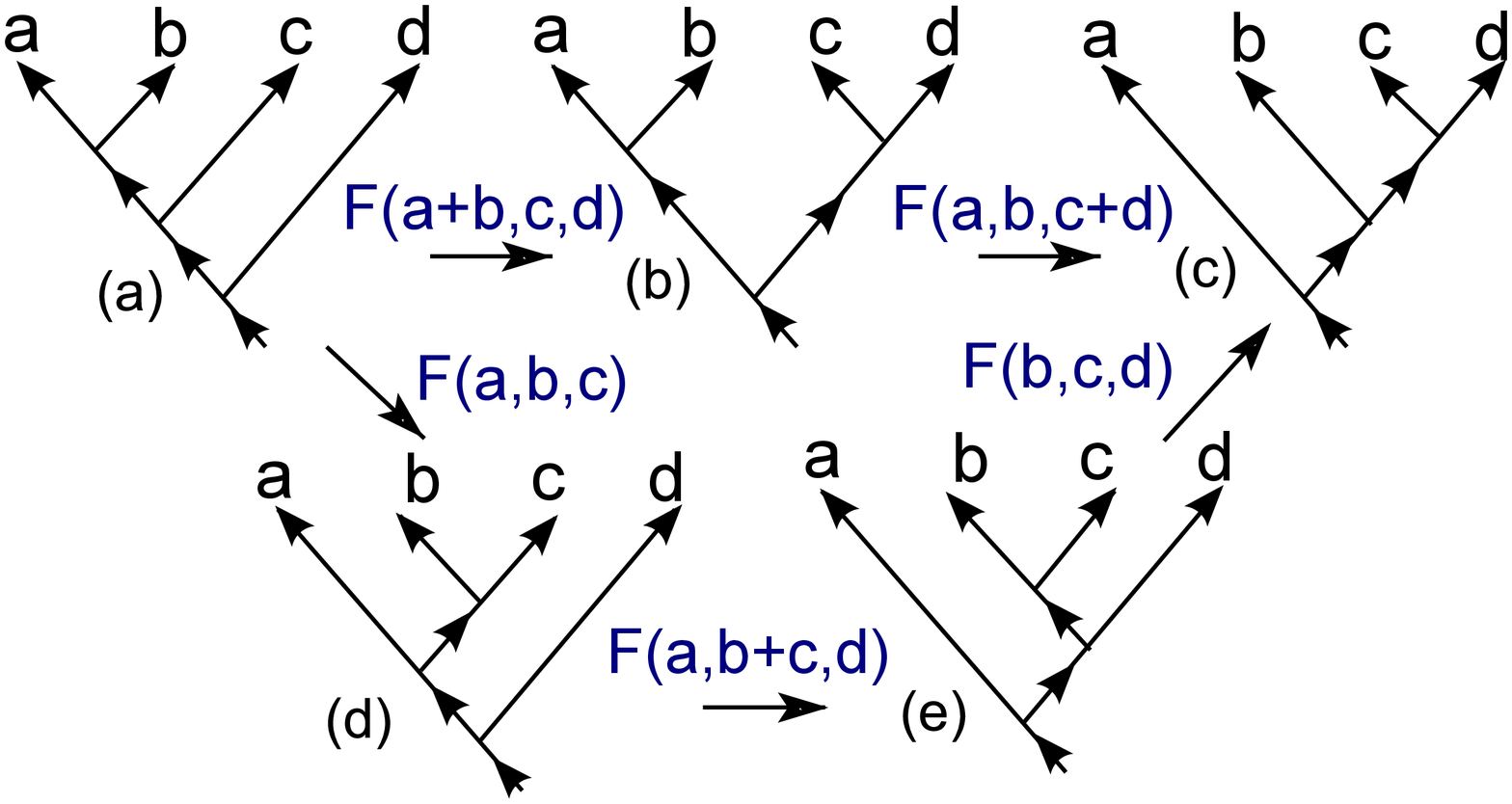}
\end{center}
\caption{
The amplitude of $(a)$ can be related to the amplitude $(c)$ in two different ways by the fusion rule
(\ref{rule3}). Self-consistency requires the two sequences of operation result in the same linear relations
between the amplitudes of $(a)$ and $(c)$.
        }
\label{sfeq2}
\end{figure}

We have seen that the rules (\ref{rule1} - \ref{rule3}) and the conventions (\ref{nullerase} - \ref{rule4'}) uniquely specify the wave function $\Phi$. 
Accordingly, the wave function $\Phi$ is completely determined once the parameters $\{F(a,b,c),d_a,\alpha(a,b), \gamma_a\}$ are given.
However, not every choice of parameters corresponds to a well-defined wave function. The reason is that for most choices of these parameters, the local rules/constraints are \emph{not self-consistent} -- that is, there are no wave functions $\Phi$ that satisfy them. 
In fact, only those $\{F,d,\gamma,\alpha \}$ that satisfy the following algebraic equations lead to self-consistent rules and a 
well-defined wave function $\Phi$:
\begin{subequations}
\label{selfconseq}
\begin{align} 
	F(a+b,c,d) &\cdot F(a,b,c+d) = 			\label{pentid} \\ 
	&F(a,b,c)  \cdot F(a,b+c,d) \cdot F(b,c,d),    \nonumber \\
	F(a,b,c) &= 1 \text{ if }a\text{ or }b\text{ or }c=0, \label{F0} \\
	d_ad_b &= d_{a+b} \label{dcons}, \\
	\gamma_{a} &= F(a^{*},a,a^{*})d_{a} \label{gammaF}, \\
	\alpha(a,b) &= F(a,b,(a+b)^{\ast}) \gamma_{a+b}.	    \label{alphaF}  
\end{align}
\end{subequations}
We now explain why these conditions are necessary for self-consistency; we show that they are sufficient in appendix \ref{consistent}.
We begin with the first equation (\ref{pentid}). The origin of this condition can be understood by considering the sequence 
of manipulations shown in Fig. \ref{sfeq2}. We can see that the amplitudes of the string-net configurations (a) and (c) can be related to one another in two different ways: 
$(a) \rightarrow (b) \rightarrow (c)$ and $(a) \rightarrow (d) \rightarrow (e) \rightarrow (c)$. 
In order for these two relations to be consistent with one another, $F$ must satisfy equation (\ref{pentid}), known as the ``pentagon identity.''
The other conditions can be derived from similar consistency requirements (see appendix \ref{consistent}).

In addition to equations (\ref{selfconseq}), we will need to impose one more constraint on $\{F,d,\gamma\,\alpha \}$ in order to construct a consistent string-net model:
\begin{equation}
|F(a,b,c)| = 1.
\label{unit}
\end{equation}
This constraint has a different origin from equations (\ref{selfconseq}): it is not necessary for constructing a well-defined wave function $\Phi$,
but rather for constructing an exactly soluble Hamiltonian with $\Phi$ as its ground state. More specifically, we will see that (\ref{unit}) is important
in ensuring that our exactly soluble Hamiltonians are \emph{Hermitian}. (see appendix \ref{property}) 

\subsection{Gauge transformations \label{gaugesec}}
In general, it is not so easy to find solutions to the conditions (\ref{selfconseq}) and (\ref{unit}). However, once we have one solution
$\{F, d, \gamma, \alpha\}$, we can construct an infinite class of other solutions $\{\tilde{F}, \tilde{d}, \tilde{\gamma}, \tilde{\alpha}\}$ 
by defining
\begin{eqnarray} 
\tilde{F}(a,b,c) &=&F(a,b,c) \cdot
\frac{f(a,b+c) f(b,c)}{f(a,b) f(a+b,c)}, \label{gauge} \\
\tilde{d}_a &=& d_a, \nonumber \\
\tilde{\gamma}_a &=& \gamma_a \cdot \frac{f(a,a^*)}{f(a^*,a)}, \nonumber \\
\tilde{\alpha}(a,b) &=& \alpha(a,b) \cdot \frac{f(b,(a+b)^*) f(a,a^*)}{f(a,b) f((a+b)^*,a+b)}. \nonumber 
\end{eqnarray}
Here $f(a,b)$ is any complex function with
\begin{align*}
|f(a,b)| &= 1 \ \ , \ \ f(a,b) =1 \ \ \ \text{ if }a\text{ or }b=0. 
\end{align*}

Similarly, we can construct solutions by defining
\begin{eqnarray}
\tilde{F}\left( a,b,c\right) &=&
F \left( a,b,c\right), \label{gauge2} \\
\tilde{d}_a &=& d_a \cdot g(a), \nonumber \\
\tilde{\gamma}_a &=& \gamma_a \cdot g(a), \nonumber \\
\tilde{\alpha}(a,b) &=& \alpha(a,b) \cdot g(a+b) \nonumber
\end{eqnarray}
where $g(a)$ is any complex function with
\begin{align}
|g(a)| &= 1 \ \ , \ \ g(a+b) = g(a) \cdot g(b). 
\label{gi}
\end{align}
We will refer to (\ref{gauge}),(\ref{gauge2}) as ``gauge transformations'' and we will say that 
$\{F, d, \gamma, \alpha\}$ and $\{\tilde{F}, \tilde{d}, \tilde{\gamma}, \tilde{\alpha}\}$ are ``gauge equivalent'' if they differ by
such a transformation. 
As the name suggests, gauge equivalent solutions are closely related to one another. In fact, it is possible to show that
if $\{F, d, \gamma, \alpha\}$ and $\{\tilde{F}, \tilde{d}, \tilde{\gamma}, \tilde{\alpha}\}$ are gauge equivalent solutions 
to (\ref{selfconseq}), (\ref{unit}), then the corresponding wave functions $\Phi, \tilde{\Phi}$ can be transformed into one another by a \emph{local unitary transformation} (See appendix \ref{gaugeapp}). 
Here, by a local unitary transformation we mean a unitary transformation that can be generated by the time evolution of a local Hamiltonian over a finite 
period of time. The existence of this local unitary transformation has an important physical 
meaning: it implies that $\Phi$ and $\tilde{\Phi}$ belong to the same quantum phase. \cite{ChenGuWen10,GuWangWen10} Therefore, if we are primarily interested in constructing 
different topological phases, then we only need to consider one solution to (\ref{selfconseq},\ref{unit}) within each gauge equivalence class.

This freedom to make gauge transformations can be quite useful. For example, using the first gauge transformation (\ref{gauge}), we can always 
transform $\gamma$ so that 
\begin{align}
\gamma_a = \begin{cases}
			\pm 1, & \text{if $a = a^*$} \\
			 1, & \text{otherwise}
	    \end{cases}. 
\end{align}
The second gauge transformation, parameterized by $g(a)$ is also quite useful. As we show in section \ref{examples}, this 
transformation allows us, in many cases, to transform $\gamma, \alpha$ so that $\gamma_a = \alpha(a,b) = 1$.

Another gauge transformation which can simplify our models can be obtained by generalizing the
conventions (\ref{dot1}) and (\ref{dot2}) to
\begin{eqnarray}
\left\< \raisebox{-0.18in}{\includegraphics[height=0.45in]{vertex3a.eps}} \right|
&=&
\beta(a,b) \left\< \raisebox{-0.18in}{\includegraphics[height=0.45in]{vertex3.eps}} \right|,   \nonumber \\
\left\< \raisebox{-0.18in}{\includegraphics[height=0.45in]{vertex4a.eps}} \right|
&=&
\beta(a,b) \left\< \raisebox{-0.18in}{\includegraphics[height=0.45in]{vertex4.eps}} \right| \label{dot2'}
\end{eqnarray}
where $|\beta(a,b)| =1 $ are complex numbers with modulus $1$. With this modification, the self-consistency condition
(\ref{alphaF}) becomes
\begin{equation}
\alpha(a,b) = \frac{\beta(b,c)}{\beta(a,b)} F(a,b,(a+b)^{\ast}) \gamma_{a+b}.
\end{equation}
It is not hard to show that by choosing $\beta$ appropriately, we can always transform $\alpha$ so that 
\begin{align}
\alpha(a,b) &= \begin{cases}
			1 \text{ or } \omega \text{ or } \omega^2, & \text{if $a = b = (a+b)^*$ } \\
			1, & \text{otherwise}
	    \end{cases}
\end{align}
where $\omega$ is a 3rd root of unity.

\section{String-net Hamiltonians} \label{fixedh}
In this section, we will construct a large class of exactly soluble lattice Hamiltonians that have the wave functions $\Phi$ as their ground states.
The basic input for our construction is a finite abelian group $G$ and a solution $\{F(a,b,c),d_a,\gamma_a,\alpha(a,b)\}$ 
to the self-consistency conditions, (\ref{selfconseq},\ref{unit}). Given this input, we will construct an exactly soluble Hamiltonian whose
ground state $|\Phi_{latt}\>$ obeys the local rules (\ref{rule1} - \ref{rule3}) and (\ref{nullerase} - \ref{rule4'}) on the lattice.
 
\subsection{Definition of the Hamiltonian}
Let us first specify the Hilbert space for our model. The model is a generalized spin system, where the spins are located
on the links of the 2D honeycomb lattice. Each spin can be $|G|$ different states which are labeled by elements of the group: 
\{$|a\> : a \in G\}$. When a spin is in state $|a\>$, we regard the link as being occupied by a string of type-$a$, oriented in certain direction.
If the spin is in state $|0\>$, we think of the link as being occupied by the null string.

\begin{figure}
\begin{center}
\includegraphics[height=2in,width=2.5in]{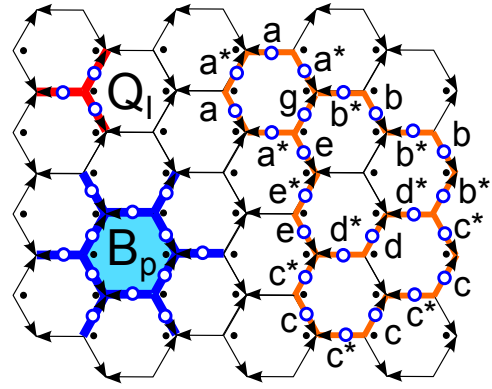}
\end{center}
\caption{Lattice spin model (\ref{h}). 
The $Q_I$ operator acts on three spins on the links connected to vertex $I$,
while the $B_{p}$ operator acts on 12 spins on the links adjacent to the hexagonal plaquette $p$. 
The $Q_{I}$ term constrains the string-nets to satisfy the branching rules while the $B_{p}$ term provides dynamics for the string-nets.
On the right is the typical ground state configuration.
The $a=0$ spin state corresponds to the vacuum or null string.
}
\label{lattice}
\end{figure} 

The Hamiltonian for our model is of the form
\begin{equation} 
H=-\sum_{I}Q_{I}-\sum_{p}B_{p}.
\label{h}
\end{equation}
Here, the two sums run over the sites $I$ and plaquettes $p$ of the honeycomb lattice. The operator $Q_{I}$ acts on the $3$ spins adjacent to the 
site $I:$
\begin{equation}
Q_{I}
\left | \raisebox{-0.1in}{\includegraphics[height=0.4in]{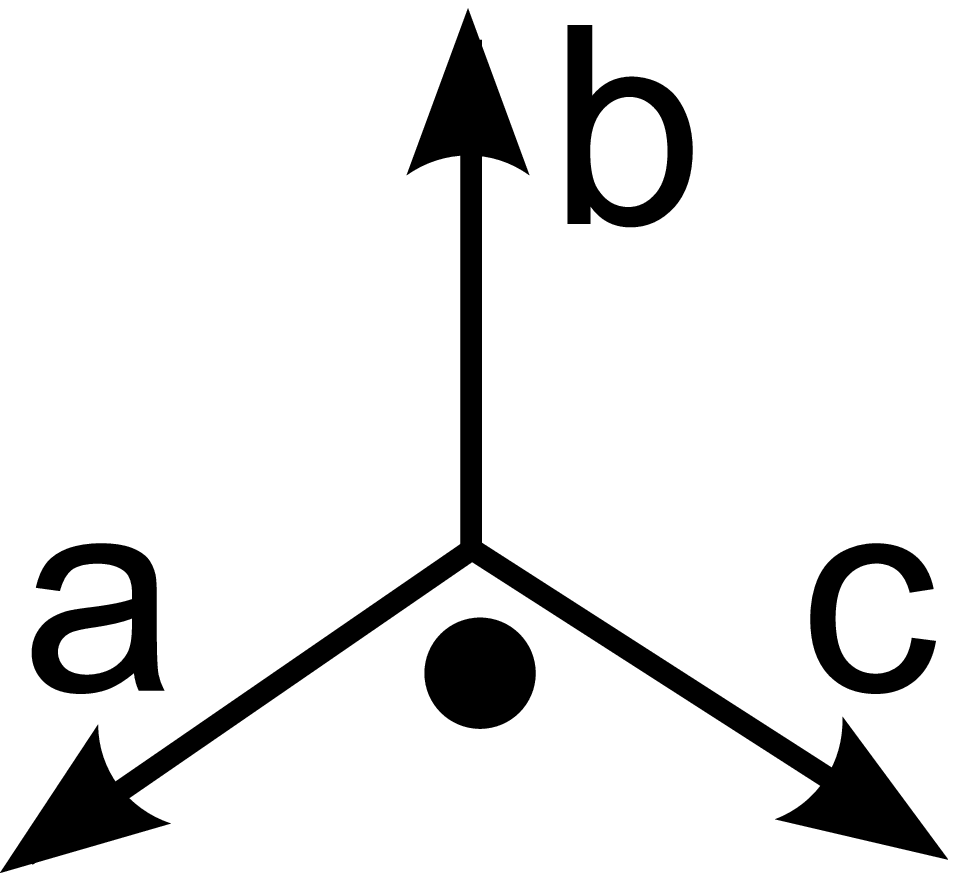}} \right\> 
=\delta_{abc}
\left | \raisebox{-0.1in}{\includegraphics[height=0.4in]{QI.eps}}\right\> 
\label{Q}
\end{equation}
where 
\begin{equation}
\delta_{abc}= \begin{cases}
                     1, & \text{if $a+b+c = 0$} \\
                     0, & \text{otherwise}
            \end{cases} 
\end{equation}
(See Fig. \ref{lattice}).
We can see that the $Q_I$ term penalizes states that don't satisfy the branching rules.

The operator $B_p$ provides dynamics for the string-net configurations and makes them condense. The definition of this operator is more complicated. It can be written as a linear combination
\begin{equation}
B_{p}=\sum_{s \in G} a_{s}B_{p}^{s}
\end{equation}
where $B_{p}^{s}$ describes a $12$ spin interaction 
involving the spins on the $12$ links that are adjacent to the vertices of the hexagon $p$ (See Fig. \ref{lattice}) and 
where $a_s$ are some complex coefficients satisfying $a_{s^{\ast }}=a_{s}^{\ast }$. 
The operator $B_p^s$ has a special structure, which we now describe. 
First, it annihilates any state that does not obey the branching rules at the $6$ vertices surrounding the plaquette. 
Second, while it acts non-trivially on the inner $6$ spins along the boundary of $p$, it does not affect the outer $6$ spins at all. The outer spins are still important, however, because the matrix element of $B_p^s$ between two inner spin configurations,
$\<g,h,i,j,k,l|$ and $|g',h',i',j',k',l'\>$ depends on the state of the outer spins, $(a,b,c,d,e,f)$. 
These matrix elements are defined by
\[
\left\< 
\raisebox{-0.26in}{\includegraphics[height=0.6in]{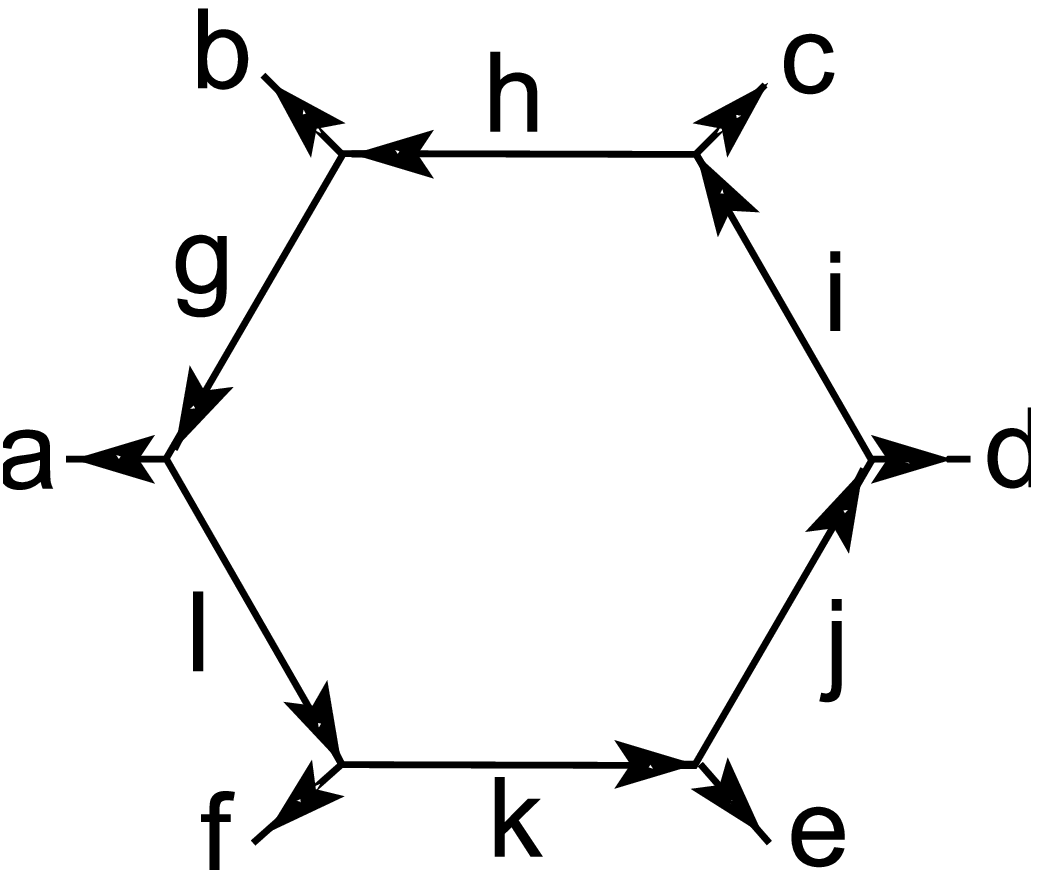}} 
\right| B_{p}^{s} \left| \raisebox{-0.26in}{\includegraphics[height=0.6in]{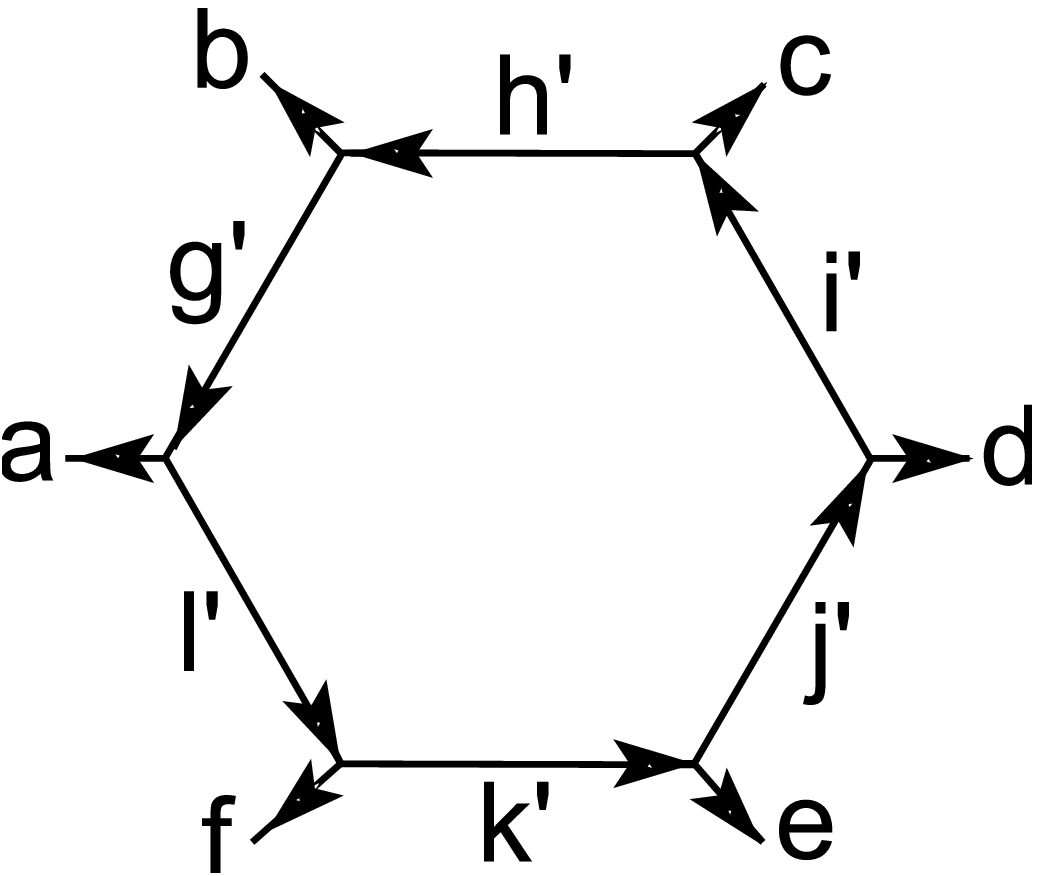}}  \right\>
=B_{p,g'h'i'j'k'l'}^{s,ghijkl}(abcdef) 
\]
where 
\begin{flalign}
	&B_{p,g'h'i'j'k'l'}^{s,ghijkl}(abcdef) = \delta_{g'}^{g+s} \delta_{h'}^{h+s} \delta_{i'}^{i+s} \delta_{j'}^{j+s} \delta_{k'}^{k+s} \delta_{l'}^{l+s} \notag \\
	& \cdot F_{s^*g'b} \cdot F_{s^*h'c} \cdot F_{s^*i'd} \cdot F_{s^*j'e} \cdot F_{s^*k'f} \cdot F_{s^*l'a}.  
\label{b}
\end{flalign} 
and $F_{abc} \equiv F(a,b,c)$. 
Note that the above expression is only valid if the initial and final states obey the branching rules, i.e. $h=b+g$, $h'=b+g'$, etc. If either state doesn't obey the branching rules, the matrix element of $B_p^s$ vanishes.

An important point is that the above matrix elements are calculated for a particular orientation configuration in which the inner links are oriented cyclically. This choice of orientations leads to simple matrix elements, but unfortunately there is no way to extend this orientation configuration to the whole honeycomb lattice while preserving translational symmetry. If we instead choose the translationally invariant orientation configuration shown in Fig. \ref{lattice}, the matrix elements are modified as 
\begin{align*}
\left\< 
\raisebox{-0.26in}{\includegraphics[height=0.6in]{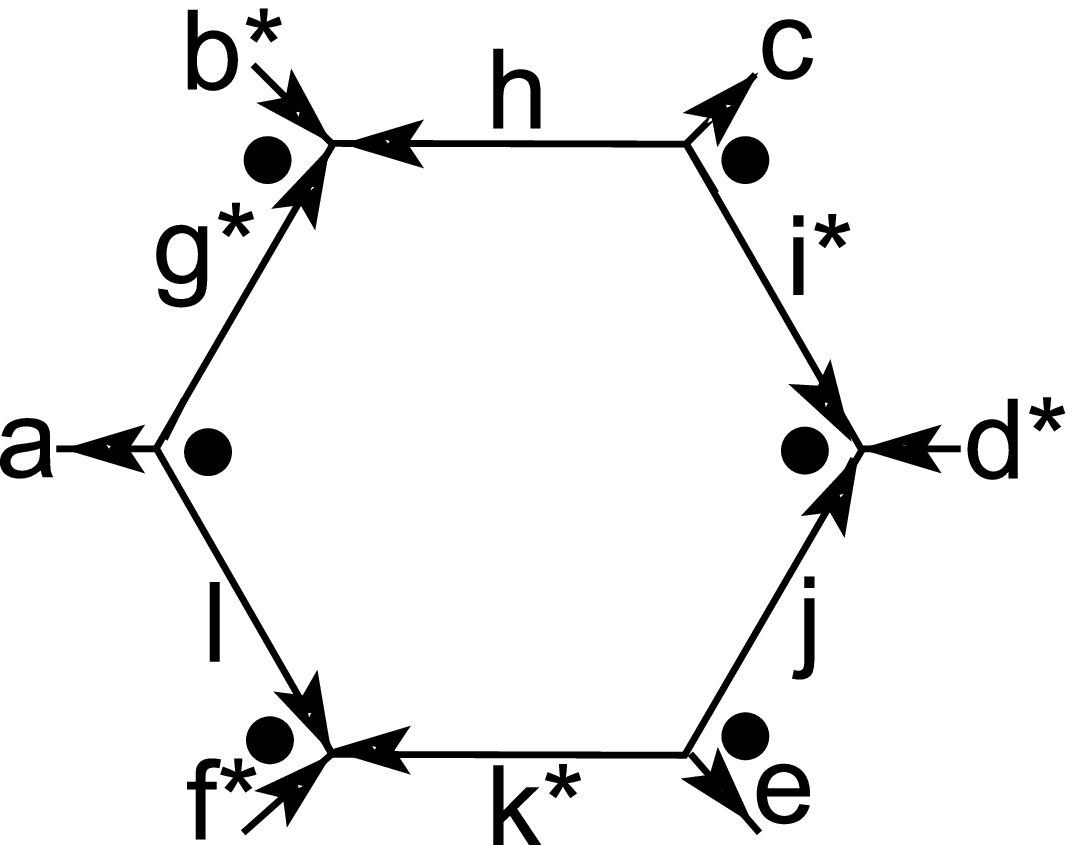}} 
\right| B_{p}^{s} \left | \raisebox{-0.26in}{\includegraphics[height=0.6in]{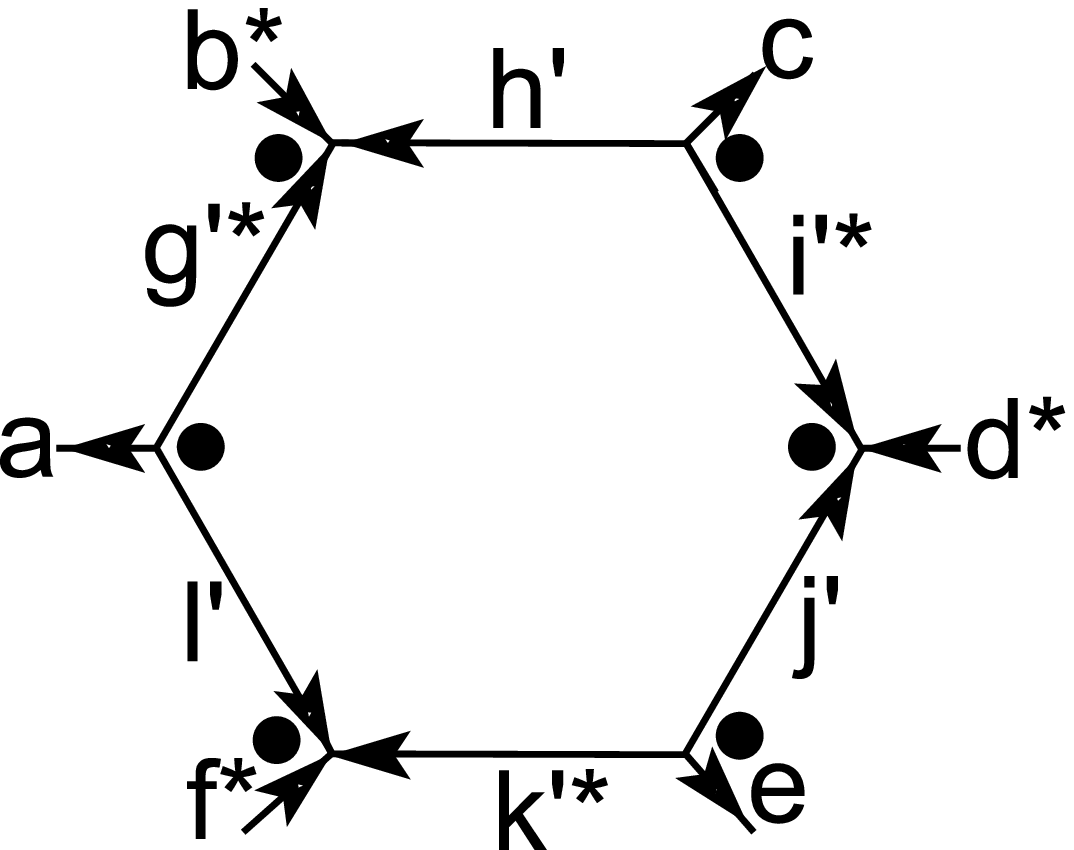}} \right\>
&= \nonumber \\
\mathbf{B}_{p,g'h'i'j'k'l'}^{s,ghijkl}&(abcdef) 
\end{align*}
with
\begin{align}
\mathbf{B}_{p,g'h'i'j'k'l'}^{s,ghijkl}(abcdef) 
&= B_{p,g'h'i'j'k'l'}^{s,ghijkl}(abcdef) & \nonumber \\
&\cdot \frac{\alpha_{g^{*}l}\alpha_{h'c}\alpha_{j^{*}i}\alpha_{k'f}\gamma_{g}\gamma_{i}\gamma_{k}}
	{\alpha_{g'^{*}l'}\alpha_{hc}\alpha_{j'^{*}i'}\alpha_{kf}\gamma_{g'}\gamma_{i'}\gamma_{k'}}.
\label{bgenorient}
\end{align}
The additional factors come from reversing the orientations on the $g,i,k$ links.

Although the algebraic definition of $B_p^s$ is complicated, there is an alternative graphical representation for this operator which is
much simpler. It is convenient to describe this graphical representation in terms of the action of $B_p^s$ on a bra $\<X|$ rather than 
describing its action on a ket $|X\>$. In the graphical representation, the action of $B_p^s$ 
can be understood as adding a loop of the type-$s$ string around the boundary of $p$: 
\begin{equation}
\left\< 
\raisebox{-0.26in}{\includegraphics[height=0.6in]{B1.eps}} 
\right| B_{p}^{s} =\left\< 
\raisebox{-0.26in}{\includegraphics[height=0.6in]{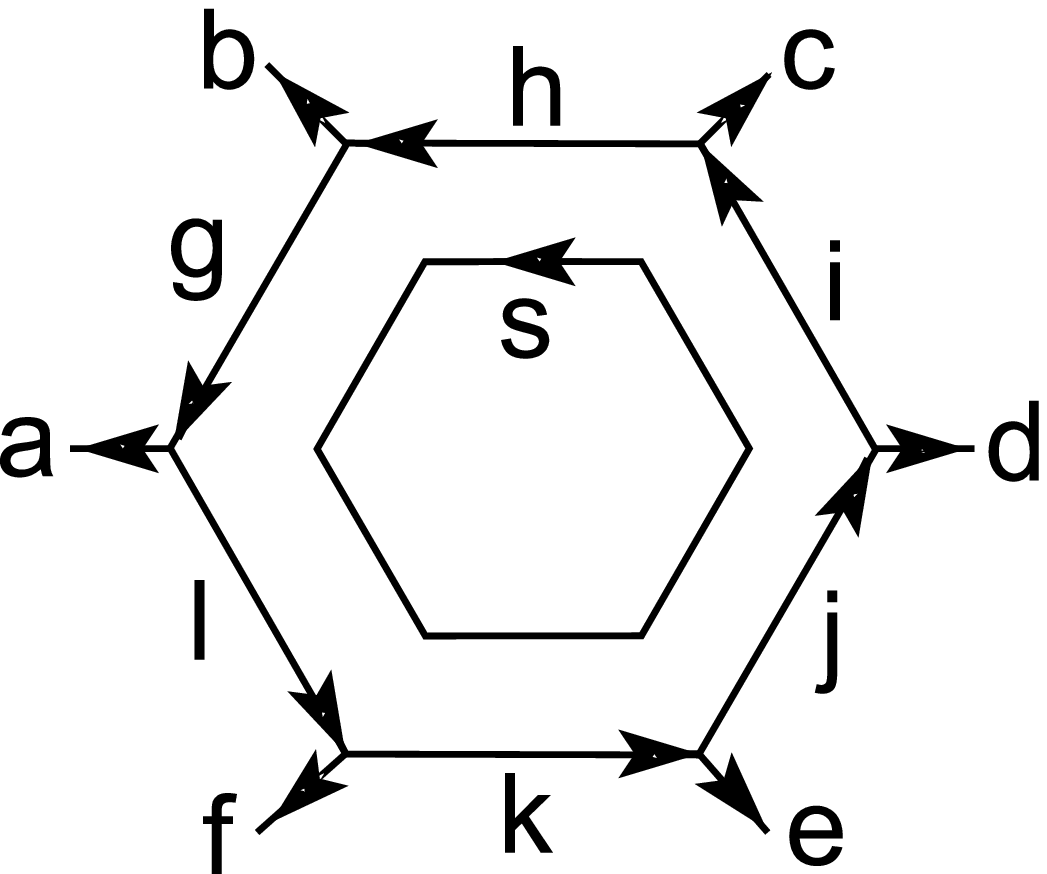}} 
\right|.
\label{graphreph}
\end{equation}
To obtain matrix elements of $B_p^s$, we use the local rules (\ref{rule1} - \ref{rule3}) and (\ref{nullerase} - \ref{rule4'}) to
``fuse'' the string $s$ onto the links along the boundary of the plaquette. That is, using these rules,
we express
$\left\< 
\raisebox{-0.26in}{\includegraphics[height=0.6in]{B3.eps}} 
\right| $ 
as the state
$\left\< 
\raisebox{-0.26in}{\includegraphics[height=0.6in]{B2.eps}} 
\right| $
multiplied by some constant. This constant tells us the matrix element of $B_p^s$ between the state
$\left\< \raisebox{-0.26in}{\includegraphics[height=0.6in]{B1.eps}} \right| $ and the state
$\left |\raisebox{-0.26in}{\includegraphics[height=0.6in]{B2.eps}} \right\> $. In appendix \ref{app_h}, we show that this prescription reproduces
the formula in equation (\ref{b}).

It is worth clarifying how exactly we use the local rules, especially since the first three rules (\ref{rule1} - \ref{rule3}) involve the 
wave function $\Phi$. To be precise, we replace every rule of the form $\Phi(X) = \sum_i c_i \Phi(X_i)$ (i.e. \ref{rule1} - \ref{rule3}) 
with a linear relation $\<X| = \sum_i c_i \<X_i|$. We then use these linear relations along with the linear relations (\ref{nullerase} - \ref{rule4'}) 
to fuse the string $s$ onto the boundary of plaquette, and obtain the required matrix elements. 

\subsection{Properties of the Hamiltonian \label{propertyh}}
Assuming $\{F(a,b,c),d_a,\gamma_a,\alpha(a,b)\}$ satisfy conditions (\ref{selfconseq}) and (\ref{unit}), the Hamiltonian has many nice
properties. 
The first property is that the Hamiltonian is Hermitian as long as $a_{s^*} = a_s^*$. This result follows from the identity
\begin{equation}
(B_p^s)^\dagger = B_p^{s^*} 
\end{equation}
which we derive in appendix \ref{property}.

The second property is that the $Q_I$ and $B_p^s$ operators commute with each other:
\begin{align}
[Q_{I},Q_{J}] = 0 \ , \ [Q_{I},B_{p}^{s}] = 0 \ , \ [B_{p}^{s},B_{p'}^{s'}] = 0.
\label{commrel}
\end{align}
The first two equalities follow easily from the definitions of $Q_I, B_p^s$. The third equation is less trivial.
We give an algebraic derivation of this identity in appendix \ref{app_commute}. 

Equations (\ref{commrel}) tell us that every term in the Hamiltonian (\ref{h}) commutes with every other term so that the model is exactly soluble.
This exact solubility holds for any value of the coefficients $a_s$. However, in what follows, we will focus on a particular value for
these coefficients, for which the mathematical structure of the model is especially simple. In particular, we consider the case where
\begin{equation}
a_s = \frac{d_s}{|G|}.
\label{as}
\end{equation}
It can be shown that $d_{s^*} = d^*_s$ so that this choice of $a_s$ is compatible with the requirement $a_s^* = a_{s^*}$. 

The third property of the Hamiltonian (which holds for the above choice of $a_s$) is that the $Q_I$ and $B_p$ are
projection operators -- i.e. they have eigenvalues $0,1$. It is easy to derive this result for $Q_I$; the derivation for
$B_p$ is given in appendix \ref{property}. 

Putting these results together, we can now derive the low energy properties of $H$. Let $|q_I, b_p\>$ denote the simultaneous eigenstates of $Q_I, B_p$:
\begin{align}
Q_I |q_I, b_p\> = q_I |q_I, b_p\> \ , \ B_p |q_I, b_p\> = b_p |q_I, b_p\> 
\end{align}
Then the corresponding energies are 
\begin{equation}
E = - \sum_I q_I - \sum_p b_p.
\end{equation}
Since the eigenvalues $q_I, b_p$ can be either $0,1$, it is clear that the ground state(s) have $q_I = b_p = 1$,
while the excited states have $q_I = 0$ or $b_p = 0$ for at least
one site $I$ or plaquette $p$. In particular, we see that there is a finite energy gap separating the ground state(s) from the excited
states. All that remains is to determine the ground state degeneracy. This degeneracy depends
on the global topology of our system. In appendix \ref{GSD}, we show that for a disk-like geometry with open boundary conditions
(see Fig. \ref{edge}), there is a unique state with $q_I = b_P = 1$. In other words, the ground state is non-degenerate. On the
other hand, in a periodic torus geometry, we find that there are $|G|^2$ degenerate ground states. This degeneracy on a torus
is a consequence of the topological order in our system.\cite{WenReview,WenBook,Einarsson}

The final property of our model which we establish in appendix \ref{property} is that the (unique) ground state of the lattice model in a disk geometry, $|\Phi_{latt}\>$, obeys the local rules (\ref{rule1} - \ref{rule3}) and (\ref{nullerase} - \ref{rule4'}). 
Therefore, since the local rules determine the ground state wave function uniquely, we conclude that $|\Phi_{latt}\>$ is identical to the
continuum wave function $\Phi$, restricted to string-net configurations on the honeycomb lattice. In other words, 
we have successfully constructed an exactly soluble Hamiltonian whose ground state is $|\Phi\>$. From now on, we will
use $|\Phi\>$ to denote both the lattice ground state and the continuum wave function, since the two are effectively identical.
 
\section{Quasiparticle excitations \label{qpsection}}
In this section, we will derive the topological properties of the quasiparticle excitations of the string-net Hamiltonian (\ref{h}). In particular,
we will find all the topologically distinct types of quasiparticles and compute their braiding statistics with one another.
Similarly to Ref. [\onlinecite{LevinWenStrnet}], our analysis proceeds in two steps: first, we construct ``string'' operators that create the 
quasiparticle excitations, and then we use the commutation algebra of the string operators to derive the quasiparticle braiding statistics.

\subsection{String operator picture}
Previously, it has been argued that quasiparticle excitations with nontrivial braiding statistics cannot be created by applying 
local operators to the ground state. Instead, these excitations are naturally created using extended \emph{string-like} operators.\cite{LevinWenHop}
The basic picture is as follows. For each topologically distinct quasiparticle excitation $\alpha$, there is a corresponding string 
operator. We denote this string operator by $W_\alpha(P)$ where $P$ is the path along which the string operator acts. In general,
$W_\alpha(P)$ is defined for both open and closed paths $P$; in the former case, we refer to $W_\alpha(P)$ as an open string
operator, while in the latter case, we say that $W_\alpha(P)$ is a closed string operator. 

The string operator $W_\alpha(P)$ is characterized by several properties. First, if $P$ is an open path, then when $W_\alpha(P)$ is 
applied to the ground state $\|\Phi \>$, the result is
\begin{equation}
W_\alpha(P) |\Phi\> = |\Phi_{ex}\>
\end{equation}
where $|\Phi_{ex}\>$ is an excited state containing a quasiparticle $\alpha$ at one end of $P$, and the antiparticle of $\alpha$
at the other end of $P$. Second, the excited state created by an open string operator does not depend on the path of the string:
\begin{equation}
W_\alpha(P) |\Phi\> = W_\alpha(P') |\Phi\>
\label{pathind}
\end{equation}
for any two paths $P, P'$ that have the same endpoints. Third, if $P$ is a closed path, then $W_\alpha(P)$ does not
create any excitations at all: $W_\alpha(P) |\Phi\> \propto |\Phi\>$.

Physically, one may think of an open string operator as describing a process in which 
a particle-antiparticle pair is created out of the ground state and the two particles are brought to the two ends of the string. 
Likewise, a closed string operator describes a process in which a pair of quasiparticles is created, and then one of them is moved around the path of the 
string until it returns to its original position, where it annihilates its partner. Note that throughout this discussion, 
we assume that the system is defined in a topologically trivial geometry, such as a disk. In topologically
non-trivial geometries, there are additional complications coming from the existence of multiple degenerate ground states, 
which we will not discuss here.

\subsection{Constructing the string operators}
We now construct string operators that create each of the different quasiparticle excitations of the string-net Hamiltonian (\ref{h}).
We follow the same strategy as Ref. [\onlinecite{LevinWenStrnet}]. 
First, we describe a particular ansatz for defining string operators. 
Next we search for the special choices of $(s, \omega, \bar{\omega})$ that lead to string operators satisfying the path independence condition (\ref{pathind}). 
Finally, we argue that the set of string operators that we find is complete in the sense that it allows us to create \emph{all} of the topologically distinct quasiparticle types.

We begin by describing our ansatz for constructing string operators. This ansatz allows us to build a string operator $W(P)$ given some data
$(s, \omega, \bar{\omega})$, where $s \in G$, and $\omega$, $\bar{\omega}$ are two complex-valued functions defined on the group $G$ with
$\omega(0) = \bar{\omega}(0) = 1$.
Here we suppress the string label $\alpha$ until later discussion of the quasiparticles.
First, suppose that $P$ is a closed path. In order to define $W(P)$ we need to specify how $W(P)$ acts on each string-net configuration. Here, we find it convenient to define the action of $W(P)$ on a bra $\<X|$ rather than describing its action on a ket $|X\>$. We describe the action of $W(P)$ using a graphical representation. More 
specifically, when $W(P)$ is applied to a string-net 
state $\<X|$, it simply adds a ``dashed'' string along the path $P$ under the preexisting string-nets: 
\begin{equation}
\left \langle \raisebox{-0.12in}{\includegraphics[height=0.35in]{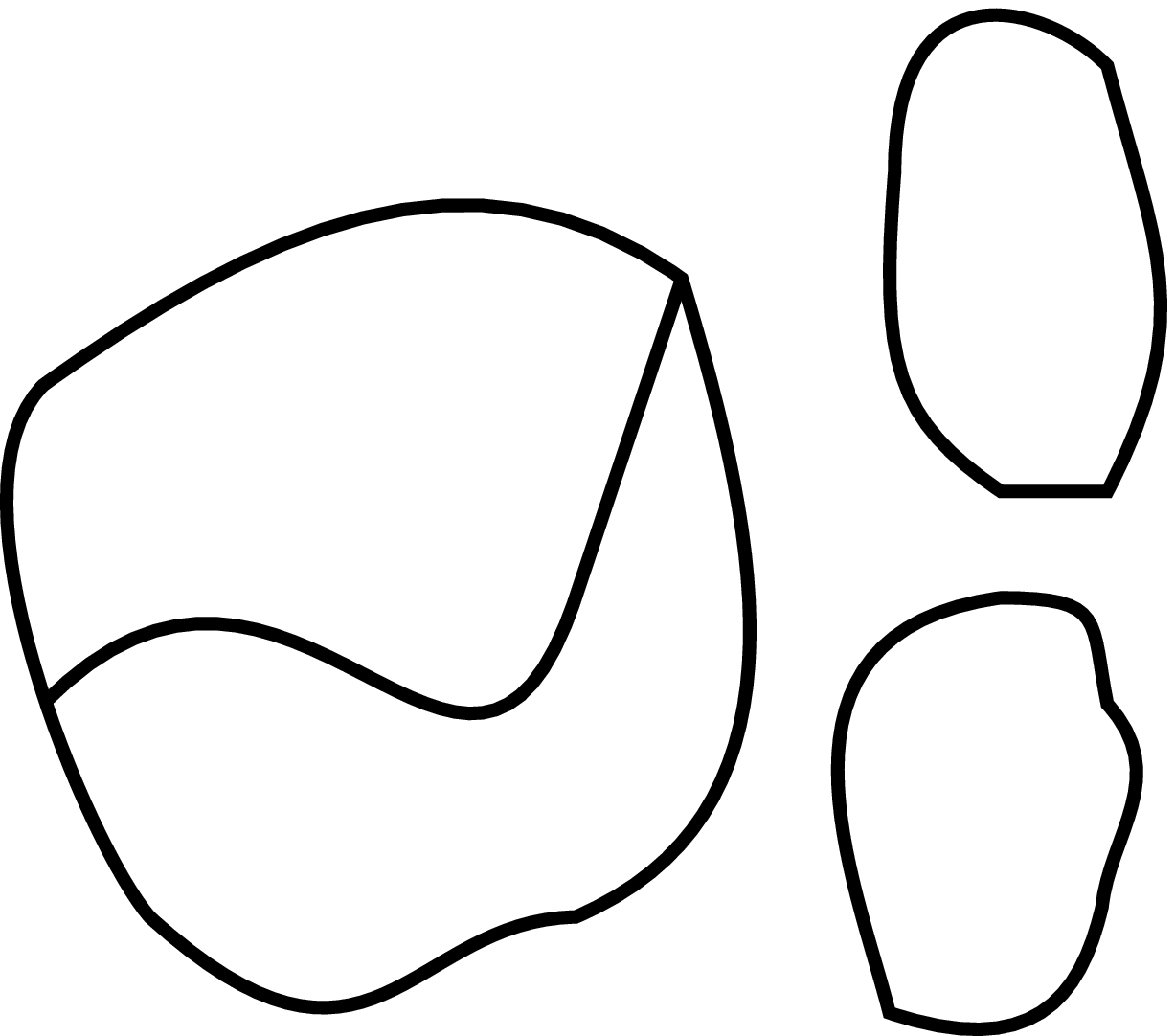}} \right\vert W(P) 
=\left \langle \raisebox{-0.12in}{\includegraphics[height=0.35in]{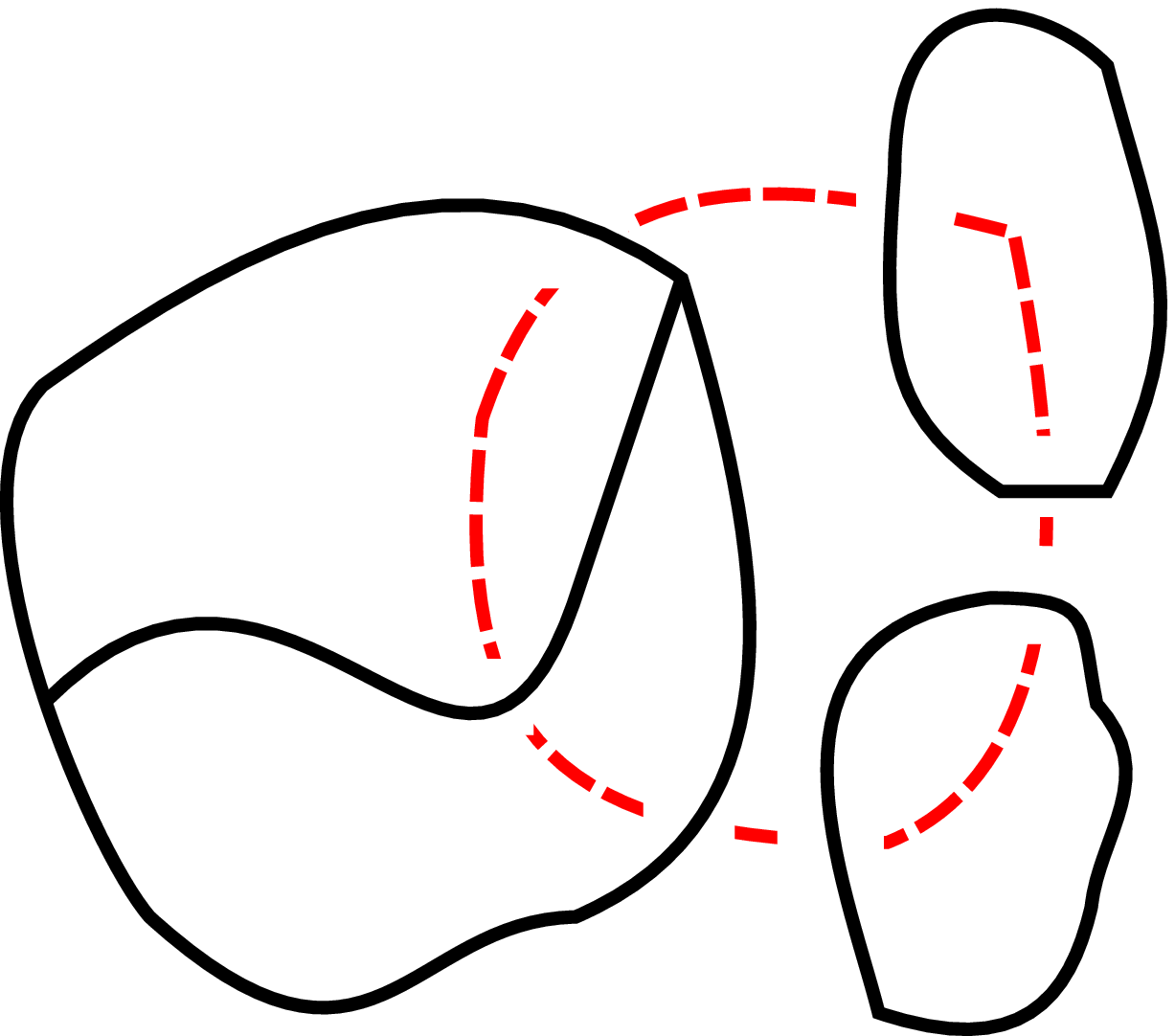}} \right\vert.
\end{equation}
We then replace the dashed string with a type-$s$ string, and we replace 
every crossing using the rules
\begin{align}
\left \langle \raisebox{-0.17in}{\includegraphics[height=0.4in,width=0.4in]{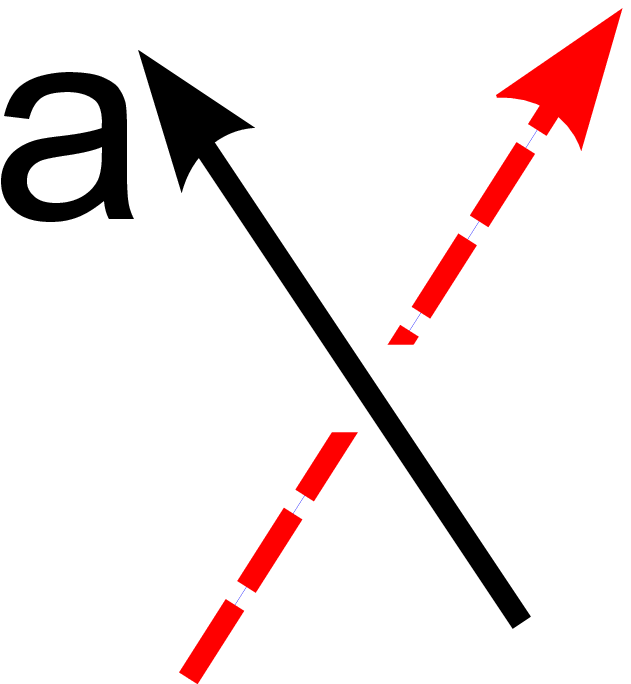}}\right\vert & =
\omega(a)\left\langle \raisebox{-0.17in}{\includegraphics[height=0.4in,width=0.4in]{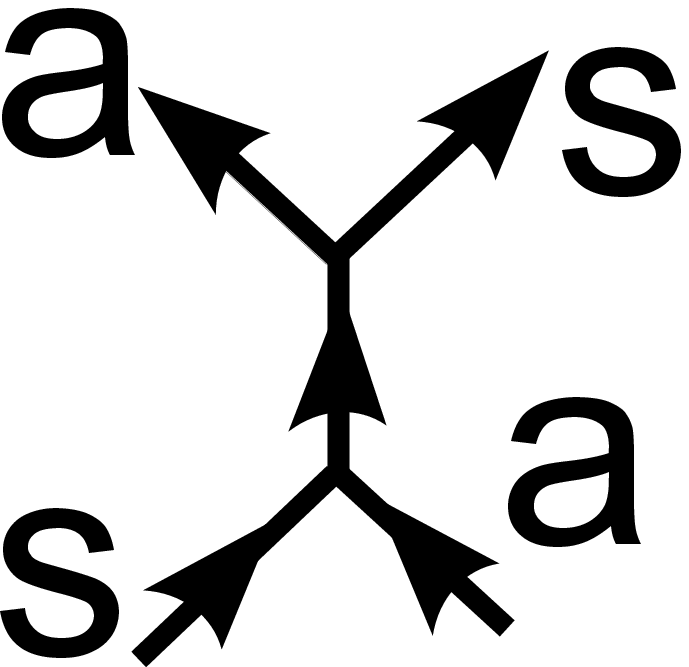}}\right\vert,  \label{rule7}
\\
\left\langle \raisebox{-0.17in}{\includegraphics[height=0.4in,width=0.4in]{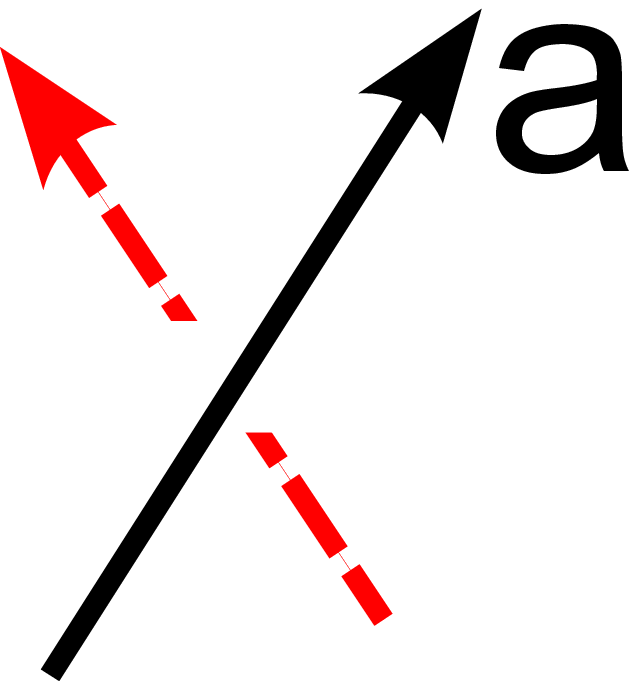}}\right\vert & =
\bar{\omega}(a)\left\langle \raisebox{-0.17in}{\includegraphics[height=0.4in,width=0.4in]{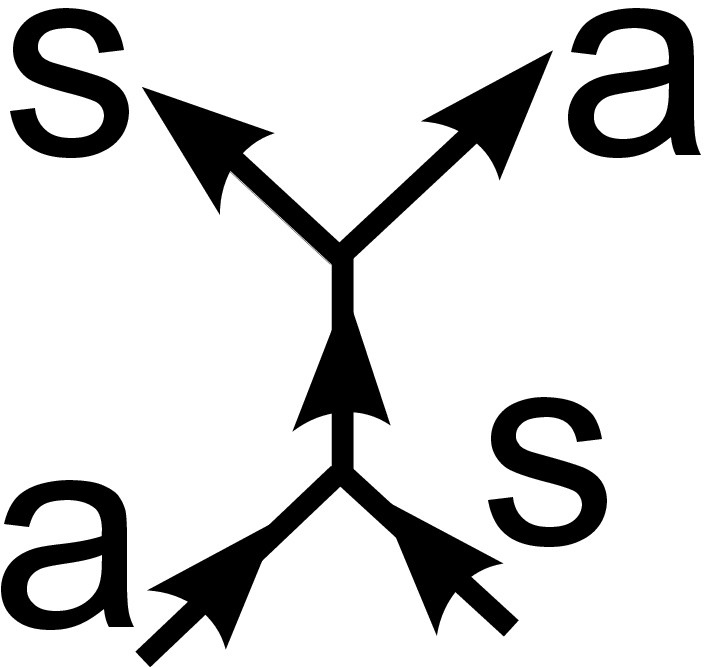}}\right\vert.  \notag
\end{align}
After making these replacements, the end result is a new string-net state $\<X'|$, multiplied by a product of complex numbers 
$\omega(a)$, $\bar{\omega}(a)$ -- one for each crossing. This construction defines the action of the operator $W(P)$. 
That is, if $\<X|$ is the initial string-net state then $\<X|W(P) = c \<X'|$ where $\<X'|$ is the final string-net state obtained by
adding a dashed string using the above rules and $c$ is a product of the $\omega(a)$ and $\bar{\omega}(a)$ factors along 
the path $P$.

\begin{figure}
\begin{center}
\includegraphics[height=1in,width=3in]{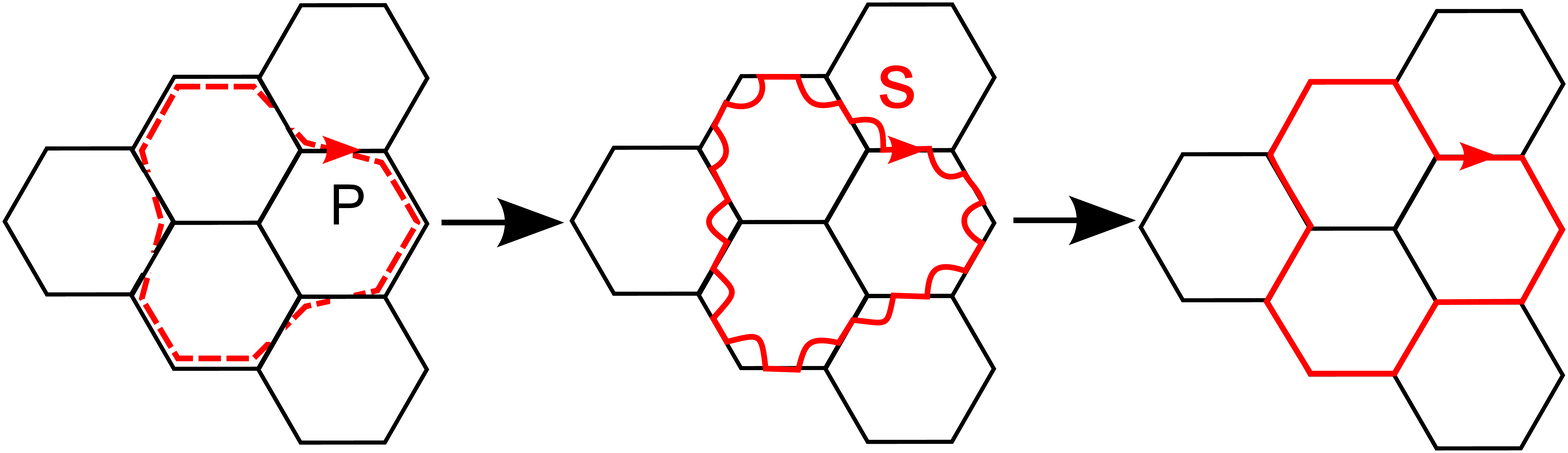}
\end{center}
\caption{
	The action of the string operator $W(P)$ along a path $P$ on the honeycomb lattice.
	First, the $W(P)$ operator adds a dashed line along the path $P$ under the preexisting string-nets.
	Then we replace the dashed line with a type-$s$ string and use (\ref{rule7}) to resolve the crossings.
	Finally, we use the local rules (\ref{rule1} - \ref{rule3}) and (\ref{nullerase} - \ref{rule4'}) to fuse the string $s$ onto the links along the path $P$. 
}
\label{wp}
\end{figure}

The above ansatz allows us to define string operators in the continuum; we now explain how to define the string operators on the \emph{lattice}. 
Let $P$ be a closed path on the honeycomb 
lattice. The corresponding string operator $W(P)$ is defined as follows. First, we shift the path $P$ slightly so that it no longer lies 
exactly on the honeycomb lattice. The way in which the path $P$ is shifted is not especially 
important, but for concreteness, we follow a particular prescription for shifting the path $P$, shown in Fig. \ref{wp}. 
The action of the operator $W(P)$ can then be described by a four step process. In the first three steps, we add a dashed string along the path $P$, replace the dashed string with a type-$s$ string, and replace each crossing using the rules (\ref{rule7}). In the final step, we use the local rules (\ref{rule1} - \ref{rule3}) and (\ref{nullerase} - \ref{rule4'}) to ``fuse'' the string $s$ onto the links along the path $P$. 
In the above discussion, we have implicitly assumed that the initial string-net state $\<X|$
obeys the branching rules at all the vertices along the path $P$; if $\<X|$ does not obey the branching rules at any of these vertices, then
we define $\<X|W(P) = 0$.

The reader may notice that the definition of the closed string operators is very similar to the graphical representation of the plaquette term in
Hamiltonian, $B_p^s$ (\ref{graphreph}). \footnote{In fact, $B_p^s$ can be thought of as a special case of a closed string operator where the string runs along 
the boundary of a single plaquette $p$.} Just like $B_p^s$, it is possible to construct an explicit algebraic formula for the matrix elements of 
$W(P)$. We will not write out the explicit formula here, but the basic structure is very similar to $B_p^s$: one finds that the string operator 
$W(P)$ only affects the spin states along the path $P$, and the matrix elements between these spin states are a function of the spins on the edges 
adjacent to $P$. 

We can also define \emph{open} string operators on the lattice. We use a graphical representation very similar to the closed string case: the action of
an open string operator $W(P)$ is defined by adding a dashed string along the path $P$ (shifted slightly)
and then using the local rules (\ref{rule1} - \ref{rule3}) and (\ref{nullerase} - \ref{rule4'}), (\ref{rule7}) to resolve crossings and fuse the 
string onto the links along $P$. There is some arbitrariness in defining the action of the string operator near the endpoints of $P$ since the 
local rules are not defined for string-net configurations that violate the branching rules. However, it does not matter how exactly we define the action of
the string operator near its endpoints, since this choice only affects the \emph{local} properties of the quasiparticle excitation created by 
$W(P)$, and does not affect the \emph{topological} properties which are our main concern here.

\subsection{Path independence constraint} \label{stropsec}
The above ansatz allows us to define a string operator $W(P)$ for each choice of $(s,\omega, \bar{\omega})$. 
However, we are only interested in the special class of string operators that create deconfined quasiparticle excitations
when we apply them along an open path. As discussed above, these string operators must satisfy path independence (\ref{pathind}):
$W(P) |\Phi\> = W(P') |\Phi\>$ for any two paths $P, P'$ that have the same endpoints. Below we search for the special values of 
$(s,\omega, \bar{\omega})$ that lead to path independent string operators.

To this end, we note that the path independence condition can be equivalently written as
\begin{equation}
\<X| W(P)|\Phi\> = \<X | W(P')|\Phi\>
\end{equation}
where $\<X|$ is an arbitrary string-net state. Furthermore, we observe that we only need to check path independence for ``elementary'' 
deformations $P \rightarrow P'$, since larger deformations can be built out of elementary ones. In this way, we can see that $W$
will satisfy path independence if and only if
\begin{equation}
	 \left\langle \raisebox{-0.17in}{\includegraphics[height=0.4in,width=0.4in]{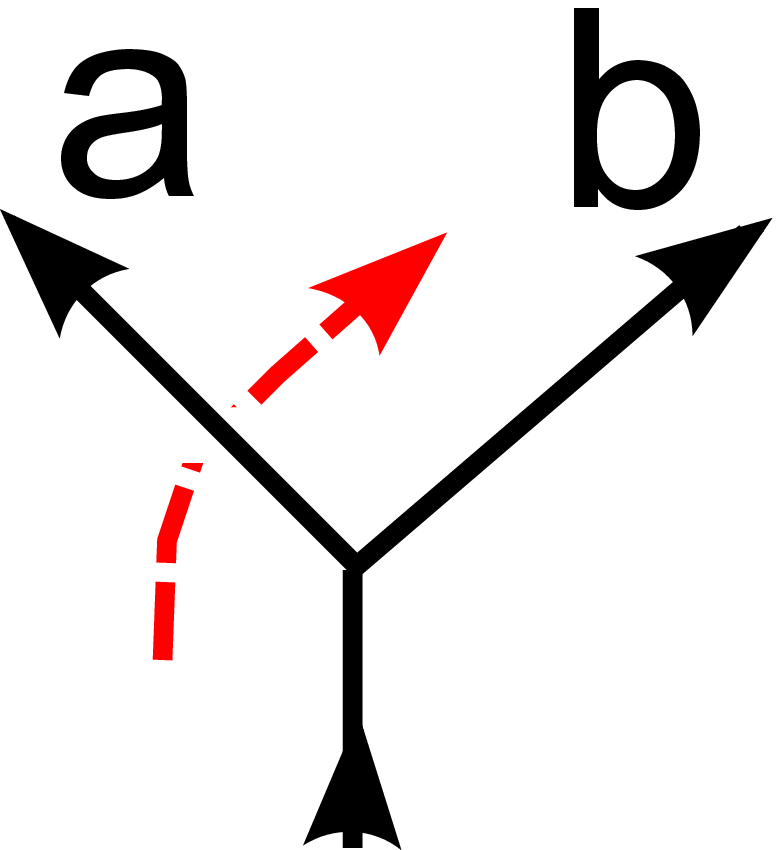}}\right\vert \Phi \bigg\rangle 
=  \left\langle \raisebox{-0.17in}{\includegraphics[height=0.4in,width=0.4in]{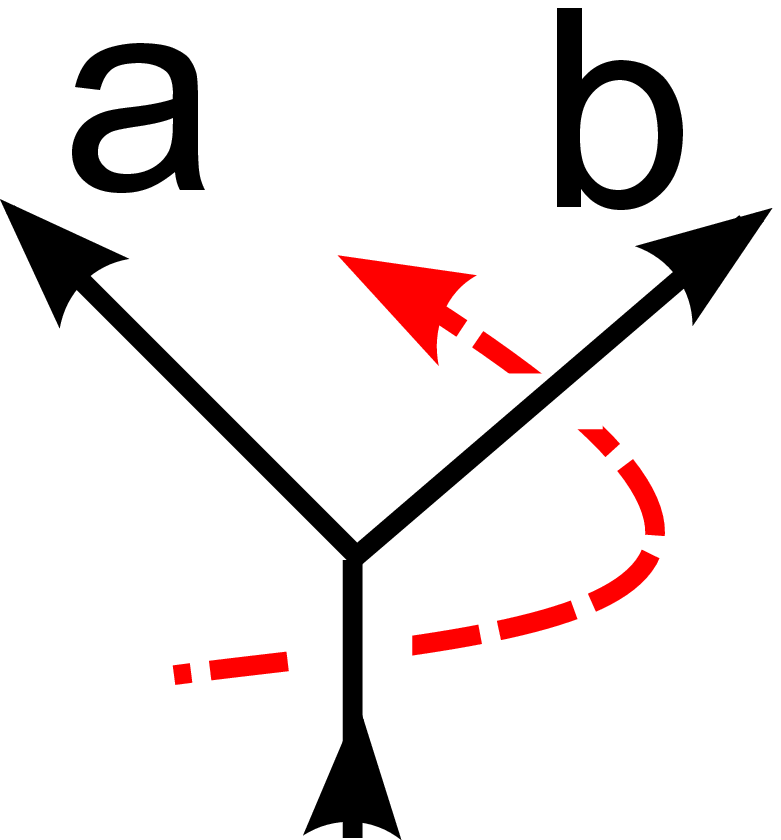}}\right\vert \Phi \bigg\rangle 
\label{pathind1}
\end{equation}
for any $a,b$ and
\begin{equation}
\left\langle
\raisebox{-0.17in}{\includegraphics[height=0.4in,width=0.4in]{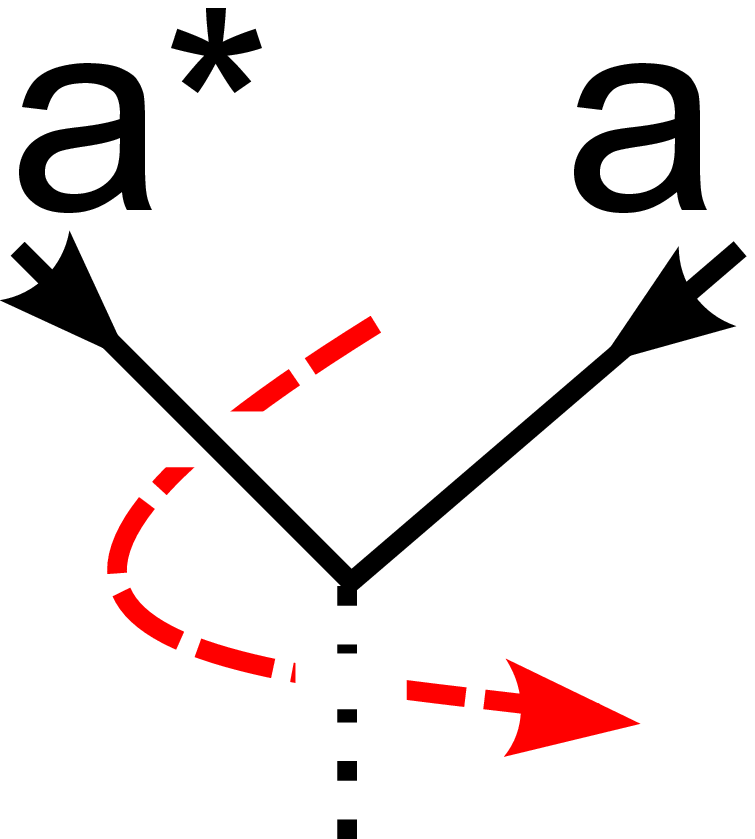}}
\right\vert \Phi \bigg \rangle
= \left\langle 
\raisebox{-0.17in}{\includegraphics[height=0.4in,width=0.4in]{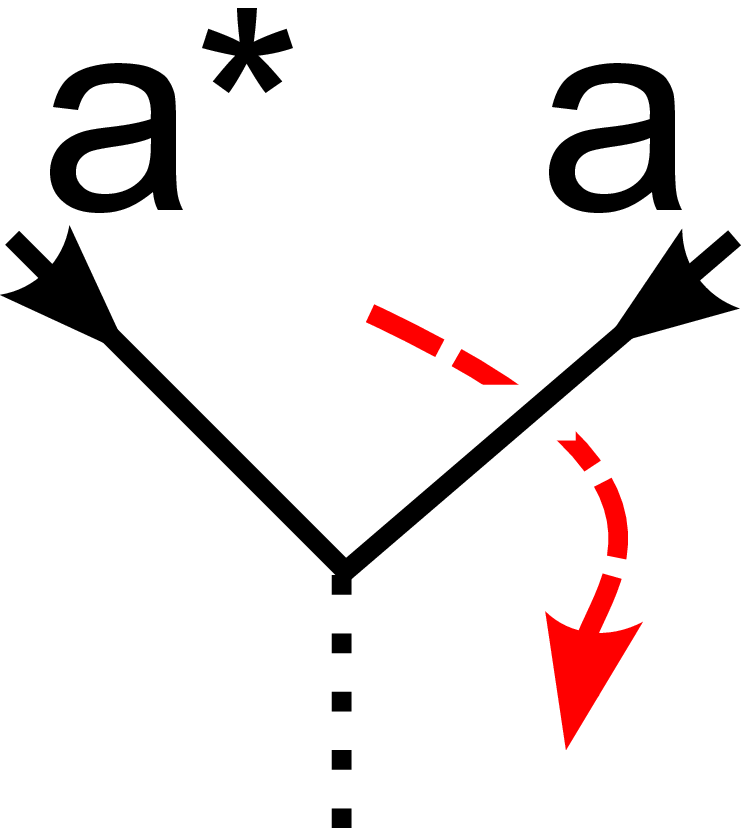}}
\right\vert \Phi \bigg \rangle
\label{pathind2}
\end{equation}
for any $a$. (The reason that the above two elementary deformations are sufficient to establish general path independence 
is that any vertex with any set of orientations can be built out of the above two vertices according to (\ref{nullabs}), (\ref{dot1}), (\ref{dot2}); 
thus the above two conditions imply path independence with respect to every vertex).

To proceed further, we translate the above graphical relations (\ref{pathind1}-\ref{pathind2}) into algebraic conditions on $(s,\omega, \bar{\omega})$ using
the local rules (\ref{rule1} - \ref{rule3}), (\ref{nullerase} - \ref{rule4'}) and (\ref{rule7}). The result is (see appendix \ref{hexagon}):
\begin{eqnarray}
\omega(a) \omega(b) &=& \omega(a+b) \cdot \frac{F_{abs} F_{sab}}{F_{asb}} \frac{F_{s(a+b)(a+b)^*}}{F_{saa^*} F_{sbb^*}}, \label{wbar1} \\
\bar{\omega}(a) &=& \omega(a)^{-1} \cdot F_{ass^*}^{-1} F_{saa^*}^{-1}. \label{wbar2}
\end{eqnarray}
Next, we define
\begin{gather}
	\mathbf{w}(a)=\omega(a)F(s,a,a^{*}) \ \ , \ \ 
        \mathbf{\bar{w}}(a)=\bar{\omega}(a)F(a,s,s^{*}), \nonumber\\
	c_{s}(a,b)=\frac{F(a,b,s)F(s,a,b)}{F(a,s,b)}. \label{cs}
\end{gather}
The above relations can then be written in the simple form
\begin{gather}
	\mathbf{w}(a)\mathbf{w}(b)=c_{s}(a,b) \mathbf{w}(a+b), \label{seq} \\
	\mathbf{\bar{w}}(a)= \mathbf{w}(a)^{-1} \label{seq2}.
\end{gather}

We wish to find all complex valued functions $\mathbf{w}, \mathbf{\bar{w}}: G \rightarrow \mathbb{C}$ that satisfy (\ref{seq}), (\ref{seq2}). Clearly, it is sufficient to
find $\mathbf{w}$ satisfying (\ref{seq}), as $\mathbf{\bar{w}}$ can be obtained immediately from Eq. (\ref{seq2}).

To solve Eq. (\ref{seq}), we observe that the self-consistency condition (\ref{pentid}) implies that $c_s(a,b)$ obeys the identity
\begin{equation}
	c_{s}(a,b)c_{s}(a+b,c)=c_{s}(b,c)c_{s}(a,b+c).
	\label{2cocyl}
\end{equation}
Equation (\ref{2cocyl}) means that $c_s(a,b)$ is a well-known mathematical object, namely the ``factor system'' of a projective representation.\cite{ChenGuWenSPT} 
From this point of view, the problem of solving equation (\ref{seq}) is equivalent to the problem of finding a (1D) projective representation $\mathbf{w}: G \rightarrow \mathbb{C}$ corresponding to the factor system $c_s(a,b)$.

There are two cases to consider: $c_s(a,b)$ may be symmetric in $a,b$ or it may be non-symmetric. First, suppose $c_s(a,b)$ is non-symmetric. 
In this case, we can see that Eq. (\ref{seq}) has no nonzero solutions, since the left-hand side is manifestly symmetric in $a,b$ while the right-hand side is non-symmetric. 
Hence, our ansatz does not yield any path independent string operators of type $s$. 
To build a path independent string operator, we have to use a more general ansatz\cite{LevinWenStrnet} where the parameters $\mathbf{w}, \mathbf{\bar{w}}$ are \emph{matrices} rather than scalars. 
Equivalently, we need to look for higher dimensional projective representations with factor system $c_s(a,b)$. 
We will not discuss this construction here since the resulting particles have \emph{non-abelian} statistics,\cite{PropitiusThesis} and our focus is on models with purely abelian statistics. In fact, throughout this paper we will restrict to choices of $F(a,b,c)$ such that $c_s(a,b)$ is symmetric for all $s$.

If $c_s(a,b)$ is symmetric, Eq. (\ref{seq}) can be solved as follows.
Since every finite abelian group is isomorphic to a direct product of cyclic groups,
we can assume without loss of generality that the group is $G = \mathbb{Z}_{N_1} \times ... \times \mathbb{Z}_{N_k}$.
Let $a_1, a_2,..., a_k$ be the generators of $G$. Once we find the value of $\mathbf{w}(a_k)$ for
each generator $a_k$, then $\mathbf{w}$ is fully determined by equation (\ref{seq}). To find the value of $\mathbf{w}(a_k)$,
we first rewrite equation (\ref{seq}) as
\begin{equation}
\frac{\mathbf{w}(a)\mathbf{w}(b)}{\mathbf{w}(a+b)} = c_s(a,b).
\end{equation}
Setting $a=a_k$ and $b=y a_k$ where $y$ is some integer, we obtain
\begin{equation}
\frac{\mathbf{w}(a_k) \mathbf{w}(y a_k)}{\mathbf{w}((y+1) a_k)} = c_s(a_k,y a_k).
\end{equation}
We then take the product of the above equations over $y = 0,1,...,N_k-1$. After canceling terms on the left hand side, and using the
fact that $N_k a_k = 0$, we find
\begin{equation}
\mathbf{w}(a_k)^{N_k} = \prod_{y=0}^{N_k-1} c_{s}(a_k, y a_k).
\label{solseq}
\end{equation}
We can see that $\mathbf{w}(a_k)$ can take $N_k$ different values for each $k$. Hence, there
are $\prod_k N_k = |G|$ solutions to Eqs. (\ref{seq}), (\ref{seq2}) for each choice of $s$. The parameter $s$ can also
take $|G|$ different values, so altogether we find $|G|^2$ solutions, corresponding to $|G|^2$ path independent string
operators.

At this point we have constructed $|G|^2$ string operators. Since these string operators satisfy path independence, we know that
when we apply them to the ground state (along an open path) they will create quasiparticle excitations at the
ends of the string. Thus, the above operators will allow us to construct $|G|^2$ different quasiparticle excitations. 
The next question is to determine whether this set of excitations is complete, i.e. whether it contains every 
topologically distinct quasiparticle. To address this question, we recall that the ground state
degeneracy of the model on a torus is $|G|^2$, and hence by general arguments we expect that the system supports a total of 
$|G|^2$ topologically distinct excitations. Furthermore, we will show later that the above quasiparticle excitations are all topologically 
distinct. Putting these two facts together, we conclude that the above set of quasiparticles is indeed 
complete.

\subsection{Labeling scheme for quasiparticle excitations \label{label}}
Having constructed all the quasiparticle excitations, we now describe a scheme for labeling these excitations.
To begin, we note that it is sufficient to define a labeling scheme for the 
solutions to Eq. (\ref{seq}) since these solutions are in one-to-one correspondence with the 
different excitations. Next, we recall that Eq. (\ref{seq}) has $|G|$ different solutions $\mathbf{w}$ 
for each string type $s \in G$. Thus, an obvious way to label the different solutions is to use a string type 
index $s \in G$, together with an integer index that runs over $1,...,|G|$, e.g. 
$\mathbf{w}_{(s,1)}, \mathbf{w}_{(s,2)},...,\mathbf{w}_{(s,|G|)}$. 

While the above labeling scheme is perfectly adequate, we will see below that it is actually more natural to label the 
solutions to Eq. (\ref{seq}) by a string type index $s \in G$ together with
an index $m$ that runs over the different $1D$ linear representations of $G$. We note that this alternative scheme
is sensible since every abelian group has exactly $|G|$ different $1D$ representations. 

Following the latter approach, we will label each type-$s$ solution to Eq. (\ref{seq}) by an ordered pair 
$\alpha = (s,m)$ where $s \in G$ and $m$ is a $1D$ representation of $G$. We will denote this solution by 
$\mathbf{w}_\alpha$. Similarly we will denote the corresponding string operator by $W_\alpha$, and we will refer to
the quasiparticle excitation created by $W_\alpha$ as $\alpha$.

In order to make this labeling scheme well-defined, we need to specify a particular type-$s$ solution $\mathbf{w}_{(s,m)}$ to 
Eq. (\ref{seq}) for each ordered pair $(s,m)$. We begin with the special case where $s=0$. In this case, equation (\ref{seq}) 
takes a simple form since $c_0(i,j) = 1$ (by Eqs. (\ref{cs}),(\ref{F0})):
\begin{equation}
\mathbf{w}(a)\mathbf{w}(b)=\mathbf{w}(a+b).
\label{seqs0}
\end{equation}
We can see that the above equation is precisely the condition for $\mathbf{w}$ to be a $1D$ linear representation of $G$. Hence, 
there is a very natural way to define a solution $\mathbf{w}_{(0,m)}$ for each $m$: we simply define 
\begin{equation}
\mathbf{w}_{(0,m)} = \rho_m
\label{fluxdef}
\end{equation}
where $\rho_m$ is the $1D$ representation corresponding to $m$.

Next we explain how $\mathbf{w}_{(s,m)}$ is defined when $s \neq 0$. Here, we proceed in two steps. In the first step, we
choose some arbitrary type-$s$ solution to Eq. (\ref{seq}) and we define $\mathbf{w}_{(s,0)}$ to be this solution; the
particular solution we choose is a matter of convention -- we will give some examples of conventions in sections \ref{zn},\ref{zksection}.
After choosing $\mathbf{w}_{(s,0)}$, we then define
\begin{equation}
\mathbf{w}_{(s,m)} = \mathbf{w}_{(s,0)} \cdot \rho_m.
\label{gendef}
\end{equation}
To understand why this definition is sensible, note that $\mathbf{w}_{(s,0)} \cdot \rho_m$ will always solve Eq. (\ref{seq}) if 
$\mathbf{w}_{(s,0)}$ solves Eq. (\ref{seq}). Hence, (\ref{gendef}) gives a complete parameterization of the different type-$s$ 
solutions to Eq. (\ref{seq}).

At this point, it is useful to introduce some terminology. We will call the $(s,0)$ excitations 
``fluxes'' and the $(0,m)$ excitations ``charges.'' Likewise, we will think of a general excitation $(s,m)$ as a composite of a 
flux and a charge. We think of the parameter $s$ as describing the amount of flux carried by the excitation, while
$m$ describes the amount of charge. (The motivation for this terminology is that the mutual statistics between the $(s,0)$ and $(0,m)$ 
excitations is identical to the mutual statistics between fluxes and charges in lattice gauge theory, as we will demonstrate in 
sections \ref{zn},\ref{zksection}). 

In this language, the basic idea behind our labeling scheme is to define the ``pure'' charge excitations $(0,m)$ using
Eq. (\ref{fluxdef}), and to define the pure fluxes $(s,0)$ in some arbitrary way -- the definition of the pure fluxes
is a matter of convention. All the other excitations can then be labeled as a composite of a charge and a flux, as in Equation (\ref{gendef}). 

\begin{figure}[tb]
        \centering
        \includegraphics[height=1.3in,width=1.8in]{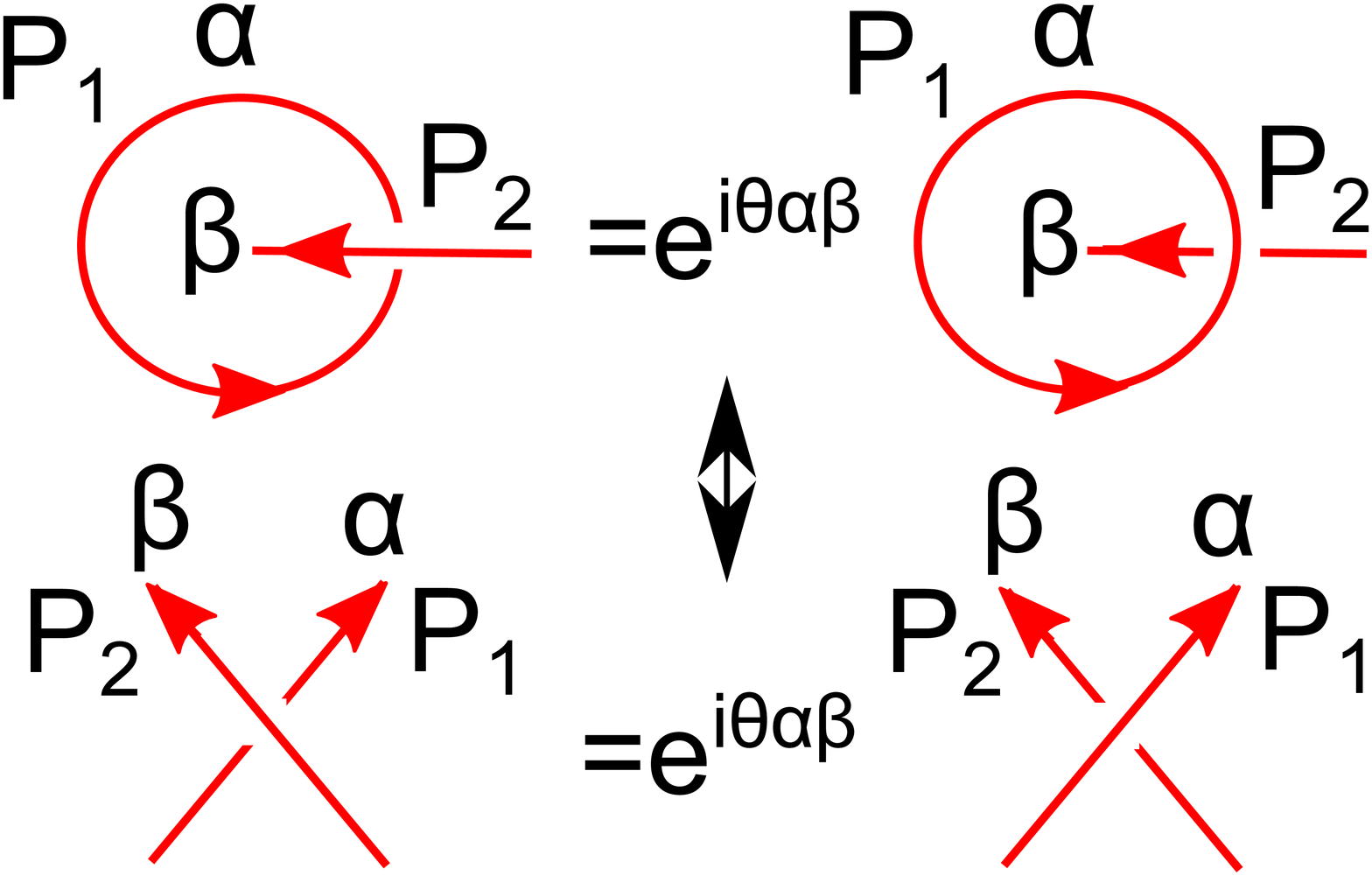}
        \caption{
                Computation of mutual statistics from string operators.
        On the top, we compare the action of $W_{\beta}(P_2)W_{\alpha}(P_1)$ with its reverse action
$W_{\alpha}(P_1)W_{\beta}(P_2)$. The phase difference between the two products is equal to the mutual
statistics $e^{i\theta_{\alpha\beta}}$ between the two quasiparticles created by $W_\alpha,W_\beta$.
In the bottom, we zoom in around the intersection point since the commutation algebra only depends on
the properties of the string operators near the intersection point.}
        \label{thetaab}
\end{figure}

\subsection{Braiding statistics of quasiparticles}
In the previous section, we constructed a string operator $W_\alpha$ for each ordered pair $\alpha = (s,m)$ where $s \in G$, and $m$ is a 1D representation of $G$. 
We argued that these $|G|^2$ string operators serve as creation operators for $|G|^2$ different quasiparticle excitations, which we label by $\alpha$.
In this section, we will compute the braiding statistics of these quasiparticle excitations. More specifically, we will compute the mutual statistics 
$\theta_{\alpha \beta}$ for every pair of quasiparticles $\alpha, \beta$, as well as the exchange statistics $\theta_\alpha$ for every
quasiparticle $\alpha$. Here we use the convention that $\theta_{\alpha \beta}$ and $\theta_\alpha$ are associated with clockwise braiding of particles.

To begin, we review the general relationship between braiding statistics and the string operator algebra in abelian topological phases. 
First we explain how the mutual statistics $\theta_{\alpha \beta}$ is encoded in the 
string algebra; afterwards, we will discuss the exchange statistics $\theta_\alpha$.
Let $\alpha, \beta$ be two (abelian) quasiparticle excitations with mutual statistics $\theta_{\alpha \beta}$, and let $W_\alpha, W_\beta$
be the corresponding string operators. Then, for any two paths $P_1, P_2$ that intersect one another as in Fig. \ref{thetaab}, the corresponding string operators $W_\alpha(P_1), W_\beta(P_2)$ obey the
commutation algebra\cite{KitaevToric,LevinWenHop,LevinWenStrnet}
\begin{equation}
W_\beta(P_2) W_\alpha(P_1) |\Phi\> = e^{i\theta_{\alpha \beta}}  W_\alpha(P_1) W_\beta(P_2) |\Phi\>
\label{mutstat}
\end{equation}
where $|\Phi\>$ denotes the ground state of the system.
A simple way to derive this result is to consider the case where $P_2$ is an open path, and $P_1$ forms a closed loop (see Fig. \ref{thetaab}. 
In this case, the two string operators $W_\alpha(P_1)$, $W_\beta(P_2)$ have different physical interpretations. The operator $W_\beta(P_2)$ describes a process in which $\beta$ and its antiparticle are created and then moved to opposite endpoints of the path $P_2$. 
On the other hand, $W_\alpha(P_1)$ describes a three step physical process in which (1) $\alpha$ and its antiparticle are created out of the ground state $|\Phi\>$, (2) $\alpha$ is moved all the way around the closed loop $P_1$, and then finally (3) $\alpha$ and 
its antiparticle are annihilated. Given these interpretations, we can see that the left hand side of (\ref{mutstat}) describes a process in which $\alpha$ is first
moved around $P_1$ and then $\beta$ and its antiparticle are moved to the endpoints of $P_2$, while the right hand side describes a process 
in which $\beta$ and its antiparticle are first moved to the endpoints of $P_2$ and then $\alpha$ is moved around $P_1$. 
By the definition of mutual statistics, we know that these two processes 
differ by the statistical Berry phase $e^{i\theta_{\alpha \beta}}$, thus implying Eq. (\ref{mutstat}). More generally, one can argue that Eq. (\ref{mutstat}) holds for 
\emph{any} two paths $P_1, P_2$ that intersect each other once, since the commutation algebra depends only on the properties of the string operators near the intersection point.

The exchange statistics $\theta_\alpha$ is also encoded in the string operator algebra. In the most general case, $\theta_\alpha$ can be extracted by examining the 
commutation relations between three string operators, as discussed in Ref. [\onlinecite{LevinWenHop}]. 
However, the calculation of $\theta_\alpha$ can be simplified, using the fact that $W_\alpha$ obeys the path independence conditions (\ref{pathind1}-\ref{pathind2}), and that $W_\alpha(P \cup P') = W_\alpha(P) W_\alpha(P')$ for any two paths $P, P'$ 
that share an endpoint. Using these properties, it can be shown that $W_\alpha(P_1), W_\alpha(P_2), W_\alpha(P_3), W_\alpha(P_4)$ obey the algebra
\begin{equation}
W_\alpha(P_2) W_\alpha(P_1) |\Phi\> = e^{i\theta_\alpha} W_\alpha(P_4) W_\alpha(P_3) |\Phi\>
\label{exstat}
\end{equation}
for any four paths $P_1, P_2, P_3, P_4$ with the geometry of Fig. \ref{fig:theta} (See appendix \ref{thetaapp} for a derivation).

With the above relations in hand, we are now ready to compute the statistics of the quasiparticle excitations in our model. We begin with the exchange statistics.
Let $\alpha = (s,m)$ be any quasiparticle excitation. We wish to find the exchange statistics of $\alpha$. The first step is to multiply both sides of
Eq. (\ref{exstat}) by the ``no-string'' (vacuum) ket $\<0|$:
\footnote{Strictly speaking, instead of choosing the bra $\left< 0 \right|$ to be the global vacuum state, we should choose it to be a \emph{local} vacuum state which contains two strings ending at the four
endpoints of $P_1, P_2$. This choice ensures that the matrix elements in (\ref{exchangeop}) are nonzero.}
\begin{equation}
\<0| W_\alpha(P_2) W_\alpha(P_1) |\Phi\> = e^{i\theta_\alpha} \<0|  W_\alpha(P_4) W_\alpha(P_3) |\Phi\>.
\label{exchangeop}
\end{equation}
Next, we note that the graphical definition of $W_\alpha$ tells us that the action of $W_\alpha(P_2)$ on $\<0|$
is simply to add a type-$s$ string along the path $P_2$ (there are no additional phase factors since $\omega_\alpha(0) = 1$ by assumption).
The same is true for $W_\alpha(P_4)$, so we obtain:
\begin{equation}
\left\< \raisebox{-0.16in}{\includegraphics[height=0.4in]{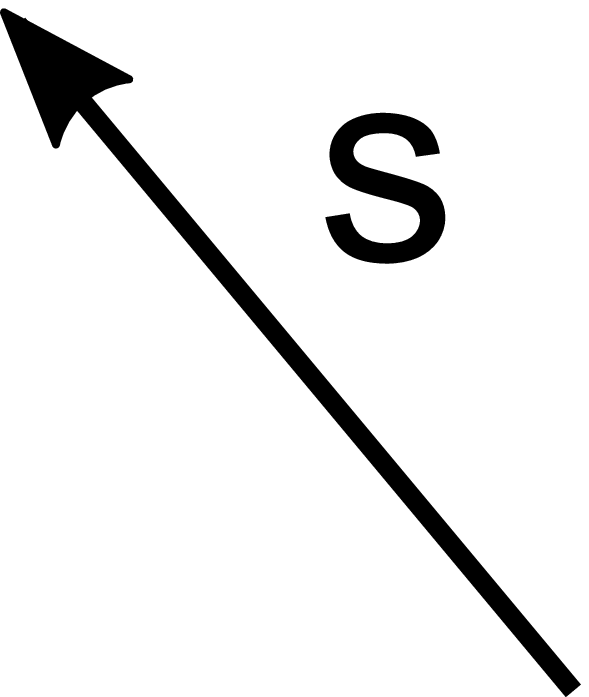}}\right| W_\alpha(P_1) \bigg |\Phi \bigg \rangle 
= e^{i\theta_\alpha} 
\left\<\raisebox{-0.16in}{\includegraphics[height=0.4in]{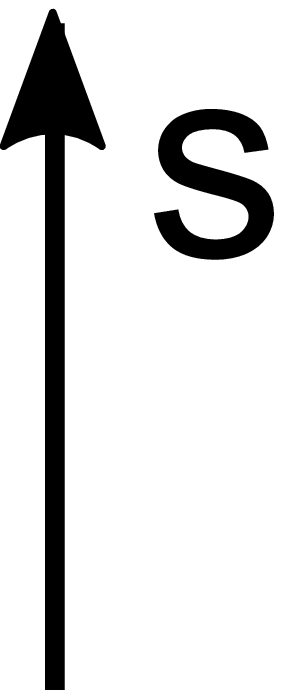}}\right| W_\alpha(P_3) \bigg|\Phi \bigg \rangle.
\end{equation}
Applying the definition of $W_\alpha$ once more, we derive
\begin{equation}
\omega_\alpha(s) \left\<\raisebox{-0.16in}{\includegraphics[height=0.4in]{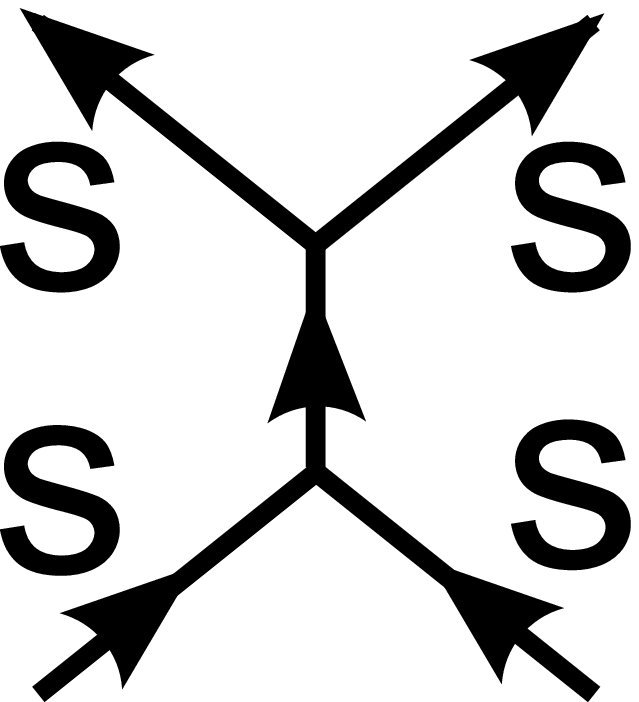}}\right| \Phi \bigg \rangle 
= e^{i \theta_\alpha} 
\left\<\raisebox{-0.16in}{\includegraphics[height=0.4in]{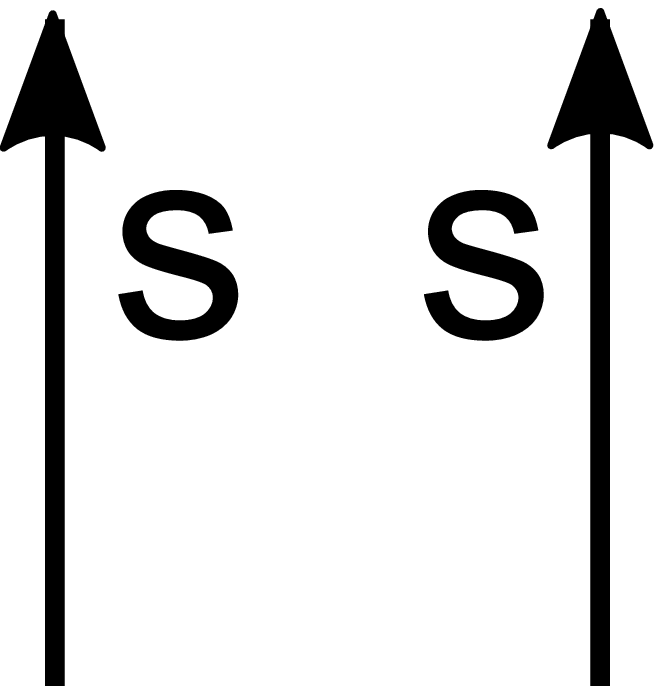}} \right| \Phi \bigg \rangle.
\end{equation}
At the same time, using the local rules (\ref{rule3}), we have
\begin{equation}
\left\<\raisebox{-0.16in}{\includegraphics[height=0.4in]{wp1wp2.eps}}\right| \Phi \bigg \rangle 
= F(s,s,s^*) 
\left\<\raisebox{-0.16in}{\includegraphics[height=0.4in]{wp3wp4.eps}}\right| \Phi \bigg \rangle.
\end{equation}
We conclude that
\begin{eqnarray}
e^{i\theta_\alpha} &=& \omega_\alpha(s) F(s,s,s^*) \nonumber \\
&=& \mathbf{w}_\alpha(s).
\label{theta}
\end{eqnarray}

\begin{figure}[tb]
        \centering
        \includegraphics[height=0.6in,width=1.7in]{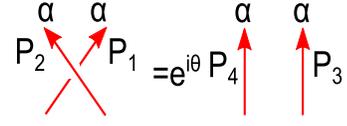}
        \caption{Computation of exchange statistics from string operators.
                The commutation relation of string operators
$W_{\alpha}(P_2)W_{\alpha}(P_1)=e^{i\theta}W_{\alpha}(P_4)W_{\alpha}(P_3)$ gives the exchange statistics of
quasiparticle $\alpha$ provided $W_\alpha$ is path independent and $W_\alpha(P \cup P')=W_\alpha(P) W_\alpha(P')$.}
        \label{fig:theta}
\end{figure}

Next, we find the mutual statistics between two excitations $\alpha = (s,m)$, $\beta = (t,n)$. Similarly to above, the first step is
to multiply both sides of Eq. (\ref{mutstat}) by the ket $\<0|$:
\begin{equation}
\<0| W_\beta(P_2) W_\alpha(P_1) |\Phi\> = e^{i\theta_{\alpha \beta}} \<0| W_\alpha(P_1) W_\beta(P_2) |\Phi\>.
\end{equation}
Evaluating the action of the string operators on both sides as above, we derive
\begin{equation}
\left\<\raisebox{-0.16in}{\includegraphics[height=0.4in]{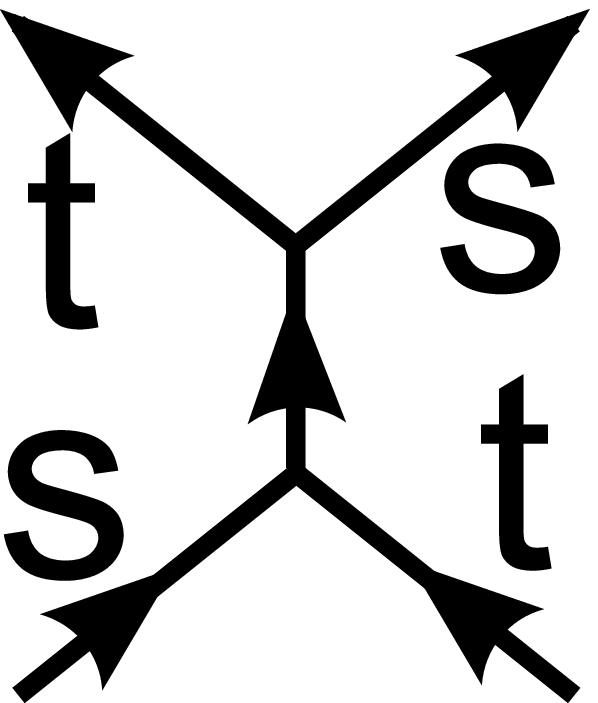}} \right|\Phi \bigg \rangle \omega_\alpha(t) 
= e^{i\theta_{\alpha \beta}} \bar{\omega}_\beta(s) 
\left\<\raisebox{-0.16in}{\includegraphics[height=0.4in]{wpwp1.eps}} \right|\Phi \bigg \rangle.
\end{equation}
We conclude that
\begin{eqnarray}
e^{i\theta_{\alpha \beta}} &=& \frac{\omega_\alpha(t)}{\bar{\omega}_\beta(s)}  \nonumber \\
&=& \frac{\mathbf{w}_\alpha(t)}{\bar{\mathbf{w}}_\beta(s)} \nonumber \\
&=& \mathbf{w}_\beta(s) \mathbf{w}_\alpha(t).
\label{sst}
\end{eqnarray}
As a consistency check, note that when $\alpha =\beta$, the above expression simplifies to
\begin{equation}
e^{i\theta_{\alpha \alpha}} = \mathbf{w}_\alpha(s)^2 = e^{2i\theta_\alpha}
\label{sst1}
\end{equation}
as it should.

The above formulas (\ref{theta}), (\ref{sst}) allow us to compute the complete set of quasiparticle braiding statistics given
any solution to the self-consistency conditions, $(F(a,b,c), d_a, \alpha(a,b), \gamma_a)$. The computation requires several steps. First,
we need to compute $c_s(a,b)$ from $F(a,b,c)$. Second, we need to solve Eq. (\ref{seq}). Third, we need to define our labeling scheme 
by specifying which solutions we label by $\mathbf{w}_{(s,0)}$. Finally, after taking these steps, we can obtain the exchange statistics 
of $\alpha = (s,m)$ and the mutual statistics between $\alpha = (s,m)$ and $\beta = (t,n)$ using Eqs. (\ref{theta}) and (\ref{sst}). 

In sections \ref{zn},\ref{zksection}, we will follow the above procedure to find the complete quasiparticle braiding statistics for every abelian string-net model.
However, before proceeding to the general case, it is illuminating to consider a few special cases where the quasiparticle braiding statistics 
are particularly simple. We begin with the exchange statistics of the charge excitations $(0,m)$. From equation (\ref{theta}) we can see that 
the exchange statistics of $\alpha = (0,m)$ is given by
\begin{equation}
\exp(i\theta_{(0,m)}) = \mathbf{w}_{(0,m)}(0) = \rho_m(0) = 1
\label{fluxex}
\end{equation}
where the second equality follows from equation (\ref{fluxdef}). We conclude that the charge excitations
are all \emph{bosons}. 

It is also simple to derive the mutual statistics between two charge excitations $\alpha = (0,m)$, $\beta = (0,n)$.
From equation (\ref{sst}) we have
\begin{equation}
\exp(i\theta_{(0,m) (0,n)}) = \mathbf{w}_{(0,n)}(0) \mathbf{w}_{(0,m)}(0)  = 1
\label{fluxmut}
\end{equation}
implying that the charge excitations are \emph{mutually bosonic}. 

Finally, it is easy to compute the mutual statistics between a charge excitation $\alpha = (0,m)$ and a 
flux excitation $\beta = (t,0)$:
\begin{equation}
\exp(i\theta_{(0,m) (t,0)}) = \mathbf{w}_{(t,0)}(0) \mathbf{w}_{(0,m)}(t) = \rho_m(t)
\label{fluxchargemut}
\end{equation}
where the second equality follows from equation (\ref{fluxdef}). The above result (\ref{fluxchargemut}) is exactly equal to the Aharonov-Bohm phase associated
with braiding a charge $m$ around a flux $t$ in conventional lattice gauge theory. This similarity is not a coincidence, since these models are in fact realizations of
the topological lattice gauge theories of Dijkgraaf and Witten\cite{DijkgraafWitten}, as discussed in the introduction.

The results (\ref{fluxex}), (\ref{fluxmut}), and (\ref{fluxchargemut}) reveal an important feature of the quasiparticle statistics in abelian string-net
models: we can see that both the mutual/exchange statistics of the charges
(\ref{fluxex}), (\ref{fluxmut}), and the mutual statistics between fluxes and charges (\ref{fluxchargemut}), depend only on the group $G$. In other words, these
quantities depend only on the branching rules, not on the parameters $(F(a,b,c), d_a, \alpha(a,b), \gamma_a)$ that describe the more detailed structure of the
string-net condensate. Hence, these quantities are not useful for distinguishing different types of string-net condensates with the same branching rules; to 
make these distinctions, we need to examine the exchange/mutual statistics of the \emph{flux} excitations $(s,0)$. 

\subsection{Expressing braiding statistics in terms of \texorpdfstring{$F(a,b,c)$}{F(a,b,c)}}
In this section, we obtain formulas that directly relate the quasiparticle braiding statistics to $F(a,b,c)$. In some cases, these
formulas may provide the easiest approach for computing the quasiparticle statistics. In other cases, it may be
more convenient to use the more indirect approach outlined below equation (\ref{sst1}).
 
To begin, we derive a formula for the exchange statistics
of a pure flux quasiparticle $\alpha = (s,0)$. To obtain the exchange statistics of $\alpha = (s,0)$, we need to find $\mathbf{w}_\alpha(s)$. 
Following the same steps as in the derivation of Eq. (\ref{solseq}), it is easy to show that
\begin{equation}
(\mathbf{w}_\alpha(s))^{p} = \prod_{y=0}^{p-1} c_s(s,ys) = \prod_{y=0}^{p-1} F(s,ys,s) 
\end{equation}
where $p$ is the smallest positive integer such that $p \cdot s = 0$. Applying the formula for exchange statistics (\ref{theta}), we conclude that
\begin{equation}
\exp(pi\theta_{(s,0)}) = \prod_{y=0}^{p-1} F(s,ys,s) \label{ex0}.
\end{equation}

Naively, one might think that the above result (\ref{ex0}) only provides \emph{partial} information about the exchange statistics since it only tells us $\theta_{(s,0)}$ modulo $2\pi/p$. 
However this partial information is the best we can hope for, unless we specify a particular convention for which quasiparticle is labeled by
$(s,0)$. To see this, note that Eq. (\ref{gendef}) implies that
\begin{equation}
\exp(i\theta_{(s,m)}) = \exp(i\theta_{(s,0)}) \cdot \rho_m(s) \label{ex0b}
\end{equation}
where $\rho_m$ is the $1D$ representation corresponding to $m$. It is not hard to show that as $m$ runs over the set of $1D$ representations,
$\rho_m(s)$ runs over the set of $p$th roots of unity, $e^{2\pi i k/p}$. Therefore if we change our labeling convention so that $(s,0) \rightarrow (s,m)$,
the exchange statistics $\theta_{(s,0)}$ can shift by any multiple of $2\pi/p$. 

In light of this observation, we can see that equations (\ref{ex0}) and (\ref{ex0b}) actually give us complete information about 
the exchange statistics of every quasiparticle excitation. To compute the exchange statistics, we simply set $e^{i\theta_{(s,0)}}$ equal to
one of the $p$th roots of $\prod_{y=0}^{p-1} F(s,ys,s)$ for each $s$. We can choose whatever $p$th root of unity that we like -- different choices correspond
to different definitions of what constitutes a ``pure'' flux $(s,0)$. Once we have $e^{i\theta_{(s,0)}}$, we can then compute $e^{i\theta_{(s,m)}}$ for any $m$ using Eq. (\ref{ex0b}).

Like the exchange statistics, it is also possible to express the mutual statistics $\theta_{\alpha \beta}$ directly in terms of $F(i,j,k)$. The simplest way to
do this is to use the relationship between mutual statistics and exchange statistics, namely
\begin{equation}
\theta_{\alpha \beta} = \theta_\gamma - \theta_\alpha - \theta_\beta
\label{thetast}
\end{equation}
where $\gamma$ is the quasiparticle obtained by fusing $\alpha$ with $\beta$.

\section{Quasiparticle statistics of general abelian string-net models} \label{qpsection2}
So far we have derived a general framework for constructing abelian string-net models and computing their quasiparticle braiding statistics. 
We now use this machinery to derive the quasiparticle statistics of all possible abelian string-net models. First, in section \ref{zn}, we find the quasiparticle
statistics for the simplest class of models, namely the models corresponding to the group $G = \mathbb{Z}_N$. Then, in section \ref{zksection}, we will analyze the general case 
$G = \mathbb{Z}_{N_1} \times ... \times \mathbb{Z}_{N_k}$. 

\subsection{\texorpdfstring{$\mathbb{Z}_N$}{ZN} string-net models \label{zn}} 
We begin by analyzing the abelian string-net models with group $G = \mathbb{Z}_N$. We note that a similar analysis of $\mathbb{Z}_N$ string-net models, using a different formalism,
was given in Ref. [\onlinecite{LanWen13}]. Also an analysis of $\mathbb{Z}_N$ string-net models with parity invariance was given in Ref. [\onlinecite{HungWan12}].  

To begin, we recall that in the $\mathbb{Z}_N$ models, the string type $a$ can be thought of as group element $a \in \mathbb{Z}_N$
or equivalently integers $a \in \{0,1,...,N-1\}$. The dual string type is defined by $a^{*}=-a$ (mod $N$), while the allowed branchings
are the triplets $(a,b,c)$ that satisfy $a+b+c=0$ (mod $N$). 
 
The first step is to construct all possible string-net models with the above $\mathbb{Z}_N$ structure. To do this, we need to find all solutions to the 
self-consistency equations (\ref{selfconseq}), (\ref{unit}). In fact, it
suffices to find all solutions to (\ref{selfconseq}), (\ref{unit}) \emph{up to} gauge transformations, since gauge equivalent solutions 
correspond to string-net models in the same quantum phase. 

Given this freedom to make gauge transformations, we can assume without loss of generality that $d_a \equiv 1$
since we can clearly gauge transform any solution to (\ref{dcons}) to $d_a = 1$ using a $g$-gauge transformation (\ref{gauge2}) with $g(a) = d_a^{-1}$.
It is then clear that $\gamma_a, \alpha(a,b)$ are fully determined by $F$ according to (\ref{gammaF}), (\ref{alphaF}):
\begin{align}
d_a &= 1 \ \ \ , \ \ \ \gamma_a = F(a^*,a,a^*) \ \ , \nonumber \\
\alpha(a,b) &= F(a,b,(a+b)^*) \gamma_{a+b}.
\label{dalphagamma}
\end{align}
The only remaining parameter is $F$, so our problem reduces to finding all $\{F(a,b,c)\}$ satisfying
\begin{align}
        F(a+b,c,d) & F(a,b,c+d) = \label{cocycl} \\ 
	&F(a,b,c) F(a,b+c,d) F(b,c,d),    \nonumber \\
        F(a,b,c) &= 1 \ \ \text{ if }a\text{ or }b\text{ or }c=0 \label{F0repeat} 
\end{align}
modulo the gauge transformation
\begin{equation}
\tilde{F}(a,b,c) = F(a,b,c) \cdot \frac{f(a,b+c) f(b,c)}{f(a,b) f(a+b,c)} \label{cobdry} 
\end{equation}
where $|f(a,b)| = 1$.

Fortunately, the problem of finding all solutions to equation (\ref{cocycl}), modulo the gauge transformations (\ref{cobdry}) is a well-known
mathematical question from the subject of group cohomology\cite{PropitiusThesis,ChenGuWenSPT,MooreSeiberg}. In the context of group cohomology, the
solutions to equation (\ref{cocycl}) are known as ``$3$-cocycles'', while the ratio $\frac{f(a,b+c) f(b,c)}{f(a,b) f(a+b,c)}$ is
known as a ``$3$-coboundary.'' The question of finding all cocycles modulo coboundaries is exactly the problem of computing the cohomology
group $H^{3}(G,U(1))$. 

In this section we are interested in the special case $G = \mathbb{Z}_N$. In this case, it is known that $H^3(\mathbb{Z}_N, U(1)) = \mathbb{Z}_N$. In particular, it is
known that there are $N$ distinct solutions to (\ref{cocycl}) up to gauge transformations. In addition, an explicit form for the $N$ distinct 
solutions is known\cite{PropitiusThesis,MooreSeiberg}:
\begin{equation}
        F(a,b,c) = e^{2\pi i \frac{p a}{N^{2}} (b+c-[b+c])}  \label{sol}.
\end{equation}
Here, the integer parameter $p = 0\dots N-1$ labels the different solutions. The arguments 
$a,b,c$ can take values in the range $0,1,...,N-1$ and the square bracket $[b+c]$ denotes $b+c$ (mod $N$) with values also taken in
the range $0,1,...,N-1$. Notice that all of these solutions satisfy the additional constraint (\ref{F0repeat}). 

For each of the above $N$ solutions, we can construct a corresponding string-net model, namely the exactly soluble lattice Hamiltonian
defined in Eq. (\ref{h}). Our next task is to derive the topological properties of these lattice models. In particular, we would like
to determine the braiding statistics of the quasiparticle excitations in these models. To this end, let us recall that  
for a general abelian group $G$, the corresponding string-net model has $|G|^2$ topologically distinct quasiparticle excitations. These 
excitations can be labeled by ordered pairs $\alpha = (s,m)$ where $s$ runs over the group elements $s \in G$, and $m$ runs over the 
$1D$ representations of $G$. In the case of $G = \mathbb{Z}_N$, the group elements $s \in G$ can be parameterized by integers $s = 0,1,...,N-1$ and the 
$1D$ representations can also be parameterized by integers $m = 0,1,..., N-1$, where the representation $\rho_m: G \rightarrow \mathbb{C}$ is 
defined by
\begin{equation*}
\rho_m(a) = e^{2\pi i \frac{m a}{N}}.
\end{equation*}

We now compute the braiding statistics of the quasiparticle excitations $\alpha = (s,m)$. We follow the approach outlined
below Eq. (\ref{sst}). First, we compute $c_s(a,b)$:
\begin{equation*}
c_s(a,b) = \frac{F(a,b,s) F(s,a,b)}{F(a,s,b)} = e^{2\pi i \frac{p s}{N^{2}} (a+b-[a+b])}.
\end{equation*}
Next, we solve Eq. (\ref{seq}) for each $s$. By inspection, we can see that one solution of Eq. (\ref{seq}) is given by 
$\mathbf{w}(a) = \exp(2 \pi i \frac{p s a}{N^2})$. We choose the convention where the above solution is labeled by 
$\alpha = (s,0)$, i.e., we define
\begin{equation*}
\mathbf{w}_{(s,0)}(a) = e^{2\pi i \frac{p s a}{N^2}}.
\end{equation*}
Then, by Eq. (\ref{gendef}) we have
\begin{equation}
\mathbf{w}_{(s,m)}(a) = e^{2\pi i \left(\frac{p s a}{N^2} + \frac{m a}{N} \right)}.
\label{wzn}
\end{equation}

We are now ready to derive the exchange statistics and braiding statistics of the quasiparticle excitations. Substituting
(\ref{wzn}) into (\ref{theta}), we find that the exchange statistics of $\alpha = (s,m)$ is given by
\begin{equation}
\theta_{(s,m)}= 2\pi \left(\frac{ps^2}{N^2}+\frac{ms}{N} \right).
\label{exz}
\end{equation}
Evidently there are two contributions to the exchange statistics of $(s,m)$ quasiparticle: the first term can be interpreted as coming
from flux-flux exchange statistics while the second term can be thought of charge-flux Aharonov-Bohm phase.

Similarly, substituting (\ref{wzn}) into (\ref{sst}), we find that the mutual statistics of $\alpha = (s,m)$ and $\beta = (t,n)$ is
\begin{equation}
\theta_{(s,m) (t,n)}= 2\pi \left(\frac{2 p st}{N^2}+\frac{ns + mt}{N} \right).
\label{mutz}
\end{equation}
Again, we can see that there are two contributions to the statistics: one coming from the flux-flux statistics and the
other from moving $s$ and $t$ fluxes around $n$ and $m$ charges respectively.

\subsection{\texorpdfstring{$\mathbb{Z}_{N_{1}}\times \dots \times \mathbb{Z}_{N_{k}}$}{ZN1 times ... times ZNk} string-net models \label{zksection}}
We now generalize the discussion to abelian string-net models with group $G = \mathbb{Z}_{N_{1}}\times\dots\times \mathbb{Z}_{N_{k}}$. Since every finite
abelian group is isomorphic to a direct product of cyclic groups, this case is sufficiently general to cover all abelian string-net models.

To begin, let us recall the structure of these models. In these models,
the strings are labeled by group elements $a \in \mathbb{Z}_{N_{1}}\times\dots\times \mathbb{Z}_{N_{k}}$. Equivalently, the string types can be
parameterized by $k$-component integer vectors $a=(a_{1},a_{2},...,a_{k}) $ where $0 \leq a_i \leq N_i - 1$. 
The dual string is defined by $a^{*}_{i}=-a_{i}$ (mod $N_{i}$) for each $i$, while the allowed branchings are the
triplets $(a, b, c)$ that satisfy $a_i + b_i + c_i = 0$ (mod $N_{i}$) for each $i$.

The next step is to find all possible $\mathbb{Z}_{N_1} \times \dots \times \mathbb{Z}_{N_k}$ string-net models. As in the $\mathbb{Z}_N$ case discussed above, the
latter problem reduces to finding all $F(a,b,c)$ satisfying (\ref{cocycl}), (\ref{F0repeat}) modulo the gauge transformation (\ref{cobdry}).
This problem is closely related to the problem of computing the cohomology group $H^{3}(\prod_{i} \mathbb{Z}_{N_{i}},U(1))$. This cohomology
group has been calculated previously and is given by\cite{PropitiusThesis,MooreSeiberg}
\begin{eqnarray}
	H^{3} \left(\prod_{i} \mathbb{Z}_{N_{i}},U(1) \right)
	&=& \prod_i \mathbb{Z}_{N_i} \cdot \prod_{i < j} \mathbb{Z}_{(N_i, N_j)} \nonumber \\
	&\cdot& \prod_{i < j < k} \mathbb{Z}_{(N_i, N_j, N_k)}
\end{eqnarray}
where $(N_i, N_j)$ denotes the greatest common divisor of $N_i$ and $N_j$, and similarly for $(N_i, N_j, N_k)$.

Now, as explained in section \ref{stropsec}, it is important to distinguish between two types of solutions to (\ref{cocycl}), (\ref{F0repeat}):
solutions with $c_s(a,b) = c_s(b,a)$, and solutions with $c_s(a,b) \neq c_s(b,a)$ where $c_s$ is defined as in Eq. (\ref{cs}). In the former case, the corresponding string-net
model has only abelian quasiparticle excitations, while in the latter case, the model supports non-abelian excitations.\cite{PropitiusThesis} In this
paper, we focus entirely on models with abelian excitations, so we will restrict ourselves to solutions that satisfy 
$c_s(a,b) = c_s(b,a)$. It is known\cite{PropitiusThesis} that the solutions with this property are classified by a subgroup of the cohomology group $H^3(\prod_i \mathbb{Z}_{N_i}, U(1))$, namely $\prod_i \mathbb{Z}_{N_i} \prod_{i < j} \mathbb{Z}_{(N_i, N_j)}$. In particular, the total number of gauge inequivalent solutions to (\ref{cocycl}) is
\begin{equation}
\mathcal{N} = \prod N_i \cdot \prod_{i < j} (N_i, N_j).
\end{equation}

An explicit form for the $\mathcal{N}$ distinct solutions is known:\cite{PropitiusThesis,MooreSeiberg}
\begin{equation}
F(a,b,c) =e^{2\pi i a^{T}\mathbf{N}^{-1} \mathbf{P}\mathbf{N}^{-1}(b+c-[b+c]) }.
\label{gensol}
\end{equation}
Here, the arguments $a,b,c$ are $k$ component integer vectors with $0 \leq a_i \leq N_i -1$ for each $i$, and the square bracket $[b+c]$ denotes a vector whose $i$-th component is 
$(a_i+b_i)$ (mod $N_{i}$) with values taken in the range $0,1,...,N_i -1$. The matrix $\mathbf{N}$ is the $k \times k$ diagonal matrix $\mathbf{N} = diag(N_1,...,N_k)$ and $\mathbf{P}$ is a $k \times k$ 
upper-triangular integer matrix that parameterizes the different solutions. The diagonal elements of $\mathbf{P}$ are restricted to the range $0 \leq \mathbf{P}_{ii} \leq N_i - 1$, while the elements above the diagonal are restricted to the range $0 \leq \mathbf{P}_{ij} \leq (N_i, N_j) - 1$. As in the $\mathbb{Z}_N$ case, we can see that the $F(a,b,c)$ not only obey (\ref{cocycl}), but also satisfy the additional constraint (\ref{F0repeat}).

The $\mathcal{N}$ solutions defined in (\ref{gensol}) can be used to construct $\mathcal{N}$ different lattice models. Our next task is to determine the braiding statistics of the quasiparticle excitations in these models. These excitations can be labeled by ordered pairs $\alpha = (s,m)$ where $s$ runs over the group elements $s \in G$, and $m$ runs over the 
$1D$ representations of $G$. In the case of $G = \mathbb{Z}_{N_1} \times \dots \times \mathbb{Z}_{N_k}$, the group elements $s \in G$ can be parameterized by $k$ component integer vectors
$s^T=(s_1,s_2,\dots,s_k)$ with $0 \leq s_i \leq N_i -1$, and the $1D$ representations can also be parameterized by $k$ component integer vectors $m^T=(m_1,m_s,\dots,m_k)$ with $0 \leq m_i \leq N_i - 1$. In this parameterization,
the representation $\rho_m: G \rightarrow \mathbb{C}$ is defined by
\begin{equation*}
\rho_m(b) = e^{2 \pi i m^T \mathbf{N}^{-1} b}.
\end{equation*}

To find braiding statistics of the quasiparticle excitations $\alpha = (s,m)$, we follow the same steps as in the $\mathbb{Z}_N$ case. First, we compute $c_s(a,b)$:
\begin{equation*}
c_s(a,b) = e^{2\pi i s^{T}\mathbf{N}^{-1} \mathbf{P}\mathbf{N}^{-1}(a+b-[a+b]) }.
\end{equation*}
Next, we solve Eq. (\ref{seq}) for each $s$. By inspection, we can see that one solution of Eq. (\ref{seq}) is given by 
$\mathbf{w}(a) = \exp(2\pi i s^T\mathbf{N}^{-1} \mathbf{P}\mathbf{N}^{-1} a)$.
We choose the convention where the above solution is labeled by $\alpha = (s,0)$, i.e., we define
\begin{equation*}
\mathbf{w}_{(s,0)}(a) = e^{2\pi i s^T \mathbf{N}^{-1} \mathbf{P}\mathbf{N}^{-1} a}.
\end{equation*}
Then, by Eq. (\ref{gendef}) we have
\begin{equation}
\mathbf{w}_{(s,m)}(a) = e^{2\pi i (s^T \mathbf{N}^{-1} \mathbf{P}\mathbf{N}^{-1} a + m^T \mathbf{N}^{-1} a)}.
\label{wzngen}
\end{equation}

We are now ready to derive the exchange statistics and braiding statistics of the quasiparticle excitations. Substituting
(\ref{wzngen}) into (\ref{theta}), we find that the exchange statistics of $\alpha = (s,m)$ is given by
\begin{equation}
\theta_{(s,m)} =2\pi (s^T \mathbf{N}^{-1} \mathbf{P}\mathbf{N}^{-1} s + m^T \mathbf{N}^{-1} s).
\label{exzk}
\end{equation}
Similarly, substituting (\ref{wzngen}) into (\ref{sst}), we find that the mutual statistics of $\alpha = (s,m)$ and $\beta = (t,n)$ is
\begin{eqnarray}
\theta _{( s,m) (t,n)} &=& 2\pi (s^T \mathbf{N}^{-1} (\mathbf{P} + \mathbf{P}^T) \mathbf{N}^{-1} t \nonumber \\
&+& n^T \mathbf{N}^{-1} s + m^T \mathbf{N}^{-1} t).
\label{mutzk}
\end{eqnarray}

\section{Chern-Simons description} \label{cssec}
So far we have derived the basic topological properties of abelian string-net models, including their quasiparticle braiding statistics and
their ground state degeneracy on a torus. It is natural to wonder: can these properties be described by some low energy effective theory? In this section, we
find a field theoretic description in terms of multicomponent $U(1)$ Chern-Simons theory.

Before explaining this effective field theory description, we first (briefly) review the basic formalism of $U(1)$ Chern-Simons theory.\cite{WenBook,WenReview,WenKmatrix}
It is believed that the topological properties of any abelian gapped many-body system can be described by some $M$ component $U(1)$ Chern-Simons
theory of the form
\begin{equation}
L=\frac{K_{IJ} }{4\pi }\varepsilon^{\mu \nu \lambda }a_{I\mu }\partial _{\nu }a_{J\lambda } \label{k1}
\end{equation}
where $K$ is an $M \times M$ symmetric, non-degenerate integer matrix. In this formalism, the different quasiparticle excitations are described by 
coupling $L$ to bosonic particles that carry integer gauge charge $q_I$ for
each gauge field $a_I$. Hence, the quasiparticles are parameterized by $M$ component integer vectors $q$.
The mutual statistics between two quasiparticles $q$ and $q'$ is
\begin{equation}
\theta_{qq'} = 2\pi q^T K^{-1} q'
\end{equation}
while the exchange statistics of quasiparticle $q$ is $\theta_q = \frac{1}{2} \theta_{qq}$. Two quasiparticles $q$ and $q'$ are
said to be ``topologically equivalent'' if $q - q' = K \Lambda$ where $\Lambda$ is an integer $M$ component vector. The number
of topologically distinct quasiparticles is given by $|\det{K}|$, as is the ground state degeneracy on the torus. 

With this background, we are now ready to give a Chern-Simons description for the abelian string-net models. Recall
that in the previous section, we explicitly constructed all possible abelian string-net models. These models are parameterized
by two pieces of data: (1) a finite abelian group $G = \mathbb{Z}_{N_1} \times \dots \mathbb{Z}_{N_k}$, and (2) a $k \times k$ upper triangular integer matrix $\mathbf{P}$ with
$0 \leq \mathbf{P}_{ii} \leq N_i - 1$ and $0 \leq \mathbf{P}_{ij} \leq (N_i, N_j) - 1$. Our basic task is to find a $K$-matrix for each choice of $G$ and $\mathbf{P}$
that captures the topological properties of the associated abelian string-net model. We will argue that the following $K$-matrix does the job:
\begin{equation}
K = \bpm \mathbf{0} & \mathbf{N} \\ \mathbf{N} & \mathbf{\tilde{P}} \epm.
\label{kmat2}
\end{equation}
Here, $\mathbf{N}$ is a $k \times k$ diagonal matrix, $\mathbf{N} = diag(N_1,...,N_k)$, and $\mathbf{\tilde{P}}$ is a $k \times k$ symmetric, integer matrix defined by $\mathbf{\tilde{P}} = -\mathbf{P} - \mathbf{P}^T$. The '$\mathbf{0}$' is meant to denote $k \times k$ matrix with vanishing elements. 
Thus, $K$ has dimension $2k \times 2k$. 

To show that the abelian string-net models (\ref{gensol}) are described by the above Chern-Simons theory (\ref{kmat2}), 
we need to establish a one-to-one correspondence between the quasiparticle excitations of each system. 
In the abelian string-net model, the quasiparticle excitations are parameterized by ordered pairs
$(s,m)$, where $s,m$ are $k$ component integer vectors with $0 \leq s_i, m_i \leq N_i - 1$ for each $i$. On the other hand, in the Chern-Simons theory (\ref{kmat2}), 
the topologically distinct quasiparticles are parameterized by $2k$ component integer vectors $q$ (modulo $K \mathbb{Z}^{2k}$). Thus, we need a one-to-one correspondence between 
ordered pairs $(s,m)$ and $2k$ component integer vectors $q$. It is easy to see that the following mapping does the job: 
\begin{equation}
q = \bpm s \\ m \epm \ \ (\text{modulo }K \mathbb{Z}^{2k}).
\end{equation}
In fact, not only is this mapping a one-to-one correspondence, but it also preserves exchange statistics: i.e. the exchange statistics of $q$ in the
Chern-Simons theory is the same as the exchange statistics of $(s,m)$ in the abelian string-net model. To see this, note that
\begin{eqnarray}
\theta_{q} &=& \pi q^T K^{-1} q \nonumber \\
&=& \pi \bpm s & m \epm \cdot \bpm -\mathbf{N}^{-1} \mathbf{\tilde{P}} \mathbf{N}^{-1} & \mathbf{N}^{-1} \\ \mathbf{N}^{-1} & \mathbf{0} \epm \cdot \bpm s \\ m \epm \nonumber \\
&=& \pi (-s^T \mathbf{N}^{-1} \mathbf{\tilde{P}} \mathbf{N}^{-1} s + 2 s^T \mathbf{N}^{-1} m) \nonumber \\
&=& \theta_{(s,m)}
\end{eqnarray}
where the last line follows from (\ref{exzk}). Likewise, the mapping preserves mutual statistics: letting $q = \bpm s \\ m \epm$ and $q' = \bpm t \\ n \epm$, we have
\begin{eqnarray}
\theta_{qq'} &=& 2\pi q^T K^{-1} q' \nonumber \\
&=& 2\pi \bpm s & m \epm \cdot \bpm -\mathbf{N}^{-1} \mathbf{\tilde{P}} \mathbf{N}^{-1} & \mathbf{N}^{-1} \\ \mathbf{N}^{-1} & \mathbf{0} \epm \cdot \bpm t \\ n \epm \nonumber \\
&=& 2\pi (-s^T \mathbf{N}^{-1} \mathbf{\tilde{P}} \mathbf{N}^{-1} t +  s^T \mathbf{N}^{-1} n + m^T \mathbf{N}^{-1} t) \nonumber \\
&=& \theta_{(s,m) (t,n)}
\end{eqnarray}
where the last line follows from (\ref{mutzk}). We conclude that the quasiparticle braiding
statistics of the Chern-Simons theory (\ref{kmat2}) exactly agree with the statistics of the abelian string-net model.

In addition to quasiparticle statistics, there is one other topological quantity we need to compare to be sure that the string-net models are described by the Chern-Simons theory (\ref{kmat2}).\cite{KapustinTopbc}
This quantity is the thermal Hall conductance\cite{KaneFisherThermal} (also known as the ``chiral central charge''\cite{KitaevHoneycomb}).
In general, the thermal Hall conductance of a multicomponent $U(1)$ Chern-Simons theory is given by the signature of the $K$ matrix.\cite{KaneFisherThermal} 
It is easy to check that the signature of (\ref{kmat2}) is $0$, so we conclude that (\ref{kmat2}) has vanishing thermal Hall conductance. At the same
time, we know that the abelian string-net model has vanishing thermal Hall conductance, since the Hamiltonian is a sum of commuting projectors.\textsuperscript{6}
Thus, the thermal Hall conductances also match.

In summary, we have shown that the Chern-Simons theory (\ref{kmat2}) correctly captures the quasiparticle statistics and 
thermal Hall conductance of the abelian string-net model (\ref{gensol}). In this sense, (\ref{kmat2}) provides a good low energy effective theory for the string-net model
(\ref{gensol}).

\section{Characterizing the phases that can be realized by string-net models} \label{phases}
Having systematically constructed all the abelian string-net models, we are faced with an important question: 
which topological phases can be realized by abelian string-net models, and which phases cannot be realized? In this 
section, we answer this question in three different ways: we describe three different criteria for characterizing 
the set of topological phases that can be realized by abelian string-net models. The first criterion is based on braiding statistics, 
the second based on the $K$-matrix formalism, and the third based on properties of the edge.

\subsection{Braiding statistics and \texorpdfstring{$K$}{K}-matrix characterizations}\label{kmatchar}
Our first criterion for determining whether a phase is realizable is based on braiding statistics and thermal Hall conductance. 
Specifically, we will show that an abelian topological phase is realizable if and only if (i) it has
a vanishing thermal Hall conductance, and (ii) it has a least one \emph{Lagrangian subgroup}. Here, a ``Lagrangian subgroup''\cite{KapustinTopbc} $\mathcal{M}$ is a subset of quasiparticles with two properties. First, all the quasiparticles in $\mathcal{M}$ are bosons and have trivial mutual statistics with one another:
$e^{i\theta_m} = e^{i\theta_{mm'}} = 1$ for all $m, m' \in \mathcal{M}$. Second, if $l$ is a quasiparticle that is not contained in $\mathcal{M}$, then it has nontrivial
mutual statistics with at least one quasiparticle $m \in \mathcal{M}$:  $e^{i\theta_{lm}} \neq 1$.

Before deriving the above result, we find it useful to explain another criterion for characterizing the set of realizable phases. We will then show that the two criteria are equivalent and prove them both simultaneously. The second criterion is based on the Chern-Simons $K$-matrix formalism. Here, the starting point for our analysis is the statement that every abelian topological phase can be described by a multicomponent $U(1)$ Chern-Simons theory with a symmetric, non-degenerate integer matrix $K$. The only constraint on $K$ is that it must have even elements on the diagonal if the corresponding topological phase is built out of bosons, as we assume here. 

Given this Chern-Simons framework, we should be able to find a criterion that tells us which $K$-matrices can be realized by abelian string-net models and which $K$-matrices cannot be realized.
Naively, one might think that the answer to the question follows from equation (\ref{kmat2}): one might think that the realizable $K$-matrices are exactly the matrices of the form (\ref{kmat2}). However, this is not quite correct. The problem is that the correspondence between $K$-matrices and topological phases is not one-to-one: multiple $K$-matrices can correspond to the same topological phase. For example, if $K, K'$ differ by a change-of-basis transformation of the form $K' = W^T K W$, where $W$ is an integer matrix with determinant $\pm 1$, then $K, K'$ describe the same topological 
phase. \footnote{One way to see that $K,K'$ describe the same phase is to note that the corresponding Chern-Simons
theories can be mapped onto one another by a field redefinition $a_I' = W_{IJ} a_j$.}
Given this equivalence between different $K$-matrices, we have to work a little bit harder to derive a criterion that distinguishes the realizable $K$-matrices from the non-realizable $K$-matrices. We will show that the following criterion does the job: a $K$-matrix is realizable if and only if it has even dimension, $2k \times 2k$, and there exist $k$ linearly independent integer vectors $\Lambda_1,\dots,\Lambda_k$ such that
\begin{equation}
\Lambda_i^T K \Lambda_j = 0 \label{null}
\end{equation}
for all $i,j$. The $\Lambda_i$ are known as ``null vectors.''

We prove these results in three steps. First, we show that the braiding statistics and $K$-matrix criteria are equivalent to one another. Next, we show that the braiding statistics criterion is \emph{necessary} for a phase to be realizable. Finally, we show that the $K$-matrix criteria are \emph{sufficient} for a phase to be realizable. Once we establish these three claims, both results follow immediately.

The first claim --- i.e. the fact that the braiding statistics and $K$-matrix criteria are equivalent to one another --- follows immediately from Ref. [\onlinecite{LevinProtedge}]. Indeed, that work showed explicitly that a $K$-matrix satisfies (\ref{null}) if and only if it satisfies conditions (i), (ii). Next, consider the second claim, namely the statement that the braiding statistics criteria are necessary for a phase to be realizable. To establish this result, it suffices to show that every string-net model obeys conditions (i) and (ii). This can be done straightforwardly: indeed, we know that all string-net models have a vanishing thermal Hall conductance, since their Hamiltonians are sums of commuting projectors.\textsuperscript{6} Also, it is easy to see that the set of charge excitations $\{(0,m)\}$ always constitutes a Lagrangian subgroup (see Eqs. (\ref{fluxex}), (\ref{fluxmut}), and (\ref{fluxchargemut})).

To complete the argument, we now show that the $K$-matrix criterion is sufficient for a phase to be realizable. In other words, we show that every $K$ matrix that obeys (\ref{null}) can be realized by an appropriate abelian string-net model. Our strategy will be to show that every $K$-matrix that obeys (\ref{null}) can be written in the form (\ref{kmat2}) after making an appropriate change-of-basis transformation $K \rightarrow W^T K W$, where $W$ is an integer matrix with determinant $\pm 1$. To this end, we observe that any $K$-matrix that obeys (\ref{null}) can be written in the form
\begin{equation*}
K = \bpm \mathbf{0} & \mathbf{A} \\ \mathbf{A}^T & \mathbf{B} \epm
\end{equation*}
after a suitable change of basis (see appendix A1b  of Ref. [\onlinecite{WangLevin13}] for a detailed proof). Next, we use the Smith normal form to find matrices $\mathbf{S}, \mathbf{T}$ with unit determinant such that $\mathbf{S} \mathbf{A} \mathbf{T} = \mathbf{N}$, where $\mathbf{N}$ is a diagonal matrix, $\mathbf{N} = diag(N_1,\dots,N_k)$. We then make the change of basis $K \rightarrow K' = W_1^T K W_1$, where
\begin{equation*}
W_1 = \bpm \mathbf{S}^T & \mathbf{0} \\ \mathbf{0} & \mathbf{T} \epm.
\end{equation*}
After this change of variables, we have
\begin{equation*}
K' = \bpm \mathbf{0} & \mathbf{N} \\ \mathbf{N} & \mathbf{B'} \epm
\end{equation*}
where $\mathbf{B'} = \mathbf{T}^T \mathbf{B} \mathbf{T}$. To complete the argument, we make one more change of basis: we define $K'' = W_2^T K' W_2$, where
\begin{equation*}
W_2 = \bpm \mathbf{1} & \mathbf{U}^T \\  \mathbf{0} & \mathbf{1} \epm.
\end{equation*}
The result is
\begin{equation}
K'' = \bpm \mathbf{0} & \mathbf{N} \\ \mathbf{N} & \mathbf{\tilde{P}} \epm
\end{equation}
where $\mathbf{\tilde{P}} =  \mathbf{B'} + \mathbf{U} \mathbf{N} + \mathbf{N} \mathbf{U}^T$. It is not hard to see that we can always choose $\mathbf{U}$ so that $\mathbf{\tilde{P}}$ obeys the inequalities $0 \leq -\mathbf{\tilde{P}}_{ii} \leq 2(N_i -1)$ and $0 \leq -\mathbf{\tilde{P}}_{ij} \leq (N_i, N_j) - 1$. This completes the argument.

\subsection{Examples of realizable and non-realizable phases}\label{exreal}
We now describe a few representative examples of realizable and non-realizable phases. A simple example of a realizable phase is the phase corresponding to the $K$-matrix
\begin{equation}
K_1 = \bpm 0 & 2 \\ 2 & 0 \epm.
\end{equation}
To see that $K_1$ is realizable, note that it has a null vector $\Lambda_1^T = \bpm 1 & 0 \epm$ and it has dimension $2 \times 2$, so it satisfies the two conditions from section \ref{kmatchar}. This Chern-Simons theory can equivalently be thought of as $Z_2$ gauge theory.\cite{KouLevinWen}

More generally, it is easy to see that any $K$-matrix of the form $\bpm \mathbf{0} & \mathbf{N} \\ \mathbf{N} & \mathbf{0} \epm$ is realizable where 
$\mathbf{N} = diag(N_1,\dots,N_k)$. Like $K_1$, these Chern-Simons theories are equivalent to discrete gauge theories where the gauge group is $G = \mathbb{Z}_{N_1} \times \dots \times \mathbb{Z}_{N_k}$.

Another example of a realizable phase is:
\begin{equation}
K_2 = \bpm 2 & 0 \\ 0 & -2 \epm.
\end{equation}
Again, it is easy to see that $K_2$ is realizable, since it has a null vector $\Lambda_1^T = (1, 1)$. This phase is an 
example of a ``doubled Chern-Simons theory'': it is a sum of two Chern-Simons theories with opposite chiralities. It is easy
to see that general doubled Chern-Simons theories of the form $\bpm \mathbf{A} & \mathbf{0} \\ \mathbf{0} & - \mathbf{A} \epm$
are also realizable. Here, $\mathbf{A}$ is a symmetric, integer, non degenerate matrix, with even elements on the diagonal.

A third example of a realizable phase is given by
\begin{equation}
K_3 = \bpm 2 & 0 \\ 0 & - 8 \epm.
\end{equation}
Here, an appropriate null vector is given by $\Lambda_1^T = (2, 1)$. On other other hand, an example of a phase that is not realizable is
\begin{equation}
K_4 = \bpm 2 & 0 \\ 0 & -4 \epm.
\end{equation}
One way to see this is to note that $K_4$ does not have any null vectors since the equation $2 x^2 - 4 y^2 = 0$ does not 
have any (nonzero) integer solutions. 

The $K_3, K_4$ examples are interesting because they show that the simplest expectations for which topological phases can 
and cannot be realized by string-net models are incorrect. For example, an optimist might have guessed that the string-net 
models can realize all phases with vanishing thermal Hall conductance: indeed, this scenario is the best one could hope 
for since ground states of Hamiltonians which are sums of commuting projectors always have vanishing thermal Hall 
conductance, as discussed in the introduction. The $K_4$ example disproves this conjecture: $K_4$ has vanishing signature and therefore 
vanishing thermal Hall conductance, yet it apparently cannot be realized by any string-net model. Similarly, a pessimist 
might have guessed that the string-net models can only realize phases that are compatible with time-reversal symmetry since
all the models realized in Ref. [\onlinecite{LevinWenStrnet}] were compatible with time-reversal symmetry. The $K_3$ example disproves this 
conjecture as well. To see this, note that the phase corresponding to $K_3$ is manifestly 
incompatible with time-reversal symmetry: indeed, this phase contains a particle with exchange statistics $\theta = -\pi/8$, 
but it does not contain any particle with the opposite exchange statistics, $+\pi/8$. We conclude that the set of realizable 
phases lies somewhere in between the pessimistic and optimistic scenarios: the string-net models can realize more than just 
the time-reversal symmetric phases, but they cannot realize all the phases with vanishing thermal Hall conductance. 

\subsection{Characterization in terms of gapped edges}\label{edgechar}
In light of the $K_3, K_4$ examples discussed above, it is clear that the difference between realizable and non-realizable 
phases can be quite subtle. Thus, it is natural to wonder whether there is a \emph{simple physical property} that distinguishes 
the two types of phases. Remarkably, such a property does exist: it is possible to show that an abelian topological phase is realizable
if and only if the boundary between the phase and the vacuum can be fully gapped by suitable 
edge interactions. In other words, the realizable phases are exactly the phases that support \emph{gapped edges}.

To derive this edge state characterization of realizable phases, we again make use of the results of Ref. [\onlinecite{LevinProtedge}]. 
In particular Ref. [\onlinecite{LevinProtedge}] proved that 
an abelian phase can support a gapped edge if and only if (i) it has a vanishing thermal Hall conductance and (ii) it has at least 
one Lagrangian group. In other words, the results of Ref. [\onlinecite{LevinProtedge}] imply that the edge state characterization is \emph{equivalent}
to the braiding statistics and $K$-matrix characterizations from section \ref{kmatchar}. Therefore, since we have already established the 
braiding statistics and $K$-matrix results (\ref{null}), the edge state characterization follows immediately.

Before concluding, we would like to point out that it is actually quite easy to see that all realizable phases support a gapped edge: indeed
in appendix \ref{gapedge}, we explicitly construct interactions that gap the boundary of a general abelian string-net model. Thus, we really only need
Ref. [\onlinecite{LevinProtedge}] to prove the converse statement --- i.e. to prove that all abelian phases with a gapped edge are realizable.

The above result has a number of interesting implications. One implication is that we can now show that even if we generalized our string-net construction to other types of branching rules, we would not realize any new abelian topological phases beyond those that we have realized here. To see this, we note that quite generally, \emph{all} string-net models support a gapped edge, independent of their branching rules. Indeed, in appendix \ref{gapedge}, we construct explicit interactions that gap the edge of abelian string-net models. An identical construction can be used to gap the edge of general string-net models. \cite{KitaevKong} This construction means that the only topological phases that are accessible to
string-net models are those that support gapped edges. Thus, since we have already realized every abelian phase with a gapped edge, we cannot possibly realize any new abelian phases, even if we consider other branching rules. (In fact, we believe that other types of branching rules always give \emph{non-abelian} topological phases, though we do not have a proof of this conjecture).

\section{Examples} \label{examples}
In this section, we work out some illustrative examples, namely $\mathbb{Z}_{2}$, $\mathbb{Z}_{3}$, $\mathbb{Z}_4$ and $\mathbb{Z}_2\times \mathbb{Z}_2$ string-net models. 
For each example, we determine whether $\alpha$ and $\gamma$ can be gauged away by $f$ and $g$ gauge transformations.
For the $\mathbb{Z}_2$ and $\mathbb{Z}_4$ models, we find a gauge where all $\gamma = \alpha = 1$. On the other hand, we find that for some of the $\mathbb{Z}_3$ models, 
every gauge has at least one $\alpha \neq 1$, while for some of the $\mathbb{Z}_2 \times \mathbb{Z}_2$ models, every gauge has at least one $\gamma \neq 1$. 
Thus, these examples show that the dot and null string structures are
essential to realizing some, but not all, string-net models.

\subsection{\texorpdfstring{$\mathbb{Z}_{2}$}{Z2} string-net model}
First we specify the Hilbert space of the $\mathbb{Z}_2$ string-net models. 
The string types are labeled by $\mathbb{Z}_2$ group elements $\{0,1\}$. The dual string types are $0^* = 0$, $1^* = 1$
(since $-1 \equiv 1$ (mod $2$)). The branching rules are 
$\{(0,0,0),(0,1,1)\}$ which requires the strings form closed loops. Thus the Hilbert space is the set of all possible closed loops.

Next we construct the Hamiltonians and wave functions for the different $\mathbb{Z}_2$ string-net models. To do this, 
we have to solve the self-consistency conditions (\ref{selfconseq}-\ref{unit}) for $\{F,d,\gamma,\alpha\}$. According to the general solution
given in (\ref{sol}),(\ref{dalphagamma}), there are two distinct solutions, parameterized by an integer $p=0,1$:
\begin{align*}
& F(1,1,1) = (-1)^p, \ \ \gamma_1 = \alpha(1,0) = \alpha(0,1) = (-1)^p, \nonumber \\
& \text{other } F, d,\gamma,\alpha = 1.
\end{align*}
While the above solutions are perfectly sufficient for constructing exactly soluble models, it is
desirable to have solutions with as many $\gamma$'s and $\alpha$'s equal to $1$ as possible, since this will lead to simpler models with more
symmetry and topological invariance. 
Therefore, we now discuss how to simplify these solutions using the $f,g$ gauge transformations (\ref{gauge}, \ref{gauge2}). 

The first step in this simplification process
is to choose a $g$-gauge transformation to make as many $\gamma$'s equal to $1$ as possible. In the above case, we can
make every $\gamma = 1$ using $g(0) = 1, g(1) = (-1)^p$. After this gauge transformation we have
\begin{align}
& F(1,1,1) = (-1)^p, \ \ d_1 = (-1)^p, \nonumber \\
& \text{other } F, d,\gamma,\alpha = 1.
\label{z2sol}
\end{align}
The second step in the simplification process is to use an $f$-gauge transformation to make as many $\alpha$'s equal to $1$ as possible. However, 
in Eq. (\ref{z2sol}), we already have all $\alpha=1$ and thus there is no need for an $f$-gauge transformation. We can see that in the above gauge, 
$\gamma = \alpha = 1$ so null strings and dots are irrelevant. Thus, the resulting solutions (\ref{z2sol}) can be constructed using the original 
formalism of Ref. [\onlinecite{LevinWenStrnet}]. 

With the solutions (\ref{z2sol}) in hand, we can construct the wave functions and Hamiltonian using (\ref{rule1} - \ref{rule3}), (\ref{nullerase} - \ref{rule4'}) and (\ref{h}). For the $p=0$ solution, the wave function is 
\begin{equation}
	\Phi(X)=1 
	\label{z2wf1a}
\end{equation}
for any closed string-net configuration $X$. The corresponding Hamiltonian is the toric code.\cite{KitaevToric,LevinWenStrnet} On the other hand, for the $p=1$ solution, the wave function is
\begin{equation}
	\Phi(X)=(-1)^{\text{loop}(X)}
	\label{z2wf2a}
\end{equation} with loop($X$) meaning the total number of closed loops in the configuration $X$. The corresponding Hamiltonian is the ``doubled semion model.'' \cite{LevinWenStrnet}  

What if we work in other gauge, say $g(0) = 1, g(1) = -(-1)^p$? After this gauge transformation, we will get the solutions
\begin{align*}
	& F(1,1,1)=(-1)^p, \ d_1 = -(-1)^p,  \\
	& \gamma_1 = \alpha(1,0) =\alpha(0,1)=-1,\text{ others} =1.
\end{align*}
In this gauge, we need to keep track of the null strings as well as the dot structure. However, for simplicity let us consider an orientation configuration on the
honeycomb lattice where the orientations chosen so that there are no vertices with $3$ incoming or $3$ outgoing strings. In this case, there are no dots in the lattice
model, and $\alpha$ can be safely ignored. The wave function for the $p=0$ case can then be written as 
\begin{equation}
	\Phi(X)=(-1)^{\text{loop}(X)}(-1)^{(\text{null-in}(X)-\text{null-out}(X))/2}.
	\label{z2wf1b}
\end{equation}
Here ``null-in(X)'' means the total number of vertices along the loops in $X$ in which (1) two oppositely oriented strings and a null string meet one another,
and (2) the null string lies on the \emph{inside} of the loop. The quantity ``null-out(X)'' is similar except it counts vertices where the null string is on
the outside of the loop. Similarly, the wave function for the $p=1$ solution becomes
\begin{equation}
	\Phi(X)=(-1)^{(\text{null-in}(X)-\text{null-out}(X))/2}.
	\label{z2wf2b}
\end{equation}
We can see that the wave functions (\ref{z2wf1b}, \ref{z2wf2b}) are more complicated than they are for the other gauge choice (\ref{z2wf1a}, \ref{z2wf2a}).
This motivates our efforts to gauge away $\gamma$ and $\alpha$ as much as possible. That being said, we would like to mention that for some special orientation
configurations, different gauges can lead to identical lattice wave functions. For example, for the honeycomb lattice with the orientations shown in Fig. \ref{lattice},
the $g$-gauge transformation does not change the wave function at all (see appendix \ref{gaugeapp}).

Independent of the gauge choice, the $p=0,1$ models each have $|G|^2 =4$ distinct types of quasiparticles which are labeled by ordered pairs $(s,m)$ with $s,m=0,1$.
The braiding statistics of these particles can be read off from (\ref{exz},\ref{mutz}). Alternatively, we can describe these braiding statistics by an appropriate $U(1)$
Chern-Simons theory. In particular, according to the general result (\ref{kmat2}), these models are described by $U(1) \times U(1)$ Chern-Simons theories with
$K$-matrix    
\begin{equation*}
K=\left( 
\begin{array}{cc}
0 & 2 \\ 
2 & -2p
\end{array}
\right).
\end{equation*}

\subsection{\texorpdfstring{$\mathbb{Z}_{3}$}{Z3} string-net model}
The $\mathbb{Z}_3$ models have three types of strings $\{0,1,2\}$ with dual strings $0^{*}=0,1^{*}=2,2^{*}=1$.
The branching rules are $\{(0,0,0),(0,1,2),(1,1,1),(2,2,2)\}$.
Thus the Hilbert space consists of all possible string-nets with the above string types and branching rules.

To construct the Hamiltonians and wave functions for the $\mathbb{Z}_3$ models, we have to solve the self-consistency conditions for $\{F,d,\gamma,\alpha\}$.
According to the general solution given in (\ref{sol}),(\ref{dalphagamma}), there are three distinct solutions, parameterized by an integer $p=0,1,2$:
\begin{align*}
& F(1,1,2)=F(1,2,1)=F(1,2,2)=e^{i 2\pi p/3},  \\ 
& F(2,1,2)=F(2,2,1)=F(2,2,2)=e^{-i2\pi p/3},  \\ 
& \gamma_1=e^{-i 2\pi p/3},  \ \ \gamma_2=e^{i 2\pi p/3},  \\
& \alpha(1,0)=\alpha(0,1)=e^{-i 2\pi p/3}, \\
& \alpha(2,0)=\alpha(0,2)=\alpha(1,1)= \alpha(2,2) = e^{i 2\pi p/3},  \\
& \text{others } = 1.
\end{align*}
As in the previous example, we now try to simplify this solution as much as possible using appropriate gauge transformations.
First, we choose a $g$-gauge transformation (\ref{gauge2}) to make $\gamma=1$.
By choosing $g(a)=e^{i 2\pi p a/3}$, we have
\begin{align}
& F(1,1,2)=F(1,2,1)=F(1,2,2)=e^{i 2\pi p/3}, \nonumber \\ 
& F(2,1,2)=F(2,2,1)=F(2,2,2)=e^{-i2\pi p/3}, \nonumber \\ 
& d_1 =e^{i2\pi p/3}, \ d_2=e^{-i2\pi p/3}, \nonumber \\
& \alpha(2,2) = e^{-i2\pi p/3}, \ \text{others } = 1.
\label{z3sol}
\end{align}
Next we look for an $f$-gauge transformation (\ref{gauge}) to make $\alpha=1$. However, when $p=1,2$ no such $f$ exists.
In fact, no combination of $f,g$ can gauge away $\alpha$ completely. To see this, notice that if 
$a+a+a=0$, the quantity $\alpha(a,a) \cdot \alpha(a^*, a^*)$ is gauge invariant 
under both $f$ and $g$ transformations. Hence, if this quantity is not equal to unity, then we have no hope of gauging away $\alpha$. 
In the above case, we have $\alpha(1,1) \cdot \alpha(2,2) \neq 1$, so it is indeed not possible to find a gauge where $\alpha = 1$.
We note that since $\alpha$ cannot be gauged away, the above $p=1,2$ solutions are not accessible to the original string-net 
construction of Ref. [\onlinecite{LevinWenStrnet}]. Thus, we have an explicit example of a string-net model that requires the 
dot structure to be realized.

As for the low energy excitations, the $p=0,1,2$ models each have $|G|^2 = 9$ distinct quasiparticles labeled by ordered pairs $(s,m)$
with $s,m = 0,1,2$. The quasiparticle statistics are described by the $U(1) \times U(1)$ Chern-Simons theory (\ref{kmat2}) with $K$-matrix
\begin{equation*}
K=\left( 
\begin{array}{cc}
0 & 3 \\ 
3 & -2p
\end{array}
\right). 
\end{equation*}

\subsection {\texorpdfstring{$\mathbb{Z}_4$}{Z4} string-net model}
As discussed in section \ref{abstrnet}, the Hilbert space for the $\mathbb{Z}_4$ models involves four string types labeled by $\{0,1,2,3\}$.
The dual string types are $0^{*}=0,1^{*}=3,2^{*}=2,3^{*}=1.$
The branching rules are $\{(0,0,0),(0,1,3),(0,2,2),(1,1,2),(3,3,2)\}.$

To construct the Hamiltonians and wave functions for the $\mathbb{Z}_4$ models, we solve the self-consistency conditions for $\{F,d,\gamma,\alpha\}$. 
As in the previous examples, we use the general solution (\ref{sol}),(\ref{dalphagamma}), which tells us that there are $4$ solutions
labeled by $p = 0,1,2,3$. We then try to gauge away $\gamma$ and $\alpha$ as 
much as possible using $f,g$. Interestingly we find a combination of gauge transformations which makes all $\gamma=\alpha=1$ (more generally, 
such a gauge exists for any $\mathbb{Z}_N$ string-net model where $N$ is not divisible by $3$). To accomplish this, we first apply a $g$-gauge 
transformation (\ref{gauge2}) with $g(a) = i^{pa}$ followed by an $f$-gauge transformation with $f(3,2) = (-i)^{p}$, $f(3,3) = (-1)^p$. The
result is:
\begin{align}
        & F_{113} = F_{331} = F_{232} = F_{212} = F_{131} = i^p, \nonumber \\
	& F_{133} = F_{311} = F_{123} = F_{321} = F_{313} = (-i)^p, \nonumber \\
	& F_{122} = F_{231} = F_{223} = F_{312} = F_{222} = F_{333} = (-1)^p, \nonumber \\
	& d_1 = i^p, \ d_2=(-1)^p, \ d_3=(-i)^p, \ \text{others } =1.
        \label{}
\end{align}
Since $\alpha = \gamma = 1$, these models can be defined without worrying about dots or null strings, as in Ref. [\onlinecite{LevinWenStrnet}].
(However, the reflection symmetry constraint in Ref. [\onlinecite{LevinWenStrnet}] $d_s=d_{s^*}$ suppresses the $p=1,3$ solutions.)
As a result, the local rules are simple and highly symmetric. For example, the nontrivial local rules corresponding to $F$ are given by: 
\begin{align*}
\Phi \left( \raisebox{-0.16in}{\includegraphics[height=0.4in]{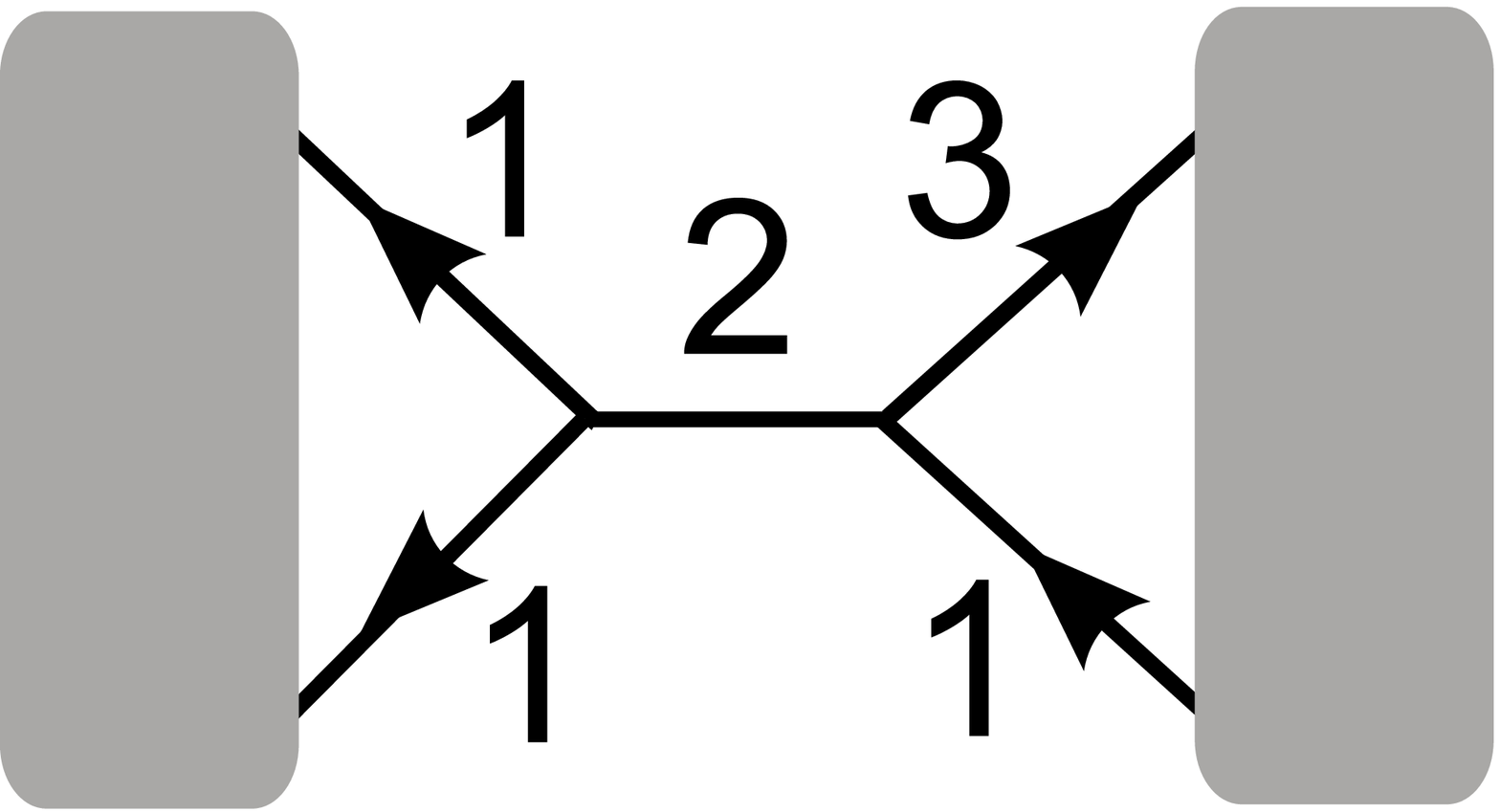}}\right)=
i^p \Phi \left( \raisebox{-0.16in}{\includegraphics[height=0.4in]{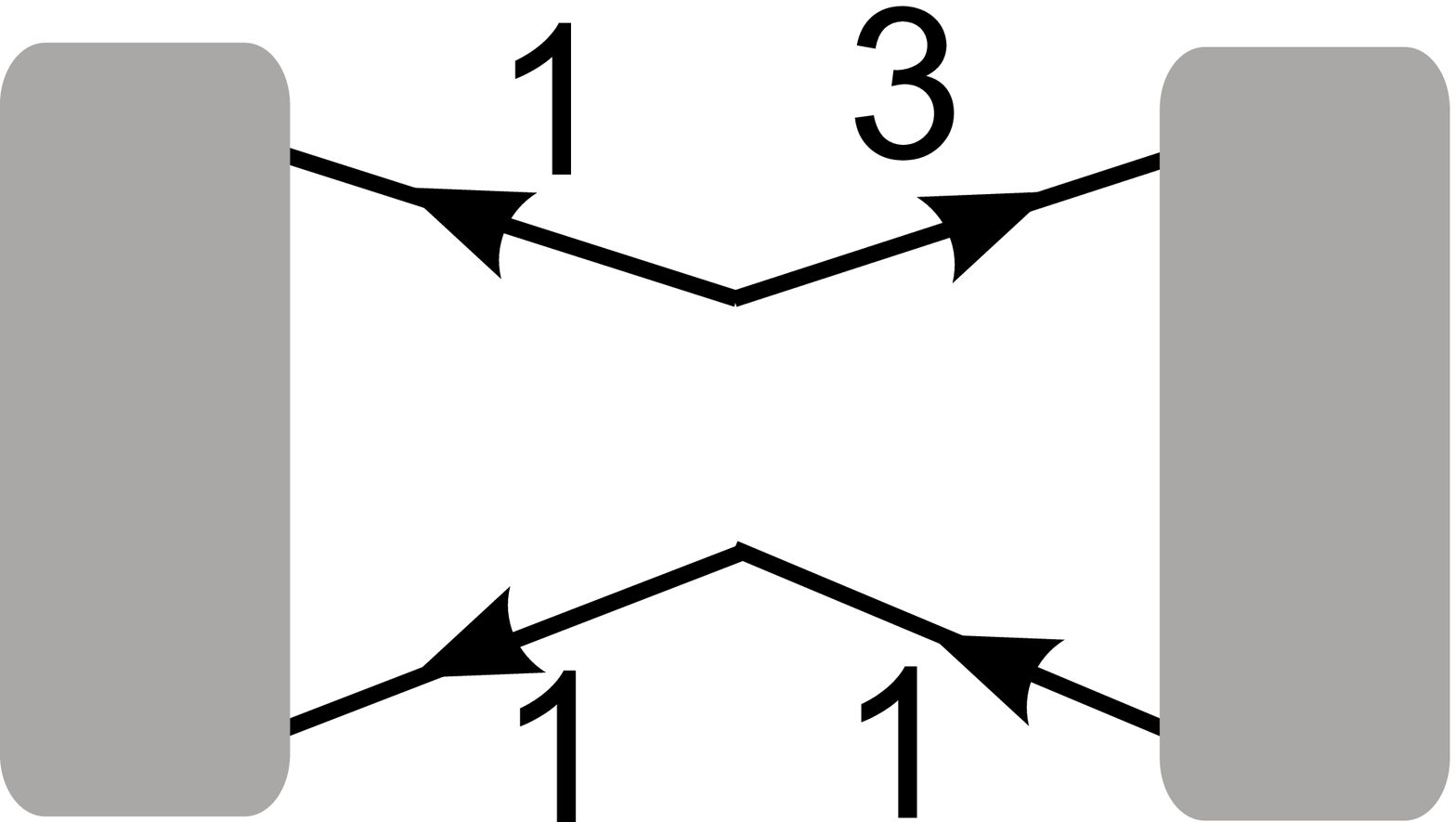}}\right), \\
\Phi \left( \raisebox{-0.16in}{\includegraphics[height=0.4in]{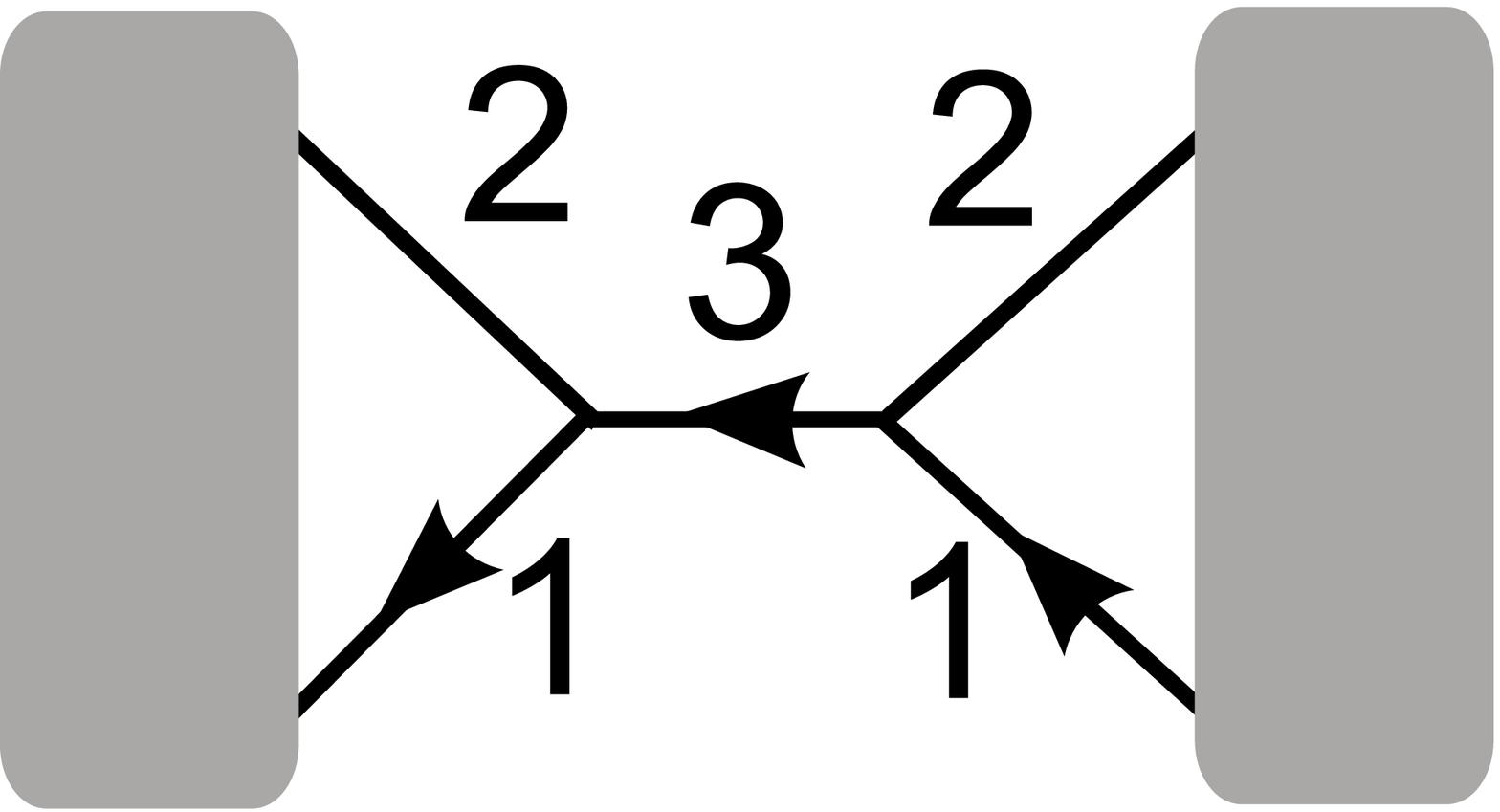}}\right)=
(-1)^p \Phi \left( \raisebox{-0.16in}{\includegraphics[height=0.4in]{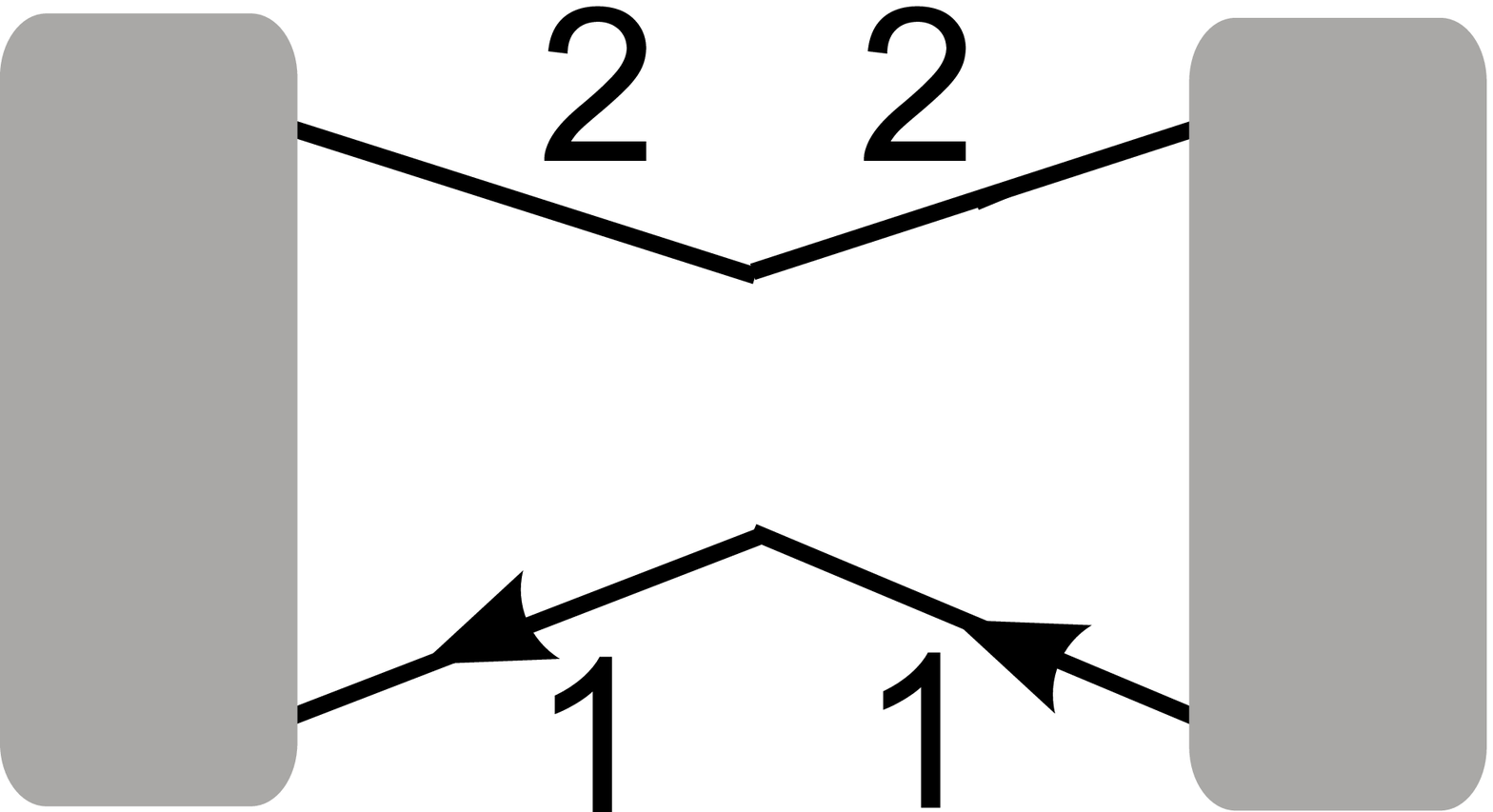}}\right), \\
\Phi \left( \raisebox{-0.16in}{\includegraphics[height=0.4in]{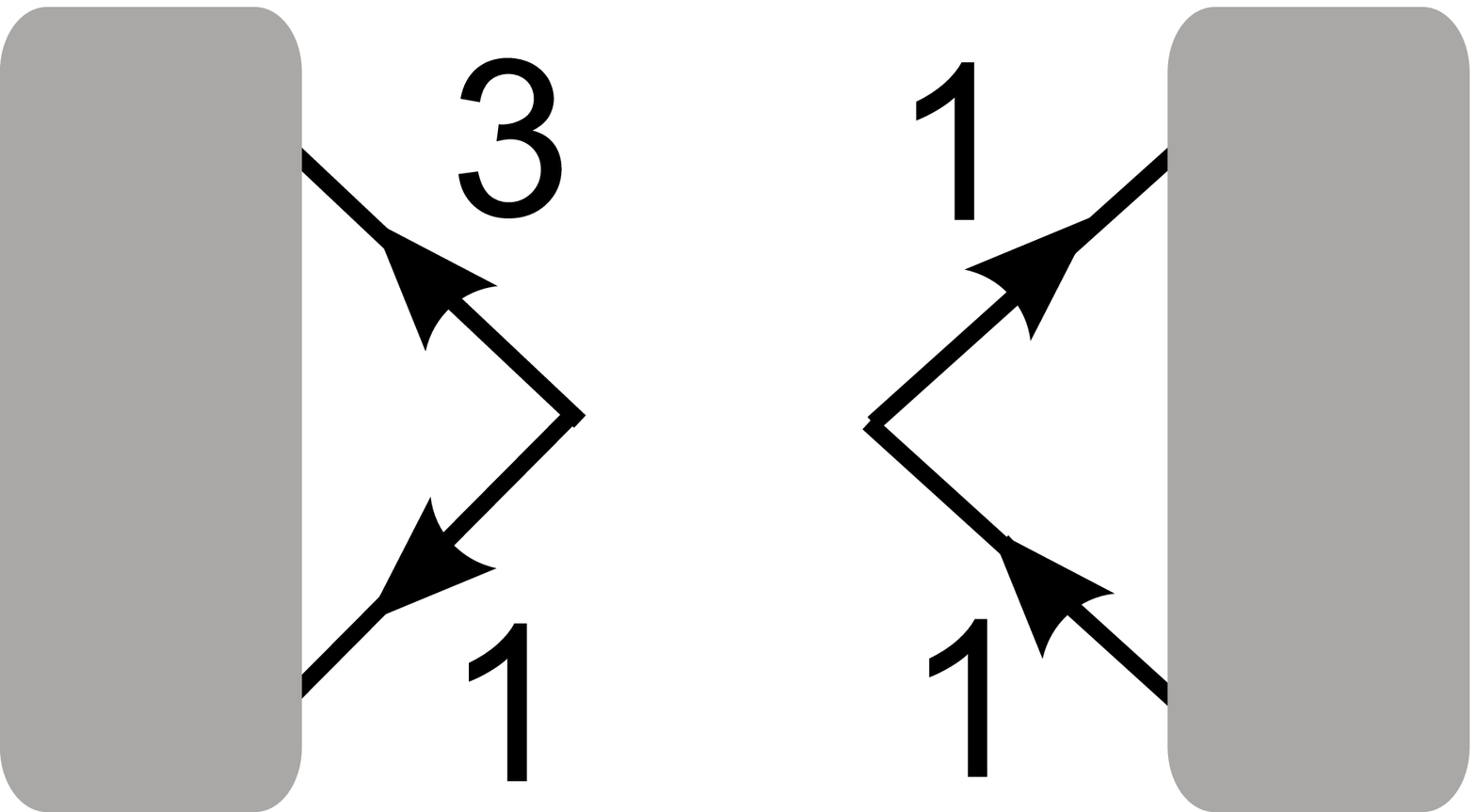}}\right)=
i^p \Phi \left( \raisebox{-0.16in}{\includegraphics[height=0.4in]{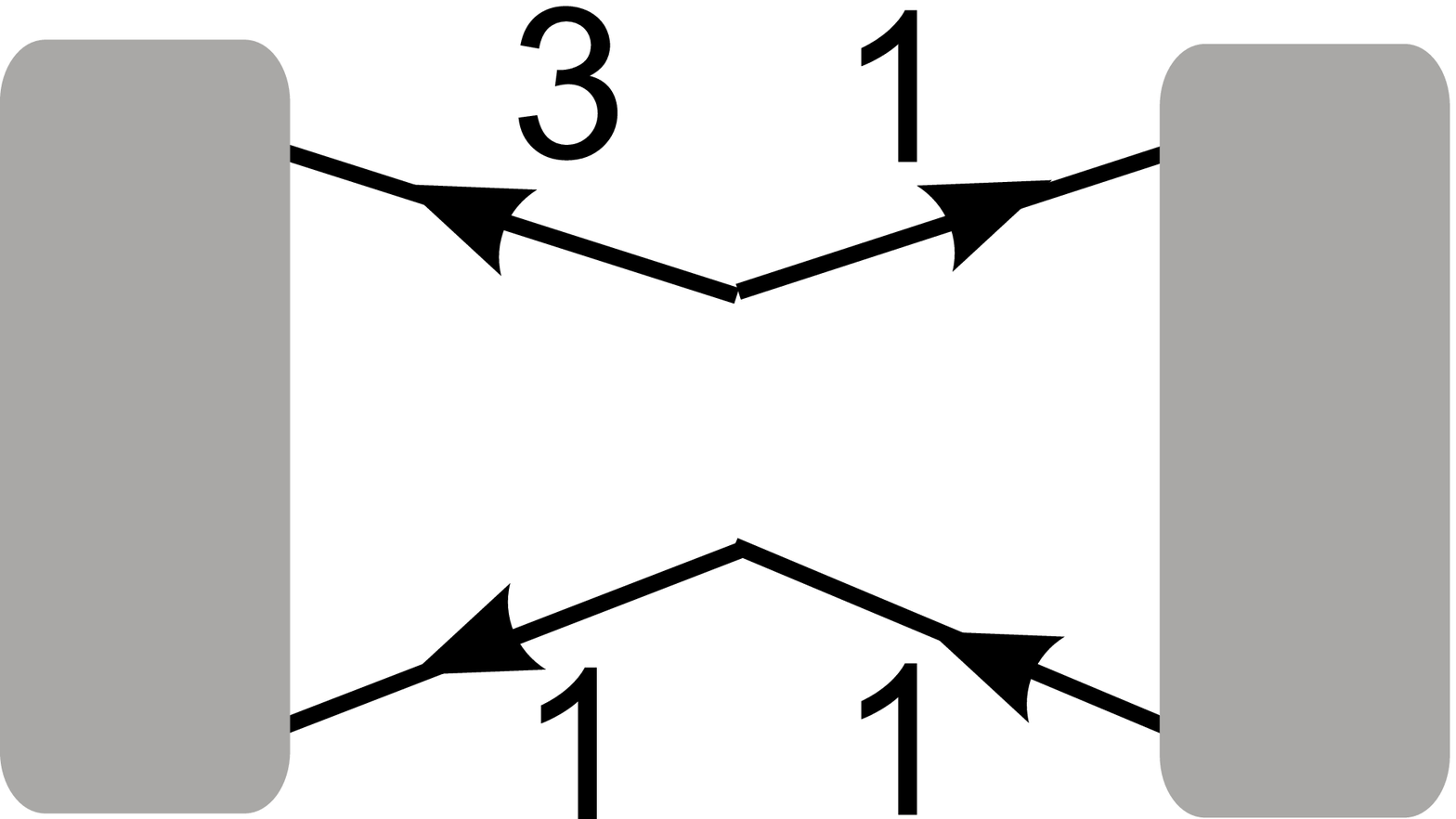}}\right), \\
\Phi \left( \raisebox{-0.16in}{\includegraphics[height=0.4in]{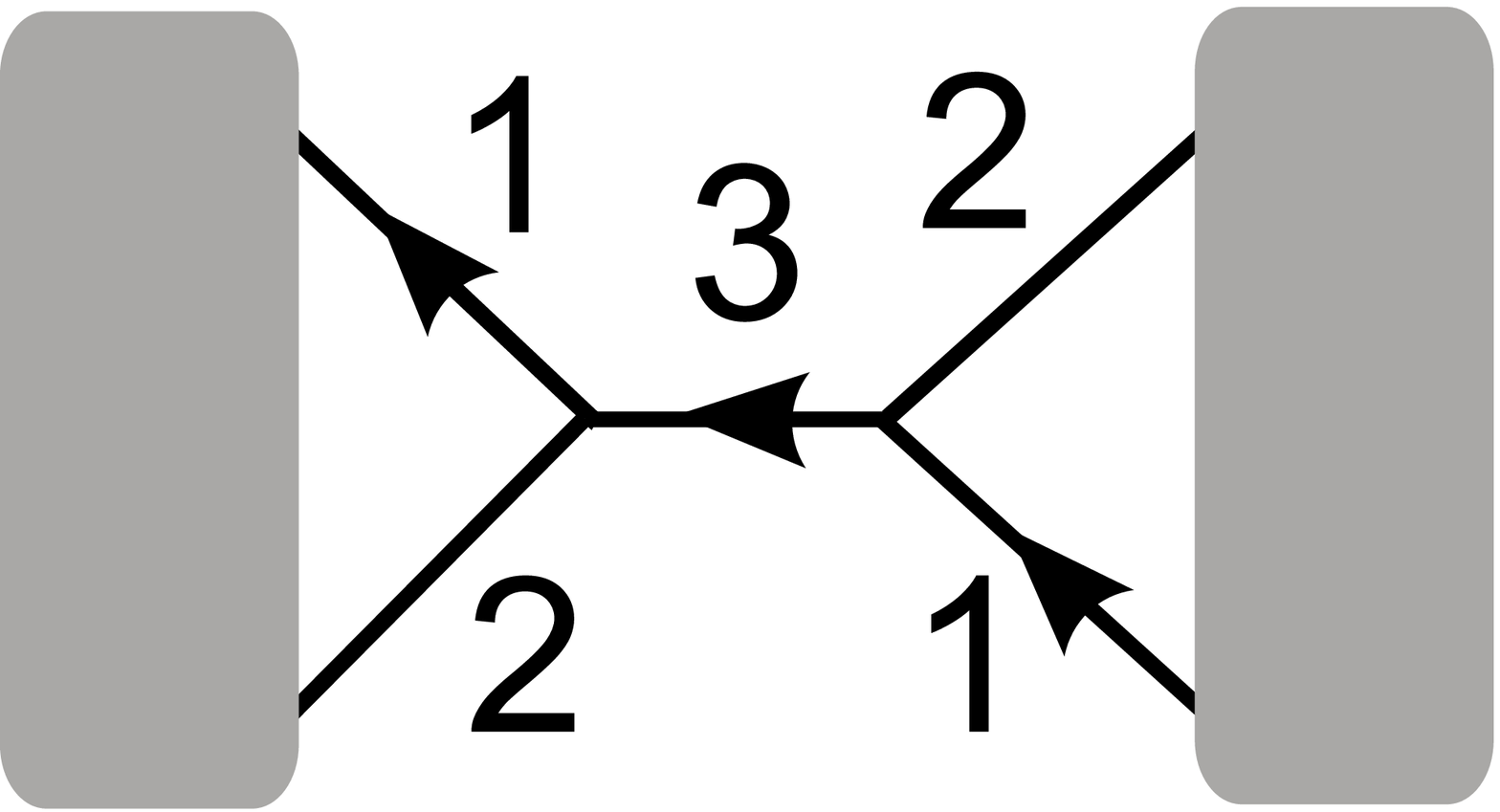}}\right)=
i^p\Phi \left( \raisebox{-0.16in}{\includegraphics[height=0.4in]{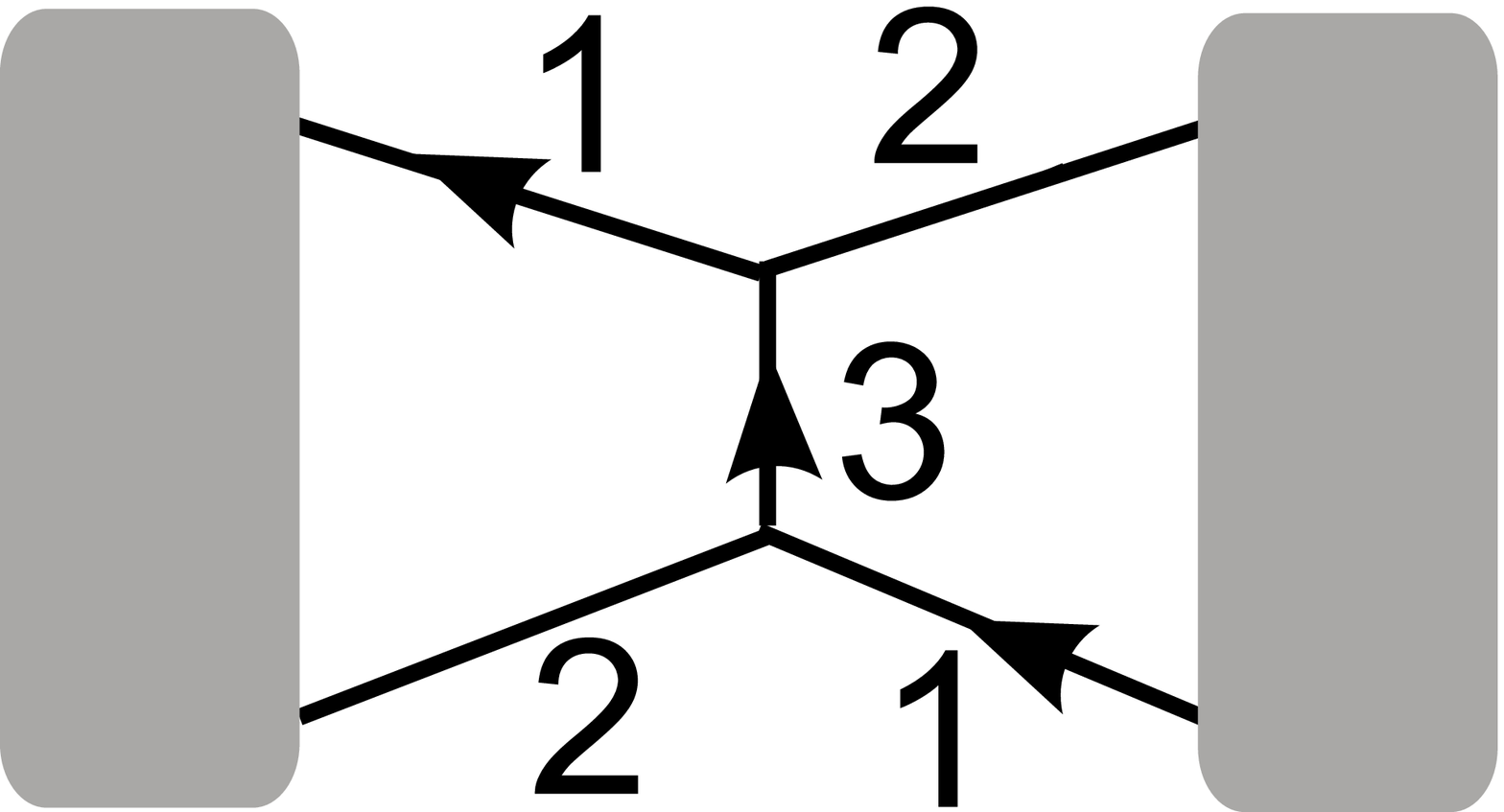}}\right), \\
\Phi \left( \raisebox{-0.16in}{\includegraphics[height=0.4in]{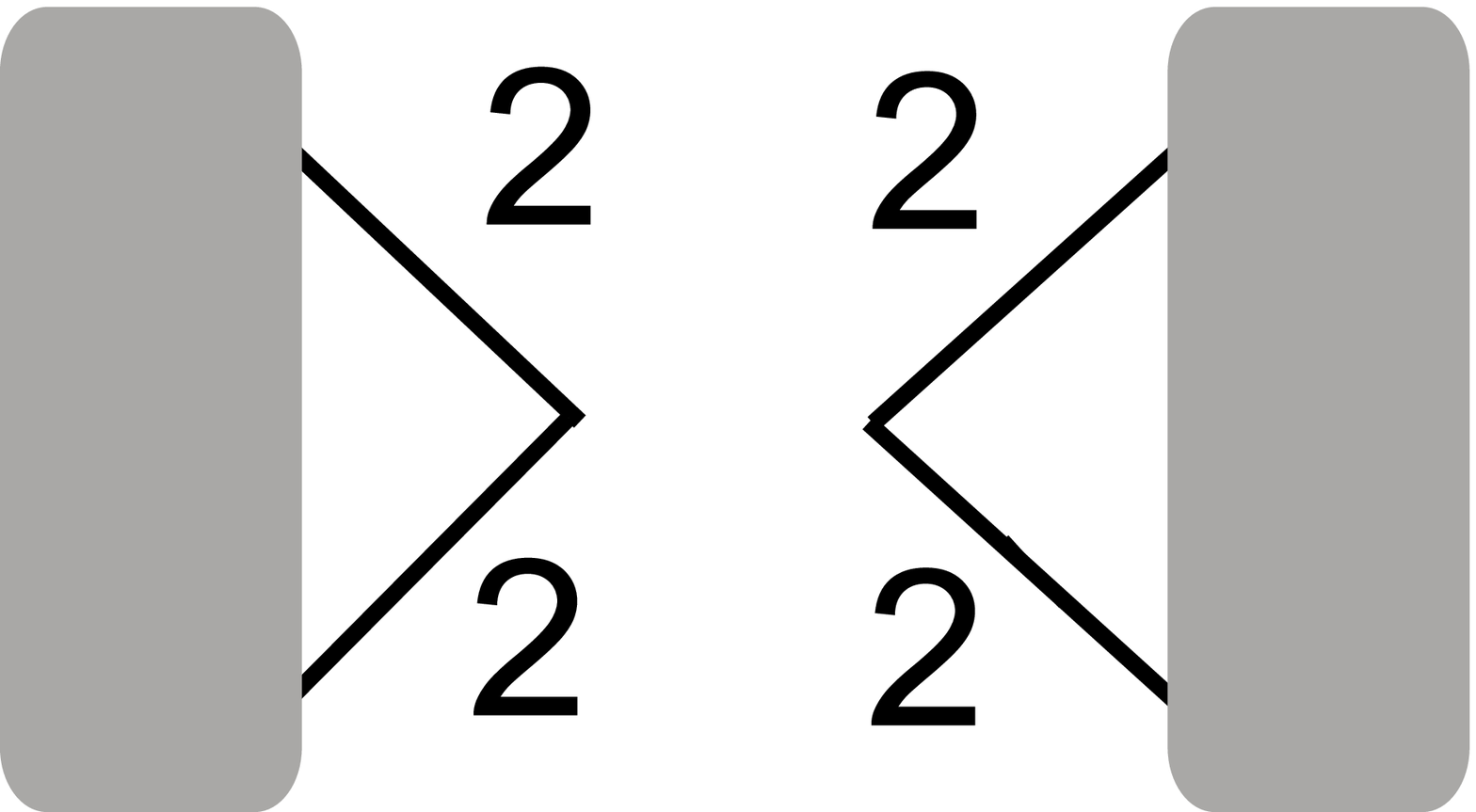}}\right)=
(-1)^p\Phi \left( \raisebox{-0.16in}{\includegraphics[height=0.4in]{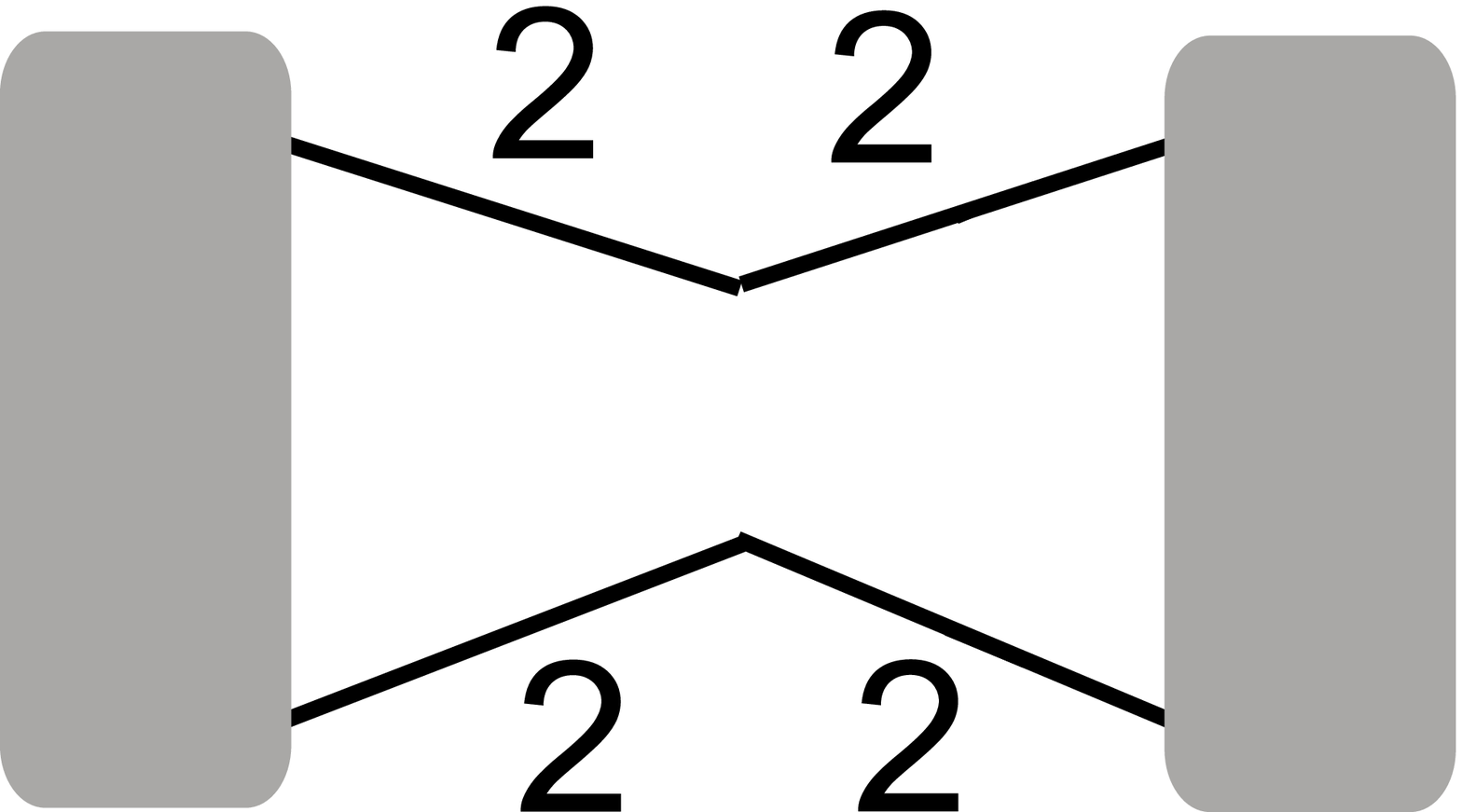}}\right), \\
\Phi \left( \raisebox{-0.16in}{\includegraphics[height=0.4in]{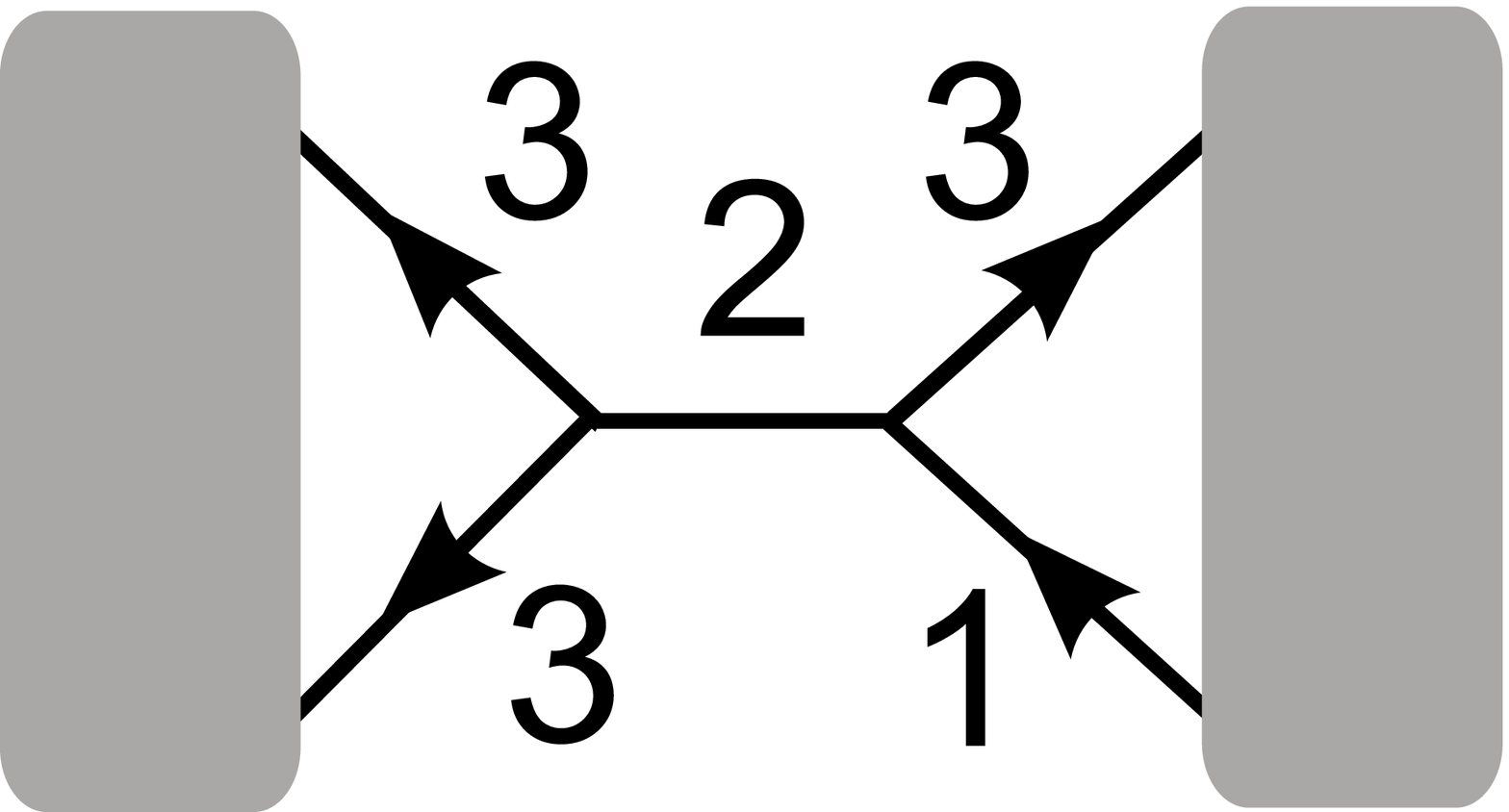}}\right)=
(-1)^p\Phi \left( \raisebox{-0.16in}{\includegraphics[height=0.4in]{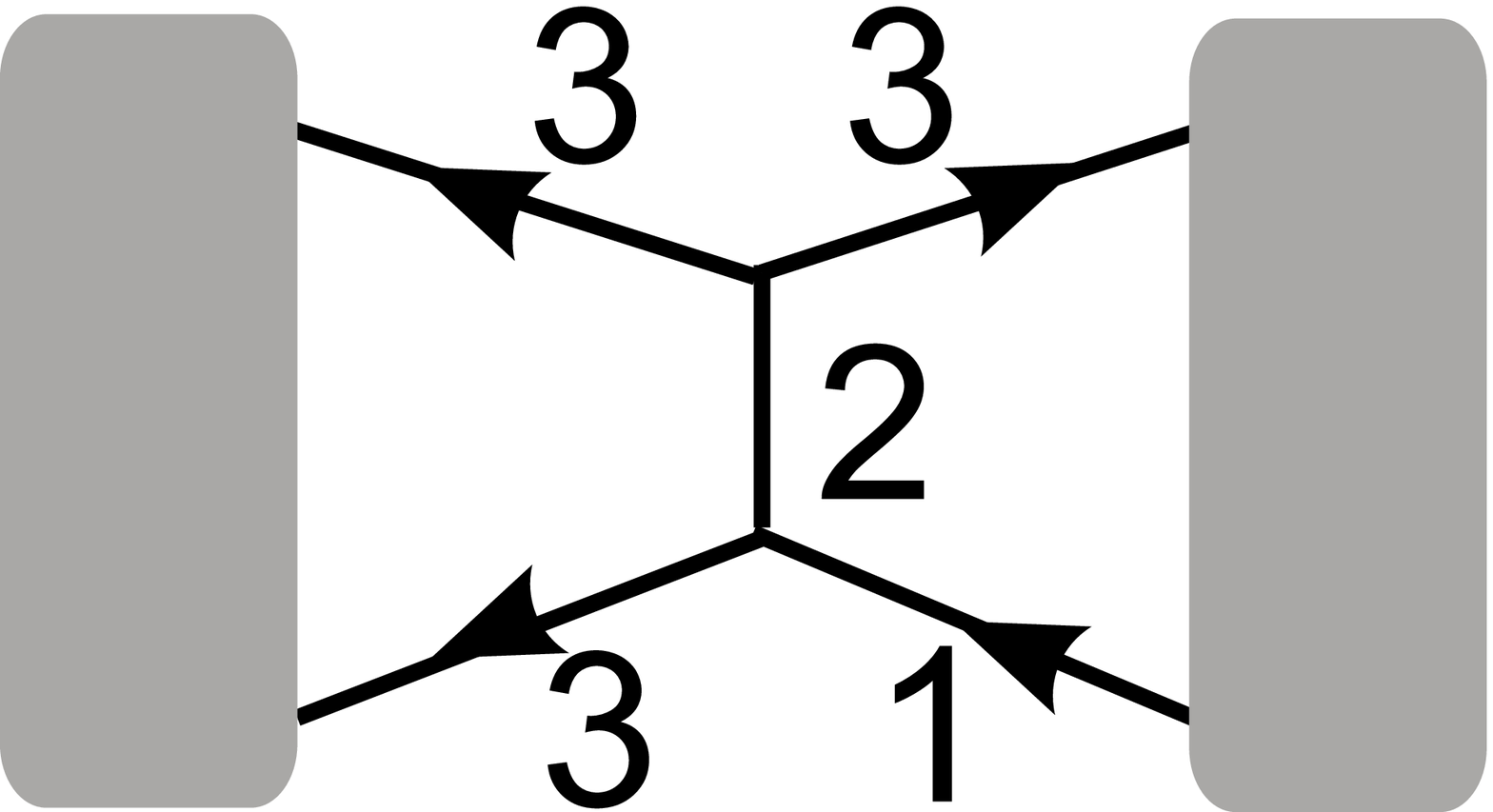}}\right).
	\label{}
\end{align*}
Here, each rule summarizes a group of nontrivial $F$ which are related by $90$, $180$ and $270$ degree rotational symmetry.

Each of these models have $|G|^2 = 16$ distinct quasiparticles which are labeled by ordered pairs $(s,m)$ with $s,m=0,1,2,3$.
The braiding statistics are described by the $U(1)\times U(1)$ Chern-Simons theory (\ref{kmat2})  
with $K$-matrix
\begin{equation*}
K=\left( 
\begin{array}{cc}
0 & 4 \\ 
4 & -2p
\end{array}
\right). 
\end{equation*}

Interestingly, the $p=1,3$ phases break time-reversal symmetry. To see this, note that the $p=1$ phase has a quasiparticle 
$(1,0)$ with exchange statistics $\theta = \pi/8$, but no quasiparticle with statistics $-\pi/8$. Similarly, the $p=3$ phase 
has a quasiparticle with statistics $\theta = -\pi/8$, but no quasiparticle with statistics $\pi/8$. (In fact, the $\mathbb{Z}_3$ 
string-net models with $p=1,2$ also break time-reversal symmetry by the same reasoning). Thus, we can see explicitly
that string-net models can realize time-reversal breaking topological phases. 

\subsection{\texorpdfstring{$\mathbb{Z}_2\times \mathbb{Z}_2$}{Z2 times Z2} string-net model}
Our final example is the two-component $\mathbb{Z}_{2}\times \mathbb{Z}_{2}$ string-net model.
These models have four string types $\{(0,0),(1,0),(0,1),(1,1)\}$ and we abbreviate them as $\{0,r,b,g\}$.
Each string is self-dual, $0=0^{*},r=r^{*},b=b^{*},g=g^{*}$, and the branching rules 
are $\{(0,r,r),(0,b,b),(0,g,g),(r,g,b)\}$.

To construct the wave functions and Hamiltonians for these models, we have to solve the self-consistency conditions for 
$\{F,d,\gamma,\alpha\}$. To this end, we use the general solution (\ref{gensol}),(\ref{dalphagamma}), which tells us
that there are eight different solutions, with each solution labeled by a $2 \times 2$ upper triangular matrix
\begin{equation}
\mathbf{P} = 
\bpm
p & q \\
0 & r
\epm.
\label{ab}
\end{equation}
with $p,q,r=0,1$. The solutions corresponding to $q=0$ are less interesting since they 
give rise to models that are products of two decoupled $\mathbb{Z}_2$ string-net models. Therefore, here we focus on the
case where $q=1$. More specifically, we focus on the case $q=1,p=r=0$ as this illustrates the basic features
of the $q=1$ solutions. In this case, (\ref{gensol}) gives
\begin{align*}
& F_{rbb} = F_{rbg} = F_{rgb} = F_{rgg} = -1, \\
& F_{gbb} = F_{gbg} = F_{ggb} = F_{ggg} = -1, \\
& \gamma_g = -1 , \ \alpha(0,g) = \alpha(g,0) = \alpha(r,g) = -1, \text{others } = 1.
\end{align*}
We can gauge away $\alpha$ using an appropriate $f$-gauge transformation, $f(b,r)=f(g,b)=-1$. The result is
\begin{align*}
        &F_{rbr}=F_{rbg}=F_{rgr}=F_{rgg}=F_{brr}=F_{brg}=F_{bbr}= \\
        &F_{bbg}=F_{grr}=F_{grg}=F_{ggr}=F_{ggg}=-1, \\
	& \gamma_g = -1, \text{others } = 1.         
\label{}
\end{align*}

Ideally, we would also like to gauge away $\gamma$.
However, $\gamma$ cannot be gauged away using any combination of $f$ and $g$ gauge transformations. One way
to see this is to note that for any abelian string-net model and any $a=a^*, b=b^*$, the quantity
$\gamma(a) \gamma(b) \gamma(a^*+b^*)$ is gauge invariant under $f,g$ gauge transformations. Thus
if this quantity is different from unity, then we can never gauge away $\gamma$. In the above case,
we have $\gamma(r)\gamma(b)\gamma(g) = -1$, so there is no gauge where $\gamma = 1$.
Since $\gamma$ cannot be gauged away, we have an example of a string-net model that
requires our more careful treatment of the null string to be realized.

Each of the $\mathbb{Z}_2 \times \mathbb{Z}_2$ string-net models has $|G|^2 = 16$ quasiparticles, which are labeled by ordered pairs $(s,m)$
where $s = (s_1, s_2), m = (m_1, m_2)$, and $s_i, m_i = 0,1$. The quasiparticle statistics are described by the four-component 
$U(1)$ Chern-Simons theory (\ref{kmat2}) with $K$-matrix
\begin{equation*}
K= \bpm 0 & 0 & 2 & 0 \\
	0 & 0 & 0 & 2 \\
	2 & 0 & -2p & -q \\
	0 & 2 & -q & -2r \epm.
\end{equation*}
 
\section{Conclusion} \label{conclusion}

In this paper, we have derived several criteria for determining which abelian topological phases can and cannot be realized by string-net models. 
The simplest of these criteria states that an abelian topological phase can be realized by a string-net model if and only if it supports a gapped
edge. This result is interesting because it shows that the string-net models realize the most general abelian phases that one could
possibly hope for: indeed one could never expect string-net models to realize a phase with a protected gapless edge, since we can show explicitly
that the edge of these models can be gapped by suitable interactions (see appendix \ref{gapedge}). 

One direction for future work would be to extend our analysis to the non-abelian case. 
A natural conjecture is that the realizable phases are exactly the phases with gapped edges, just like in 
the abelian case:
\begin{conjecture}\label{conj}
A general 2D topological phase can be realized by a string-net model 
if and only if it supports a gapped edge. 
\end{conjecture}
As in the abelian case, it is easy to establish the ``only if'' direction in this conjecture, since it can be shown explicitly that string-net models always support a gapped edge.\cite{KitaevKong}
Proving the other direction --- i.e. proving that the string-net models realize \emph{all} topological phases with a gapped edge --- requires more work. 
In particular, this result seems to require a complete characterization of 
which topological phases support a gapped edge; currently such a characterization only exists in the abelian case.\cite{LevinProtedge,KapustinTopbc} Several works\cite{Kong13,FuchsSchweigert12} have proposed abstract classifications of gapped edges, though it has not been shown that these classifications include
every possible gapped edge that can occur in microscopic models. That being said, these proposals seem to be consistent with the above conjecture. 

Another motivation for considering the non-abelian case is to develop the most general possible formulation of string-net models.
As we mentioned in the introduction, our construction of abelian string-net models is more general than the original construction of
Ref. [\onlinecite{LevinWenStrnet}], due to two new ingredients, $\gamma, \alpha$, which are related to the $\mathbb{Z}_2$ and $\mathbb{Z}_3$ Frobenius-Schur indicators respectively.
It would interesting to understand how to incorporate $\gamma, \alpha$ into the construction of general (non-abelian) string-net models. Other
generalizations of Ref. [\onlinecite{LevinWenStrnet}], such as those described in Refs. [\onlinecite{Kong12} - \onlinecite{LanWen13} ], have avoided including $\gamma, \alpha$, at the cost of breaking
the rotational symmetry of the lattice models. Our alternative formulation involving $\gamma, \alpha$, preserves as much symmetry and
topological invariance as possible, and therefore may give simpler models in some cases.

A final question is to determine whether string-net models can be generalized further. In particular, are there
any 2D commuting projector spin Hamiltonians that realize topological phases beyond the reach of string-net models? An example of such a Hamiltonian (or a proof that
no such Hamiltonian exists) would advance our understanding of both exactly soluble models and topological phases.

\appendix

\section*{Acknowledgement}
C.-H. Lin thanks C. Wang and M. Cheng for useful discussions. C. -H. Lin and M. Levin acknowledge support from the Alfred P. Sloan foundation and NSF DMR-1254721.

\section{Motivation for \texorpdfstring{$\gamma$}{gamma} and \texorpdfstring{$\alpha$}{alpha}}\label{dot}
In this section, we explain the motivation for introducing $\gamma$ and $\alpha$. Our explanation is based on the 
string-net formulation of lattice gauge theory\cite{KogutSusskind,BanksMyersonKogut,LevinWenStrnet} or equivalently, 
quantum double models\cite{BuerschaperAguado09}, so we first briefly review how this string-net formulation works. 
Consider a quantum double model\cite{KitaevToric} associated with some finite group $G$ (in this discussion $G$ can 
be \emph{any} finite group and need not be abelian). The string-net description of this model goes as follows: the 
string types are labeled by irreducible representations $\mu$ of $G$, and the null string is labeled by the trivial 
representation $\mathbf{1}$. The dual string type of $\mu$ is defined to be the dual representation $\bar{\mu}$, and 
the branching rules allow $\{\mu_1, \mu_2, \mu_3\}$ to meet at a vertex if and only if the tensor product 
$\mu_1 \otimes \mu_2 \otimes \mu_3$ contains the trivial representation $\mathbf{1}$. If the trivial representation 
appears more than once in this tensor product, i.e. $\mu_1 \otimes \mu_2 \otimes \mu_3 = n \cdot \mathbf{1} + ...$ 
then the vertex carries an additional label $\nu$ that runs over $n$ values. 

Our motivation for introducing $\alpha, \gamma$ can be seen by considering the ground state wave function $\Phi$ in these models. This wave function can be written down explicitly once we fix some conventions. In particular, for every allowed branching $\mu_1, \mu_2, \mu_3$ with vertex label $\nu$, we fix a corresponding three index tensor $T_{m_1 m_3 m_3}$ that belongs to the invariant subspace within $\mu_1 \otimes \mu_2 \otimes \mu_3$ --- i.e. the subspace corresponding to the identity representation $\mathbf{1}$. Here, $m_i$ is an index which transforms under $G$ according to the representation $\mu_i$. Also, $m_i$ is an upper or lower index depending on whether the corresponding string points outward or inward.

Once we fix these invariant tensors, the wave function $\Phi$ can be computed as follows. Consider some string-net configuration $X$. To calculate the amplitude $\Phi(X)$, we assign tensors $T$ to every vertex in $X$ following the above prescription. We then contract all the indices $m_1,m_2,...$ of these tensors along the adjoining strings (every string will have one lower index and one upper index). The complex number that results from this tensor contraction gives $\Phi(X)$.

We now carefully examine the above formula for $\Phi$ and we show that we need to introduce $\alpha$ and $\gamma$ to 
make everything work out. We begin with $\alpha$. The need for $\alpha$ appears when we consider vertices that are attached to three outgoing (or incoming) strings, each of which are labeled by the same representation $\mu$. In this case, the associated invariant tensor $T$ has $3$ lower indices $m_1, m_2, m_3$ that transform under the same representation $\mu$. The problem occurs if the tensor $T$ is not cyclically symmetric, i.e. $T_{m_1 m_2 m_3} \neq T_{m_3 m_1 m_2}$. In this case, there is an ambiguity in our prescription for calculating $\Phi(X)$, since we don't know which index $m_i$ should be assigned to which string. To resolve this ambiguity, we introduce a ``dot'' at each vertex. We then assign $m_1$ to the string that is adjacent to the dot in the clockwise direction, and then assign $m_2$ and $m_3$ to the remaining strings, ordered in the clockwise direction. With this convention, the value of $\Phi(X)$ depends on the position of the dot. To see how this dependence comes about, note that if we cyclically symmetrize our tensors, we can assume without loss of generality that $T_{m_1 m_2 m_3} = T_{m_3 m_1 m_2} \cdot \alpha(\mu)$ where $\alpha(\mu) = 1, \omega, \omega^2$ is a third root of unity. If we then move the dot in the clockwise direction, we have
$\Phi(X) \rightarrow \Phi(X) \cdot \alpha(\mu)$ similarly to Eqs. (\ref{rule4} - \ref{rule4'}).

The motivation for $\gamma$ is similar. Consider a vertex that adjoins one null string and two outgoing (incoming) strings labeled with the same representation $\mu$. 
Since the identity representation is one dimensional, thus we can write $T$ as $T_{m_1 m_2}$. If $T$ is not symmetric, i.e. $T_{m_1 m_2} \neq T_{m_2 m_1}$,
then we again have an ambiguity in our prescription for calculating $\Phi(X)$. To resolve this ambiguity, we use the position of the null string to break the symmetry: we adopt
the convention that $m_1$ is assigned to the string that is in the clockwise direction from the null string, and $m_2$ is assigned to the other string. In this
way, the null string serves the same function as the dot. Now, by symmetrizing our tensors appropriately, we can assume without loss of generality that
$T$ is either symmetric or antisymmetric: $T_{m_1 m_2} = T_{m_2 m_1} \cdot \gamma(\mu)$ where $\gamma(\mu) = \pm 1$. If we then flip the position of the null string
then $\Phi(X) \rightarrow \Phi(X) \cdot \gamma(\mu)$ similarly to Eqs. (\ref{rule5} - \ref{rule6}). With this picture, we can also motivate the rules (\ref{normal} - \ref{dot2}): these rules 
correspond to conventions for how to raise and lower indices of our tensors $T_{m_1 m_2 m_3}$.  

The reader may notice that the $\alpha, \gamma$ formalism in this paper differs slightly from what we have motivated above: in this paper, we include $\alpha$ factors 
for \emph{all} vertices with three outgoing or incoming strings, even if they have different labels. Likewise, we include $\gamma$ factors for all vertices
with one null string and two incoming or outgoing strings, independent of their labels. While this observation is correct, we only use $\alpha, \gamma$ in these
more general situations because it simplifies our notation: it is always possible to gauge away the $\alpha$ and $\gamma$ factors in these cases, as discussed
in section \ref{gaugesec}.

To summarize, we have shown that the $\alpha$ and $\gamma$ factors naturally appear in the string-net formulation of quantum double models. It is this observation that motivates us to include $\alpha, \gamma$ in our construction.

\section{Derivation of self-consistency conditions \label{consistent}}

\begin{figure}
\begin{center}
\includegraphics[height=2in,width=2.2in]{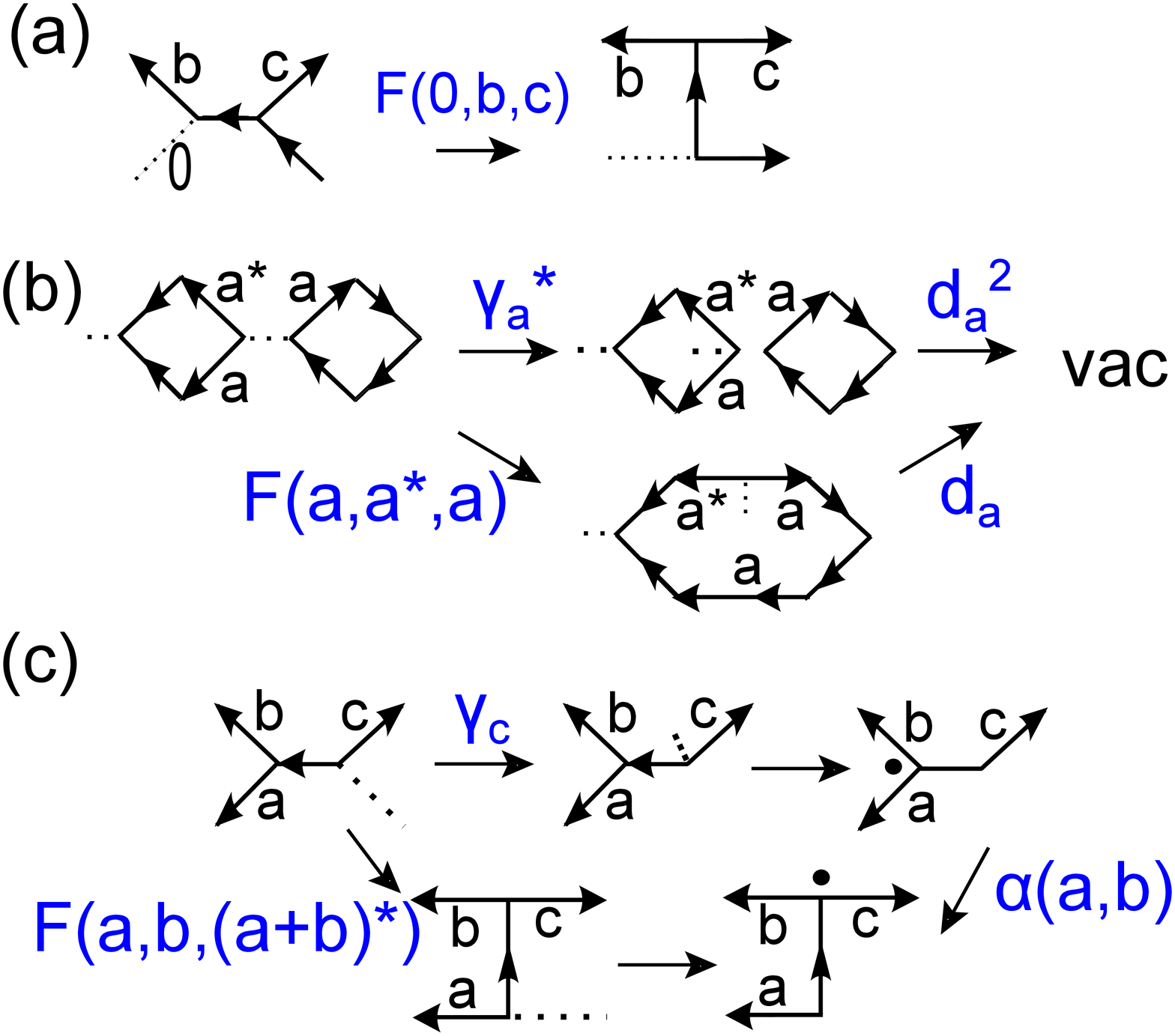}
\end{center}
\caption{
Derivation of self-consistency conditions.
Self-consistency in sequences (a),(b) and (c) requires the conditions (\ref{F0app}),(\ref{gammaFapp}) and 
(\ref{alphaFapp}), respectively.
}
\label{sfeq1}
\end{figure}

\begin{figure}[tb]
\begin{center}
\includegraphics[height=1.in,width=1.4in]{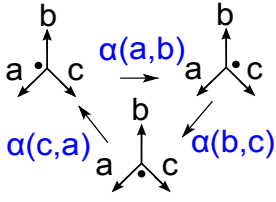}
\end{center}
\caption{
A sequence of three $\alpha$ moves will return to the original configuration. Thus self-consistency requires
$\alpha(a,b)\alpha(b,c)\alpha(b,a)=1$ with $a+b+c=0$ as shown in (\ref{alpha3}). }
\label{sfeq3}
\end{figure}

In this section, we show that the local rules (\ref{rule1} - \ref{rule3}) and the conventions (\ref{nullerase} - \ref{rule4'}) are self-consistent if and only if the
parameters $\{F(a,b,c), d_a, \gamma_a, \alpha(a,b)\}$ satisfy the following conditions 
\begin{subequations}
\label{selfconseqapp}
\begin{align}
	F(a+b,c,d)& F(a,b,c+d) = 	\label{pentidapp}\\
	&F(a,b,c) F(a,b+c,d) F(b,c,d),   \nonumber \\
	F(a,b,c) &= 1 \ \text{ if }a\text{ or }b\text{ or }c=0, \label{F0app} \\
	d_a d_b &= d_{a+b}, \label{dconsapp} \\
	\gamma_{a} &= F(a^{*},a,a^{*})d_{a} \label{gammaFapp}, \\
	\alpha(a,b) &= F(a,b,(a+b)^{*}) \gamma_{a+b}.	    \label{alphaFapp}
\end{align}
\end{subequations}
First, we show that the above conditions are \emph{necessary} for self-consistency.
The first condition (\ref{pentidapp}) was derived previously in the text (see section \ref{scc}).
The second condition (\ref{F0app}) can be understood by considering the rule (\ref{rule3}) for the special case where $a =0$, the null string. 
As shown in Fig. \ref{sfeq1}(a), equation (\ref{F0app}) must be satisfied in order for the rule (\ref{rule3}) to be consistent with the topological invariance rule (\ref{rule1}) and the rule for erasing null strings (\ref{nullsame}). Similarly, the conditions (\ref{gammaFapp}) and (\ref{alphaFapp}) follow from considering the string-net configurations shown in Fig. \ref{sfeq1}(b), and Fig. \ref{sfeq1}(c) and demanding consistency between the two sequences of moves shown there. 
To derive (\ref{dconsapp}), we note that
\begin{equation}
	\alpha(a,b)\alpha(b,c)\alpha(c,a) = 1 \ \text{if }a+b+c=0
	\label{alpha3}
\end{equation}
as can be shown by considering a sequence of three $\alpha$ moves as in Fig. \ref{sfeq3}.
It then follows from (\ref{alphaFapp}) that
\begin{equation}
	1=\alpha_{ab}\alpha_{bc}\alpha_{ca}=\frac{F_{abc}F_{bca}F_{cab}}{F_{aa^*a}F_{bb^*b}F_{cc^*c}}\frac{1}{d_a d_b d_c}. 
	\label{alpharel}
\end{equation}
Here we use the abbreviation $F_{abc}=F(a,b,c)$ and $\alpha_{ab}=\alpha(a,b)$.
To proceed further, we note that (\ref{pentidapp}) implies the identity
\begin{equation}
\frac{F_{abc}F_{bca}F_{cab}}{F_{aa^*a}F_{bb^*b}F_{cc^*c}} = 1.
\label{Fidentity}
\end{equation}
Substituting (\ref{Fidentity}) into (\ref{alpharel}), we derive (\ref{dconsapp}).

The above arguments show that Eqs. (\ref{selfconseqapp}) are \emph{necessary} conditions for the rules to be self-consistent. 
To prove that they are \emph{sufficient}, suppose that $\{F(a,b,c), d_a, \gamma_a, \alpha(a,b)\}$ obey Eqs. (\ref{selfconseqapp}). Then, using
$\{F,d,\gamma,\alpha\}$, we can construct an exactly soluble lattice Hamiltonian $H$ (\ref{h}). The Hamiltonian $H$ has a number of interesting properties, 
which are discussed in section \ref{propertyh} and proven in the appendices. 
Here we concentrate on one of these properties, namely the fact that the ground state of $H$ satisfies the local rules (\ref{rule1} - \ref{rule3}) on the lattice:
\begin{align*}
	\Phi_{latt}\left( \raisebox{-0.17in}{\includegraphics[height=0.42in,width=0.5in]{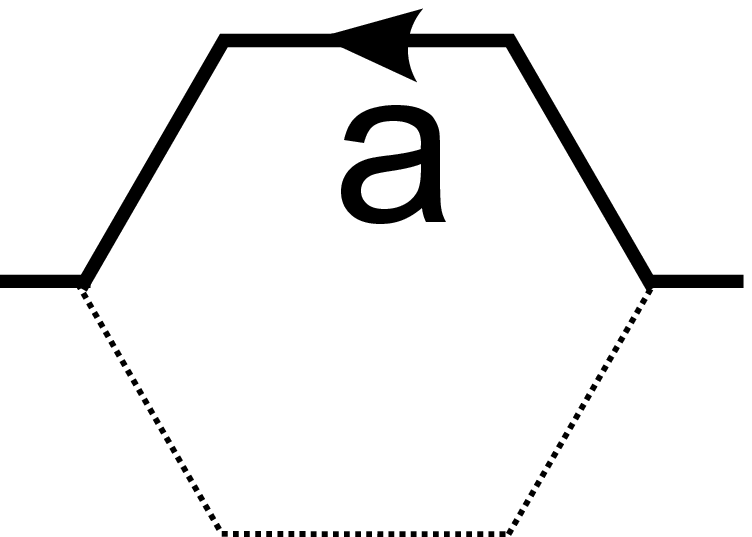}} \right)
	& =\Phi_{latt}\left( \raisebox{-0.17in}{\includegraphics[height=0.42in,width=0.5in]{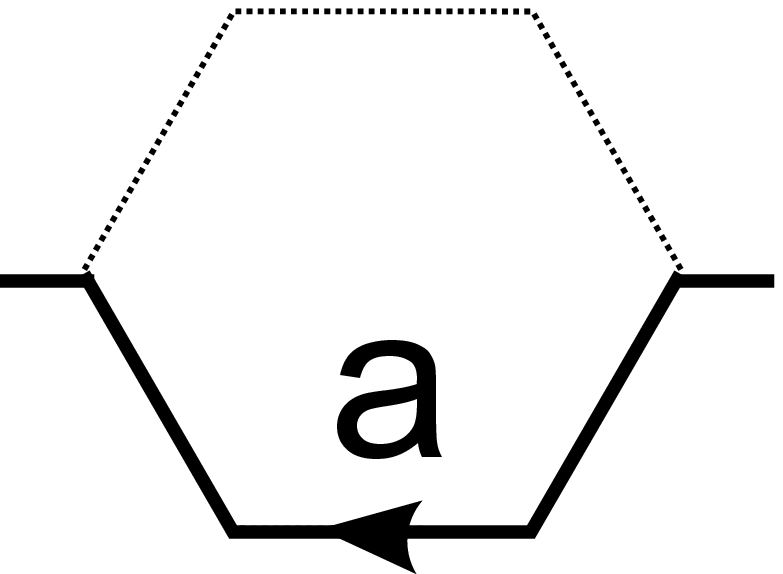}} \right), \\
	\Phi_{latt}\left( \raisebox{-0.17in}{\includegraphics[height=0.4in,width=0.46in]{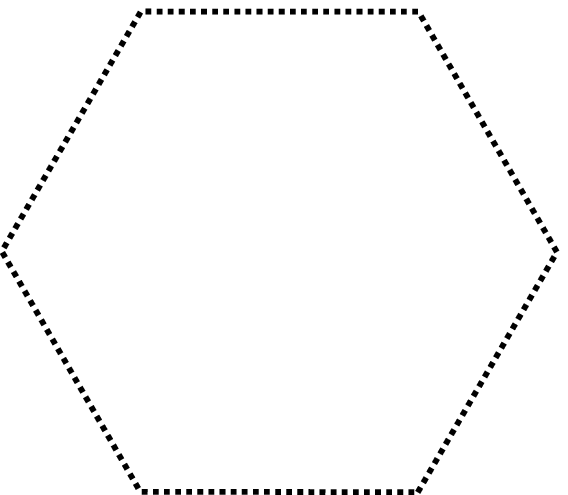}}\right) 
	& =d_{s}\Phi_{latt}\left( \raisebox{-0.17in}{\includegraphics[height=0.4in,width=0.46in]{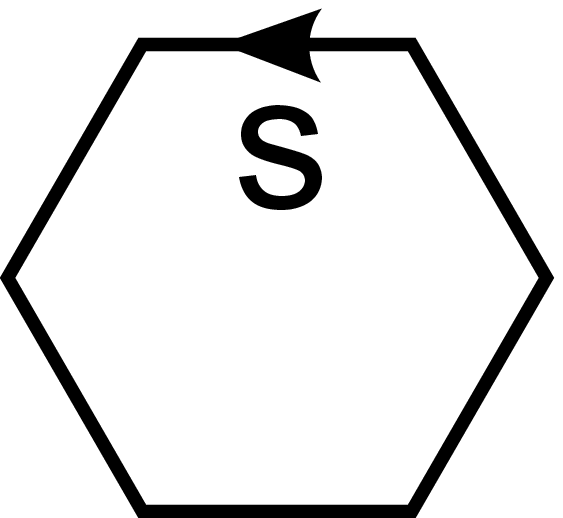}} \right), \\
	\Phi_{latt}\left( \raisebox{-0.2in}{\includegraphics[height=0.5in,width=0.51in]{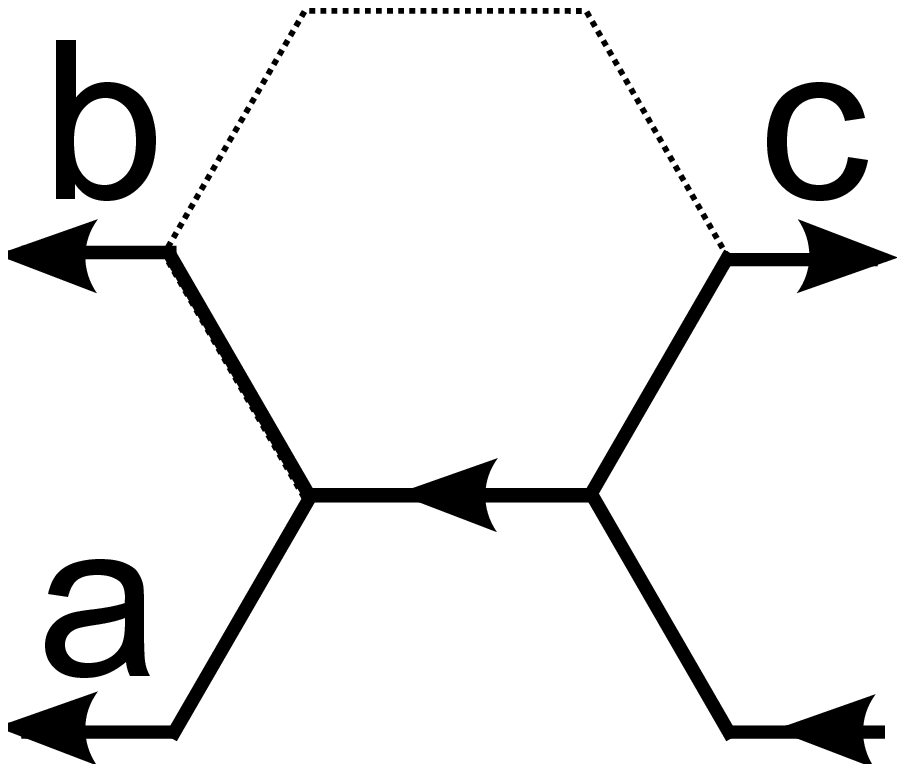}}\right) 
	& =F(i,j,k) \Phi_{latt}\left( \raisebox{-0.2in}{\includegraphics[height=0.5in,width=0.51in]{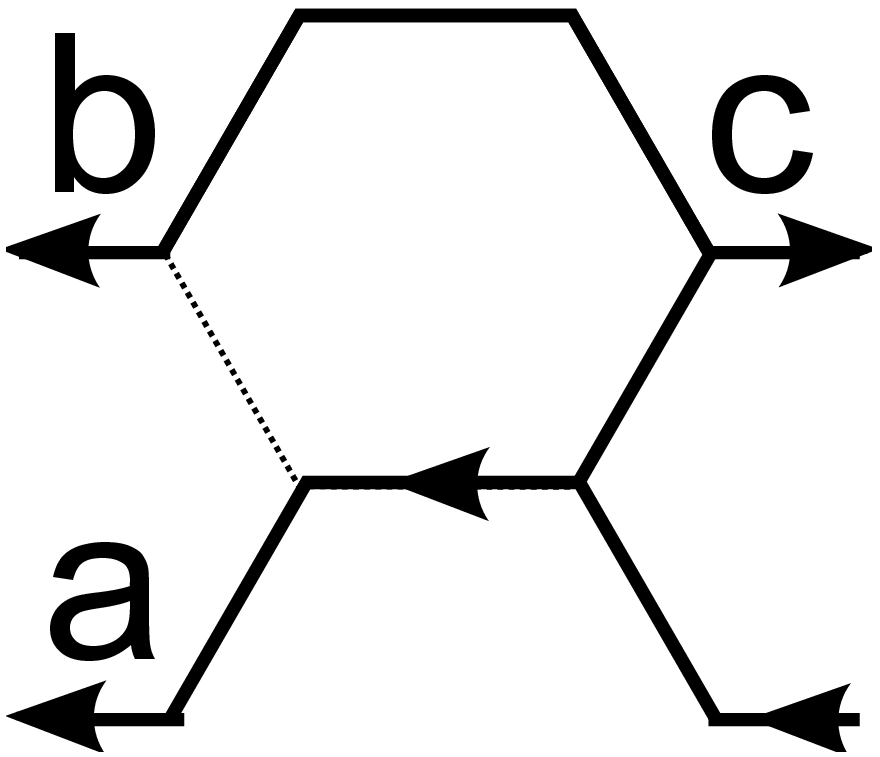}} \right).
\end{align*}
Given this fact, it is now easy to prove that the rules are self-consistent: suppose to the contrary that the rules (\ref{rule1} - \ref{rule3}) 
are not self-consistent. Then there exists two sequences of ``moves'' which relate the same initial and final (continuum) string-net states $X_1, X_2$, but with different proportionality constants: i.e. $\Phi(X_1) = c \Phi(X_2)$ and $\Phi(X_1) = c' \Phi(X_2)$ with $c \neq c'$. Clearly, these two sequences of moves can be adapted from the continuum to the honeycomb lattice if we make the lattice sufficiently fine. But this leads us to an immediate contradiction
since the ground state $\Phi_{latt}$ of $H$ gives an explicit example of a wave function that obeys the rules on the lattice. We conclude that our assumption is false: the rules must be self-consistent.

\section{Local unitary transformations and gauge equivalence \label{gaugeapp}}
In this section, we study the two gauge transformations $f,g$ (\ref{gauge},\ref{gauge2}) which act on the parameters $\{F,d,\gamma,\alpha\}$. We show that both
$f$ and $g$ can be implemented by a local unitary transformation.$U$\cite{ChenGuWen10} More precisely, we show that, for both $f$ and $g$, the lattice Hamiltonians and ground state wave functions corresponding to $\{F,d,\gamma, \alpha\}$ and $\{\tilde{F},d,\tilde{\alpha},\tilde{\gamma}\}$ are related to one another by 
\begin{align*}
\tilde{H} = U H U^\dagger, \ \ \ 
|\tilde{\Phi}\> = U |\Phi\>.
\end{align*}

We begin with the $f$-transformation. 
In this case, the local unitary transformation $U$ that connects the wave functions $\Phi$ and $\tilde{\Phi}$ is defined by:
\begin{align*}
\left\< \raisebox{-0.18in}{\includegraphics[height=0.4in,width=0.41in]{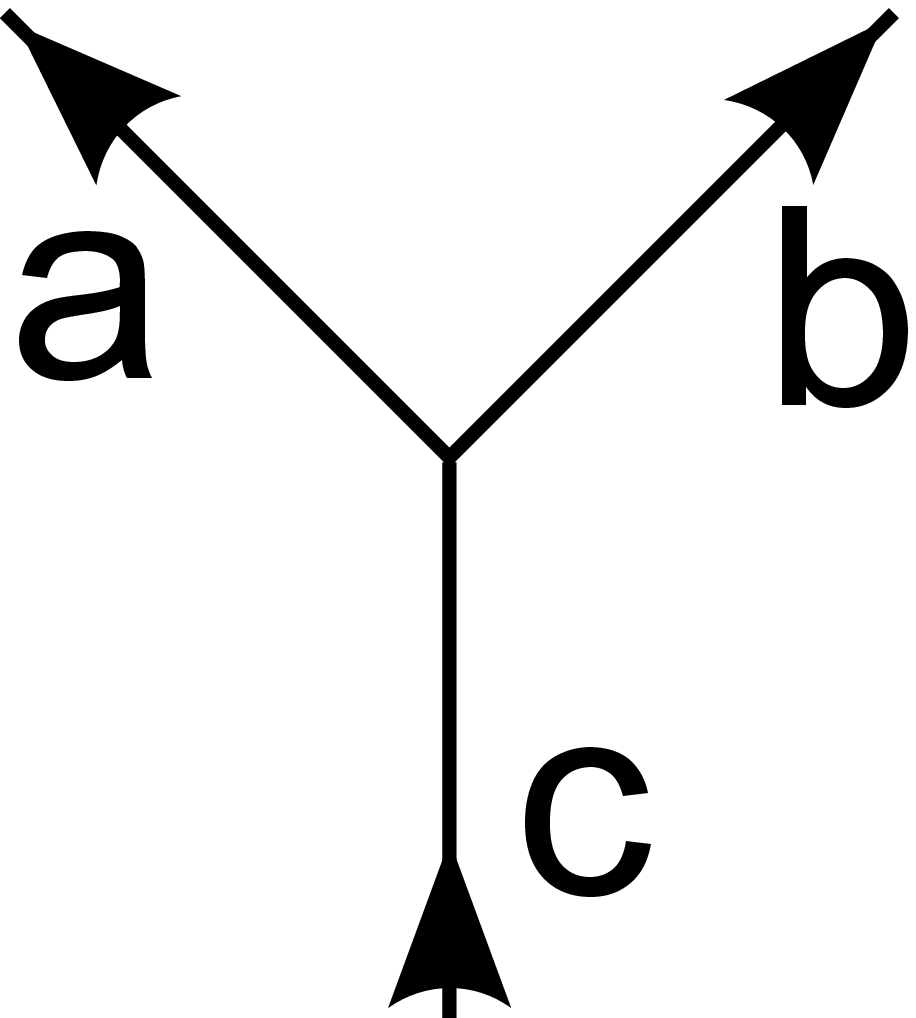}} \right|U
&=\frac{1}{f(a,b)}
\left\< \raisebox{-0.18in}{\includegraphics[height=0.4in,width=0.41in]{gauge1.eps}} \right|, \\
\left\< \raisebox{-0.1in}{\includegraphics[height=0.3in,width=0.41in]{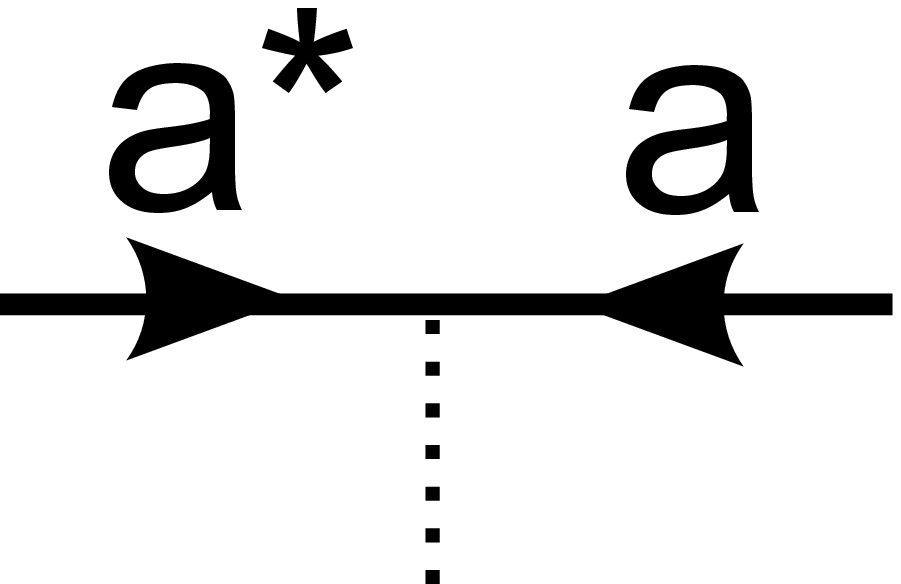}} \right|U
&=f(a^*,a)
\left\< \raisebox{-0.1in}{\includegraphics[height=0.3in,width=0.41in]{gauge3.eps}} \right|,\\
\left\< \raisebox{-0.18in}{\includegraphics[height=0.4in,width=0.41in]{vertex2.eps}} \right|U 
&=\frac{f(b,b^*)}{f(a,b^*)}
\left\< \raisebox{-0.18in}{\includegraphics[height=0.4in,width=0.41in]{vertex2.eps}} \right|,\\
\left\< \raisebox{-0.18in}{\includegraphics[height=0.4in,width=0.41in]{vertex3.eps}} \right|U 
&=\frac{1}{f(a,b)f(c,c^*)}
\left\< \raisebox{-0.18in}{\includegraphics[height=0.4in,width=0.41in]{vertex3.eps}} \right|,\\
\left\< \raisebox{-0.18in}{\includegraphics[height=0.4in,width=0.41in]{vertex4.eps}} \right|U
&=\frac{f(a^*,a)f(b^*,b)}{f(a^*,b^*)}
\left\< \raisebox{-0.18in}{\includegraphics[height=0.4in,width=0.41in]{vertex4.eps}} \right|.
\end{align*}
Here, by this notation, we mean that $U$ multiplies each string-net basis state by the above phase factors for each vertex of
the above form. 

It is not hard to check that $U$ transforms $\Phi$ to $\tilde{\Phi}$. Indeed, if we substitute the
above transformation into the local rules (\ref{rule1} - \ref{rule3}) and (\ref{nullerase} - \ref{rule4'}), we can see that 
if $|\Phi\>$ obeys the local rules corresponding to $\{F, d, \gamma, \alpha\}$, then
$U|\Phi\>$ obeys the local rules corresponding to $\{\tilde{F}, \tilde{d}, \tilde{\gamma}, \tilde{\alpha}\}$. Therefore, since
the local rules determine the wave function uniquely, we must have $U|\Phi\> = |\tilde{\Phi}\>$.
Similarly, it is easy to check that the gauge transformed Hamiltonian $\tilde{H}$ is unitarily equivalent to the original Hamiltonian, i.e. $UHU^{\dagger}= \tilde{H}$, by applying $U$ to (\ref{bgenorient}), (\ref{Q})
We conclude that two solutions related by $f$-transformation describe the same phase since the two wave functions and Hamiltonians are connected by a local unitary transformation $U$.

As for the $g$-transformation, one can show that $\tilde{d_s}\tilde{B}_{p}^{s}=d_s B_{p}^{s}$ by substituting the transformation (\ref{gauge2}) into (\ref{bgenorient})
and noting that all the factors of $g$ cancel. It then follows that the Hamiltonian is invariant under the $g$-transformation, i.e. $H = \tilde{H}$.
Accordingly, the wave function is also invariant under the transformation: $|\Phi\> = |\tilde{\Phi}\>$. In other words, the unitary transformation that
relates the two models is simply the identity map, $U = \mathbf{1}$. We would like to mention that the triviality of the $g$-transformation is special to the
particular orientation choice shown in Fig. \ref{lattice}: for other orientation choices, $|\Phi\> \neq |\tilde{\Phi}\>$. That being said, it is
clear that $|\Phi\>$ and $\tilde{\Phi}\>$ are still related by local unitary transformations for other orientation choices. Indeed, this result follows
from the fact that a change in the orientation configuration can be implemented by a local unitary transformation (made up of a product of $\alpha$ and $\gamma$ factors).

\section{Graphical representation of the Hamiltonian \label{app_h}}
In this section we demonstrate the graphical representation of $B_p^s$ leads to the same matrix elements as in equation (\ref{b}). 

According to the graphical representation, the action of the operator $B_{p}^{s}$ on the string-net state 
$\left\< \raisebox{-0.2in}{\includegraphics[height=0.6in]{B1.eps}} \right| $ is equivalent to adding a loop of type-$s$ string.
This allows us to obtain the matrix elements of $B_{p}^{s}$:
\begin{widetext}
\begin{align*}
&
\left\<
\raisebox{-0.2in}{\includegraphics[height=0.6in]{B1.eps}}
\right| B_{p}^{s}
=\left\<
\raisebox{-0.27in}{\includegraphics[height=0.65in]{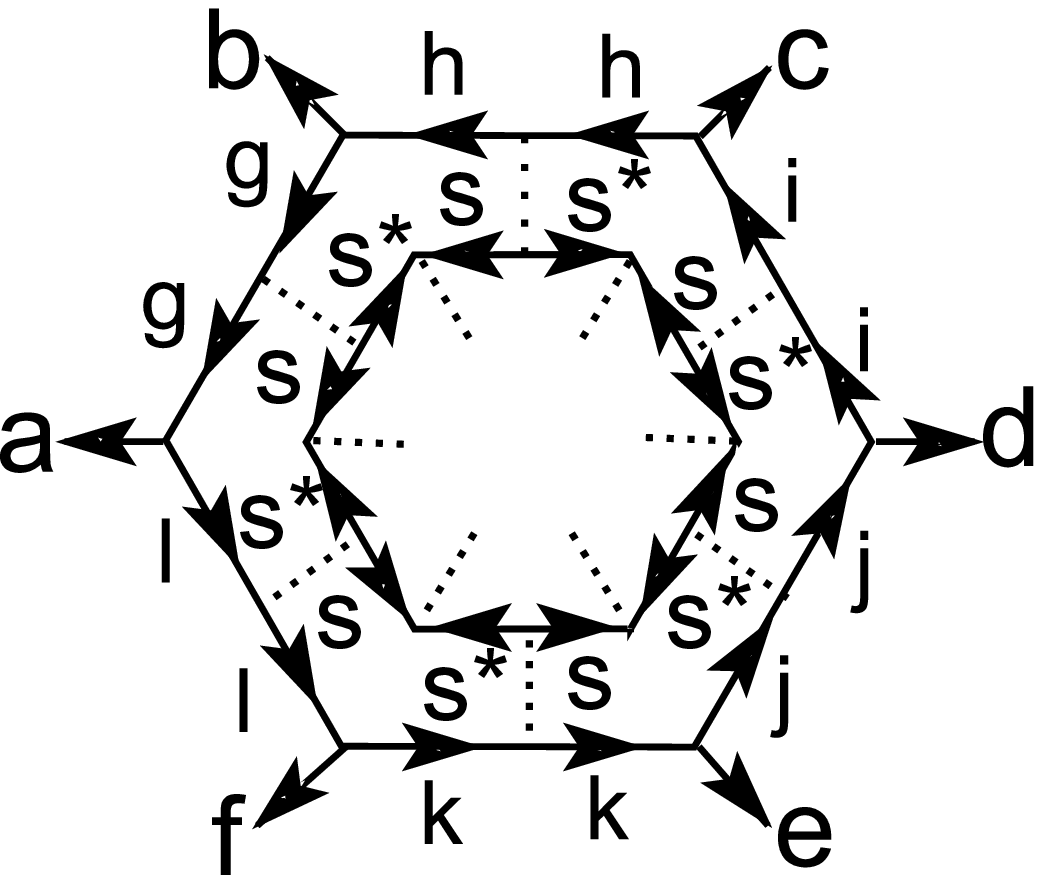}}
\right| 
={\textstyle\prod\limits_{x=\{g,h,i,j,k,l\}}}
F_{s^*sx} 
\left\<
\raisebox{-0.27in}{\includegraphics[height=0.65in]{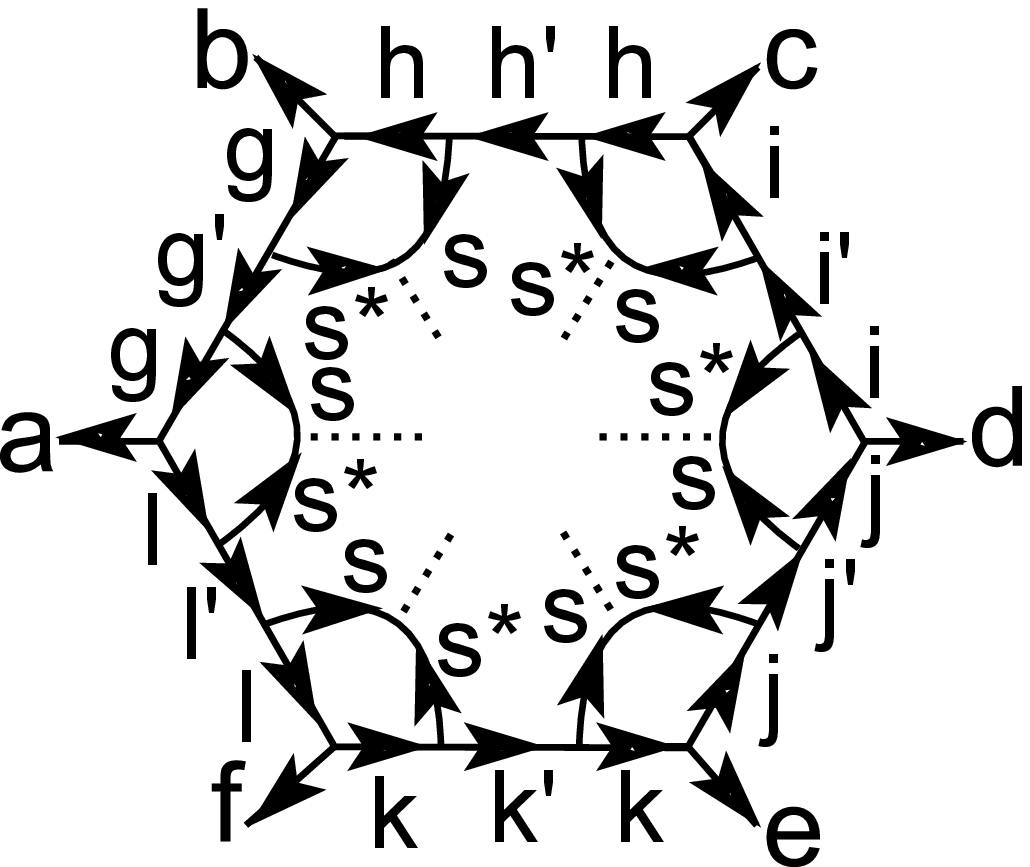}}
\right| 
\\
&=
\left(
\textstyle\prod\limits_{x=\{g,h,i,j,k,l\}}
F_{s^*sx} 
\right)
F_{s^{*}g'b}  
F_{s^{*}h'c} 
F_{s^{*}i'd} 
F_{s^{*}j'e}  
F_{s^{*}k'f}  
F_{s^{*}l'a}
\left\<
\raisebox{-0.27in}{\includegraphics[height=0.65in]{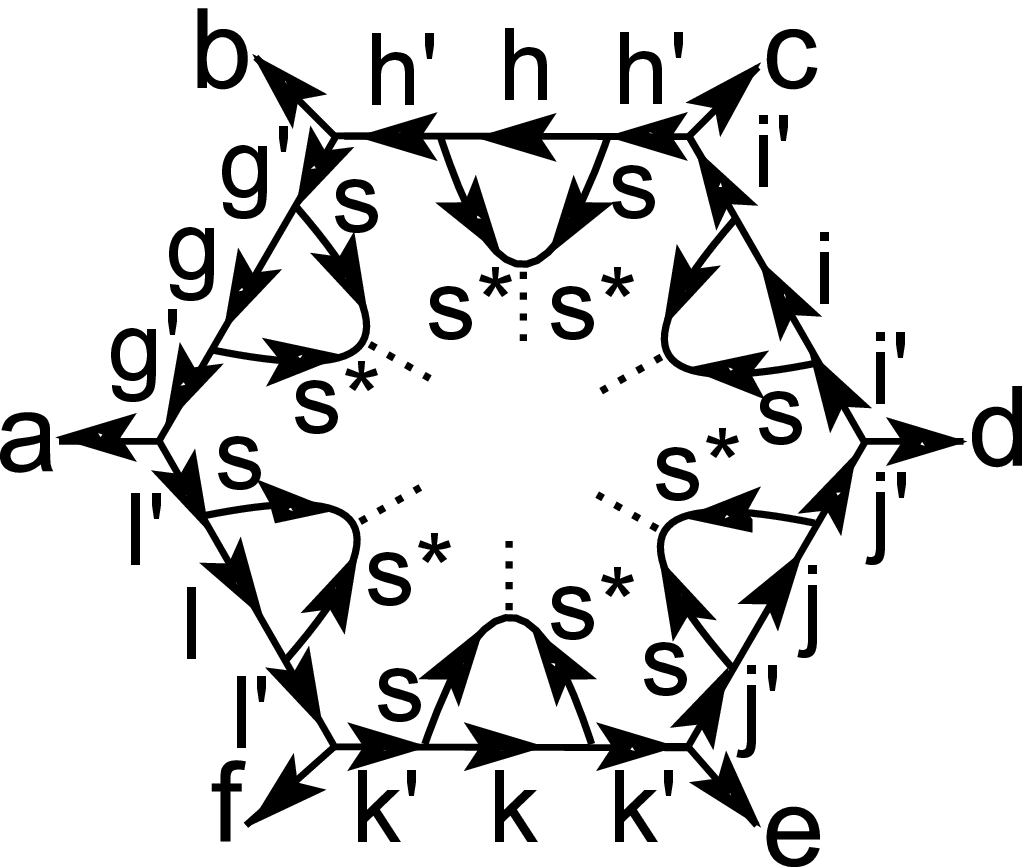}}
\right| 
\\
&=
\left(
\textstyle\prod\limits_{x=\{g,h,i,j,k,l\}}
F_{s^*sx}F_{s^*x'x'^*}\alpha_{sx}\gamma_{x'^*}\gamma_{s^*} d_s \right)
F_{s^{*}g'b}  
F_{s^{*}h'c}  
F_{s^{*}i'd} 
F_{s^{*}j'e}  
F_{s^{*}k'f}  
F_{s^{*}l'a}
\left\<
\raisebox{-0.2in}{\includegraphics[height=0.6in]{B2.eps}}
\right| \\
&=
F_{s^{*}g'b}  
F_{s^{*}h'c}  
F_{s^{*}i'd} 
F_{s^{*}j'e}  
F_{s^{*}k'f}  
F_{s^{*}l'a}
\left\<
\raisebox{-0.2in}{\includegraphics[height=0.6in]{B2.eps}}
\right| \equiv B_{p,g'h'i'j'k'l'}^{s,ghijkl}(abcdef)  
\left\<
\raisebox{-0.2in}{\includegraphics[height=0.6in]{B2.eps}}
\right|,
\end{align*}
\end{widetext}
with $x'=x+s$.
Here the fourth equality follows from
\begin{align*}
\left\< \raisebox{-0.1in}{\includegraphics[height=0.3in]{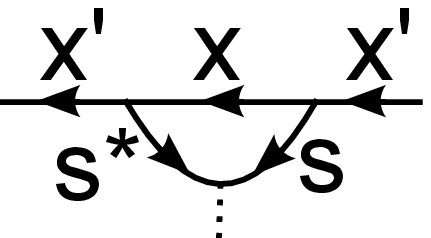}} \right| &=
\alpha_{sx}
\left\< \raisebox{-0.1in}{\includegraphics[height=0.3in]{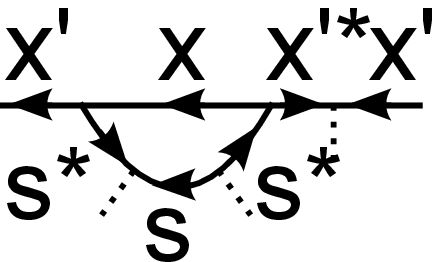}} \right|  \\
&=\alpha_{sx}F_{s^*x'x'^{*}}\gamma_{x'^{*}}\gamma_{s^*}d_s
\left\< \raisebox{-0.04in}{\includegraphics[height=0.15in]{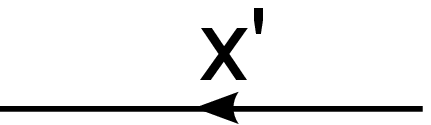}} \right|
\end{align*}
while the fifth equality follows from the identity
\begin{equation*}
	F_{s^*sx}F_{s^*x'x'^*}\alpha_{sx}\gamma_{x'^*}\gamma_{s^*} d_s=1
	\label{}
\end{equation*}
with $x'=x+s$.
This identity can be derived from the self-consistency conditions (\ref{selfconseq}), by expressing $\alpha$,$\gamma$ in terms of $F$, and then using the pentagon identity (\ref{pentid}) together with (\ref{F0}) to simplify the resulting expression.

Our derivation is now complete: we can see that the above expression agrees with the matrix elements of Eq. (\ref{b}), as we wished to show.

\section{Showing \texorpdfstring{$B_{p_1}^{s_1}$}{Bp1s1} and \texorpdfstring{$B_{p_2}^{s_2}$}{Bp2s2} commute \label{app_commute}}
In this section we will show that the operators $B_{p_{1}}^{s_{1}}$ and $B_{p_{2}}^{s_{2}}$ commute with one another. 
We only have to consider the case where the two plaquettes are the same, $p_{1} = p_{2}$, or the case where $p_{1}$ and $p_{2}$ are adjacent
since it is clear that the two operators will commute if $p_{1}$ and $p_{2}$ are further apart.

The first case is when the two $B_p$ operators act on the same plaquette, i.e. $p_{1}=p_{2} = p$. We need to show $B_{p}^{s}$ and $B_{p}^{t}$ commute.
One can prove this by writing down the matrix elements of $B_{p}^{s}B_{p}^{t}:$
\begin{widetext}
\begin{align*}
	( B_{p}^{s}B_{p}^{t})_{g''h''i''j''k''l''}^{ghijkl}( abcdef)=
	F_{s^{*}g'b} 
	F_{s^{*}h'c}
	F_{s^{*}i'd}
	F_{s^{*}j'e}
	F_{s^{*}k'f}
	F_{s^{*}l'a}
	F_{t^{*}g''b} 
	F_{t^{*}h''c}
	F_{t^{*}i''d}
	F_{t^{*}j''e}
	F_{t^{*}k''f}
	F_{t^{*}l''a}
\end{align*} 
where $x'=x+s$ and $x''=x'+t$ with $x\in\{ g,h,i,j,k,l\}.$
Using the pentagon identity, 
$F_{s^{*},x',y}F_{t^{*}x''y}=F_{(s+t)^{*}x''y}F_{s^{*}t^{*}(x''+y)}/F_{s^{*}t^{*}x''}$,
the above expression can be rewritten as
\begin{align*}
	( B_{p}^{s}B_{p}^{t})_{g''h''i''j''k''l''}^{ghijkl}(abcdef)
	&=F_{(s+t)^{*}g''b}F_{(s+t)^{*}h''c}F_{(s+t)^{*}i''d}F_{(s+t)^{*}j''e}F_{(s+t)^{*}k''f}F_{(s+t)^{*}l''a}\\
	&=( B_{p}^{s+t})_{g''h''i''j''k''l''}^{ghijkl}(abcdef).
\end{align*}
\end{widetext}
We conclude that 
\begin{equation}
B_{p}^{s}B_{p}^{t}=B_{p}^{t}B_{p}^{s}=B_{p}^{s+t}.
\label{bcommute1}
\end{equation}
proving commutativity when $p_1=p_2=p$.
\begin{figure}[ptb]
\begin{center}
\includegraphics[height=1.5in,width=1.1in]{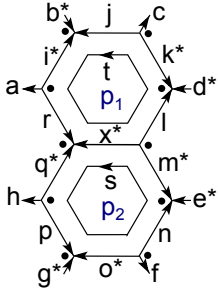}
\end{center}
\caption{Two plaquette operators $B_{p_1}^t$ and $B_{p_2}^s$ act on two adjacent plaquettes and add two loops in two different orders.
	If we apply $B_{p_2}^s$ first and then $B_{p_1}^t$, the resulting matrix elements are (\ref{bprod1}).
	Conversely, if we apply $B_{p_1}^t$ first and then ${B_{p_2}^s}$, we will get (\ref{bprod2}).
}
\label{com}
\end{figure}

The second case is when the two $B_p$ operators act on two adjacent plaquettes $p_1, p_2$. We want to show $B_{p_1}^t B_{p_2}^s=B_{p_2}^s B_{p_1}^t$ (see Fig. \ref{com}).
To prove this, we use (\ref{bgenorient}) to write down the matrix elements of operators on each side and then show they are equal.

The matrix elements of $B_{p_2}^s B_{p_1}^t$ are
\begin{align}
	&B_{p_2,q'x^{*'}m'n'o'p'}^{s,qx^{*}mnop}(hr^*lefg) \ \times \nonumber \\
	&B_{p_1,i''j''k''l''(x-s)''r''}^{t,ijkl(x-s)r}(abcdm^{\prime *}q').
	\label{bprod1}
\end{align}
Here we use the notation $y'=y+s$ and $y''=y+t$ for any string type-$y$. 
Similarly, the matrix elements of $B_{p_1}^t B_{p_2}^s$ are
\begin{align}
	&B_{p_1,i''j''k''l''x''r''}^{t,ijklxr}(abcdm^*q) \ \times \nonumber \\
	&B_{p_2,q'(x+t)^{*'}m'n'o'p'}^{s,q(x+t)^{*}mnop}(hr^{\prime \prime*}l''efg). 
	\label{bprod2}
\end{align}
Writing out (\ref{bprod1}) and (\ref{bprod2}) explicitly using (\ref{bgenorient}), we see that many of the factors of $F, \alpha, \gamma$
are the same in both expressions. Thus, to prove $(\ref{bprod1}) = (\ref{bprod2})$, it is sufficient to compare the factors which are different.
Specifically, we need to show the product
\begin{align}
	F_{s^* q' r^*} F_{s^* x^{*'} l} F_{t^* l'' m'^*} F_{t^* (x+t-s) q'} \ \times \nonumber \\ 
	\alpha_{x^{*\prime} l} \alpha_{x''^* l''} \alpha_{x q} \alpha_{(x+t-s) q'} \gamma_{x+t} \gamma_{x-s}
	\label{bprod3}
\end{align}
which appears in (\ref{bprod1}) is equal to the product 
\begin{align}
	F_{s^*q'r^{''^*}} F_{x^*(x+t-s)^* l''} F_{t^* l'' m^*} F_{t^* x'' q} \ \times \nonumber \\
	\alpha_{(x+t-s)^* l''} \alpha_{x^* l} \alpha_{(x-s) q'} \alpha_{x'' q} \gamma_x \gamma_{x+t-s}
	\label{bprod4}
\end{align}
which appears in (\ref{bprod2}). 

Equivalently, we need to show that the ratio of (\ref{bprod3}) and (\ref{bprod4}) is equal to unity. To this end, we divide the
$F, \alpha$ factors in the ratio $(\ref{bprod3})/(\ref{bprod4})$ into two groups which are associated with the
left and right ends of the common link, respectively. Combining all the factors from the left end of the link gives
\begin{equation*}
	C_{L} \equiv \frac{F_{s^* q' r^*} F_{t^* (x+t-s) q'} \alpha_{x q} \alpha_{(x+t-s) q'} }
	{F_{s^*q'r^{''^*}} F_{t^* x'' q} \alpha_{(x-s) q'} \alpha_{x''q}}.
	\label{}
\end{equation*}
Similarly, the factors from the right end of the link give
\begin{equation*}
	C_R \equiv \frac{F_{s^* x^{*'} l} F_{t^* l'' m'^*} \alpha_{x^{*\prime} l} \alpha_{x''^* l''}}
	{F_{x^*(x+t-s)^* l''} F_{t^* l'' m^*} \alpha_{(x+t-s)^* l''} \alpha_{x^* l}}.
	\label{}
\end{equation*}
Showing $(\ref{bprod3})/(\ref{bprod4}) = 1$ is then equivalent to proving that
\begin{equation}
	C_L C_R \cdot \frac{\gamma_{x+t}\gamma_{x-s}}{\gamma_x\gamma_{x+t-s}} = 1.
	\label{c12}
\end{equation}
We notice that $C_L$ and $C_R$ are related by $C_R=C_L^{-1}(x\rightarrow -x,s\leftrightarrow t, q\rightarrow l, r\rightarrow m)$. 

The next step is to express all the $\alpha$ factors in $C_1, C_2$ in terms of $F$ using (\ref{alphaF}) and then simplify the resulting expressions
using the pentagon identity (\ref{pentid}). The result is:
\begin{align}
	C_L=\frac{F_{x s^* x^{*'}}}{F_{t^* x'' s^*}F_{x'' s^* (x+t-s)^*}}, \label{cl} \\
	C_R=\frac{F_{s^* x^{*'} t^*} F_{x^{*'} t^* (x+t-s)}}{F_{x^* t^* x''}}. \label{cr}
\end{align}
Furthermore, using (\ref{gammaF}) (\ref{dcons}), we derive
\begin{align}
\frac{\gamma_{x+t}\gamma_{x-s}}{\gamma_x\gamma_{x+t-s}} = \frac{F_{(x+t)(x+t)^*(x+t)}F_{(x-s)(x-s)^*(x-s)}}{F_{xx^*x}F_{(x+t-s)(x+t-s)^*(x+t-s)}}.
\label{gammasimp}
\end{align}
In the final step, we take the product of (\ref{cl}),(\ref{cr}),(\ref{gammasimp}). The expression on the right hand side can then be reduced to $1$,
using the pentagon identity (\ref{pentid}) together with (\ref{F0}). This establishes the identity (\ref{c12}) and completes the proof that the
$B_p^s$ terms commute with one each other.

\section{Properties of the Hamiltonian (\ref{h}) \label{property}}
In this section, we will establish the following properties of the Hamiltonian (\ref{h}):
\begin{enumerate}
\item{$(B_p^{s})^\dagger = B_p^{s^*}$.}
\item{$B_p$ is a projection operator, that is $B_p^2 = B_p$.}
\item{The ground state wave function on the honeycomb lattice satisfies the local rules (\ref{rule1} - \ref{rule3}).}
\end{enumerate}
Let us show them in order. To prove the first result, we use the pentagon identity (\ref{pentid}) and (\ref{F0}) to derive
\begin{eqnarray}
F_{s x' y} F_{s^* x y} &=& F_{0 x y} F_{s s^* (x+y)}/F_{s s^* x} \nonumber \\
&=& F_{s s^* (x+y)}/F_{s s^* x}
\end{eqnarray}
where $x' = x+s$. We then set
$x = g,h,i,j,k,l$ and $y=b,c,d,e,f,a$, and take the product of these six equations, thereby deriving
\begin{equation}
B_{p,ghijkl}^{s,g'h'i'j'k'l'}(abcdef)^{-1} = B_{p,g'h'i'j'k'l'}^{s^*,ghijkl}(abcdef).
\label{bpid}
\end{equation}
Finally, we use $F_{ijk}^{-1}=F_{ijk}^*$ (\ref{unit}) to rewrite (\ref{bpid}) as 
\begin{equation}
B_{p,ghijkl}^{s,g'h'i'j'k'l'}(abcdef)^* = B_{p,g'h'i'j'k'l'}^{s^*,ghijkl}(abcdef).
\label{bpherm2}
\end{equation}
This establishes the first property.

To prove the second result, we use the identity (\ref{bcommute1}) established in appendix \ref{app_commute} to derive
\begin{align}
B_p^2 = \sum_{s,t} \frac{d_s d_t}{|G|^2}B_p^s B_p^t = \sum_{s,t} \frac{d_{s} d_{t}}{|G|^2} B_p^{s+t} .
\end{align} 
We then use the self-consistency condition (\ref{dcons}) to write $d_s d_t = d_{s+t}$. Changing variables to $s' = s+t$, we derive
\begin{equation}
B_p^2 = \sum_{s'} \frac{d_{s'}}{|G|} B_p^{s'} = B_p.
\end{equation}

Finally we show that the ground state $\Phi_{latt}$ of $H$ obeys the local rules 
(\ref{rule1} - \ref{rule3}). 
To see this, we use the fact that $B_{p}| \Phi_{latt}\>=|\Phi_{latt}\>$ together
with the following relations:
\begin{align}
\left\< \raisebox{-0.17in}{\includegraphics[height=0.4in,width=0.5in]{wf1.eps}}\right| B_{p} 
& =\left\< \raisebox{-0.17in}{\includegraphics[height=0.4in,width=0.5in]{wf2.eps}} \right| B_{p}, \nonumber \\
\left\< \raisebox{-0.17in}{\includegraphics[height=0.4in,width=0.45in]{wf3.eps}}\right| B_{p}
& =d_{s}\left\< \raisebox{-0.17in}{\includegraphics[height=0.4in,width=0.45in]{wf4.eps}} \right| B_{p}, \nonumber \\
\left\< \raisebox{-0.17in}{\includegraphics[height=0.4in,width=0.45in]{wf5.eps}}\right| B_{p}
& =F\left( a,b,c\right) \left\< \raisebox{-0.17in}{\includegraphics[height=0.4in,width=0.45in]{wf6.eps}} \right| B_{p}. \label{bprel}
\end{align}
Multiplying these equations by the ground state ket $|\Phi_{latt}\>$, we can immediately see that the ground state wave function $\Phi_{latt}(X)=\<X|\Phi_{latt}\>$ satisfies the local rules (\ref{rule1} - \ref{rule3}). 

The relations (\ref{bprel}) can be proved using the expression for the matrix elements of $B_{p}^{s}$ in (\ref{b}) together with the pentagon identity.
For example, to prove the last equation, we expand out the left hand side as
\begin{align*}
	&\left< \raisebox{-0.2in}{\includegraphics[height=0.5in,width=0.51in]{wf5.eps}}\right| B_{p}
	=\sum_{s}\frac{d_s}{|G|}\left< \raisebox{-0.2in}{\includegraphics[height=0.5in,width=0.51in]{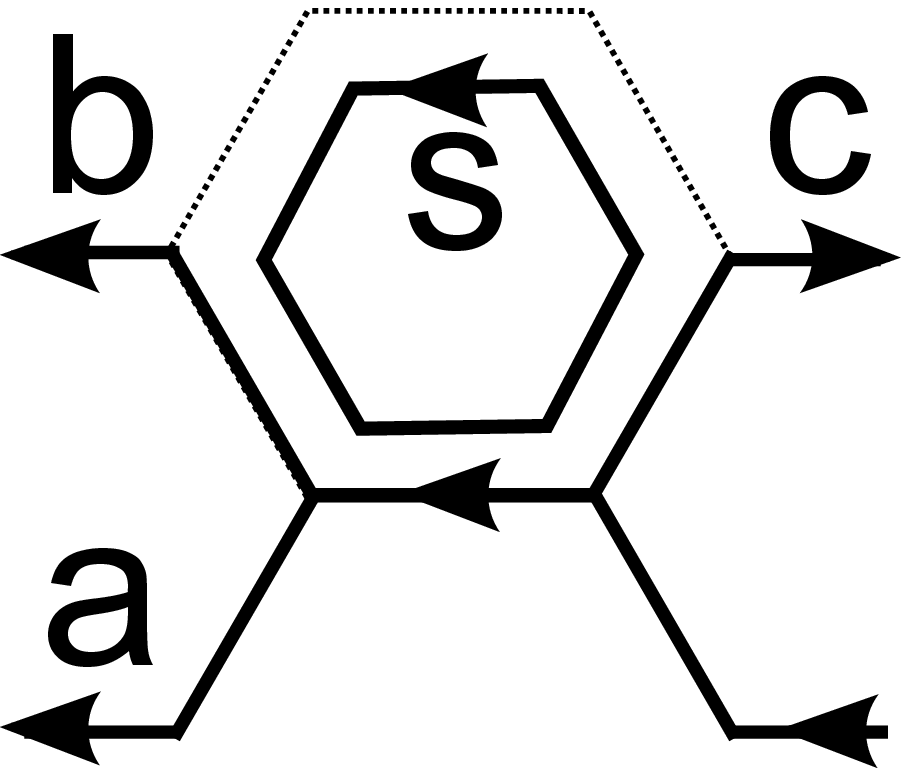}} \right| \\
	&=\sum_{s}\frac{d_s}{|G|}\alpha^{-1}_{(a+b)^{*}a}\alpha_{(a+b)c}\gamma_{b^{*}}\gamma_{(a+b)^{*}}
	\left< \raisebox{-0.2in}{\includegraphics[height=0.5in,width=0.51in]{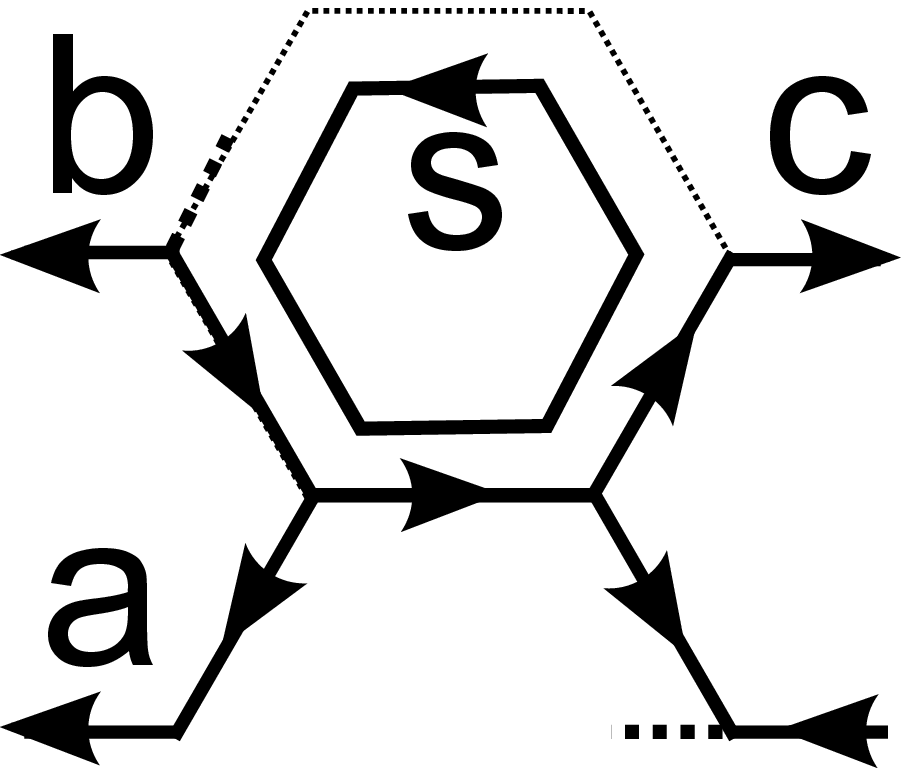}} \right| \\
	&=\sum_{s}\frac{d_s}{|G|}\alpha^{-1}_{(a+b)^{*}a}\alpha_{(a+b)c}\gamma_{b^{*}}\gamma_{(a+b)^{*}}\times\\
	&B^{s,000c(a+b)^{*}b^{*}}_{sssc'(a+b)^{*'}b^{*'}}(b00c(a+b+c)^{*}a)
	\left< \raisebox{-0.2in}{\includegraphics[height=0.5in,width=0.51in]{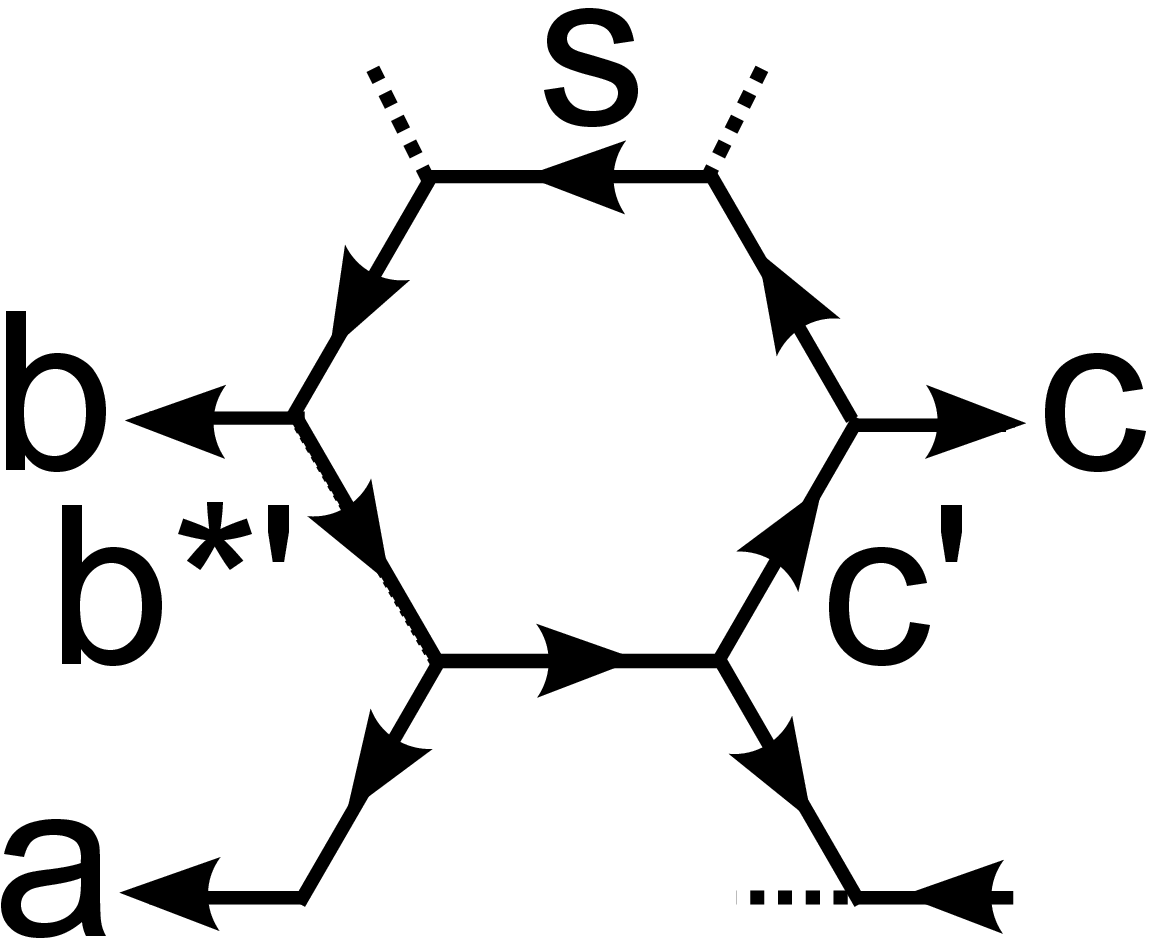}} \right|. 
\end{align*}
Here $B$ is defined in (\ref{b}), $x'=x+s$.
Similarly, on the right hand side, we find
\begin{align*}
	&\left< \raisebox{-0.2in}{\includegraphics[height=0.5in,width=0.51in]{wf6.eps}}\right| B_{p}^{s}
	=\sum_{s}\frac{d_s}{|G|}\left< \raisebox{-0.2in}{\includegraphics[height=0.5in,width=0.51in]{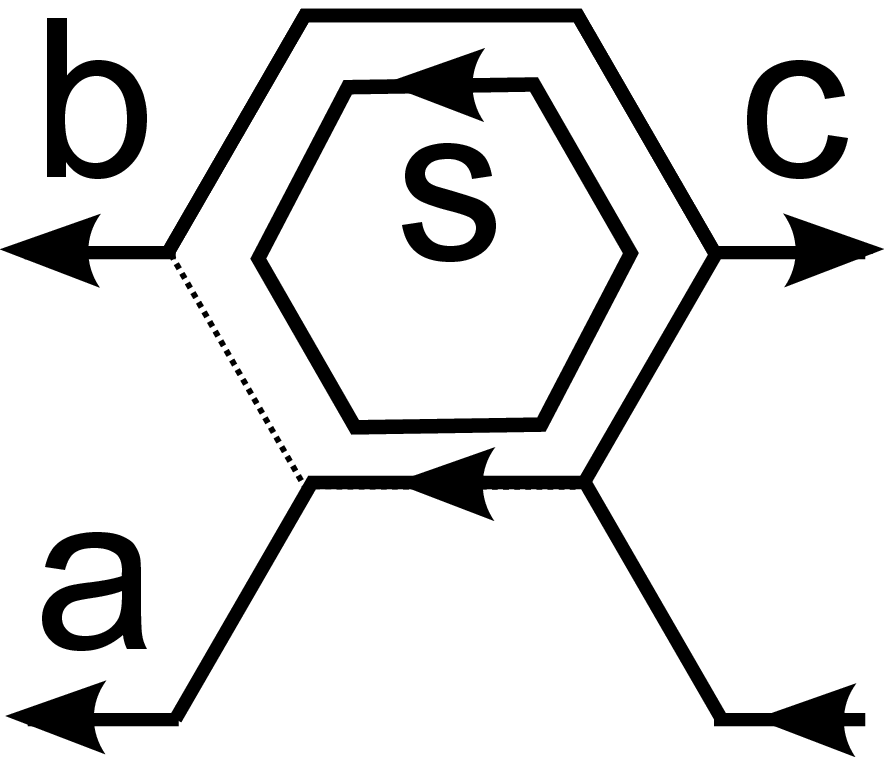}}\right|  \\
	&=\sum_{s}\frac{d_{s}}{|G|}\alpha_{a(b+c)}\gamma_{a^{*}} \times \\
	&B^{s,bbb(b+c)a^{*}0}_{b'b'b'(b+c)'a^{*'}s}(b00c(a+b+c)^{*}a)
	\left< \raisebox{-0.2in}{\includegraphics[height=0.5in,width=0.61in]{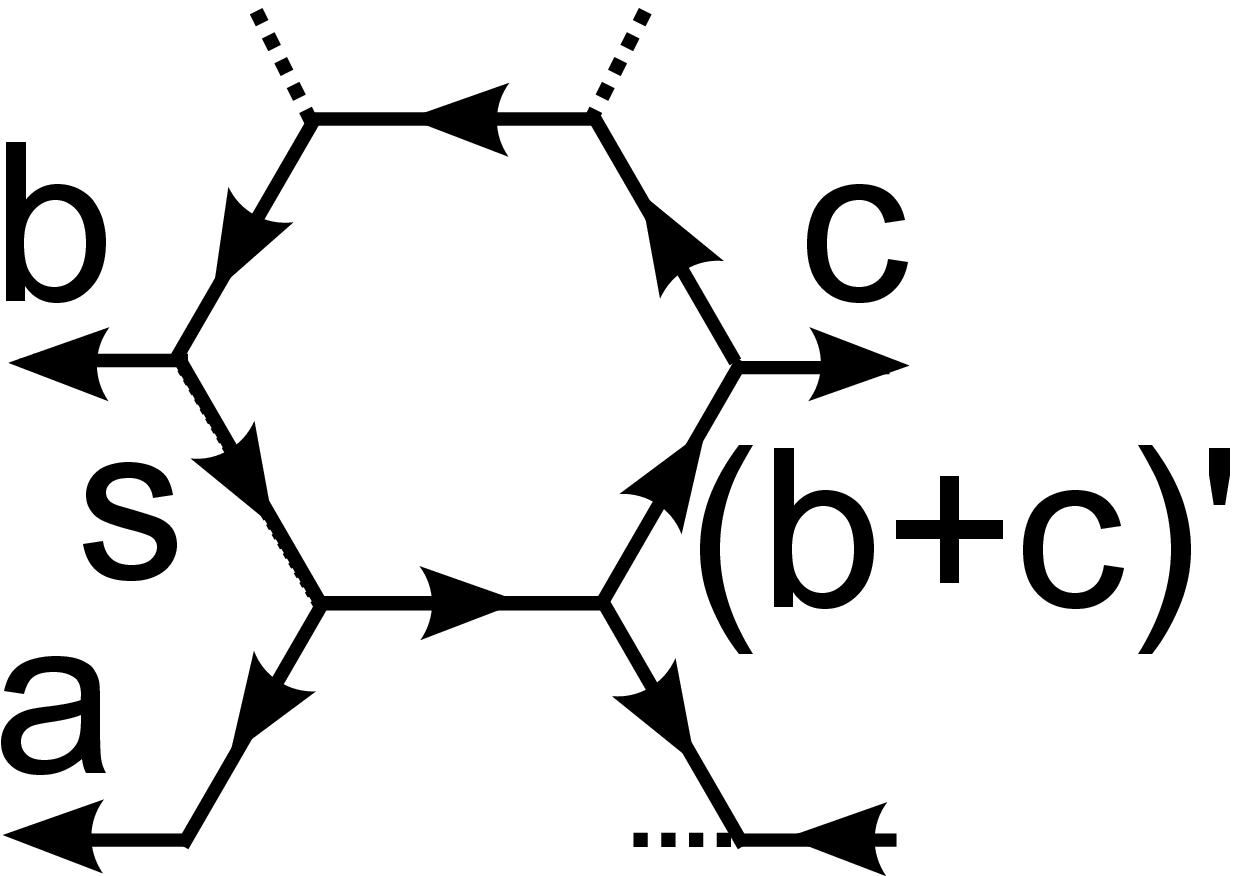}}\right| . 
\end{align*}
Changing the dummy variable $s$ to $s+b$ in the first expression, we can see that the resulting final state is the same as the second final one. 
We then compute the ratio of the two corresponding amplitudes. Using (\ref{selfconseq}), this ratio can be simplified to $F(a,b,c)$.
This justifies the third equation in (\ref{bprel}). In addition, by taking $a=0,b=c^*$, we obtain the first equation. The second equation can be shown in a similar manner.

\section{Gapped edge states \label{gapedge}}
In this section, we analyze the abelian string-net models in a disk geometry, and we show that the boundary of the disk
can be gapped for an appropriate choice of edge interactions. We note that a more general analysis of gapped boundaries of string-net models
was given in Ref. [\onlinecite{KitaevKong}].

\begin{figure}[tb]
        \begin{center}
                \includegraphics[height=1.8in,width=2.1in]{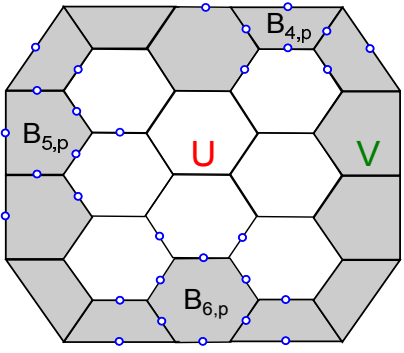}
        \end{center}
                \caption{
                        The lattice spin model in a disk geometry.
                        The gray region denotes the edge part $V$ and the white is the bulk $U$.
                        The total Hamiltonian has the form $H=H_{bulk}+H_{edge}$
                        where the bulk Hamiltonian $H_{bulk}$ is defined by (\ref{hbulk})
                        and the edge Hamiltonian $H_{edge}$ contains three kinds of spin interaction terms
                        $B_{4,p},B_{5,p}$, and $B_{6,p}$ defined in (\ref{bnp}).
                }
        \label{edge}
\end{figure}

For concreteness, let us consider the geometry shown in Fig. \ref{edge}. Let us assume that in the bulk of the disk, the Hamiltonian is defined 
by (\ref{h}), that is,
\begin{equation}
H_{bulk} = - \sum_{I \in U} Q_I - \sum_{p \in U} B_p.
\label{hbulk}
\end{equation}
Here the second sum runs over all plaquettes $p$ in region $U$ in Fig. \ref{edge}, while the first sum runs over all vertices $I$
that are inside or on the boundary of region $U$.

Our task is to find an edge Hamiltonian $H_{edge}$ acting on the spins near the boundary of the disk such that 
$H = H_{bulk} + H_{edge}$ has an energy gap and either a unique or finitely degenerate ground state. Once we find such an edge Hamiltonian, we will have proven 
explicitly that the edge of the abelian string-net models can be gapped.

Fortunately, it is easy to construct the desired $H_{edge}$. The key point is that, although we have focused on 
the string-net Hamiltonian (\ref{h}) in the context of the honeycomb lattice, the Hamiltonian (\ref{h}) can be readily generalized 
to any planar trivalent graph such as the graph shown in Fig. \ref{edge}. Hence there is a natural way to define 
$H_{edge}$ that maintains the exact solubility and other properties of the bulk Hamiltonian. To be specific, we define
\begin{equation}
H_{edge} = - \sum_{I \in V} Q_I - \sum_{p \in V} {B_{n,p}} \label{hedge}
\end{equation}
where the second sum runs over all plaquettes $p$ in region $V$ and the first sum runs over all vertices on the boundary of the disk.
Each of the plaquette terms $B_{n,p}$ is a $2n$-spin interaction term with $n = 4,5,6$ (see Fig. \ref{edge}). Similarly to the bulk plaquette
terms $B_p$, we define
\begin{equation}
B_{n,p}=\sum_{s \in G} \frac{d_s}{|G|} \cdot B_{n,p}^{s}
\label{bnp}
\end{equation}
where
\begin{align*}
        \left\<
        \raisebox{-0.2in}{\includegraphics[height=0.5in]{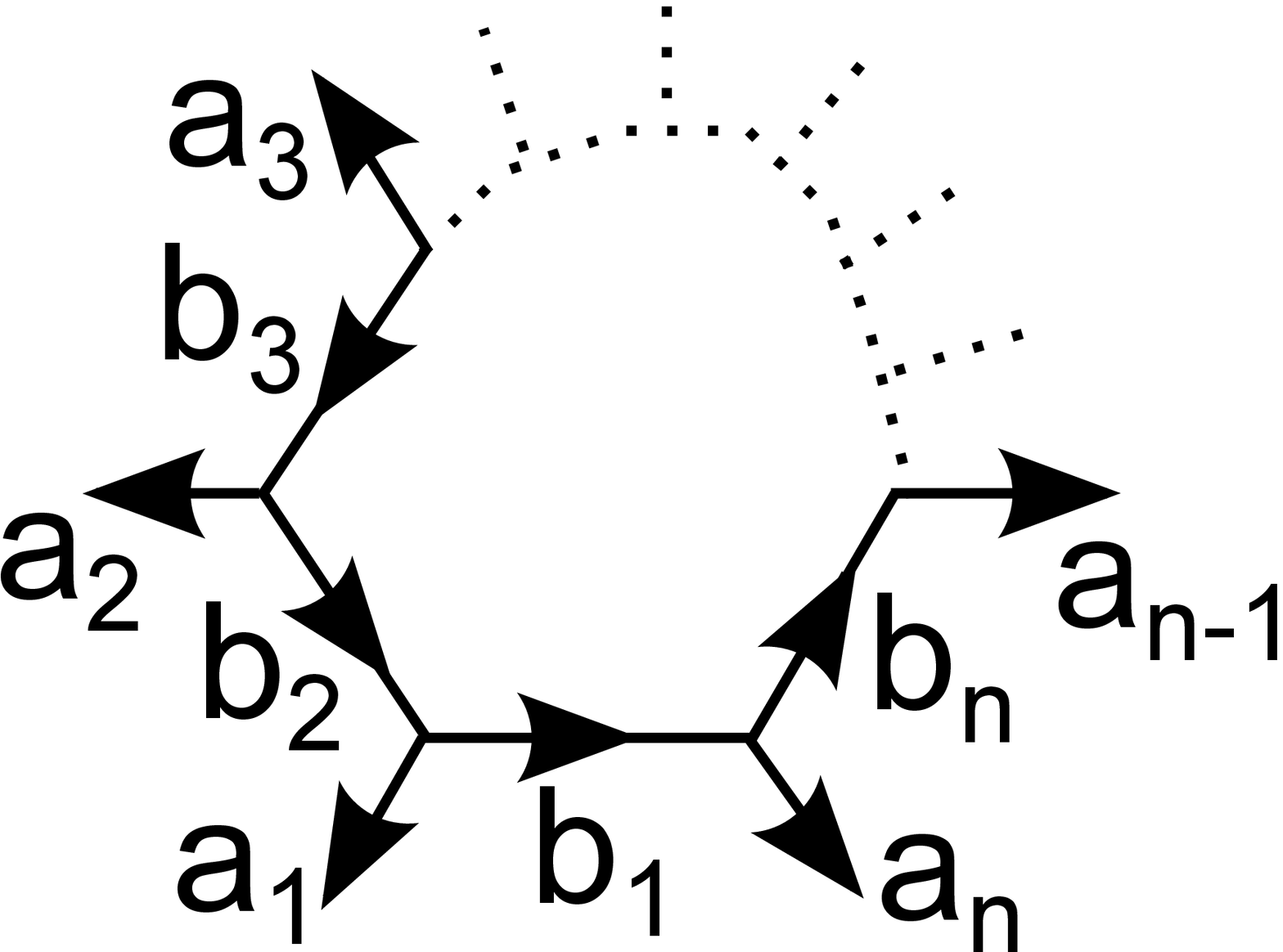}}
        \right| B_{n,p}^{s} \left| \raisebox{-0.2in}{\includegraphics[height=0.5in]{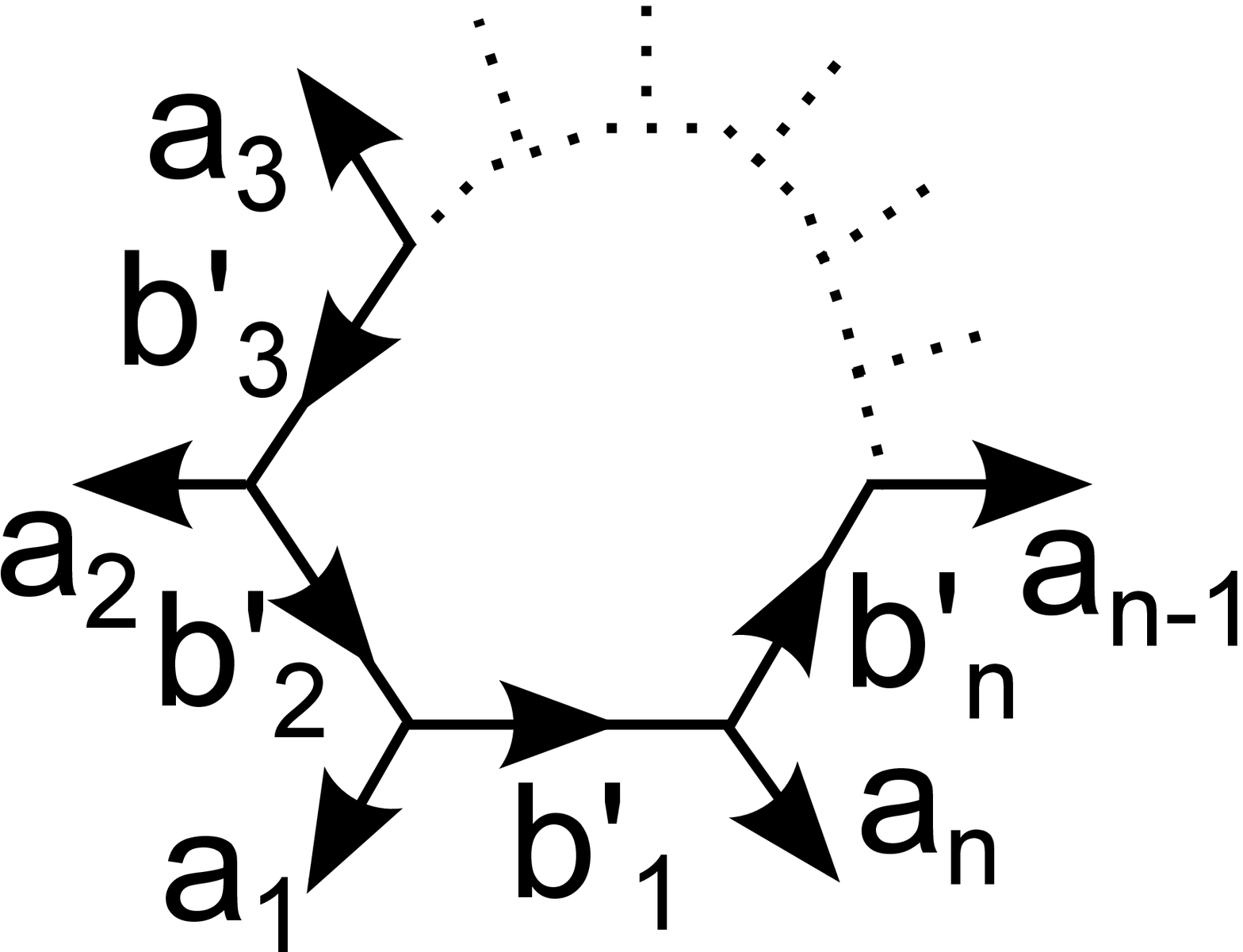}} \right\>
        &= \nonumber \\
	B_{n,p,b'_{1}b'_{2}\dots b'_{n}}^{s,b_{1}b_{2}\dots b_{n}}&(a_{1}a_{2}\dots a_{n})
\end{align*}
and
\begin{equation}
        B_{n,p,b'_{1}b'_{2}\dots b'_{n}}^{s,b_{1}b_{2}\dots b_{n}}(a_{1}a_{2}\dots a_{n})
	= \prod_{i=1}^n \delta_{b_i'}^{b_i+s} \cdot \prod_{i=1}^{n}F(s^{*},b'_{i},a_{i}).
        \label{bpgen}
\end{equation}
(Note that Eq. (\ref{bpgen}) only defines the plaquette operators for a particular choice of 
orientations. The definition of $B_p$ for other orientations will generally include additional factors of $\gamma$ and 
$\alpha$ as in Eq. (\ref{bgenorient}).)

Now, using exactly the same arguments as in the original honeycomb lattice case (\ref{h}), one can show that every
term in $H = H_{bulk} + H_{edge}$ commutes with every other term:
\begin{equation}
	[Q_{I},B_{n,p}^{s}]=[B_{p}^{s},B_{n,p}^{s}]=[B_{n,p}^{s},B_{n',p'}^{s'}]=0.
	\label{}
\end{equation}
Also, one can show that every term in the Hamiltonian is a projection operator, i.e. 
$Q_I, B_p, B_{n,p}$ all have eigenvalues $0,1$. Putting these facts together, we deduce that the ground state 
has eigenvalues $q_I = b_p = b_{n,p} = 1$, and that the lowest excited state is separated from the ground state by
an energy gap of at least $1$.

At this point, we have almost proven that the abelian string-net models support a gapped edge: 
all that remains is to show that the ground state is unique or finitely degenerate;
in fact we show the ground state is \emph{unique} on a disk.
We establish this result in appendix \ref{GSD}.

\section{Ground state degeneracy \label{GSD}}
In this  section, we calculate the ground state degeneracy of the string-net model (\ref{h}).  
We consider two geometries: a periodic torus geometry and a disk geometry, where the latter is defined in appendix \ref{gapedge}. We show that the degeneracy on a 
torus is $D = |G|^2$, while the degeneracy on a disk is $D=1$.

Our calculation, which is similar to that of Ref. [\onlinecite{LevinBurnell}], is based on the commuting projectors $Q_{I}$ and $B_{p}$ from Sec.\ref{fixedh}. 
Let us consider the product of these projectors over all vertices and plaquettes on a lattice.
Since $Q_I$ and $B_p$ have eigenvalues $0$ and $1$, this product will select out the states with eigenvalues $q_{I}=b_{p}=1$ for all $I$ and $p$. 
In other words, it will select out the ground states. Hence the ground state degeneracy can be calculated by taking the trace of the product of all the 
projectors:
\begin{equation}
	D=\text{Tr}\left(\prod_{p}B_{p} \prod_{I}Q_{I} \right).
\end{equation}
Here the trace is over all states in the Hilbert space.

We notice that the product of $Q_{I}$'s constrains the trace to configurations obeying the branching rules.
Thus $D$ reduces to the trace of the product of $B_p$ over allowed string-net configurations $X$.
Plugging in the expression for $B_{p}$, we have
\begin{align}
	D &= \sum_{\text{strnet X}}\<X|\prod_{p}\sum_{s}\frac{d_{s}}{|G|}B_{p}^{s}|X\> \nonumber \\
	&= \frac{1}{|G|^{N_{plaq}}}\sum_{\{s\}}\sum_{X}\<X|\prod_{p}d_{s}B_{p}^{s}|X\>. \label{d1}
\end{align}
Here $\{s\}$ means sets of $N_{plaq}$ numbers which specify $B_{p}^{s}$ over all plaquettes and $N_{plaq}$ is the total number of plaquettes. 

The next step is to calculate the expectation value $\<X| \prod_p B_{p}^{s} |X\>$. Let us first consider the disk geometry.
For a disk, this expectation value is nonzero only if $s=0$ for all plaquettes. 
The reason is that to get a nonzero expectation value, the final state after the action of $\prod_{p}B_{p}$ has to be the same as the original state. 
Since the disk has a boundary, this forces $s=0$ on the boundary and thus $s=0$ in the interior of the disk as well. Hence, in this case, (\ref{d1}) simplifies
to
\begin{equation}
	D=\frac{1}{|G|^{N_{plaq}}}\sum_{X}1=\frac{1}{|G|^{N_{plaq}}}N_{config}
	\label{d2}
\end{equation}
where $N_{config}$ is the number of string-net configurations on a disk.
To count these configurations, we can compare the number of free parameters to the number of constraints.
The number of free parameters is the number of links $N_{param}=N_{link}$.
As for the number of constraints, there is a constraint at each site coming from the branching rules. However,
not all the branching constraints are independent because if we sum up all the constraint equations, we get an 
equality which is automatically satisfied. Thus, the total number of constraints is $N_{const} = N_{site} - 1$.
Combining these calculations, the total number of string-net configurations on a disk is 
\begin{equation}
	N_{config}=|G|^{N_{param}-N_{const}}=|G|^{N_{link}-N_{site}+1}.
\end{equation}
Substituting this expression into (\ref{d2}), we derive
\begin{equation}
	D=\frac{1}{|G|^{N_{plaq}}}N_{config}=|G|^{N_{link}-N_{plaq}-N_{site}+1}.
\end{equation}
The final step is to note that a disk has an Euler characteristic of $1$, so 
\begin{equation}
	N_{site}-N_{link}+N_{plaq}=1.
\end{equation}
Thus we conclude $D=1$ on a disk.
\begin{figure}
\begin{center}
\includegraphics[height=1.1in,width=1.8in]{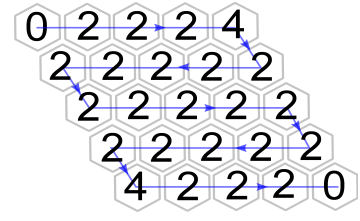}
\end{center}
\caption{
	A $5\times 5$ lattice on torus. We want to apply the $B_p^s$ operators to the whole lattice. 
	We start by applying the $B_p^s$ operators to the upper left hand corner and then following a zig-zag path to fill the lattice. 
	When a $B_p^s$ acts on each plaquette, it creates a closed loop-$s$ around the boundary of the plaquette. 
	Two adjacent loops will merge to a bigger loop, multiplied by a phase factor of $d_{s^*}$. 
	In general, each application of $B_p^s$ comes with a factor $d_{s^*}^{N_{ext}/2}$ 
	with $N_{ext}$ being the number of external legs on the boundary of the plaquette $p$ that are occupied by strings. 
	These numbers $N_{ext}$ are shown above in each plaquette. 
}
\label{gsdeg}
\end{figure}

Next we consider a torus geometry. Again, we need to calculate the expectation value $\<X| \prod_p B_{p}^{s} |X\>$ and substitute into (\ref{d1}). In the case of a torus, this expectation value is nonzero for any configuration in which $s$ is constant. This is because a torus does not have boundary and therefore as long as $s$ is constant, each link is acted on by two adjacent $B_{p}^{s}$ in opposite directions, so that $\prod_p B_p^s |X\> \propto |X\>$. Hence, in this case 
(\ref{d1}) simplifies to
\begin{equation}
D = \frac{d_s^{N_{plaq}}}{|G|^{N_{plaq}}}\sum_{s \in G}\sum_{X}\<X|\prod_{p}B_{p}^{s}|X\>.
\label{d3}
\end{equation}
To proceed further, we use the identity
\begin{equation}
	\<X|\prod_{p}B_{p}^{s}|X\>= d_{s^*}^{N_{plaq}}
\label{bpprodid}
\end{equation}
for constant $s$. We prove this identity in two steps. First, we note that every string-net state $|X\>$ can be written as 
$|X\> \propto \prod B_p^s \cdot \prod W_\alpha(P) |0\>$
where $|0\>$ denotes the vacuum state, and the product runs over some arbitrary set of plaquette operators $B_p^s$ and closed string operators $W_\alpha(P)$. Using this fact, together with the fact that $B_p^s, W_\alpha(P)$ commute with one another, it follows that 
\begin{equation}
\<X| \prod_p B_p^s |X\> = \<0| \prod_p B_p^s |0\>.
\end{equation}
We then calculate the latter expectation value by brute force. To this end, we order the plaquette operators in a particular way, starting with the upper left
hand corner of Fig. \ref{gsdeg} and then following a zig-zag path. When we apply the first plaquette operator to the vacuum state $\<0|$, we get
a single closed loop around the boundary of that plaquette. When we apply the second plaquette operator, the result is a larger closed loop, multiplied by
a phase factor of $d_{s^*}$. In general, it can be shown that each plaquette operator $B_p^s$ comes with phase factor of $d_{s^{*}}^{N_{ext}/2}$ with $N_{ext}$ being the number of external legs on the boundary of the plaquette $p$ that are occupied by strings. The values of $N_{ext}$ for the zig-zag ordering are shown in Fig. \ref{gsdeg}. We can
see that after applying all the plaquette operators, the total number of factors of $d_{s^*}$ is $(N_{plaq}-4) \cdot 1 + 2 \cdot 2 = N_{plaq}$. Hence,
$\<0| \prod_p B_p^s |0\> = d_{s^*}^{N_{plaq}}$. This proves the identity (\ref{bpprodid}). 

Substituting (\ref{bpprodid}) into (\ref{d3}), and using the fact that $d_s \cdot d_{s^*} = d_0 = 1$, we derive
\begin{align}
	D &=\frac{1}{|G|^{N_{plaq}}}\sum_{s \in G}\sum_{X}1 \nonumber \\
	&=\frac{|G|}{|G|^{N_{plaq}}}N_{config} \nonumber \\
	&=|G|^{N_{link}-N_{site}-N_{plaq}+2}
\end{align}
where the counting of string-net configurations $N_{config}$ is identical to the disk case.
Now a torus has an Euler characteristic of $0$ so
\begin{equation}
	N_{site}-N_{link}+N_{plaq}=0.
\end{equation}
Therefore, we conclude that the ground state degeneracy on a torus is $D=|G|^2$.

\begin{figure}[tb]
\begin{center}
\includegraphics[height=1.6in,width=2.3in]{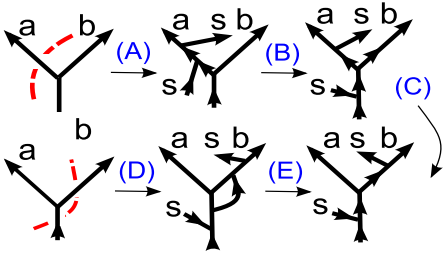}
\end{center}
\caption{
        The sequence to show Eq. (\ref{seq}). The arrows involve factors from local rules and resolving
crossings (see (\ref{phase1})).
}
\label{seqfig}
\end{figure}

\section{Derivation of Eqs. (\ref{wbar1}, \ref{wbar2}) \label{hexagon}}
\begin{figure}[tb]
\begin{center}
\includegraphics[height=1.3in,width=2.8in]{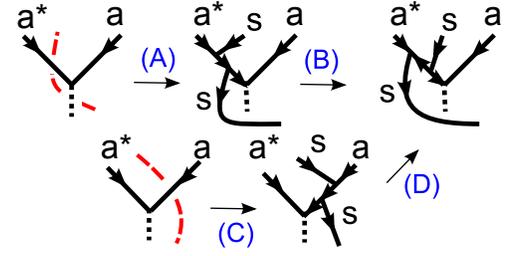}
\end{center}
\caption{
        The sequence to show Eq. (\ref{seq2}).  The arrows involve factors from local rules and resolving
crossings (see (\ref{phase2})).
}
\label{seq2fig}
\end{figure}

In this section, we show that the string operator $W(P)$ obeys the path independence conditions (\ref{pathind1}) and (\ref{pathind2}) if and only if
$\omega$ obeys Eqs. (\ref{wbar1}, \ref{wbar2}). The main steps in the argument are shown in Figs. \ref{seqfig}, \ref{seq2fig}. 

We begin with Fig. \ref{seqfig}. Clearly the string operator $W(P)$ will satisfy (\ref{pathind1}) if and only if the amplitudes of the two string-net
configurations on the left hand side of Fig. \ref{seqfig} are equal to one another. At the same time, using the local rules (\ref{rule1} - \ref{rule3}) and (\ref{nullerase} - \ref{rule4'}) and (\ref{rule7}) we can relate both of these amplitudes to the amplitude of the configuration shown in the bottom right hand corner of Fig. \ref{seqfig}. In this way, one can show that (\ref{pathind1}) is equivalent to the algebraic equation
\begin{equation}
(A)(B)(C)=(D)(E)
\label{seqalg}
\end{equation}
where
\begin{align}
	&(A)=\omega(a), \label{phase1}\\
	&(B)=F_{s^{*}(a+s)b}\alpha^{-1} _{s^{*}(a+s)}\alpha_{s^{*}(a+b+s)}\gamma_a \gamma_{(a+b)^{*}}, \nonumber\\
	&(C)=F_{asb}, \nonumber\\
	&(D)=\bar{\omega}(b)\omega(a+b), \nonumber\\
	&(E)=F_{abs}F_{(b+s)s^{*}s}d_s^* \gamma_s. \nonumber
\end{align}
Simplifying (\ref{seqalg}) using (\ref{selfconseq}), we obtain
\begin{equation}
\omega(a) = \bar{\omega}(b) \omega(a+b) \frac{F_{abs}F_{bss^*}F_{(a+s)b(a+b)^*}}{F_{ab(a+b)^*} F_{asb}}.
\label{seqalg2}
\end{equation}
Setting $a=0$ and using the fact that $\omega(0)=1$, along with (\ref{F0}), we derive (\ref{wbar2}):
\begin{equation}
\bar{\omega}(b) = \omega(b)^{-1} \cdot F_{bss^*}^{-1} F_{sbb^*}^{-1}. 
\end{equation}
Substituting the above relation back into (\ref{seqalg2}) and simplifying using the pentagon identity (\ref{pentid}) gives (\ref{wbar1}):
\begin{equation}
\omega(a) \omega(b) = \omega(a+b) \cdot \frac{F_{abs} F_{sab}}{F_{asb}} \frac{F_{s(a+b)(a+b)^*}}{F_{saa^*} F_{sbb^*}}.
\end{equation}

From the above analysis, we conclude that a string operator $W(P)$ will obey the path independence condition (\ref{pathind1}) if and only if $\omega$ obeys 
Eqs. (\ref{wbar1},\ref{wbar2}). The next question is to see whether the other path independence condition (\ref{pathind2}) gives any new constraints
on $\omega$. We will show that it does not.

To this end, we consider Fig. \ref{seq2fig}. Clearly $W(P)$ will satisfy (\ref{pathind2}) if and only if the amplitudes of the two configurations 
on the left hand side of Fig. \ref{seq2fig} are equal. At the same time, these amplitudes can both be related to the amplitude of the configuration shown in the top right hand corner of Fig. \ref{seq2fig}. Thus we see that (\ref{pathind2}) is equivalent to the equation
\begin{equation}
(A)(B)=(C)(D)
\label{seqalg3}
\end{equation}
where
\begin{align}
	&(A)=\omega(a^*), \label{phase2} \\
	&(B)=F_{a^*sa}, \nonumber\\
	&(C)=\bar{\omega}(a), \nonumber\\
	&(D)=\alpha^{-1}_{a^*(s+a)}\gamma_s. \nonumber
\end{align}
Simplifying (\ref{seqalg3}) using (\ref{selfconseq}) gives
\begin{equation}
	\omega(a^*)=\bar{\omega}(a)\frac{F_{a^*as}F_{ass^*}}{F_{a^*sa}},
	\label{}
\end{equation}
which is exactly the relation (\ref{seqalg2}) by setting $a\rightarrow -a$ and $b\rightarrow a$.
Thus it does not give rise to new constraints. This completes the proof that the string operator $W(P)$ is path independent if only if $\omega$ obeys 
(\ref{wbar1},\ref{wbar2}). 

\section{Exchange statistics \label{thetaapp}}
In this section, we derive the string algebra (\ref{exstat}) from the main text. That is, we show that the string operators $W_\alpha(P_1), W_\alpha(P_2), 
W_\alpha(P_3), W_\alpha(P_4)$ obey
\begin{equation}
W_\alpha(P_2) W_\alpha(P_1) |\Phi\> = e^{i\theta_\alpha} W_\alpha(P_4) W_\alpha(P_3) |\Phi\>
\label{exstatapp}
\end{equation}
where $\theta_\alpha$ is the exchange statistics of the quasiparticle $\alpha$ created by the string operator $W_\alpha$, and $P_1, P_2, P_3, P_4$ are
any four paths with the geometry of Fig. \ref{fig:theta}. We will derive the algebra (\ref{exstat}) given two assumptions about the string operators:
\begin{eqnarray}
	W_\alpha(P)|\Phi\>&=&W_\alpha(P')|\Phi\> \label{pathindapp}, \\
	W_\alpha(P_{1})W_\alpha(P_{2})|\Phi\>&=&W_\alpha(P_{1}\cup P_{2})|\Phi\>. \label{piececonn}	
\end{eqnarray}
Here, the first assumption (\ref{pathindapp}) is that the string operators satisfy \emph{path independence}: i.e., their action on the ground state is
the same for any two paths $P, P'$ with the same endpoints. The second assumption (\ref{piececonn}) is that the string operator are \emph{piece-connected}:
i.e. two string operators $W_\alpha(P_1)$, $W_\alpha(P_2)$ acting on paths $P_1, P_2$ that share a common endpoint, can be ``glued'' together in the natural
way. After proving (\ref{exstatapp}), we will then show that the above two assumptions are valid.

\begin{figure}
\begin{center}
\includegraphics[height=0.8in,width=2.8in]{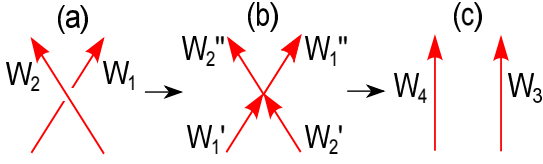}
\end{center}
\caption{
	There are two steps to connect the string operation in (a) and (c).
	First, we use piece-connectedness to decompose $W_1,W_2$ in (a) into four pieces $W_1',W_1'',W_2',W_2''$ in (b).
	Second, we use path independence to deform the $W_1''W_2'$ into $W_3$ and $W_2''W_1'$ into $W_4$ in (c).
}
\label{Wop5}
\end{figure}

Before proving (\ref{exstatapp}), we first introduce some notation: we denote $W_\alpha(P_1)$ by $W_1$ and similarly for $P_2, P_3, P_4$. In this
notation, (\ref{exstatapp}) can be written as
\begin{equation}
W_2 W_1 |\Phi\> = e^{i\theta_\alpha} W_4 W_3 |\Phi\>.
\end{equation}
The first step in the proof is to use (\ref{piececonn}) to decompose $W_1, W_2$ into two pieces, as shown in Fig. \ref{Wop5}(a,b): 
\begin{equation}
W_2 W_1 |\Phi\> = (W_2'' W_2') (W_1'' W_1') |\Phi\>.
\label{exstat1}
\end{equation}
We then use the ``hopping operator algebra'' derived in Ref. [\onlinecite{LevinWenHop}]. This algebra
connects the exchange statistics of a particle to the commutation properties of three string operators.
Applying it in our case gives:
\begin{equation}
W_2' W_1'' W_1' |\Phi\> = e^{i\theta_\alpha} W_1' W_1'' W_2' |\Phi\>
\end{equation}
where $\theta_\alpha$ is the exchange statistics of $\alpha$. Substituting this expression into (\ref{exstat1}),
we derive
\begin{equation}
W_2 W_1 |\Phi\> = e^{i\theta_\alpha} W_2'' W_1' W_1'' W_2' |\Phi\>. 
\label{exstat3}
\end{equation}
In the final step, we use path independence (\ref{pathindapp}) to write (see Fig. \ref{Wop5}(c))
\begin{eqnarray}
(W_2'' W_1') (W_1'' W_2') |\Phi\> &=&  (W_2'' W_1') W_3 |\Phi\> \nonumber \\
&=& W_4 W_3 |\Phi\>.
\label{exstat4}
\end{eqnarray}
Comparing (\ref{exstat4}) and (\ref{exstat3}), the claim (\ref{exstatapp}) follows immediately.

\begin{figure}
\begin{center}
\includegraphics[height=1.9in,width=3in]{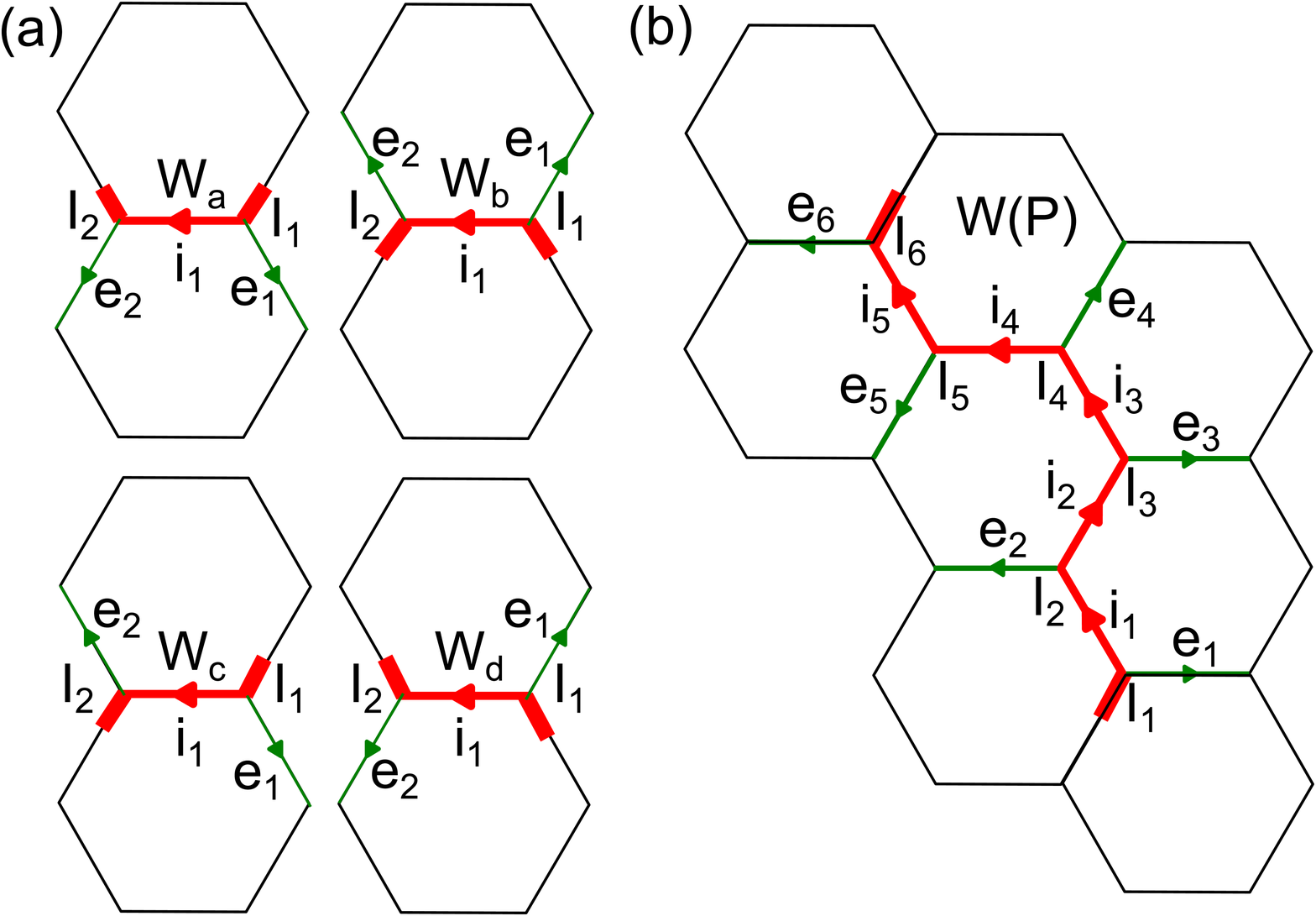}
\end{center}
\caption{
	(a) Four building blocks of any string operator defined on each link. 
	$W_{a}$ and $W_{b}$ have ends in the same side of the link while $W_{c}$ and $W_{d}$ have ends in opposite sides of the link. 
	Here $i_{1},e_{1},e_{2}$ denote the initial spin state in the link and two spin states on the external legs, respectively. 
	$I_{1}, I_{2}$ label the vertices.
	Their matrix elements are given in (\ref{Wop}).
	(b) A typical open string operator $W(P)$ along the path $P=I_{1},I_{2},\dots,I_{6}$ (the red line). 
	It can be decomposed into product of basic string blocks acting on each link along $P$.
	The matrix elements of $W(P)$ between an initial bra state $i'_{1},i'_{2},\dots,i'_{5}$ and a finial bra state $i_{1},i_{2},\dots,i_{5}$ is 
	$W_{i'_{1}\dots i'_{5}}^{i_{1}\dots i_{5}}(e_{1}\dots e_{6})=
	W_{a,i'_{5}}^{s,i_{5}}(e_{5}e_{6})
	W_{d,i'_{4}}^{s,i_{4}}(e_{4}e_{5})
	W_{b,i'_{3}}^{s,i_{3}}(e_{3}e_{4})
	W_{c,i'_{2}}^{s,i_{2}}(e_{2}e_{3})
	W_{d,i'_{1}}^{s,i_{1}}(e_{1}e_{2}).$
}
\label{stringblock}
\end{figure}

To complete the argument we need to show that the two assumptions (\ref{pathindapp},\ref{piececonn}) are valid for the
string operators $W_\alpha(P)$. The path independence assumption (\ref{pathindapp}) certainly holds since the string operators
were constructed specifically to obey this property. As for the piece-connectedness assumption (\ref{piececonn}), we can establish
this property if we can show that the open string operators can be defined in such a way that two string operators with a common
endpoint can be glued together to form a longer string. To this end, we will first introduce some basic string operators defined on 
each link of the honeycomb lattice and then show how to construct a general string operator from these building blocks. This construction
automatically gives string operators that are piece-connected.

The basic string operators act on a single link of the honeycomb lattice. For each link, there are four types of basic
operators which differ from one another at their two ends (see Fig. \ref{stringblock}(a)). 
These ends are meant to indicate how to glue two string operators together at a vertex.
Specifically, two basic string operators can be connected only if their ends \emph{match} one another. That is,
the two connecting ends must lie along the same path in the lattice. Thus we can construct a general string operator
$W(P)$ along a path $P$ on the honeycomb lattice by gluing together a sequence of basic string operators that follow
the path $P$ (see Fig. \ref{stringblock}(b)).

Now we have to define these basic string operators in such a way that they can be glued together to form longer string operators.
One way to find an appropriate definition is to start with a long string operator as defined in section \ref{stropsec}, 
and then break it up into many short string operators. In general there are many ways to do this, since there are many ways of
dividing the phase associated with a vertex between the two shorter string operators that share this vertex. Here, we
use a symmetrical convention: we split equally the phase associated with a vertex to the two connecting ends of the basic
string operators at this vertex. In this way, we obtain the following definition of the basic string operators:
\begin{align}
	W_{a,i'_{1}}^{s,i_{1}}(e_{1}e_{2})&=F(i_{1}^{*},i_{1},s)\sqrt{V_{R,s}(i_{1}+e_{1},e_{1})V_{R,s}(i_{1},e_{2})}, \notag\\
	W_{b,i'_{1}}^{s,i_{1}}(e_{1}e_{2})&=F(s^{*},s,i_{1})\sqrt{V_{L,s}(i_{1}+e_{1},e_{1})V_{L,s}(i_{1},e_{2})}, \notag \\
	W_{c,i'_{1}}^{s,i_{1}}(e_{1}e_{2})&=\bar{\omega}_{i_{1}s}\sqrt{V_{R,s}(i_{1}+e_{1},e_{1})V_{L,s}(i_{1},e_{2})}, \notag \\
	W_{d,i'_{1}}^{s,i_{1}}(e_{1}e_{2})&=\omega_{i_{1}s}\sqrt{V_{L,s}(i_{1}+e_{1},e_{1})V_{R,s}(i_{1},e_{2})}, \label{Wop}
\end{align}
with
\begin{align*}
	V_{L,s}(a,b)=\frac{F(s^{*},a-b+s,b)}{F(s^{*},s,a)}, \\
	V_{R,s}(a,b)=\frac{F(b,a-b,s)}{F(b-a,a-b,s)}.
\end{align*}
Here $V_{L,s}$ and $V_{R,s}$ are the phases associated with the left and right turning of the string at each vertex, respectively.
The phases $F$, $\omega$ and $\bar{\omega}$ are from the fusion of the string to the link without and with crossings while 
$V_{R,s},V_{L,s}$ are from the fusion at the vertices. 

With (\ref{Wop}) at hand, we define the matrix elements for general string operators $W(P)$ by taking the product of
the matrix elements of the basic string operators for each link along the path $P$ (see Fig. \ref{stringblock}(b)).
This prescription (\ref{Wop}) satisfies $W(P_{1})W(P_{2})|\Phi\>=W(P_{1}\cup P_{2})|\Phi\>$ by construction.

We would like to mention that there is one subtlety in gluing together the basic string operators:
the above matrix elements are defined for strings satisfying the branching rules at vertices, yet 
when we apply the first string operator to a string-net state $|X\>$, the branching rules will be violated at the two endpoints. 
To use (\ref{Wop}) for the second string operator connecting to the first one, 
we follow a special prescription: focusing on the vertex where the two strings meet, we pretend that the link along which the first string operator acts is in the unique state that obeys the branching rules with the other two links adjoining the vertex. With this prescription, the matrix elements of the second string operator are well-defined.

\bibliography{abelianstringnet}
\end{document}